%% file: main.tex
\documentclass[12pt]{mitthesis}
\usepackage{latexsym} % Gets \Box etc
\usepackage{amssymb}  % \gtrsim, \geqslant, etc etc: see amsguide.ps
\usepackage{amsbsy}   % \pmb and \boldsymbol
\usepackage{graphicx}       % For PostScript figures
\usepackage[dvips]{color}
\usepackage{wasysym}
\usepackage{psfrag}
\usepackage{subfigure}

\newcommand{\al}{\alpha}
\newcommand{\be}{\beta}
\newcommand{\ga}{\gamma}
\newcommand{\Ga}{\Gamma}
\newcommand{\de}{\delta}
\newcommand{\De}{\Delta}
\newcommand{\ep}{\varepsilon}
\newcommand{\eps}{\epsilon}

\newcommand{\ka}{\kappa}

\newcommand{\La}{\Lambda}
\newcommand{\ph}{\varphi}

\newcommand{\si}{\sigma}

\renewcommand{\th}{\theta}   % LaTeX: \th already defined

\newcommand{\<}{\langle} 
\renewcommand{\>}{\rangle} % LaTeX: \> already defined

\newcommand{\txt}{\textstyle}

\newcommand{\dsp}{\displaystyle}

\newcommand{\ad}{\dagger}
\newcommand\eqn[1]{(\ref{#1})}      % parentheses around the LaTex "ref" macro
\newcommand\Eqn[1]{Eq.~(\ref{#1})}  % includes ``Eq.'' in front
\newcommand{\e}{{\rm e}}
\newcommand{\beq}{\begin{equation}}
\newcommand{\eeq}{\end{equation}}
\newcommand{\ba}{\begin{array}}
\newcommand{\bea}{\begin{eqnarray}}
\newcommand{\ea}{\end{array}}
\newcommand{\eea}{\end{eqnarray}}
\newcommand{\bc}{\begin{center}}
\newcommand{\ec}{\end{center}}
\newcommand{\ben}{\begin{enumerate}}
\newcommand{\een}{\end{enumerate}}

\newcommand{\dslash}{{\partial\kern-0.55em/}}
\newcommand{\Dslash}{{D\kern-0.65em/}}

\newcommand\comment[1]{ \hbox{[{\it Comment suppressed here.}\/]} }
\newcommand\hide[1]{}

\renewcommand{\O}{{\cal O}}
\newcommand{\tr}{\hbox{tr}}

\renewcommand{\Re}{{\rm Re}\,}

\newcommand{\skipover}[1]{}

\newcommand{\half} {{\txt \frac{1}{2}}}
\newcommand{\third}{{\txt \frac{1}{3}}}

\newcommand{\quarter}{{\txt\frac{1}{4}}}

\def\phm{\phantom{-}}

\newsavebox{\eqlabel}
\makeatletter  %\catcode`\@=11
\newlength{\numblen}
\newsavebox{\eqnumb}
\def\@eqnnum{\savebox{\eqnumb}{\rm (\theequation)}%
\settowidth{\numblen}{\usebox{\eqnumb}}%
\makebox[\numblen][l]{\usebox{\eqnumb}~~~\usebox{\eqlabel}}}
\makeatother   %\catcode`\@=12

\newenvironment{equationwithlabel}[1]{ %
  \begin{equation}\label{#1} }{\end{equation}} %\savebox{\eqlabel}{~}}
\newcommand{\beql}[1]{\begin{equationwithlabel}{#1}}
\newcommand{\eeql}{\end{equationwithlabel}}
\newcommand{\eV}{{\rm eV}} 
\newcommand{\keV}{{\rm keV}} 
\newcommand{\MeV}{{\rm MeV}} 
\newcommand{\GeV}{{\rm GeV}} 
\newcommand{\vp}{{\mathbf p}}    % vector p
\newcommand{\vk}{{\mathbf k}}    % vector k
\newcommand{\vq}{{\mathbf q}}    % vector q
\newcommand{\vr}{{\mathbf r}}    % vector r
\newcommand{\vx}{{\mathbf x}}    % vector x
\newcommand{\vl}{{\boldsymbol{\ell}}}    % vector \ell
\newcommand{\dm}{{\delta\mu}} % delta mu
\newcommand{\mubar}{{\bar\mu}}% mu bar
\newcommand{\dmmax}{{\de\mu_2}}
\newcommand{\dmmin}{{\de\mu_1}}
\newcommand{\pslash}{{\not\! p}}
\newcommand{\qslash}{{\not\! q}}
\newcommand{\muslash}{{\not\! \mu}}
\newcommand{\delslash}{{\not\! \partial}}

\newcommand{\kslash}{{k\kern-0.5em/}}

\newcommand{\thrbarA}{$\bar{\bf 3}_A$}
\newcommand{\sixS}{${\bf 6}_S$}
\newcommand{\thrS}{${\bf 3}_S$}
\newcommand{\oneA}{${\bf 1}_A$}

\newcommand{\psibar}{{\bar\psi}}

\newcommand{\cfms}{{\cal S}^{ij}_{\alpha\beta}}

\newcommand{\conds}[1]{\psi{#1}{\cal S}\psi}
\newcommand{\conda}[1]{\psi{#1}{\cal A}\psi}
\newcommand{\uu}{\mid\uparrow\uparrow\rangle}
\newcommand{\ud}{\mid\uparrow\downarrow\rangle}
\newcommand{\du}{\mid\downarrow\uparrow\rangle}
\newcommand{\dd}{\mid\downarrow\downarrow\rangle}

\begin{document}

\include{cover}
\pagestyle{plain}
%\cleardoublepage
%\clearpage \phantom{blah} \clearpage
\include{contents}

%\cleardoublepage
\clearpage \phantom{blah} \clearpage
\include{chap1}
\clearpage \phantom{blah} \clearpage
\include{chap2}

\include{chap3}

\clearpage \phantom{blah} \clearpage
\include{chap4}

\clearpage \phantom{blah} \clearpage
\include{chap5}

\appendix
\include{appa}

\include{appb}

\clearpage \phantom{blah} \clearpage
\include{biblio}

\end{document}

%% file: cover.tex
% -*-latex-*-
% $Log: cover.tex,v $
% Revision 1.7  2001/02/08 18:53:16  boojum
% changed some \newpages to \cleardoublepages
%
% Revision 1.6  1999/10/21 14:49:31  boojum
% changed comment referring to documentstyle
%
% Revision 1.5  1999/10/21 14:39:04  boojum
% *** empty log message ***
%
% Revision 1.4  1997/04/18  17:54:10  othomas
% added page numbers on abstract and cover, and made 1 abstract
% page the default rather than 2.  (anne hunter tells me this
% is the new institute standard.)
%
% Revision 1.4  1997/04/18  17:54:10  othomas
% added page numbers on abstract and cover, and made 1 abstract
% page the default rather than 2.  (anne hunter tells me this
% is the new institute standard.)
%
% Revision 1.3  93/05/17  17:06:29  starflt
% Added acknowledgements section (suggested by tompalka)
% 
% Revision 1.2  92/04/22  13:13:13  epeisach
% Fixes for 1991 course 6 requirements
% Phrase "and to grant others the right to do so" has been added to 
% permission clause
% Second copy of abstract is not counted as separate pages so numbering works
% out
% 
% Revision 1.1  92/04/22  13:08:20  epeisach
\title{Color Superconducting Phases of Cold Dense Quark Matter}

\author{Jeffrey Allan Bowers}
\prevdegrees{
B.S., Physics, MIT,  1998 \\
B.S., Electrical Engineering and Computer Science, MIT, 1998 \\
M.Eng., Electrical Engineering and Computer Science, MIT, 1998 
} 
\department{Department of Physics \\ at the Massachusetts
Institute of Technology}
% If the thesis is for two degrees simultaneously, list them both
% separated by \and like this:
% \degree{Doctor of Philosophy \and Master of Science}
\degree{Doctor of Philosophy in Physics}
\degreemonth{June}
\degreeyear{2003}
\thesisdate{May 6, 2003}

%% By default, the thesis will be copyrighted to MIT.  If you need to copyright
%% the thesis to yourself, just specify the `vi' documentclass option.  If for
%% some reason you want to exactly specify the copyright notice text, you can
%% use the \copyrightnoticetext command.  
%\copyrightnoticetext{\copyright IBM, 1990.  Do not open till Xmas.}

% If there is more than one supervisor, use the \supervisor command
% once for each.
\supervisor{Krishna Rajagopal}{Associate Professor of Physics}

% This is the department committee chairman, not the thesis committee
% chairman.  You should replace this with your Department's Committee
% Chairman.
%\chairman{Patrick Lee}{Chairman, Department Committee on Graduate Students}
\chairman{Thomas Greytak}{Chairman, Department Committee on Graduate Students}

% Make the titlepage based on the above information.  If you need
% something special and can't use the standard form, you can specify
% the exact text of the titlepage yourself.  Put it in a titlepage
% environment and leave blank lines where you want vertical space.
% The spaces will be adjusted to fill the entire page.  The dotted
% lines for the signatures are made with the \signature command.
%\maketitle

% The abstractpage environment sets up everything on the page except
% the text itself.  The title and other header material are put at the
% top of the page, and the supervisors are listed at the bottom.  A
% new page is begun both before and after.  Of course, an abstract may
% be more than one page itself.  If you need more control over the
% format of the page, you can use the abstract environment, which puts
% the word "Abstract" at the beginning and single spaces its text.

%% You can either \input (*not* \include) your abstract file, or you can put
%% the text of the abstract directly between the \begin{abstractpage} and
%% \end{abstractpage} commands.

% First copy: start a new page, and save the page number.
 
%\cleardoublepage
%\clearpage
%\phantom{blah}
%\clearpage

% Uncomment the next line if you do NOT want a page number on your
% abstract and acknowledgments pages.
% \pagestyle{empty}
\setcounter{savepage}{\thepage}
\begin{abstractpage}
\input{abstract}

\end{abstractpage}

% Additional copy: start a new page, and reset the page number.  This way,
% the second copy of the abstract is not counted as separate pages.
% Uncomment the next 6 lines if you need two copies of the abstract
% page.
% \setcounter{page}{\thesavepage}
% \begin{abstractpage}
% \input{abstract}
% \end{abstractpage}

%\cleardoublepage
\clearpage
\phantom{blah}
\clearpage

\section*{Acknowledgments}
 
I cannot adequately express my gratitute to my mentor, teacher,
collaborator, and friend, \mbox{Krishna} Rajagopal, for years of
patient advice and inspiration.  Very special thanks also to Mark
Alford, for close collaboration and generous advice.  Thanks to
Frank Wilczek for helpful discussion and guidance.  Thanks to him and
to Wit Busza for serving on my thesis committee and for taking the
time to review this manuscript.  Thanks to Jack M.~Cheyne and Greig
A.~Cowan for their collaboration on the spin-one calculations of
Chapter 4 and Appendix B.  Thanks to 
Paulo Bedaque, 
%Jurgen Berges,
%Ignazio Bombaci, 
%David Blaschke, 
%Deepto Chakrabarty,
Michael Forbes,
Elena Gubankova,
Robert Jaffe,
Chris Kouvaris,
Joydip Kundu, 
Vincent Liu, 
%Jes Madsen, 
%Chetan Nayak,
%Dimitrios Psaltis, 
%Soo-Jong Rey, 
Dirk Rischke,
%Mal Ruderman, 
Thomas Sch\"afer, 
%Armen Sedrakian, 
Eugene Shuster, 
Dam Son,
%and Misha Stephanov 
%Ira Wasserman, and
%Fridolin Weber 
and Christof Wetterich
for many enlightening conversations.  This research was supported in
part by the U.S.~Department of Defense (D.O.D.) National Defense
Science and Engineering Graduate Fellowship Program, by the Kavli
Institute for Theoretical Physics (KITP) Graduate Fellowship Program,
by the U.S.~Department of Energy (D.O.E.)  under cooperative research
agreement \#DF-FC02-94ER40818, and by the National Science Foundation
under Grant No.~PHY99-07949.  I am grateful to the Kavli Institute for
Theoretical Physics (KITP) and the Institute for Nuclear Theory (INT)
at the Univeristy of Washington for their hospitality and support
during the completion of much of this work.

\clearpage
\phantom{blah}
\clearpage

%%%%%%%%%%%%%%%%%%%%%%%%%%%%%%%%%%%%%%%%%%%%%%%%%%%%%%%%%%%%%%%%%%%%%%
% -*-latex-*-

%% file: abstract.tex
% $Log: abstract.tex,v $
% Revision 1.1  93/05/14  14:56:25  starflt
% Initial revision
% 
% Revision 1.1  90/05/04  10:41:01  lwvanels
% Initial revision
% 
%
%% The text of your abstract and nothing else (other than comments) goes here.
%% It will be single-spaced and the rest of the text that is supposed to go on
%% the abstract page will be generated by the abstractpage environment.  This
%% file should be \input (not \include 'd) from cover.tex.

We investigate color superconducting phases of cold quark matter at
densities relevant for the interiors of compact stars.  At these
densities, electrically neutral and weak-equilibrated quark matter can
have unequal numbers of up, down, and strange quarks.  The QCD
interaction favors Cooper pairs that are antisymmetric in color and in
flavor, and a crystalline color superconducting phase can occur which
accommodates pairing between flavors with unequal number densities.
In the crystalline color superconductor, quarks of different flavor
form Cooper pairs with nonzero total momentum, yielding a condensate
that varies in space like a sum of plane waves.  Rotational and
translational symmetry are spontaneously broken.  We use a
Ginzburg-Landau method to evaluate candidate crystal structures and
predict that the favored structure is face-centered-cubic.  We predict
a robust crystalline phase with gaps comparable in magnitude to those
of the color-flavor-locked phase that occurs when the flavor number
densities are equal.  Crystalline color superconductivity will be a
generic feature of the QCD phase diagram, occurring wherever quark
matter that is not color-flavor locked is to be found.  If a very large
flavor asymmetry forbids even the crystalline state, single-flavor
pairing will occur; we investigate this and other spin-one color
superconductors in a survey of generic color, flavor, and spin pairing
channels.  Our predictions for the crystalline phase may be tested in
an ultracold gas of fermionic atoms, where a similar crystalline
superfluid state can occur.  If a layer of crystalline quark matter
occurs inside of a compact star, it could pin rotational vortices,
leading to observable pulsar glitches.

%% file: contents.tex
  % -*- Mode:TeX -*-
%% This file simply contains the commands that actually generate the table of
%% contents and lists of figures and tables.  You can omit any or all of
%% these files by simply taking out the appropriate command.  For more
%% information on these files, see appendix C.3.3 of the LaTeX manual. 
\tableofcontents
%\newpage
%\listoffigures
%\newpage
%\listoftables

%% file: chap1.tex
%% This is an example first chapter.  You should put chapter/appendix that you
%% write into a separate file, and add a line \include{yourfilename} to
%% main.tex, where `yourfilename.tex' is the name of the chapter/appendix file.
%% You can process specific files by typing their names in at the 
%% \files=
%% prompt when you run the file main.tex through LaTeX.
\chapter{Introduction}

\section{Overview}

In this thesis we shall discuss the behavior of cold quark matter at
densities that are relevant for the interiors of compact stars.  It is
well known that cold dense quark matter is unstable to the formation
of a condensate of quark Cooper pairs, making it a color
superconductor.  Various phases of color superconductivity have been
proposed, and in section \ref{sec:phasediagram} we review the phase
diagram of QCD to provide a larger context for a discussion of the
color superconducting phases.  In section \ref{sec:highdensity} we
discuss the color-flavor-locked (CFL) color superconductor, the ground
state of cold quark matter at very high densities.  In section
\ref{sec:unlocking} we describe how the CFL phase can be disrupted at
intermediate densities that are relevant for compact stars.  At these
intermediate densities, neutral quark matter has unequal numbers of
up, down, and strange quarks, and a crystalline color superconducting
phase is favored.  The crystalline color superconductor has the
remarkable virtue of allowing pairing between quarks with unequal
Fermi surfaces.  Cooper pairs with nonzero total momentum are favored;
the condensate spontaneously breaks translational and rotational
invariance, leading to gaps which vary periodically in a crystalline
pattern as a superposition of plane waves.  In section
\ref{sec:crystalline} we discuss the crystalline phase and look ahead
to the detailed calculations of chapters 2 and 3 where we investigate
single-plane-wave and multiple-plane-wave crystalline phases,
respectively.  In section \ref{sec:singleflavor} we discuss
single-flavor color superconductivity, which might occur when there is
a very large flavor asymmetry that forbids the crystalline state. We
also preview chapter 4, which presents a larger survey of various
single-color and single-flavor color superconductors.  Many of these
are spin-one phases with unusual spectra of elementary excitations.
Finally, in section \ref{sec:applications} we discuss physical
contexts in which the crystalline phase may occur with observable
consequences.  In \ref{subsec:compactstars} we review the
astrophysical implications of color superconductivity for compact
stars. If a layer of crystalline quark matter occurs inside of a
compact star, it could pin rotational vortices, leading to observable
pulsar glitches.  In \ref{subsec:atomicphysics}, we describe how a
crystalline superfluid might be created and detected in a trapped gas
of ultracold fermions.  These are previews of more detailed
discussions of glitches and cold atoms that appear in chapter 5.

\section{The phase diagram of QCD}
\label{sec:phasediagram}

In recent years much theoretical and experimental effort has been
devoted to understanding the behavior of quantum chromodynamics (QCD)
in extreme conditions of very high temperature or density.  Because
QCD is asymptotically free~\cite{asymfree}, its high temperature and
high density phases are readily described in terms of quark and gluon
degrees of freedom~\cite{collinsperry}.  At high temperature, the
familiar hadronic phase of QCD gives way to a deconfined plasma of
quarks and gluons, in which all the symmetries of the QCD Lagrangian
are unbroken~\cite{highTreviews}.  This phase preceded hadronization in the early
universe, and efforts are underway to produce and probe this phase in
thermalized collisions of relativistic heavy ions at Brookhaven and
CERN laboratories~\cite{qm2001}.  At low temperature and high
density, on the other hand, the hadronic phase gives way to a
degenerate Fermi system of quarks.  Under the influence of the QCD
interaction, quarks near the Fermi surface can bind together as Cooper
pairs, which condense by the BCS mechanism~\cite{BCS} to form a color
superconductor~\cite{Barrois,BailinLove,IwaIwa,ARW2,RappEtc,ARW3,Son,PR,CD,RSSV2,SW3,ShovWij,Reviews}.
While not accessible in the laboratory, this cold dense quark matter
might occur inside compact stars, with a host of potential
astrophysical implications~\cite{supercondstars}.

\begin{figure}
\begin{center}
\psfrag{T}{$T$}
\psfrag{primordialuniverse}{\parbox{1.5in}{primordial \\ universe}}
\psfrag{heavyioncollisions}{heavy ion collisions}
\psfrag{mu}[tc][tc]{$\mu_{\mbox{\scriptsize quark}}$}
\psfrag{muqh}[tc][tc]{$\mu_{qh}$}
\psfrag{muunlock}[tc][tc]{$\mu_{\mbox{\scriptsize unlock}}$}
\psfrag{compactstars}[tc][tc]{compact stars}
\psfrag{1GeV}[tc][tc]{$\third$GeV}
\psfrag{hadrongas}{\parbox{1.5in}{hadron \\ gas}}
\psfrag{hadronliquid}{\parbox{1.5in}{hadron \\ liquid}}
\psfrag{150}[rc][rc]{\parbox{0.5in}{170 \\  MeV}}
%\psfrag{crystal}{\parbox{2in}{crystalline \\ superconductor}}
%\psfrag{cfl}{\parbox{2in}{CFL \\ superconductor}}
\psfrag{crystal}{\parbox{2in}{non-CFL \\ superconductor}}
\psfrag{cfl}{\parbox{2in}{CFL \\ superconductor}}
\psfrag{quarkgluonplasma}{\parbox{2in}{quark-gluon \\ plasma}}
\psfrag{nuclear matter}[rc][rc]{nuclei}
\includegraphics[width=5.5in]{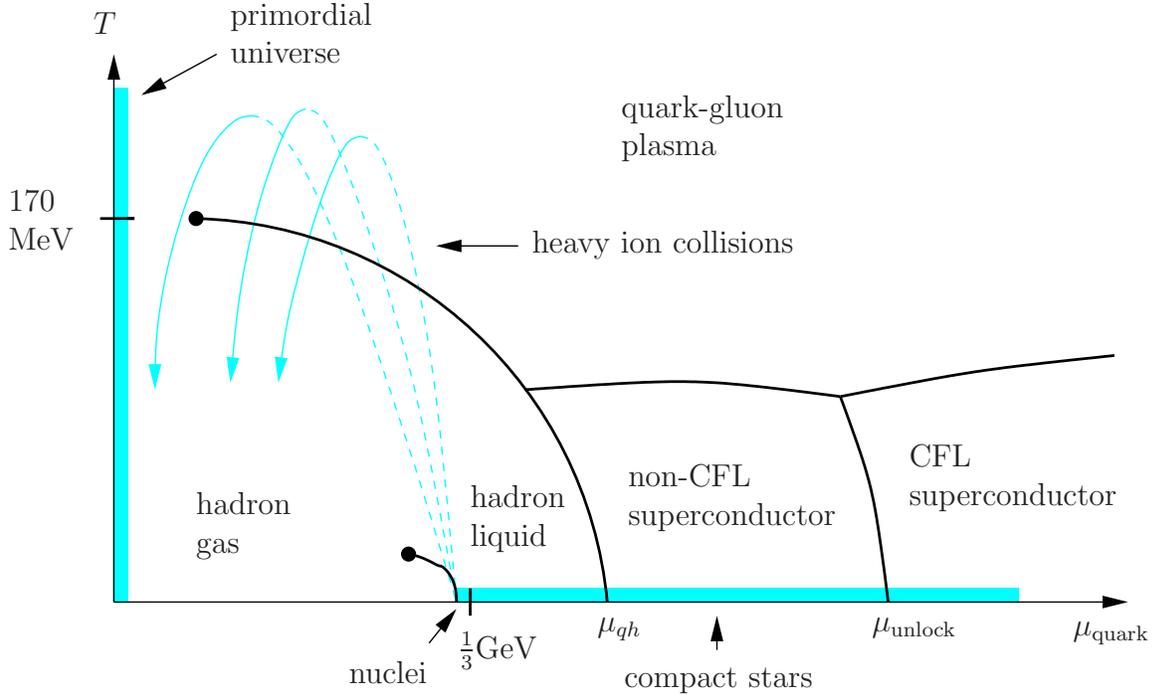}
\end{center}
\caption{
\label{phasediagram}
Schematic QCD phase diagram with hadronic, plasma, and superconducting
phases, as a function of temperature $T$ and chemical potential
$\mu_{\mbox{\scriptsize quark}} = \mu_{\mbox{\scriptsize baryon}}/3$.
The gray-shaded areas are regions of phenomenological interest.  }
\end{figure}

These various states of QCD fit together in a phase diagram as a
function of temperature and chemical potential, as shown in figure
\ref{phasediagram}.  Let us first consider the temperature axis of
this phase diagram.  The early universe moved down the vertical axis
during the first tens of microseconds after the big
bang~\cite{kolbturner}.  The temperature axis can also be studied with
lattice simulations, which reveal a transition from a quark-gluon
plasma (QGP) to a gas of hadrons at a temperature of about 170
MeV~\cite{lattice}.  With realistic quark masses ($m_s \gg m_{u,d}
\neq 0$), chiral symmetry is explicitly broken everywhere and the
transition is a rapid but smooth crossover.  Chiral symmetry is
(approximately) restored in the QGP phase and broken in the hadronic
phase.

Now let us consider the chemical potential axis.  Nuclei are droplets
of the hadron liquid phase; they sit on the horizontal axis just to
the right of the small ``curlicue'' which is a line of first-order
phase transitions between the hadron gas and the hadron liquid (this
line terminates at a second-order critical point at a temperature of
about 10 MeV)~\cite{pochodzalla}.  Moving further to the right along
the chemical potential axis, the density increases beyond nuclear
density and eventually the nucleons overlap to such an extent that a
description in terms of quark degrees of freedom is more suitable.  In
fact various model calculations suggest that on the horizontal axis
there is a first-order transition from the hadronic phase to
deconfined quark
matter~\cite{ARW2,RappEtc,BergesRajagopal,halasz,randommatrix,BuballaOertel,phasetransitions}.
Starting from this first-order transition at $\mu_{qh}$, a first-order
line separating quark and hadron phases extends upwards and to the
left, terminating at a second-order critical point before it reaches
the temperature axis~\cite{BergesRajagopal,halasz}.  
In relativistic heavy ion experiments at Brookhaven and CERN, the
collided nuclei might follow trajectories in $T$-$\mu$ space like
those shown in figure \ref{phasediagram}: the nuclei start at zero
temperature, depart from equilibrium during the first moments of
collision, then perhaps reappear on the phase diagram at high temperature,
thermalizing to create a brief fireball of quark-gluon plasma before
expanding and cooling through the quark-hadron transition to produce
hadrons.

To the right of $\mu_{qh}$ is the regime of color superconductivity.
In this regime deconfined quarks fill large Fermi seas and the
interesting physics is that of quarks near their Fermi surfaces
(quarks that are deep within Fermi seas are Pauli-blocked and
therefore essentially behave as free particles).  Pairs of quarks near
the Fermi surface that are antisymmetric in color feel an attractive
QCD interaction; this is not surprising because at low density and
temperature the quarks bind strongly together to form baryons.  This
attractive interaction makes the system unstable to the formation of a
BCS condensate of quark-quark Cooper pairs.  Because a pair of quarks
cannot be a color singlet, the BCS condensate breaks color gauge
symmetry, and the system is a color superconductor.

The phase boundary separating the QGP and color superconducting phases
in figure \ref{phasediagram} occurs at the critical temperature
$T_c(\mu)$ above which the diquark condensates vanish.  This
temperature is expected to be a few tens of MeV~\cite{Reviews}.  Hot
protoneutron stars may cool through the color superconducting phase a
few seconds after the birth of the star, with interesting implications
for neutrino transport in supernovae~\cite{CarterReddy}.  Compact
stars rapidly cool to temperatures of a few keV and then reside on the
chemical potential axis of the QCD phase diagram (at essentially zero
temperature)~\cite{Prakash}.  The chemical potential will vary with
depth and the star could be a ``hybrid star'' with a hadronic mantle
and a quark matter core~\cite{HHJ,LattimerPrakash,Weber,quarkeos}.
Because the star temperature is small compared to the critical
temperature for color superconductivity, if any quark matter is
present then it will be a color superconductor.

At very high densities (far to the right on the phase diagram) the
ground state of QCD is the color-flavor-locked (CFL) color
superconductor~\cite{ARW3,SWcont,RSSV2,Reviews}.  This is a
robust and symmetric phase that is invariant under simultaneous
$SU(3)$ rotations in color and flavor.  At very high densities the
strange quark mass can be neglected and flavor $SU(3)$ is a good
symmetry of the QCD Lagrangian.  The nonzero strange quark mass
explicitly breaks this symmetry and the effect is more pronounced at
lower densities.  Moving down the chemical potential axis from very
high density, this flavor asymmetry can eventually disrupt the CFL
phase at an ``unlocking''
transition~\cite{ABR2+1,SW2,neutrality,Bedaque,ARRW,AlfordRajagopal,BuballaOertel}.
It is uncertain whether unlocking occurs before hadronization.  If, as
in figure \ref{phasediagram}, we assume that unlocking does occur
first, then there is a window in the QCD phase diagram at intermediate
density ($\mu_{qh} < \mu < \mu_{\mbox{\scriptsize unlock}}$) for which
the ground state of QCD is deconfined quark matter that is not
color-flavor-locked.

In unlocked quark matter, pairing should still occur because there is
still an attractive interaction between quarks.  Any pairing breaks
color symmetry so the system remains a color superconductor.  In the
present work we investigate the novel pairing patterns of quark matter
that can occur within this window of non-CFL color superconductivity.
We propose that the most favorable candidate is crystalline color
superconductivity~\cite{BowersLOFF,ngloff,BowersCrystal,crystalproc,massloff,pertloff,Giannakis,LOFFphonon}.
Unlocked quark matter has different number densities of $u$, $d$, and
$s$ quarks, and we will show that the crystalline phase can accommodate
this by forming Cooper pairs with nonzero total momentum.  Condensates
of this sort spontaneously break translational and rotational
invariance, leading to gaps which vary periodically in a crystalline
pattern.

\section{High density and color-flavor-locking}
\label{sec:highdensity}

At extremely high densities, the quarks at the Fermi surface have very
large momenta and their interactions are asymptotically weak.  In this
limit, a rigorous theoretical analysis of the QCD ground state is
possible using weak-coupling, but non-perturbative, methods of BCS
theory~\cite{Barrois,Son,SW3,PR,Hong,HMSW,BLR,HS,Schaefer,BBS,ShovWij,EHHS}.
This analysis predicts that at asymptotically-high density, the
preferred ground state of QCD is the color-flavor-locked (CFL) color
superconductor involving three massless flavors of quarks (at
asymptotic densities, one can reasonably neglect the up, down, and
strange quark masses).  All nine quarks (three flavors, three colors)
together form a simple and elegant diquark condensate of the form~\cite{ARW3}
\begin{equation}
\label{CFLcondensate}
\langle \psi_{i \alpha a}(\vx) \psi_{j \beta b}(\vx) \rangle
\propto \Delta_0 \epsilon_{\alpha \beta A} \epsilon_{i j A} 
(C \gamma_5)_{ab}
\end{equation}
with indices for color ($\alpha,\beta$), flavor ($i,j$), and spin
($a,b$).  The $\epsilon_{\alpha \beta A}$ tensor indicates that the
condensate is an $SU(3)$ color antitriplet of $rg$, $rb$, and $gb$
Cooper pairs.  The color $\mathbf{\bar 3}$ channel is favored because
this is the attractive channel for perturbative single-gluon exchange.
The $\epsilon_{i j A}$ tensor indicates that the condensate is
simultaneously an $SU(3)$ flavor antitriplet of $ud$, $us$, and $ds$
Cooper pairs.  The common index $A$ is summed and therefore ``locks''
color to flavor.  The Dirac matrix $C \gamma_5$ indicates that the
condensate is both rotationally invariant ($J=0$) and parity even.
Quarks are paired with the same helicity and opposite momentum;
therefore they have opposite spin and form a spin singlet.

The value of the CFL gap $\Delta_0$ can be calculated in the
asymptotic limit, where the leading interaction is just single gluon
exchange.  Unfortunately the result can be trusted only at
unphysically large chemical potentials, of order $10^8$ MeV or
higher~\cite{RajagopalShuster}.  Extrapolation to densities of
interest for compact stars ($\mu \approx 400$ MeV) is unjustified but
yields a gap of about 10-100 MeV.  Alternatively, the gap can be
calculated by using phenomenological toy models whose free parameters
are chosen to give reasonable vacuum
physics~\cite{ARW2,RappEtc,BergesRajagopal,ARW3,CD,RSSV2,EHS,SW0}.
For example, one might use a Nambu-Jona-Lasinio (NJL) model in which
the interaction between quarks is replaced by a pointlike four-fermion
interaction (with the quantum numbers of single-gluon exchange or the
instanton interaction), and choose the coupling constant to fit the
magnitude of the vacuum chiral condensate at $\mu = 0$.  It is
gratifying that these model calculations also yield gaps of about
10-100 MeV, in agreement with the extrapolation from the asymptotic
limit.

The CFL phase has a remarkable pattern of symmetry breaking and
elementary excitations~\cite{ARW3}.  The exact microscopic symmetry
group of QCD with three massless flavors is
\begin{equation}
G_{\mbox{\scriptsize microscopic}} = SU(3)_{\mbox{\scriptsize color}} \times SU(3)_L 
\times SU(3)_R \times U(1)_B
\end{equation}
where the first factor is the color gauge symmetry, the next two
factors are the chiral flavor symmetries, and the last factor is
baryon number.  Electromagnetism is included by gauging a vector
$U(1)$ subgroup of the flavor groups.  In the CFL phase,
$G_{\mbox{\scriptsize microscopic}}$ is broken to the subgroup
\begin{equation}
G_{\mbox{\scriptsize CFL}} = SU(3)_{\mbox{\scriptsize color} + L + R} \times Z_2.
\end{equation}
The color and chiral flavor symmetries are broken to a ``diagonal''
global symmetry group of equal $SU(3)$ transformations in all three
sectors (color, left-handed flavor, and right-handed flavor).  All
nine quarks are gapped and the system has nine massive
spin-$\frac{1}{2}$ quasiquark excitations that decompose into a octet
and a singlet under the diagonal $SU(3)$.  The gluon fields produce an
SU(3) octet of spin-1 bosonic excitations that acquire a mass by the
Meissner-Higgs mechanism.  Chiral symmetry is broken and there is an
associated octet of massless psuedoscalar bosons.  Baryon number is
broken to a discrete $Z_2$ symmetry in which all the quark fields are
multiplied by $-1$; there is an associated massless scalar boson which
is the superfluid mode of the CFL phase.  The gauge symmetries are
completely broken except for a residual $U(1)$ generated by $\tilde Q
= Q + \frac{1}{\sqrt{3}}T_8$ where $Q$ is the electromagnetic charge
operator and $T_8$ is a generator in the Lie algebra of the $SU(3)$
color group.  This $U(1)_{\tilde Q}$ symmetry is a gauged subgroup of
the unbroken $SU(3)_{\mbox{\scriptsize color} + L + R}$.  There is a
corresponding $\tilde Q$ photon which is a linear combination of the
usual photon and the $T_8$ gluon (in fact there is just a small
admixture of the gluon, so an ordinary magnetic field externally
applied to a chunk of CFL matter is mostly admitted as a $\tilde Q$
magnetic field and only a small fraction of the flux is expelled by
the Meissner effect~\cite{ABRmag}).  All the elementary excitations
have integer $\tilde Q$ charges.

\section{Intermediate density and unlocking}
\label{sec:unlocking}

\begin{figure}
\begin{center}
\psfrag{mu}[r][r]{$\mu$}
\psfrag{ms}[l][l]{$m_s$}
\psfrag{pfe}[l][l]{$p_F^e$}
\psfrag{pfu}[l][l]{$p_F^u$}
\psfrag{pfd}[l][l]{$p_F^d$}
\psfrag{pfs}[l][l]{$p_F^s$}
\psfrag{pf}[l][l]{$p_F$}
\psfrag{dpf}[l][l]{$\delta p_F$}
\includegraphics[width=4in]{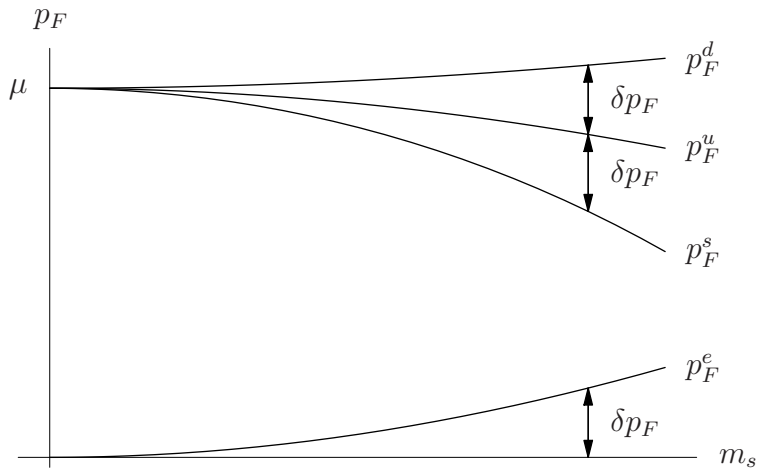}
\end{center}
\caption{
\label{fermimomenta}
Fermi momenta for noninteracting electrons and quarks, in a system
that is electrically neutral and in weak equilibrium.  All three quark
Fermi surfaces separate when the strange quark mass is nonzero, 
with $\delta p_F \approx m_s^2/4\mu$.    
}
\end{figure}

The CFL state pairs quarks with different flavors, forming a flavor
antitriplet of $ud$, $us$, and $ds$ pairs.  As in any BCS state, a
quark with momentum $\vp$ pairs with a quark of momentum $-\vp$, and
the condensate is dominated by those pairs for which each quark is in
the vicinity of its Fermi surface: $| |\vp| - p_F | \lesssim
\Delta_0$, where $\Delta_0$ is the BCS gap parameter.  Since different
flavors are paired together, the different flavors must have equal
Fermi surfaces for the BCS pairing scheme to work.  If the Fermi
momenta are different, then it is no longer possible to guarantee that
the formation of pairs lowers the free energy: the two fermions in a
pair have equal and opposite momentum, so at most one member of each
pair can be created at its Fermi surface.  This implies that the CFL
phase can be disrupted by any flavor asymmetry that would, in the
absence of pairing, separate the Fermi surfaces.  The nonzero mass of
the strange quark has precisely this effect.  To understand this,
consider noninteracting quark matter with $m_u = m_d = 0$ and $m_s
\neq 0$.  Chemical equilibrium under weak decay reactions requires
\begin{equation}
\mu_u  = \mu - \frac{2}{3} \mu_e , \ \ \  
\mu_d = \mu_s = \mu + \frac{1}{3} \mu_e 
\end{equation}
where $\mu = \third \mu_{\mbox{\scriptsize baryon}}$ is the average
quark chemical potential, and $\mu_e$ is a chemical potential for
electrons.  With these chemical potentials the noninteracting quarks
and electrons fill Fermi seas up to Fermi momenta given by
\begin{equation}
p_F^{u,d} = \mu_{u,d}, \ \ \ p_F^s = \sqrt{\mu_s^2 - m_s^2}, \ \ \ 
p_F^e = \mu_e 
\end{equation}
with corresponding number densities
\begin{equation}
N_{u,d} = \frac{1}{\pi^2} \mu_{u,d}^3, \ \ \ 
N_s = \frac{1}{\pi^2} (\mu_s^2 - m_s^2)^{3/2}, \ \ \ 
N_e = \frac{1}{3 \pi^2} \mu_e^3.
\end{equation}
Electrical neutrality then requires 
\begin{equation}
\frac{2}{3} N_u - \frac{1}{3} N_d - \frac{1}{3} N_s - N_e = 0.
\end{equation} 
These equations are readily solved and figure \ref{fermimomenta} shows
how the Fermi momenta of the quarks and electrons respond as you vary
$m_s$ at fixed $\mu$.  Notice that the effect of the strange quark 
mass, combined with the requirement of electric neutrality, is to
push $p_F^d$ up and $p_F^s$ down relative to $p_F^u$, and at the same
time induce a nonzero electron density.  To leading order in $m_s$ the
electron chemical potential is 
\begin{equation}
\mu_e = \frac{m_s^2}{4 \mu}
\end{equation}
and the Fermi momenta are 
\begin{eqnarray}
p_F^d & = & \mu + \frac{m_s^2}{12 \mu}  =  p_F^u + \frac{m_s^2}{4\mu} \nonumber \\ 
p_F^u & = & \mu - \frac{m_s^2}{6 \mu} \nonumber \\ 
p_F^s & = & \mu - \frac{5 m_s^2}{12 \mu}  =  p_F^u - \frac{m_s^2}{4\mu}
\label{UnpairedFermiMomenta}
\end{eqnarray}
The magnitude of the splitting between Fermi surfaces is $\delta p_F =
m_s^2/4 \mu$.  Notice that decreasing $\mu$ enhances this flavor
disparity so the effect is more important at intermediate densities.

The system responds differently when pairing interactions at the Fermi
surface are taken into account~\cite{neutrality}.  At asymptotically
high densities, the system is in the CFL state with equal numbers of
$u$, $d$, and $s$ quarks.  As $\mu$ decreases, the CFL state remains
``rigid'' with equal numbers of $u$, $d$, and $s$ quarks, despite the
presence of a stress which seeks to separate the Fermi surfaces.  This
rigidity maximizes the binding energy for the Cooper pairs in the CFL
phase.  The CFL state is the stable ground state of the system only
when its negative interaction energy offsets the large positive free
energy cost associated with forcing the Fermi seas to deviate from
their normal state distributions.  The free energy of the CFL state
must be compared to that of the unpaired or ``normal'' state in which
the quarks simply distribute themselves in Fermi seas as in figure
\ref{fermimomenta}.  The result is~\cite{AlfordRajagopal}
\begin{equation}
\label{CFLFreeEnergy}
\Omega_{\mbox{\scriptsize CFL}} - \Omega_{\mbox{\scriptsize normal}}
\approx - \frac{3}{\pi^2} \Delta_0^2 \mu^2 + \frac{3}{16 \pi^2} m_s^4
\end{equation}
where the negative first term represents the pairing energy gain of
the CFL phase, and the positive second term is the cost associated
with enforcing equal numbers of $u$, $d$, and $s$ quarks.  We have
neglected terms in the free energy that are of high order in
$\Delta_0$ or $m_s$.  We find that the CFL phase is favored over
unpaired quark matter only
for~\cite{ABR2+1,SW2,neutrality,ARRW,AlfordRajagopal,BuballaOertel}
\begin{equation}
\label{ApproxUnlocking}
\mu > \mu_{\mbox{\scriptsize unlock}} \approx \frac{m_s^2}{4 \Delta_0}
\end{equation}
and the CFL pairing vanishes at a first-order unlocking transition.
Notice that the unlocking occurs when the Fermi surface splitting
$\delta p_F$ in the unpaired phase becomes equal to the gap in the CFL
phase: this is consistent with the physical intuition that the CFL
phase can ``force'' the Fermi seas to deviate from their normal state
distributions only for $\delta p_F < \Delta_0$.  

In interpreting equation (\ref{ApproxUnlocking}), recall that the
value of the CFL gap is not precisely known: it is of order 10-100 MeV
and is also density-dependent.  The strange quark mass parameter $m_s$
includes the contribution from any $\langle \bar s s \rangle$
condensate induced by the nonzero current strange quark mass, making
it a density-dependent effective mass.  At densities that may occur at
the center of compact stars, corresponding to $\mu\sim 400-500$ MeV,
$m_s$ is certainly significantly larger than the current quark mass,
and its value is not well known.  In fact, $m_s$ decreases
discontinuously at the unlocking transition~\cite{BuballaOertel}.
Thus, the criterion (\ref{ApproxUnlocking}) can only be used as a
rough guide to the location of the unlocking transition in nature.  As
in figure \ref{phasediagram} we assume that unlocking occurs before
hadronization, so there is a window of intermediate densities where
non-CFL quark matter will occur.

It has been proposed~\cite{BedaqueSchaefer} that the CFL state may not
be completely rigid above the unlocking transition, but may instead
respond to the imposed stress by forming a condensate of CFL Goldstone
bosons.  To respond to the stress the system wants to reduce the
number of strange quarks and increase the number of up quarks, as in
figure \ref{fermimomenta}.  Introducing an extra up quark and a
strange hole lowers the energy by $\sim m_s^2/2\mu$, but appears to
require the breaking of a pair and therefore involves an energy cost
which is of the order of the gap $\Delta_0$.  However, a
down-particle/strange-hole pair has the quantum numbers of a kaon, so
the energy cost is actually just the mass of the kaon-like elementary
excitation in the CFL phase.  The CFL kaon is a pseudo-Goldstone boson
which acquires a small mass $m_K \sim \Delta_0 \sqrt{m_s m_d}/\mu \ll
\Delta_0$ when nonzero quark masses are introduced which explicitly
break chiral symmetry.  So the CFL vacuum may lower its free energy by
decaying into $K^0$ collective modes by the process $0 \rightarrow
(\overline{us})(du)$.  A $K^0$ condensate corresponds to a relative
rotation of left- and right-handed condensates in flavor space.  If
kaon condensation occurs, it lowers the free energy of the CFL phase
with another term of order $m_s^4$ in equation (\ref{CFLFreeEnergy}).
As a result the 4 in the denominator of (\ref{ApproxUnlocking})
increases but remains smaller than 5, so this is a relatively minor
effect as it concerns unlocking~\cite{AlfordRajagopal}.

The unlocking transition was originally studied without the
requirement of charge neutrality~\cite{ABR2+1,SW2}.  It was assumed
that $\mu_u = \mu_d = \mu_s = \mu$ and therefore in the unpaired phase
the up and down quarks would have equal Fermi momenta $p_F^u = p_F^d =
\mu$ but the strange quark would have a smaller Fermi momentum $p_F^s
= \sqrt{\mu^2 - m_s^2}$, so the system would have a net positive
charge density.  In this context, $us$ and $ds$ pairing are disrupted
at the unlocking transition, but $ud$ pairing should persist because
there is no ``stress'' trying to separate the $u$ and $d$ Fermi
surfaces.  What is left over is the simple and well-known ``2SC''
state~\cite{ARW2,RappEtc,Reviews} with a condensate of the form
\begin{equation}
\label{2SCcondensate}
\langle \psi_{i \alpha a}(\vx) \psi_{j \beta b}(\vx) \rangle
\propto \Delta_0 \epsilon_{\alpha \beta 3} \epsilon_{i j 3} 
(C \gamma_5)_{ab}.
\end{equation}
This is quite similar to equation (\ref{CFLcondensate}) for the CFL
condensate, except that there is no summation over a common color and
flavor antitriplet index $A$.  In the 2SC phase the condensate
involves only two colors and two flavors.  Four of the nine quarks are
paired: $rd$ pairs with $gu$, and $ru$ pairs with $gd$.  Five quarks
are left unpaired (the blue $u$ and $d$ quarks, and all three colors
of the strange quark).

In a charge neutral system, however, the 2SC phase is unlikely to
occur.  As we have seen, imposing charge neutrality in the unpaired
phase requires the introduction of an electron chemical potential
$\mu_e$.  This is an isospin chemical potential which separates the up
and down Fermi surfaces, so the stress that disrupts $us$ and $ds$
pairing should equally disrupt $ud$ pairing.  A detailed calculation
confirms that this physical intuition is
correct~\cite{AlfordRajagopal,Steiner:2002gx}: a calculation of the
free energies of charge-neutral CFL, 2SC, and normal (unpaired) states
reveals that the 2SC phase is nowhere favored.

Below the unlocking transition, the system is still unstable to the
formation of a condensate of Cooper pairs, but the pairing pattern
must be something other than CFL or 2SC.  Perhaps the most obvious
possibility is that the system simply abandons inter-species pairing
and forms single-flavor $\langle uu \rangle$, $\langle dd \rangle$,
and $\langle ss \rangle$
condensates~\cite{IwaIwa,ABCC,TS1flav,Schmitt:2002sc}.  Unfortunately
these condensates are rather feeble: the gaps are no larger than a few
MeV~\cite{TS1flav,ABCC}, and they could even be much smaller than
this~\cite{ABCC,ARW2}.  The single-flavor condensates are weak because
they must be either symmetric in spin (and therefore $J=1$) or
symmetric in color, whereas the QCD interaction favors condensates
that are both antisymmetric in spin ($J=0$) and antisymmetric in
color.  This is easy to understand.  The color $\mathbf{\bar 3}$
channel is the dominant attractive channel, perturbatively (with
Coulombic gluons), nonperturbatively (via the instanton interaction),
and phenomenologically (from the fact that pairs of valence quarks in
a baryon are in a color $\mathbf{\bar 3}$ state).  The $J=0$
channel is enhanced by its rotational symmetry (larger symmetry
generally implies more robust pairing).

Condensates that are antisymmetric in color and spin are also
antisymmetric in flavor (by the Pauli principle), so the QCD
interaction naturally favors inter-species (inter-flavor) pairing.
So, inter-species pairing is likely to be favored, but a novel pairing
arrangement must be proposed that can accommodate inter-species
pairing even at intermediate densities when the different species have
different Fermi momenta.  The pairing arrangement of the crystalline
color
superconductor~\cite{BowersLOFF,ngloff,BowersCrystal,crystalproc,massloff,pertloff,Giannakis,LOFFphonon}
is the most well-studied option, and it is the main emphasis of the
current work.

The crystalline phase was originally described by Larkin, Ovchinnikov,
Fulde, and Ferrell (LOFF)~\cite{LO,FF} as a novel pairing mechanism
for an electron superconductor with a Zeeman splitting between spin-up
and spin-down Fermi surfaces, neglecting all orbital effects of the
magnetic perturbation.  Quark matter is a more natural setting for the
LOFF phase, as it features a ``flavor Zeeman effect'' with no orbital
complications, as a consequence of the large strange quark mass.
Cooper pairs in the LOFF phase have nonzero total momentum: a quark
with momentum $\vp$ is paired with a quark with momentum $-\vp+2\vq$
such that each quark is near its respective Fermi surface, even though
the two Fermi surfaces are disjoint.  The magnitude $|\vq|$ is
determined by the separation between Fermi surfaces (we expect $|\vq|
\sim \delta p_F$) while the direction $\hat\vq$ is chosen
spontaneously.  This generalization of the pairing ansatz (beyond BCS
ans\"atze in which only quarks with momenta which add to zero pair) is
favored because it gives rise to a region of phase space where each
quark in a pair resides near its respective Fermi surface; as a
result, pairs can be created at a low cost in free energy and a
condensate can form.  In contrast to the BCS phase, where pairing
occurs over the entire Fermi surface, LOFF pairing has a restricted
phase space.  For a given $\vq$, the quarks that pair are only those
in ``pairing rings'', one on each Fermi surface; as explained below
and in figure \ref{ringsfig}, these circular bands are antipodal to
each other and perpendicular to $\vq$.  

As we shall see in chapter 2, the phase space is restricted to the
ring-like pairing regions by the formation of ``blocking regions''
(see figure \ref{fig:regions}) in which pairing is forbidden.
Momentum modes inside these blocking regions are occupied by one
species of quark but not the other, so pairing cannot occur.
Quasiquark excitations are gapless for momenta at the boundaries of
these blocking regions, but within the pairing rings the quasiquarks
are gapped.  Because it has both gapped and gapless quasiparticle
excitations, the crystalline state is simultaneously superconducting
and metallic.

If each Cooper pair in the condensate carries the same total momentum
$2\vq$, then in position space the condensate varies like a plane
wave: 
\begin{equation}
\label{PlaneWaveCondensate}
\langle \psi(\vx) \psi(\vx) \rangle \sim \Delta
e^{2i\vq\cdot\vx}
\end{equation}
meaning that translational and rotational symmetry are spontaneously
broken.  This justifies calling it a crystalline color superconductor.
Of course, if the system is unstable to the formation of a single
plane-wave condensate, then we expect that a superposition of multiple
plane waves is still more favorable, leading to a more complicated
spatial variation of the condensate:
\begin{equation}
\label{MultiplePlaneWaveCondensate}
\langle \psi(\vx) \psi(\vx) \rangle \sim \sum_{\vq} 
\Delta_\vq e^{2i \vq\cdot\vx}.
\end{equation}
Each $\Delta_\vq$ corresponds to condensation of Cooper pairs with
momentum $2\vq$, i.e.~another pairing ring on each Fermi surface.  As
we add more plane waves, we utilize more of the Fermi surface for
pairing, with a corresponding gain in condensation energy.  On the
other hand, the rings can ``interact'' with each other: condensation
in one mode can enhance or deter condensation in another mode.  The
true ground state of the system is obtained by exploring the 
infinite-dimensional parameter space of crystalline order parameters
$\{ \Delta_{\vq} \}$ to find that particular crystal structure which 
is a global minimum of the free energy functional 
$\Omega[ \Delta(\vx) ] = \Omega(\{\Delta_{\vq}\})$.  

As an aside, it is worth noting that crystalline phases have appeared
in other QCD contexts. In their analysis of quark matter with a very
large isospin density (with large Fermi momenta for down and {\it
anti}-up quarks) Son and Stephanov have noted that if the $d$ and
$\bar u$ Fermi momenta differ suitably, a LOFF crystalline phase will
arise~\cite{SonStephanov}.  A LOFF crystalline phase can also occur
for neutron-proton pairing in asymmetric nuclear matter with a
splitting between the neutron and proton Fermi
surfaces~\cite{NuclearLOFF}.  Moreover, the LOFF state is not the only
crystalline phase that has been investigated: at large baryon number
density, pairing between quarks and holes with nonzero total momentum
has also been discussed~\cite{DGR,ShusterSon,PRWZ,RappCrystal}.  This
results in a chiral condensate which varies in space with a wave
number equal to $2\mu$; in contrast, the LOFF phase describes a
diquark condensate which varies with a wave number $2|\vq|$ comparable
to $\delta p_F$.  Several possible crystal structures have been
analyzed for the crystalline chiral condensate~\cite{RappCrystal}, but
this phase is favored at asymptotically high densities only if the
number of colors is very large~\cite{DGR}, greater than about
$N_c=1000$~\cite{ShusterSon,PRWZ}.  It may arise at lower densities in
QCD with fewer colors, but apparently not in QCD with
$N_c=3$~\cite{RappCrystal}.

At least two alternatives to the crystalline color superconducting
phase have been proposed that also allow pairing between quarks with
unequal Fermi surfaces.  The first alternative is the deformed Fermi
sphere (DFS) superconductor~\cite{DFS}.  In this phase, the unequal
Fermi surfaces of the two paired species are deformed so that they can
intersect, and then pairing can occur in the vicinity of this
intersection.  The deformations are volume-conserving so that they do
not change particle numbers for the two species.  The larger Fermi
surface has a prolate deformation, while the smaller Fermi surface has
an oblate deformation, and pairing occurs along two bands just above
and below the equator of each spheroid.  The DFS phase breaks
rotational symmetry, but unlike the crystalline (LOFF) phase it does
not break translational symmetry because the Cooper pairs still have
zero total momentum.  
%
%In contrast to the deformed spheres of the DFS
%phase, the LOFF phase can be imagined as having ``displaced'' Fermi
%spheres: the centers of the two spheres are displaced by an amount
%equal to the total momentum of a LOFF Cooper pair.  The displacement
%allows the two spheres to intersect in a ring, where pairing occurs
%(keep in mind that while this displacement picture is useful to
%visualize the phase space for LOFF pairing, there is no physical
%adjustment of either Fermi sphere).  Whether the unequal Fermi
%surfaces are made to intersect by a displacement or by a deformation,
%the intersection has zero area measure: 
%the pairing is along circular
%bands on the Fermi surfaces and therefore the DFS and LOFF phases
%would seem to have similar condensation energies.  
%
In both the DFS and LOFF phases the pairing is along circular bands on
the Fermi surfaces and therefore the two phases would seem to have
similar condensation energies.  Meanwhile, in the DFS phase there is
also a large kinetic energy cost associated with deforming the Fermi
spheres away from their preferred spherical shapes.  There is no such
cost for the crystalline phase, so we expect the crystalline phase to
have a lower free energy.
 
The second alternative is the ``breached-pair'' color
superconductor~\cite{Sarma,Liu1,Liu2}.  The breached-pair
superconductor is translationally and rotationally invariant, in
contrast to the crystalline and DFS phases.  This state was first
encountered by Sarma~\cite{Sarma} in the context of an electron
superconductor with a Zeeman splitting between the spin-up and
spin-down Fermi surfaces, the same context in which the crystalline
(LOFF) phase was first proposed.  In this historical context it was
found that the breached-pair state was never a minimum of the free
energy, but recent developments~\cite{Liu1,Liu2} suggest that the
breached-pair state might be a stable ground state for pairing between
a light species and a heavy species with different Fermi momenta, and
therefore might accommodate $us$ and $ds$ pairing in
intermediate-density quark matter.  In the breached-pair
superconductor, the Fermi sea of heavy quarks (Fermi momentum $p_h$)
is redistributed to accommodate pairing at the light quark Fermi
surface (Fermi momentum $p_l$).  The redistribution has a small energy
cost because the heavy quark has a very flat single-particle
dispersion relation.  For $p_l < p_h$, heavy quarks are promoted from
$|\vp| \simeq p_l$ to $|\vp| \simeq p_h$, creating a ``trench'' of
unoccupied heavy quark states in the interior of the heavy quark Fermi
sea.  This trench is coincident with the light quark Fermi surface and
allows the formation a condensate of Cooper pairs at this surface, a
so-called ``interior gap''~\cite{Liu1}.  For $p_h < p_l$ (the scenario
of interest for quark matter), heavy quarks are promoted from $|\vp|
\simeq p_h$ to $|\vp| \simeq p_l$, creating a ``berm'' of occupied
heavy quark states far above the heavy quark Fermi sea.  This berm is
coincident with the light quark Fermi surface and allows the formation
of a condensate of Cooper pairs at this surface, a so-called
``exterior gap''~\cite{Liu2}.

The common mechanism in both interior and exterior gap phases is the
promotion of a shell of heavy quarks across a momentum ``breach'' of
magnitude $|p_l-p_h|$, thereby creating a second edge in the momentum
distribution of heavy quarks.  This edge behaves like a new Fermi
surface coincident with the light quark Fermi surface, accommodating
the formation of Cooper pairs.  The breach is a ``blocking region''
analogous to the aforementioned blocking regions in the crystalline
phase.  In the crystalline phase, the blocking regions restrict the
pairing to occur on rings and forbid pairing away from these rings.
In the breached-pair phase, the blocking region is spherically
symmetric and forbids pairing in the breach between $p_l$ and $p_h$.
Just as in the blocking regions of the crystal, momentum modes within
the breach are occupied by one species of quark but not the other.  In
either context, quasiparticle excitations are gapless for momenta at
the boundaries of the blocking regions.  In the breached-pair state
this means that the light quark Fermi surface is gapped while the
heavy quark Fermi surface remains ungapped, and the system is
simultaneously superconducting and metallic, like the crystalline
state.  The condensation energy must be weighed against the cost of
promotion across the breach: the cost of promotion is small only if
the heavy quark dispersion relation is sufficiently flat, so exterior
gap $\langle us \rangle$ and $\langle ds \rangle$ condensates can
occur only when the strange quark is nonrelativistic.  A breached-pair
state has been proposed for $ud$ pairing~\cite{Gapless2SC}, but the
preceding argument suggests that this is not a stable ground state.  A
gapless CFL state, also with breached pairing, has been investigated
and was found to be metastable~\cite{GaplessCFL}.  The evaluation of
the energy of breached-pair phases is subtle, and the stability is
sensitive to whether a microcanonical (fixed number density) or grand
canonical (fixed chemical potential) approach is used~\cite{Yip,Liu2}.

\section{Crystalline color superconductivity} 
\label{sec:crystalline}

Crystalline color superconductivity has only been studied in
simplified models with pairing between two quark species whose Fermi
momenta are pushed apart by a chemical potential
difference~\cite{BowersLOFF,ngloff,LOFFphonon,pertloff,Giannakis} or a
mass difference~\cite{massloff}.  We suspect that in reality, in
three-flavor quark matter whose unpaired Fermi momenta are split as in
(\ref{UnpairedFermiMomenta}), the pattern of pairing in the
crystalline phase will involve $ud$, $us$ and $ds$ pairs, with color
and flavor quantum numbers just as in the CFL phase.  However,
studying the simpler two-flavor problem should elucidate the nature of
the crystalline ground state, including its crystal structure.  We
therefore simplify the color-flavor pattern to one involving massless
$u$ and $d$ quarks only, with Fermi momenta split by introducing
chemical potentials
\begin{eqnarray}
\mu_d &=& \bar \mu + \delta\mu\nonumber\\
\mu_u &=& \bar \mu - \delta\mu \ .
\end{eqnarray}
In this toy model, we vary $\delta\mu$ by hand.  In three-flavor quark
matter, the analogue of $\delta\mu$ is controlled by the nonzero
strange quark mass and the requirement of electrical neutrality and
would be of order $m_s^2/4\mu$ as in (\ref{UnpairedFermiMomenta}).

The LOFF crystalline phase was originally studied in the context of an
electron superconductor with a Zeeman splitting between the spin-up
and spin-down Fermi surfaces~\cite{LO,FF}.  The authors considered a
magnetic perturbation $\Delta H = h \psi^\dag \sigma_z \psi$ but
disregarded any orbital effects of the magnetic field.  They were
seeking to model the physics of magnetic impurities in a
superconductor.  Magnetic effects on the motion of the
electrons~\cite{Gruenberg} and the scattering of electrons off
non-magnetic impurities~\cite{Aslamazov,Takada2} disfavor the LOFF
state. Although signs of the BCS to LOFF transition in the heavy
fermion superconductor UPd$_2$Al$_3$ have been reported~\cite{Gloos},
the interpretation of these experiments is not
unambiguous~\cite{Controversy}.  It has also been suggested that the
LOFF phase may be more easily realized in condensed matter systems
which are two-dimensional~\cite{Shimahara,2D} or
one-dimensional~\cite{1D}, both because the LOFF state is expected to
occur over a wider range of the Zeeman field $h$ than in
three-dimensional systems and because the magnetic field applied
precisely parallel to a one- or two-dimensional system does not affect
the motion of electrons therein.  Evidence for a LOFF phase in a
quasi-two-dimensional layered organic superconductor has recently been
reported~\cite{Nam}.

None of the difficulties which have beset attempts to realize the LOFF
phase in a system of electrons in a magnetic field arise in the QCD
context of interest to us.  Differences between quark chemical
potentials are generic and the physics which leads to these
differences has nothing to do with the motion of the quarks.  We
therefore expect the original analysis of LOFF (without the later
complications added in order to treat the difficulties in the
condensed matter physics context) to be a good starting point.  

In our two-flavor model, we shall take the interaction between quarks
to be pointlike, with the quantum numbers of single-gluon exchange.
This $s$-wave interaction is a reasonable starting point at accessible
densities but is certainly not appropriate at asymptotically high
density, where the interaction between quarks (by gluon exchange) is
dominated by forward scattering.  The crystalline color
superconducting state has been analyzed at asymptotically high
densities in Refs.~\cite{pertloff,Giannakis}.  We expect a
qualitatively different crystalline phase in this asymptotic regime,
but this may not be relevant for densities of interest for compact
star physics.  

With this astrophysical context in mind, it is also appropriate for us
to work at zero temperature.  Compact stars that are more than a few
minutes old are several orders of magnitude colder than the critical
temperature (of order tens of MeV) for CFL or crystalline color
superconductivity.  The crystalline color superconductor has been
studied at finite temperature, and the critical temperature is given
by $T_c/\Delta_{(T=0)} \simeq 0.44$~\cite{ngloff} (a result previously
known in the historical LOFF context~\cite{Takada1}).  It is
interesting that this differs from the usual BCS relation
$T_c/\Delta_{(T=0)} \simeq 0.57$~\cite{BCS}.  The critical
temperatures for the CFL and single-flavor color superconductors also
differ from the BCS result~\cite{Schmitt:2002sc}.

Now let us consider how the crystalline phase occurs in our two-flavor
model.  Starting at $\dm = 0$, the system forms a BCS superconductor
with gap $\Delta_0$.  In fact this BCS superconductor is precisely the
2SC phase of equation (\ref{2SCcondensate}): the Cooper pairs are
color antisymmetric (red pairs with green) and flavor antisymmetric
(up pairs with down).  The blue quarks are left unpaired.  The up and
down Fermi surfaces are coincident.  As we begin to increase $\dm$,
the system exhibits a ``rigidity'' analogous to that of the CFL phase:
despite the imposed stress $\dm$, the gap stays constant and the Fermi
surfaces remain coincident.  The BCS state is the stable ground state
of the system only when its negative interaction energy offsets the
large positive free energy cost associated with forcing the Fermi seas
to deviate from their normal state distributions.  The free energy of
the BCS state, relative to that of of the normal state in which the
quarks simply distribute themselves in Fermi seas with $p_F^u = \mu_u$
and $p_F^d = \mu_d$, is approximately
\begin{equation}
\label{BCSFreeEnergy}
\Omega_{\mbox{\scriptsize BCS}} - \Omega_{\mbox{\scriptsize normal}}
\approx - \frac{1}{\pi^2} \Delta_0^2 \bar\mu^2 + \frac{2}{\pi^2} \dm^2 \bar\mu^2
\end{equation}
where the first term is the negative pairing energy of the BCS state,
and the second term is the cost associated with enforcing equal
numbers of up and down quarks in the presence of the imposed stress
$\dm$.  This result is exact only in the weak-coupling limit in which
the gap $\Delta_0 \ll \bar\mu$.  This expression
(\ref{BCSFreeEnergy}) should be compared to equation
(\ref{CFLFreeEnergy}) for the free energy of the neutral CFL phase
with the imposed stress of a nonzero $m_s$.  When $\dm$ reaches a
critical value
\begin{equation}
\label{dm1val}
\dm_1 \approx \frac{1}{\sqrt{2}} \Delta_0 = (0.7071\cdots) \Delta_0,
\end{equation}
the BCS phase ``breaks'' and the Fermi surfaces separate (again, the
expression is exact only in the weak coupling limit in which $\Delta_0
\ll \bar\mu$).  This is the two-flavor analogue of the CFL unlocking
transition. (Bedaque~\cite{Bedaque} has investigated the mixed phase
associated with this first-order transition, where the unpaired blue
quarks also play a role.)  This result was first derived by Clogston
and Chandrasekhar~\cite{Clogston} in the context of an electron
superconductor with a Zeeman splitting.

\begin{figure}
\centering
\psfrag{muu}[r][r]{$\mu_u$}
\psfrag{mud}[l][l]{$\mu_d$}
\psfrag{p}[l][l]{$\vp$}
\psfrag{mp}[r][r]{$-\vp$}
\psfrag{p2q}[r][r]{$-\vp+2\vq$}
\psfrag{pp}[r][r]{$\vp'$}
\psfrag{mpp}[l][l]{$-\vp'$}
\psfrag{pp2q}[l][l]{$-\vp'+2\vq$}
\psfrag{2q}[c][c]{$2\vq$}
\psfrag{psiu}[c][l]{$\psi_u$}
\psfrag{psid}[c][l]{$\psi_d$}
\includegraphics[width=3.5in]{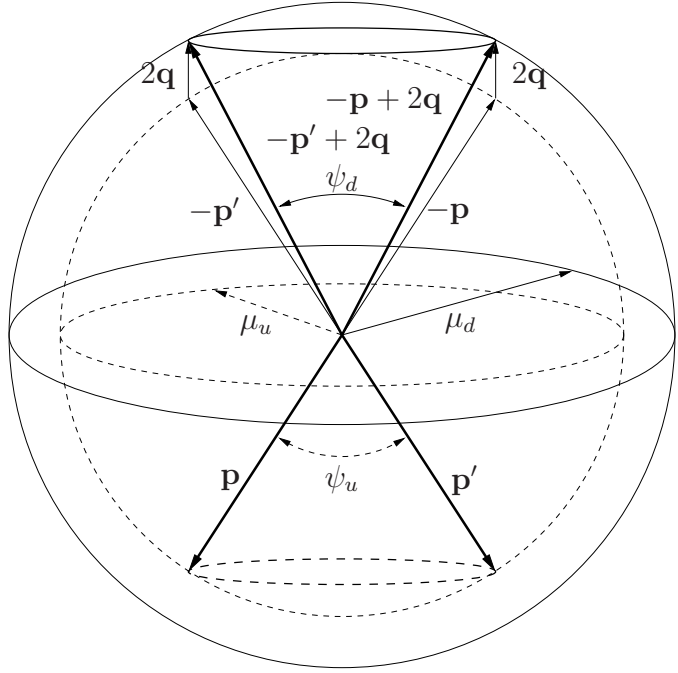}
\caption{
\label{ringsfig}
The LOFF pairing geometry for Cooper pairs with total momentum $2\vq$.
The dashed and solid spheres are the up and down quark Fermi surfaces,
respectively.  An up quark with momentum $\vp$ (or $\vp'$) near its
Fermi surface pairs with a down quark with momentum $-\vp+2\vq$ (or
$-\vp'+2\vq$) near its Fermi surface.  The pairing is strongest for up
quarks in a band centered on the dashed ring shown on the up Fermi
surface, and down quarks in a band centered on the solid ring shown on
the down Fermi surface.
}
\end{figure}

For $\dm > \dm_1$, the up and down quarks have unequal Fermi surfaces
and a crystalline state is possible.  In the simplest LOFF state, up
quarks with momentum $\vp$ are paired with down quarks with momentum
$-\vp+2\vq$.  Each Cooper pair carries the same total momentum $2{\bf
q}$.  The allowed phase space for $\vp$ is determined by the
requirement that each quark in the Cooper pair should sit near its
Fermi surface, i.e.
\begin{equation}
| |\vp| - \mu_u | \lesssim \Delta \ \ \  \mbox{and} \ \ \  | |-\vp+2\vq| - \mu_d | \lesssim \Delta
\end{equation}
where $\Delta$ is the LOFF gap parameter.  This phase space
corresponds to a circular band on each Fermi surface as shown figure
\ref{ringsfig}.  As indicated in the figure, the bands are perpendicular to
the spontaneously chosen direction $\hat\vq$ for the total momentum of
each Cooper pair.  The magnitude $|\vq| \equiv q_0$ is determined
energetically from the separation between Fermi surfaces.  We shall
find that the relation is $q_0 \approx 1.20 \dm$.  

It is useful to discuss the various scales involved in the problem.
The BCS gap $\Delta_0$ can be thought of as the fundamental energy
scale for physics at the Fermi surface.  If we consider the weak
coupling limit in which $\Delta_0 \ll \bar\mu$, then the two scales
are cleanly separated and all the other Fermi-surface energy scales in
the problem (i.e.~$|\vq|$, $\dm$, and the LOFF gap $\Delta$) should be
proportional to the fundamental scale $\Delta_0$.  We achieve this by
taking a``double scaling'' limit \label{doublescaling} in which we
choose to hold $\dm/\Delta_0$ fixed while taking the $\Delta_0/\bar\mu
\rightarrow 0$ limit.
%$\Delta_0/\bar\mu \rightarrow 0$ while $\dm/\Delta_0$ and
%$|\vq|/\Delta_0$ are held fixed and nonzero.  
In this double scaling limit, $|\vq|/\Delta_0$ and $\Delta/\Delta_0$
also stay fixed.  In fact, every dimensional quantity for the
Fermi-surface physics stays fixed as ``measured'' in units of
$\Delta_0$.  If we fail to take the double scaling limit, instead
keeping $\dm/\bar\mu$ fixed as $\Delta_0/\mu \rightarrow 0$, we would
not find crystalline color superconductivity at weak
coupling~\cite{HongLOFF}.  We will not always work in the double
scaling limit (see, for example, chapter 2), but we will often quote
analytic results that are exact in this limit (as we did in equation
\ref{dm1val}).  In the double scaling limit, $\dm \ll \bar\mu$ and the
two Fermi surfaces in figure \ref{ringsfig} are very close together.
The opening angles $\psi_u$ and $\psi_d$ of the two pairing bands
become degenerate and take on the value
\begin{equation}
\label{openingangleeqn}
\psi_0 \approx 2 \cos^{-1} \left(\frac{\dm}{|\vq|}\right) 
\approx 2 \cos^{-1} \frac{1}{1.20} =  67.1^\circ.
\end{equation}
The radial thickness of each pairing band is of order $\Delta$, while
the angular width is $\delta\psi \sim \Delta/\sqrt{|\vq|^2 - \dm^2}
\sim 1.5 \Delta/\dm$.  If we use double scaling then both $\psi_0$ and
$\delta\psi$ are constant because $|\vq|/\Delta_0$, $\dm/\Delta_0$,
and $\Delta/\Delta_0$ are all held constant while $\Delta_0
\rightarrow 0$.  Hereafter, when we speak of the ``weak coupling limit'' 
we shall always mean the double scaling limit.  

If all the Cooper pairs in the condensate have the same nonzero total
momentum $2\vq$, then the condensate varies like a single plane wave
in position space, as in equation (\ref{PlaneWaveCondensate}).  In
chapter 2 we present a careful study of this single plane wave
condensate, the simplest example of a LOFF phase.  We show that there
is a range of $\dm$ in which quark matter is unstable to the
spontaneous breaking of translational invariance by the formation of a
plane wave condensate.  Of course, once one has demonstrated an
instability to the formation of a plane wave, it is natural to expect
that the state which actually develops has a crystalline structure
consisting of multiple plane waves, as in equation
(\ref{MultiplePlaneWaveCondensate}). In chapter 3 we investigate this
possibility.  Larkin and Ovchinnikov in fact argue that the favored
configuration is a crystalline condensate which varies in space like a
one-dimensional standing wave, $\cos(2\vq\cdot\vr)$. Such a condensate
vanishes along nodal planes~\cite{LO}.  Subsequent analyses suggest
that the crystal structure may be more complicated.
Shimahara~\cite{Shimahara} has shown that in two dimensions, the LOFF
state favors different crystal structures at different temperatures: a
hexagonal crystal at low temperatures, square at higher temperatures,
then a triangular crystal and finally a one-dimensional standing wave
as Larkin and Ovchinnikov suggested at temperatures that are higher
still. In three dimensions, the question of which crystal structure is
favored was unresolved~\cite{BuzdinKachkachi}.  Our analysis, shown in
chapter 3, suggests that the favored crystal structure in three
dimensions is face-centered-cubic.

\begin{figure}
\begin{center}
\psfrag{delta}[rc][rc]{$\Delta$}
\psfrag{omega}[rc][rc]{$\Omega$}
\psfrag{deltaomega}[rc][rc]{$-\frac{\mu^2\Delta_0^2}{\pi^2}$}
\psfrag{deltabcs}{$\Delta_{\mbox{\scriptsize BCS}}$}
\psfrag{omeganormal}{$\Omega_{\mbox{\scriptsize normal}} \equiv 0$}
\psfrag{omegabcs}{$\Omega_{\mbox{\scriptsize BCS}}$}
\psfrag{omegapw}[lc][lc]{$\Omega_{\mbox{\scriptsize pw}}$}
\psfrag{omegaxtal}[lc][lc]{$\Omega_{\mbox{\scriptsize xtal}}$}
\psfrag{delta0}{$\Delta_0$}
\psfrag{deltamu}{$\dm$}
\psfrag{dm1}[tc][tc]{$\dm_1$}
\psfrag{dm2}[tc][tc]{$\dm_2$}
\psfrag{dmp}[tc][tc]{$\dm_1^\prime$}
\psfrag{dms}[tc][tc]{$\dm_*$}
\psfrag{deltapw}[lc][lc]{$\Delta_{\mbox{\scriptsize pw}}$}
\psfrag{deltaxtal}[lc][lc]{$\Delta_{\mbox{\scriptsize xtal}}$}
\includegraphics[width=5.5in]{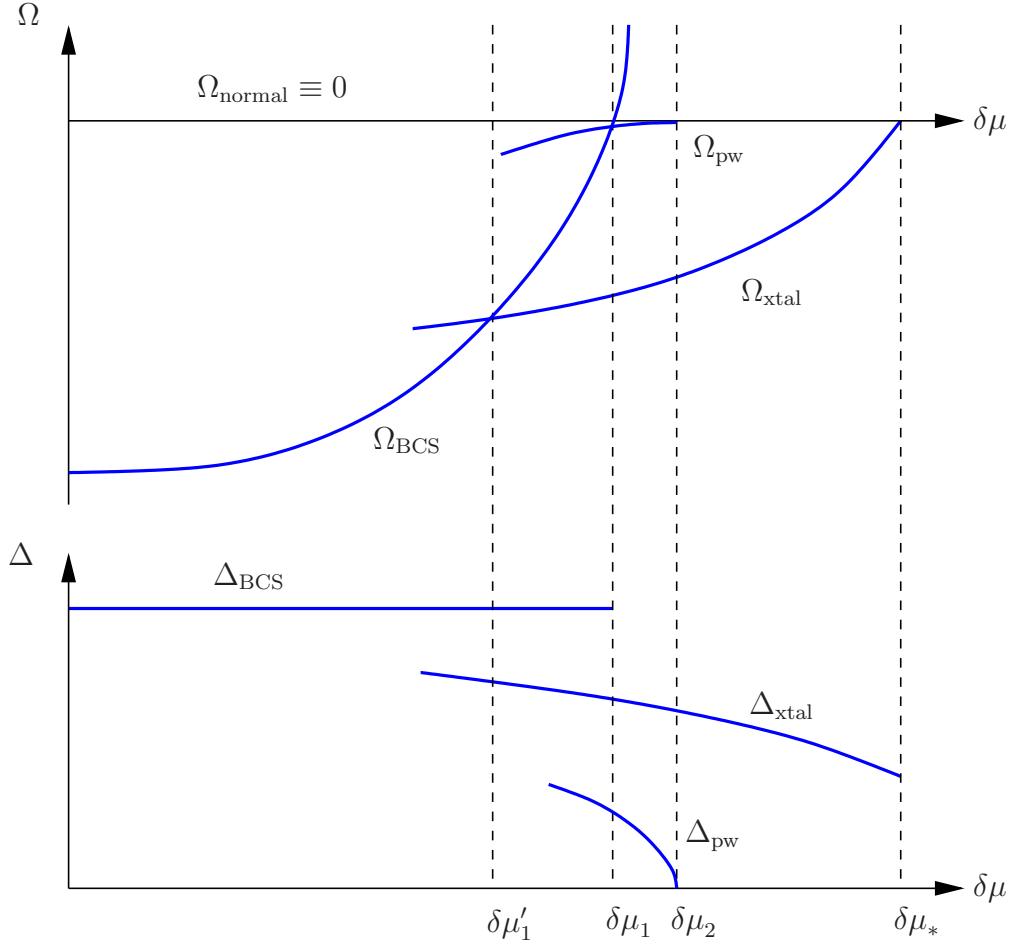}
\end{center}
\caption{
\label{omegadeltafig}
Free energies and gaps for competing states: normal, BCS, and single
and multiple plane wave crystalline phases (``pw'' and ``xtal'',
respectively).  For the plane wave state, the LOFF interval is
$[\dm_1,\dm_2] \approx [0.707 \Delta_0, 0.754 \Delta_0]$ ($\Delta_0$
is the BCS gap) and the transitions to BCS and normal states are first
and second order, respectively.  For the multiple plane wave
(crystalline) state, the LOFF interval is $[\dm_1^\prime, \dm_*]$ and
both transitions are first order.  Note that the plane wave state is
exaggerated (compare figure \ref{fig:F_plot}).  In reality $(\dm_2 -
\dm_1) \ll (\dm_*-\dm_1^\prime)$ and the crystalline state is much
more favorable than the plane wave state.  }
\end{figure}

The crystalline states appear for $\dm > \dm_1$.  In chapter 2 we will
show that the simplest LOFF state, a single plane wave condensate, can
occur in an interval $\dm_1 < \dm < \dm_2$.  At $\dm_2$ there is a
second-order transition from LOFF to the normal state (unpaired
quarks).  The second-order point occurs at
\begin{equation}
\dm_2 \approx (0.7544\cdots) \Delta_0
\end{equation}
where this relation is exact in the weak coupling limit (the numerical
coefficient is known exactly; it is the solution of a simple
transcendental equation).  At the second-order phase transition,
$\Delta/\Delta_0\rightarrow 0$ and $|{\bf q}|/\Delta_0$ tends to a
nonzero limit, which we shall denote $q_0/\Delta_0$, where $q_0\simeq
0.90 \Delta_0\simeq 1.20 \delta\mu_2$.  Near the second-order phase
transition, the quarks that participate in the crystalline pairing lie
on thin circular rings on their Fermi surfaces that are characterized
by an opening angle $\psi_0 \simeq 2 \cos^{-1} (\dm_2/q_0) \simeq
67.1^\circ$ and an angular width $\delta\psi$ that is of order
$\Delta/\dm$ and therefore tends to zero as $\Delta/\Delta_0
\rightarrow 0$.  At $\delta\mu_1$ there is a first-order phase
transition at which the LOFF solution with gap $\Delta$ is superseded
by the BCS solution with gap $\Delta_0$.  (The analogue in
three-flavor QCD would be a LOFF window in $m_s^2/\mu$, with CFL at
lower $m_s^2/\mu$ (higher density) and unpaired quark
matter at higher $m_s^2/\mu$ (lower density).)  These
results are summarized in figure \ref{omegadeltafig}, where we have
shown the free energies and gaps for the competing BCS, plane-wave
LOFF, and unpaired quark matter phases (the figure also shows the gap
and free energy for the multiple-plane-wave LOFF state, which we
discuss below).  Keep in mind that this figure is just a qualitative
sketch which exaggerates the size of the plane-wave LOFF window
$[\dm_1,\dm_2]$.  Quantitative plots are shown in figure
\ref{fig:F_plot} in chapter 2.  In the vicinity of the second-order
critical point $\dm_2$, our mean-field analysis yields a gap and free
energy for the plane-wave state that obey simple power-law relations
\begin{equation}
\Delta_{\mbox {\scriptsize pw}} \propto \sqrt{\dm_2-\dm}, 
\ \ \ \Omega_{\mbox {\scriptsize pw}} \propto - (\dm_2-\dm)^2
\end{equation}
with the expected mean field theory critical exponents for a
second-order transition.

Proceeding beyond just a single plane wave might seem to be a daunting
task. We have to explore the infinite-dimensional parameter space of
crystalline order parameters $\{ \Delta_{\vq} \}$ to find the unique
crystal structure that is a global minimum of the free energy.
However, we can exploit the fact that there is a second-order point
$\dm_2$ which indicates the onset of plane-wave instability in the
system.  In the vicinity of this second-order point, we can express
the free energy as a Ginzburg-Landau potential; the potential is
written as a series expansion of the exact free energy in powers of
the order parameters $\{\Delta_{\vq}\}$.

In chapter 3 we explicitly construct the Ginzburg-Landau potential and
apply it to a large survey of candidate crystal structures.  The
Ginzburg-Landau calculation finds many crystal structures that are
much more favorable than the single plane wave
(\ref{PlaneWaveCondensate}).  For many crystal structures, the
calculation actually predicts a strong {\em first-order} phase
transition, at some $\delta\mu_* \gg \delta\mu_2$, between unpaired
quark matter and a crystalline phase with a $\Delta$ that is
comparable in magnitude to $\Delta_0$.  Once $\delta\mu$ is reduced to
$\delta\mu_2$, where the single plane wave would just be beginning to
develop, these more favorable solutions already have very robust
condensation energies, perhaps even comparable to that of the BCS
phase.  Therefore they can even compete with the BCS phase and move
the position of the first-order transition between BCS and LOFF to a
new point $\dm_1^\prime < \dm_1$.  All of this is shown schematically
in figure \ref{omegadeltafig}.  These results are exciting, because
they suggest that the crystalline phase is much more robust than
previously thought.  However, they cannot be trusted quantitatively
because the Ginzburg-Landau analysis is only controlled in the limit
$\Delta\rightarrow 0$, and we find a first-order phase transition to a
state with $\Delta\neq 0$.  

Even though it is quite a different problem, we can look for
inspiration to the Ginzburg-Landau analysis of the crystallization of
a solid from a liquid~\cite{Chaikin}. There too, a Ginzburg-Landau
analysis predicts a first-order phase transition, and thus predicts
its own quantitative downfall.  But, qualitatively it is correct: it
predicts the formation of a body-centered-cubic crystal and experiment
shows that most elementary solids are body-centered-cubic (BCC) near
their first-order crystallization transition.

Thus inspired, let us look at how the Ginzburg-Landau calculation will
proceed.  We can start by writing down the most general expression for
the Ginzburg-Landau potential that is consistent with translational
and rotational symmetry.  We will include only the modes on the sphere
$|\vq| = q_0 = 1.2 \dm $ since these are the modes that become
unstable at $\dm_2$.  To order $\Delta^6$, the expression looks like
\begin{eqnarray}
\label{GLpotential}
\Omega(\{\Delta_\vq\}) & \propto & \sum_{\vq, |\vq| = q_0} \alpha \Delta_\vq^* \Delta_\vq
  + \frac{1}{2} \sum_\square J(\square) \Delta_{\vq_1}^* \Delta_{\vq_2} \Delta_{\vq_3}^* \Delta_{\vq_4}  \nonumber \\
 & & + \frac{1}{3} \sum_{\hexagon} K(\hexagon) \Delta_{\vq_1}^* \Delta_{\vq_2} \Delta_{\vq_3}^* \Delta_{\vq_4} \Delta_{\vq_5}^* 
\Delta_{\vq_6} + \cdots.
\end{eqnarray}
Odd powers are not allowed because the potential is invariant under
$U(1)$ baryon number (which multiples every $\Delta_\vq$ by a common
phase).  The symbol $\square$ represents a set of four equal-length
vectors $(\vq_1, \vq_2, \vq_3, \vq_4)$, $|\vq_i| = q_0$, with $\vq_1 -
\vq_2 +\vq_3 - \vq_4 = 0$, i.e.~the four vectors are joined together
to form a closed (not necessarily planar) figure. Similarly, the
symbol $\hexagon$ represents a set of six equal-length vectors
$(\vq_1, \ldots, \vq_6)$, $|\vq_i| = q_0$, with $\vq_1 - \vq_2 + \vq_3
- \vq_4 - \vq_5 + \vq_6 = 0$, i.e.~the six vectors form a closed
``hexagon''.  We sum only over closed sets of $\vq$-vectors because
otherwise the Ginzburg-Landau potential, expressed in position space
as a functional $\Omega[ \Delta(\vx) ]$, would not be translationally
invariant.  Rotational invariance implies that the coefficients
$J(\square)$ and $K(\hexagon)$ are the same for any two shapes related
by a rigid rotation.

The quadratic coefficient $\alpha$ changes sign at $\dm_2$ showing the
onset of the LOFF plane-wave instability: $\alpha \approx (\dm -
\dm_2)/\dm_2$.  If there was no interaction between the different
modes, they would just simply all condense at once, because they would
all become unstable at the second-order point.  The answer is more
complicated than this because condensation in one mode can enhance or
deter condensation in another mode.  This interaction between modes is
implemented in our Ginzburg-Landau potential by the higher order terms
involving multiple modes; thus the coefficients $J$, $K$, $\ldots$
characterize the interactions between modes and thereby determine the
crystal structure.  

As we shall see in chapter 3, these coefficients can actually be
calculated from the microscopic theory, as loop integrals in a
Nambu-Gorkov formalism.  So for a candidate crystal structures with
all $\Delta_\vq$'s equal in magnitude, we can evaluate aggregate
Ginzburg-Landau quartic and sextic coefficients $\beta$ and $\gamma$
as sums over all rhombic and hexagonal combinations of the $\vq$'s:
\begin{equation}
\beta = \sum_{\square} J(\square), \ \ \ \gamma = \sum_{\hexagon} K(\hexagon).
\end{equation}  
Then for a crystal consisting of $P$ plane waves we obtain
\begin{equation}
\label{GLpotential2}
\Omega(\Delta) \propto P \alpha \Delta^2 + \frac{1}{2} \beta \Delta^4
+ \frac{1}{3} \gamma \Delta^6 + \cdots
\end{equation} 
and we can then compare crystals by calculating $\beta$ and $\gamma$
to find the structure with the lowest $\Omega$.

Evaluating the quadratic coefficient $\alpha$ determines the location
of the plane-wave instability point, i.e.~the value of $\dm_2$.  It
also tells us that $|\vq| \simeq 1.20 \dm$, which means that each
pairing ring has an opening angle of $67.1^\circ$, as in equation
(\ref{openingangleeqn}).  As mentioned above, on its own the quadratic
term indicates that adding more plane waves (i.e.~adding more pairing
rings to the Fermi surface) always lowers the free energy.  But this 
conclusion is modified by the higher order terms in two important ways:
\begin{enumerate} 
\item
Crystal structures with intersecting pairing rings are strongly
disfavored.  Recall that each $\vq$ is associated with pairing among
quarks that lie on one ring of opening angle $\psi_0 \simeq
67.1^\circ$ on each Fermi surface. We find that any crystal structure
in which such rings intersect pays a large free energy price.
Therefore the favored crystal structures are those that feature a
maximal number of rings ``packed'' onto the Fermi surface without
intersections.  No more than nine rings of opening angle $67.1^\circ$
can be packed on a sphere without
intersections~\cite{extremal,tammes}.
\item
Crystal structures are favored if they have a set of $\vq$'s that
allow many closed combinations of four or six vectors, leading to many
terms in the $\sum_{\square}$ and $\sum_{\hexagon}$ summations in
equation (\ref{GLpotential}).  Speaking loosely, ``regular''
structures are favored over ``irregular'' structures.  All
configurations of nine nonintersecting rings are rather irregular,
whereas if we limit ourselves to eight rings, there is a regular
choice which is favored by this criterion: choose eight ${\bf q}$'s
pointing towards the corners of a cube.  In fact, a deformed cube
which is slightly taller or shorter than it is wide (a cuboid) is just
as good.
\end{enumerate}

\begin{figure}
\begin{center}
\psfrag{Omega1}[lc][lc]{$\Omega(\dm>\dm_*)$}
\psfrag{Omega2}[lc][lc]{$\Omega(\dm=\dm_*)$}
\psfrag{Omega3}[lc][lc]{$\Omega(\dm=\dm_2)$}
\psfrag{Omega4}[lc][lc]{$\Omega(\dm<\dm_2)$}
\psfrag{Delta}{$\Delta$}
\includegraphics[width=2.9in]{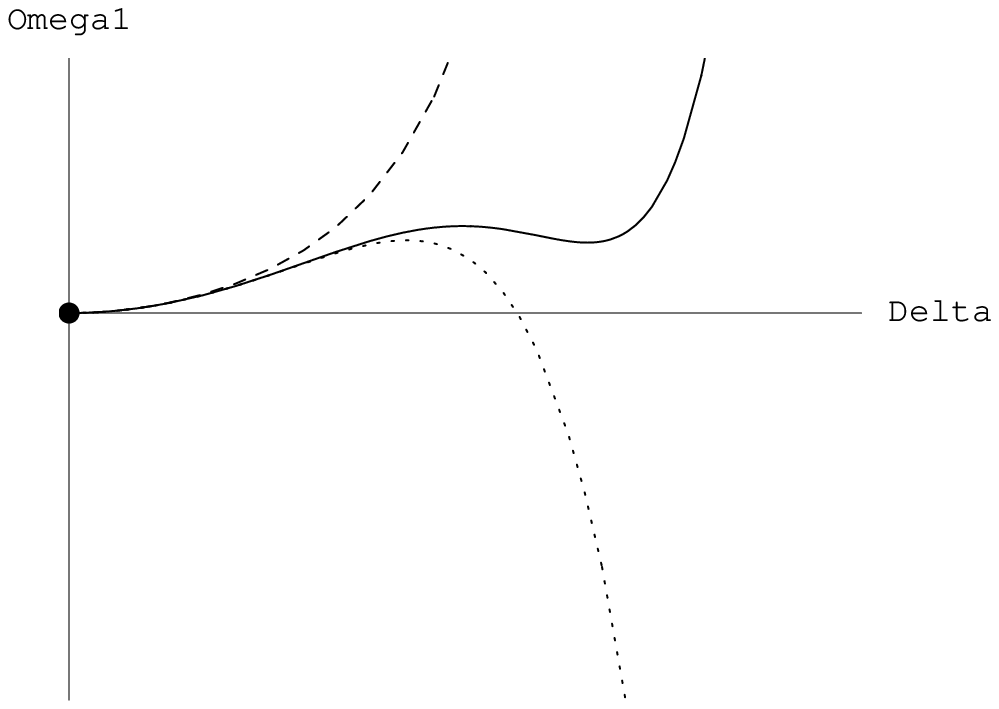}
\includegraphics[width=2.9in]{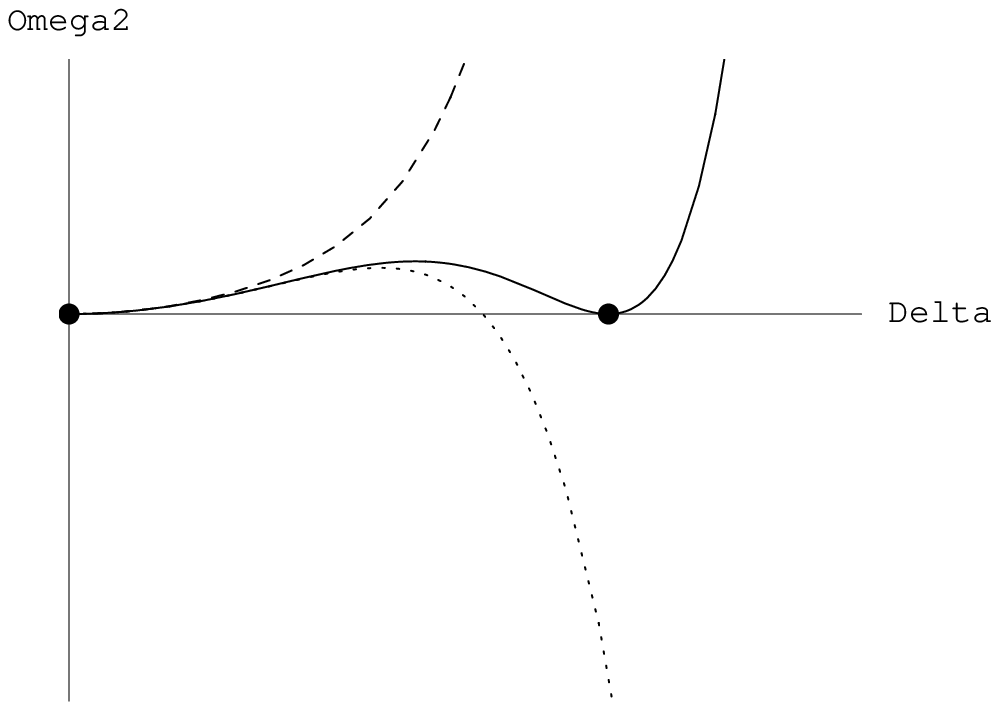}
\includegraphics[width=2.9in]{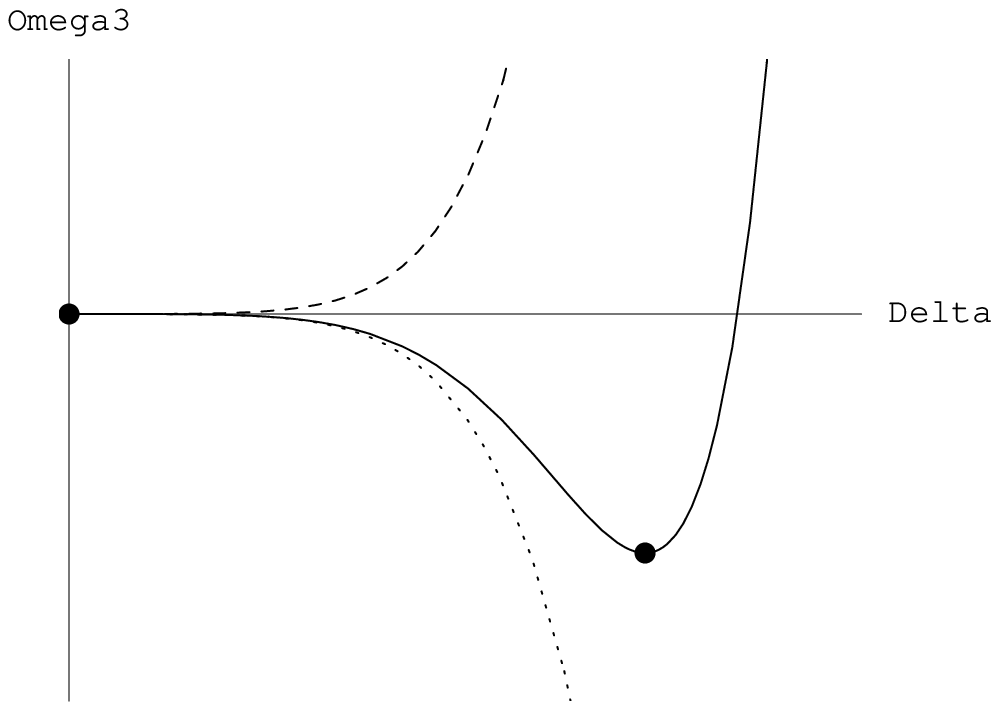}
\includegraphics[width=2.9in]{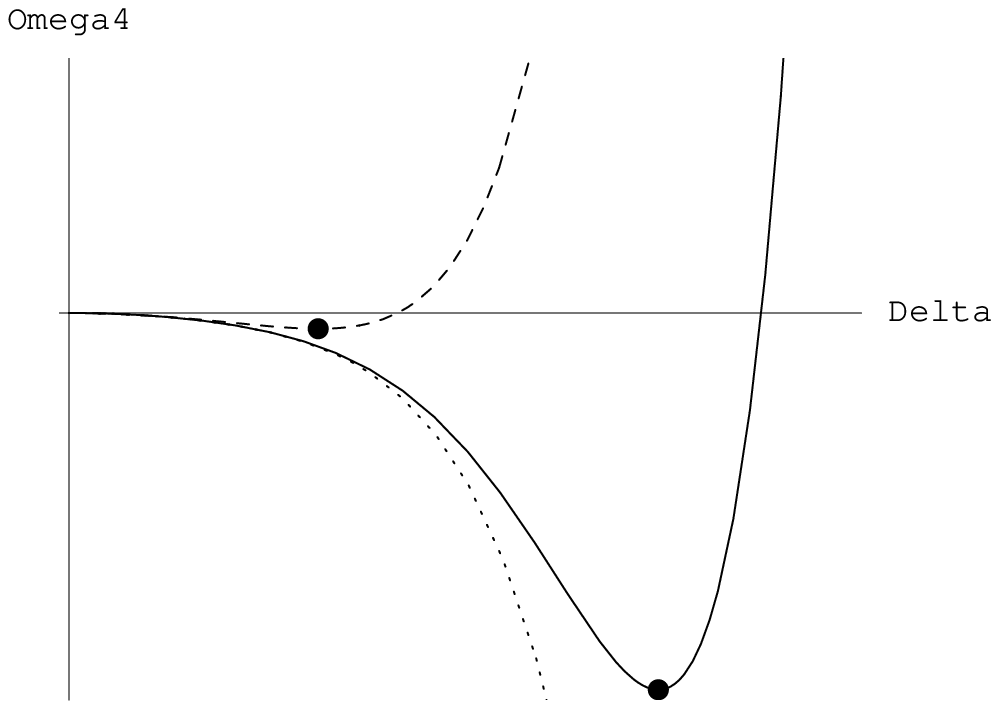}
\end{center}
\caption{
\label{eplots}
Schematic free energies for the FCC crystal (solid line) and plane
wave (dashed line), showing how they support first and second order
phase transitions, respectively.  The dotted line is the unbounded
Ginzburg-Landau expression for the free energy of the FCC crystal.}
\end{figure}

These qualitative arguments are supported by the quantitative results
of our Ginzburg-Landau analysis, which do indeed indicate that the
most favored crystal structure is a cuboid that is very close to a
cube. This crystal structure is so favorable that the coefficients
$\beta$ and $\gamma$ in the Ginzburg-Landau potential (equation
(\ref{GLpotential2})) are large and negative.  In fact, we find
several crystal structures with negative coefficients, but the cube
has by far the most negative $\beta$ and $\gamma$.  In other words,
starting at the origin in the space of crystalline order parameters
$\{ \Delta_{\vq}\}$, the ``steepest descent'' in free energy is
achieved by moving in the direction of the cube structure.  Our
Ginzburg-Landau potential is unbounded from below, so our analysis is
unable to discover the actual free energy minimum or the value of the
gap at which this minimum occurs.  But we can reasonably presume that
the lowest free energy and largest gap are achieved by moving in the
direction of steepest descent.  We could go on, to order $\Delta^8$ or
higher, until we found a Ginzburg-Landau free energy for the cube
which is bounded from below.  However, we know that this free energy
would give a strongly first-order phase transition, meaning that the
Ginzburg-Landau analysis would anyway not be under quantitative
control.  A better strategy, then, is to use the Ginzburg-Landau
analysis to understand the physics at a qualitative level, as we have
done.  We understand qualitatively what features of the
eight-plane-wave solution make it most favorable, so the next step is
to take this as an ansatz, solve the gap equation, and thus obtain a
bounded free energy without making a Ginzburg-Landau approximation.
This calculation is still in progress, but in figure \ref{eplots} we
show what the bounded free energy might look like (solid curve),
compared with the unbounded Ginzburg-Landau free energy (dotted
curve).  The series of plots shows how the unbounded free energy
indicates a first order transition: for $\dm > \dm_2$ the quadratic
coefficient $\alpha$ is positive: increasing $\Delta$ at fixed $\dm$,
the free energy should first turn upwards, then downwards under the
influence of the negative quartic and sextic terms, then eventually it
will turn upwards again because it must be bounded from below.  The
resulting curve can thus generate a first-order transition as $\dm$ is
varied, as shown in the figure.  For comparison the dashed line shows
the free energy of the plane wave crystal (with $\beta,\gamma > 0$),
which demonstrates a typical second order transition.

\begin{figure}
\centering
\psfrag{0}{\small 0}
\psfrag{0.25}{\small 0.25}
\psfrag{0.5}{\small 0.5}
\psfrag{0.75}{\small 0.75}
\psfrag{1}{\small 1}
\includegraphics[width=4in]{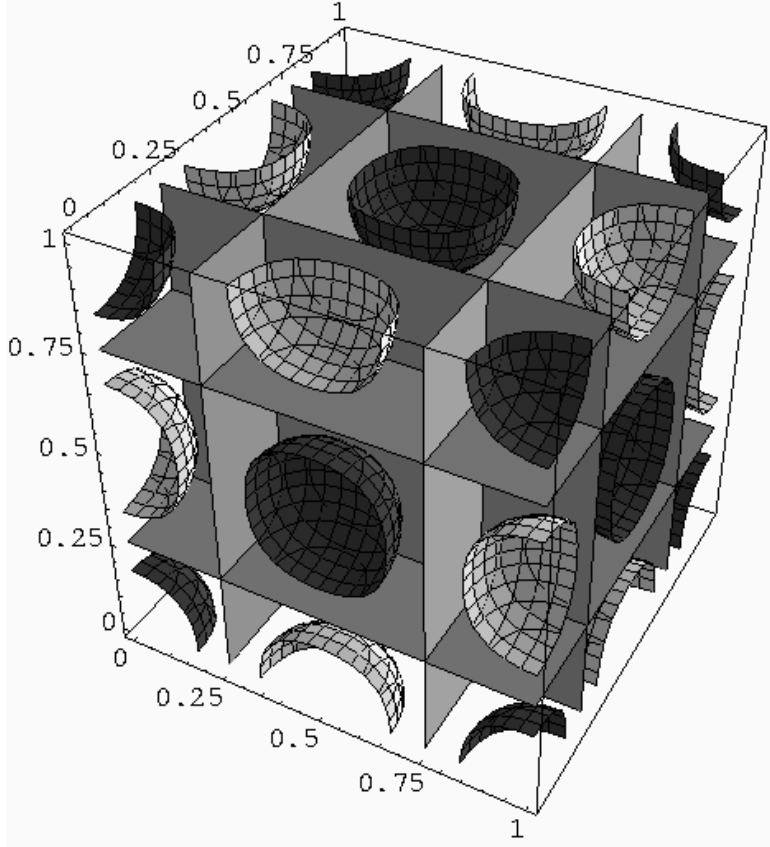}
\caption{
\label{unitcellfig}
A unit cell of the LOFF face-centered-cubic 
crystal.  The gray planes are surfaces where
$\Delta(\vx) = 0$.  The darker surfaces are contours where
$\Delta(\vx) = +4 \Delta$, and the lighter surfaces are contours where
$\Delta(\vx) = -4 \Delta$.  }
\end{figure}

The eight ${\bf q}$'s of our most-favored crystal structure are the
eight shortest vectors in the reciprocal lattice of a
face-centered-cubic crystal.  Therefore, we find that $\Delta(\vx)
\sim \langle \psi(\vx) \psi(\vx) \rangle$ exhibits face-centered-cubic
symmetry.  Explicitly,
\begin{eqnarray}
\Delta(\vx) & = & 2 \Delta \Biggl[ \cos \frac{2 \pi}{a} (x + y + z) +
\cos \frac{2 \pi}{a} (x - y + z) \nonumber \\ & & + \cos \frac{2
\pi}{a} (x + y - z) + \cos \frac{2 \pi}{a} (-x + y + z) \Biggr]\ ,
%\Delta(\vx) & = & 2 \Delta [ \cos  \frac{2 q_0}{\sqrt{3}} (x + y + z) + \cos \frac{2 q_0}{\sqrt{3}} (x - y + z) \\
%& & + \cos \frac{2 q_0}{\sqrt{3}} (x + y - z) + \cos \frac{2 q_0}{\sqrt{3}} (x - y - z) \Biggr]\ ,
\label{fcccrystal}
\end{eqnarray}
where the lattice constant ({\it i.e.}~the edge length of the unit cube) is
\begin{equation}
\label{latticeconstant}
a = \frac{\sqrt{3}\pi}{|\vq|} \simeq \frac{4.536}{\dm}
\simeq \frac{6.012}{\Delta_0}\ ,
\end{equation}
where the last equality is valid at $\dm=\dm_2$ and where $\Delta_0$
is the gap of the BCS phase that would occur at $\dm=0$.  A unit cell
of the crystal is shown in Fig.~\ref{unitcellfig}.  The figure clearly
reveals a face-centered-cubic structure.  Like any crystal, the FCC 
crystalline color superconductor should have phonon modes which are 
Goldstone bosons of spontaneously broken translation symmetry.  
Casalbuoni {\it et al} have formulated an effective theory for the 
LOFF phonons~\cite{LOFFphonon}.

\section{Single-flavor color superconductivity}
\label{sec:singleflavor}

If the Fermi momenta of the $u$, $d$, and $s$ quarks are very far
apart then the system has no choice but to abandon inter-species
pairing and form single-flavor $\langle uu \rangle$, $\langle dd
\rangle$, and $\langle ss \rangle$ condensates.  In the two-flavor
context of section \ref{sec:crystalline}, with a splitting $\dm$
between the $u$ and $d$ Fermi surfaces, single-flavor pairing will
occur for $\dm > \dm_*$.  At $\dm_*$ there is a first-order
``crystallization'' transition between the single-flavor state and the
crystal state with $ud$ pairing.  The value of the first-order point
$\dm_*$ is not known; if it were known we could estimate that an
analogous crystallization transition will occur in three-flavor
neutral quark matter when the Fermi momentum splitting $\delta p_F$
approaches $2 \dm_*$, i.e. when
\begin{equation}
\delta p_F = \delta p_F^* \approx \frac{m_s^2}{4 \mu_*} \sim 2 \dm_*.
\end{equation}
(The factor of two occurs because in our notation $p_F^d - p_F^u =
\delta p_F = 2 \dm$).  We expect that $\dm_*$ is appreciably larger
than $\Delta_0$, which implies that $\mu_*$ is appreciably
smaller than $\mu_{\mbox{\scriptsize unlock}}$.  Crystalline quark
matter occurs in the interval between $\mu_*$ and
$\mu_{\mbox{\scriptsize unlock}}$.  If $\mu_*$ is below the
hadronization point $\mu_{qh}$ then single-flavor quark matter is
unlikely to occur and the crystalline phase will occupy the entire
interval between hadronization and unlocking in the QCD phase diagram
(figure \ref{phasediagram}).  Otherwise, there may be a window just
above the hadronization point in which single-flavor pairing is
possible.

 The structure of a single-flavor $\langle uu \rangle$ condensate
(flavor index 1) is~\cite{ABCC}
\begin{equation}
\label{1flavorcondensate}
\langle \psi_{i \alpha a}(\vx) \psi_{j \beta b}(\vx) \rangle
\propto \Delta \epsilon_{\alpha \beta 3} \delta_{i 1} \delta_{j 1} 
(C \gamma_3)_{ab}
\end{equation}
with indices for color ($\alpha,\beta$), flavor ($i,j$), and spin
($a,b$).  The condensate is antisymmetric in color (as usual, the
color $\mathbf{\bar 3}$ channel is favored because this is the
attractive channel for the QCD interaction).  Only two of the three
colors pair; the choice of the index 3 for the color tensor
$\epsilon_{\alpha \beta 3}$ is arbitrary and excludes the blue quarks
from pairing.  The condensate is obviously symmetric in flavor.  By
the Pauli principle it is also symmetric in Dirac indices\footnote{The
symmetric Dirac matrices that could appear in equation
\ref{1flavorcondensate} are $C \gamma_0$, $C \sigma_{0i}$, $C
\sigma_{0i} \gamma_5$, and $C \gamma_i$.  The first is ruled out because it has
no particle-particle component.  The second and third are disfavored
by our NJL model.}.  It is a $J=1$ (vector) condensate that breaks
rotational symmetry: the Dirac matrix $C \gamma_3$ in equation
(\ref{1flavorcondensate}) indicates that the condensate has
spontaneously chosen the 3 direction in position space.  The
condensate is parity even.  The quarks are paired with opposite
helicity (LR pairing) and opposite momentum; therefore they have
parallel spin and form a ($s=1,m=\pm1$) spin triplet state.

The gap parameter $\Delta$ can be calculated using an NJL model with a
four-fermion interaction vertex~\cite{ABCC}.  These calculations are
shown in chapter 4.  Unfortunately the value of the gap is drastically
sensitive to the details of the effective interaction and to the
chemical potential. It could be as large as 1 MeV, or orders of
magnitude smaller (see the dash-dotted line in
figure~\ref{fig:Mag800}).  For $\mu$ of 400 to 500 MeV, the NJL
calculation predicts a gap that ranges from 0.1 to 10 keV (this
illustrates the sensitivity to the chemical potential).  The
calculation is also very model dependent.  In figure~\ref{fig:Mag800}
the gap is calculated using an NJL model with pointlike magnetic
gluons.  Calibrated to give the same CFL gap, a different NJL model
that includes pointlike electric and magnetic gluons predicts a much
larger gap (by about a factor of 10); with an instanton interaction,
no gap is predicted at all (the channel is flavor symmetric and there
is no interaction with an instanton vertex).

At asymptotically high density a model-independent calculation of the
gap is possible, with a perturbative gluon
interaction~\cite{TS1flav,Schmitt:2002sc,BLR}.  Extrapolating to
reasonable densities ($\mu \sim 400$ MeV), the perturbative
calculation predicts spin-one gaps of order 20 keV - 1 MeV, assuming that
the gap in the CFL phase is of order 10-100 MeV.  The perturbative
calculation also predicts that the condensate will have an additional
$C \sigma_{03}$ component: this channel, which is repulsive in an NJL
model with pointlike gluons, becomes attractive at asymptotic density
when the gluon propagator provides a form factor that strongly
emphasizes small-angle scattering.  In the $C \sigma_{03}$ channel, 
quarks are paired with the same helicity (LL or RR pairing) and 
opposite momentum; their spins are antiparallel and they form a 
($s=1, m=0$) spin triplet state.  

The elementary excitations of the single-flavor color superconductor
are quite different than those of a spin-zero phase like CFL.  The
quasiquark excitations have anisotropic dispersion relations and they
are not fully gapped.  For the $C \gamma_3$ condensate, the energy gap
to create a quasiquark goes to zero at the poles of the Fermi surface;
for the $C \sigma_{03}$ condensate, the energy gap vanishes on the
equator of the Fermi surface. When a nonzero quark mass is considered,
these modes acquire a small gap of order $m \Delta/\mu$~\cite{ABCC}.
The gapless or nearly-gapless excitations are likely to dominate
transport properties of the material (viscosities, conductivities,
etc.).  The system should also have massless spin-waves which are
Goldstone bosons of the spontaneously-broken rotational symmetry.  It
is interesting to note that, unlike the CFL phase, the single-flavor
color superconductor does not have a massless ``rotated photon'' that
can admit magnetic flux (there is no leftover $U(1)_{\tilde Q}$ gauge
symmetry).  Therefore the phase exhibits an electromagnetic Meissner
effect~\cite{Spin1Meissner}.

The condensate of equation (\ref{1flavorcondensate}) spontaneously
chooses a spatial direction and a color direction.  A lower free
energy may obtained by making the replacement $\epsilon_{\alpha\beta3}
C\gamma_3 \rightarrow \epsilon_{\alpha\beta A} C\gamma_A$ with a
summation over the common color-spin index $A$.  This is a
``color-spin locking'' (CSL) phase in which color structure is
correlated with spatial direction~\cite{TS1flav,Schmitt:2002sc}.  The
symmetry breaking pattern in the CSL phase is 
\begin{equation}
SU(3)_{\mbox{\scriptsize color}} \times SO(3)_J \ \rightarrow \ SO(3)_{\mbox{\scriptsize color} + J}
\end{equation}
i.e.~the color and rotation groups are broken to a diagonal subgroup
of simultaneous global rotations in color and space.  In the CSL phase
the quasiquark dispersion relations are isotropic and fully gapped.
This phase is analogous to the B phase of helium-3~\cite{helium3}
(whereas the condensate of equation (\ref{1flavorcondensate}) is
analogous to the A phase).  As in helium-3, there are yet other
possible phases for spin-one condensates.  Some of these other phases
have been explored in refs.~\cite{TS1flav,Schmitt:2002sc}.

In concert with the NJL model calculations for the single-flavor color
superconductor, a large catalog of color-flavor-spin channels for
diquark condensation has been surveyed~\cite{ABCC}.  This survey is
shown in chapter 4.  For the survey we use an NJL model that includes
four-fermion interactions with the quantum numbers of electric gluon
exchange, magnetic gluon exchange, and the two-flavor instanton, with
Fermi couplings $G_E$, $G_M$, and $G_I$, respectively (see equation
(\ref{kernel})).  We investigate diquark condensates that factorize
into separate color, flavor, and Dirac tensors, i.e.
\begin{equation}
\label{genericcondensate}
\langle  \psi_{i\alpha a} \psi_{j \beta b} \rangle \propto \Delta \mathfrak{C}_{\alpha\beta} \mathfrak{F}_{ij} \Gamma_{ab}.
\end{equation}
Chapter 4 shows an exhaustive survey of 24 different condensates which
have this generic form.  Many of these condensates are spin-one, with
interesting quasiquark dispersions like those discussed above.  For
all of the attractive channels, the NJL mean field theory gap
equations are solved to estimate values of the gaps.  Unfortunately,
most of these NJL gap calculations suffer from the same drastic model
dependence that inflicts the single-flavor color superconductor
calculation as described above.  The results for the five attractive
channels are shown in table \ref{channels}.  With the notable
exception of the first channel, the gap estimates in the right column
should be interpreted very cautiously.  Not only are the values of the
gaps quite model-dependent, they are also extremely sensitive to the
chemical potential: as we will see in chapter 4, they can vary by more
than two orders of magnitude when the chemical potential is changed
from 400 MeV to 500 MeV.  The numerical estimates in the table should
be interpreted as optimistic upper bounds for gaps which could be
orders of magnitude smaller.

\begin{table}
\begin{center}
\begin{tabular}{cccccccc}
channel & $N_c$ & $N_f$ & $\mathfrak{C}$ & $\mathfrak{F}$ & $\Gamma$ & $J$ & $\Delta$ \\ \hline
1 & 2 & 2 & $\epsilon_{\alpha\beta 3}$ & $\epsilon_{ij3}$ & $C\gamma_5$ & 0 & 10-100 MeV \\
2 & 2 & 2 & $\epsilon_{\alpha\beta 3}$ & $\epsilon_{ij3}$ & $C\gamma_3 \gamma_5$ & 1 & $\lesssim$ 1 MeV \\
3 & 1 & 2 & $\delta_{\alpha 1} \delta_{\beta 1}$ & $\epsilon_{ij3}$ & $C\sigma_{03}$ & 1 & $\lesssim$ 1 MeV \\
4 & 2 & 1 & $\epsilon_{\alpha\beta 3}$ & $\delta_{i1} \delta_{j1}$ & $C\gamma_3$ & 1 & $\lesssim$ 1 MeV \\
5 & 1 & 1 & $\delta_{\alpha 1}\delta_{\beta 1}$ & $\delta_{i1} \delta_{j1}$ & $C\gamma_0 \gamma_5$ & 0 & $\lesssim$ 0.01 MeV \\
\end{tabular}
\end{center}
\caption{ 
\label{channels}
Summary of attractive channels from the NJL survey of chapter 4.  The
pairing pattern is shown in equation (\ref{genericcondensate}); $N_c$
and $N_f$ are the numbers of colors and flavors that participate in
the pairing.  }
\end{table}

The first channel (2 colors, 2 flavors) in table \ref{channels} is the
familiar 2SC phase of equation (\ref{2SCcondensate}).  This phase, and
its 3-flavor, 3-color cousin (the CFL phase), have the largest gaps of
any of the color superconducting phases.  They are also the only
phases for which the NJL gap calculations are robust.  As we have
discussed previously, other color superconducting phases are only
likely to prevail when the CFL and 2SC phases are disrupted by a
flavor asymmetry, as occurs in intermediate-density neutral quark
matter.  Channels 2 and 3 (2 flavors, 1 or 2 colors) are unlikely to
be of interest: they require inter-species pairing, but have gaps that
are smaller than 1 MeV, so the same stress that disrupts the 2SC and
CFL phases will even more readily disrupt these phases (channel 3 has
been proposed to accompany the 2SC phase and allow pairing between
blue up and down quarks~\cite{ARW2,BHO}, but we have seen that the 2SC
phase is unlikely to occur in neutral quark matter).  The fourth
channel (2 colors, 1 flavor) and its 3-color cousin (the
color-spin-locking phase) are the single-flavor color superconducting
phases discussed earlier in this section.  Channel 5 (1 color, 1
flavor) vanishes for light quarks, but it may allow pairing for an
``orphaned'' color of strange quark (i.e.~a strange quark that is
neglected by all other pairing processes).

\section{Applications}
\label{sec:applications}

\subsection{Compact stars}
\label{subsec:compactstars}
 
Our current understanding of the color superconducting state of quark
matter leads us to believe that it may occur naturally within compact
stars.  The critical temperature below which quark matter is a color
superconductor is estimated to be about 10 to 50 MeV.  In compact
stars that are more than a few seconds old, the star temperature is
less than this critical temperature and any quark matter that is
present will be in a color superconducting state.  It is therefore
important to explore the astrophysical consequences of color
superconductivity~\cite{supercondstars}.

Much of the work on the consequences of quark matter within a compact
star has focussed on the effects of quark matter on the equation of
state, and hence on the mass-radius relationship~\cite{HHJ}.  The largest
contributions to the pressure of quark matter are a positive $\mu^4$
contribution from the Fermi sea, and a negative bag constant $B$.  As
a Fermi surface effect, the effect of pairing is a contribution of
order $\mu^2 \Delta^2$.  This is small compared to the two leading
terms, so the conventional wisdom has been that superconductivity has
a minor effect on the equation of state.  Recently, however, it has
been observed that if the bag constant is large enough so that nuclear
matter and quark matter have comparable pressures at some density that
occurs in compact stars, then there may be a large cancellation
between the two leading terms and the Fermi surface term can have a
large effect~\cite{quarkeos}.  Therefore mass-radius curves can be
sensitive to the presence of color superconductivity.

A gravitational wave detector could yield insight into compact star
interiors from observations of binary inspirals/mergers.  In a hybrid
star with a sharp interface between a nuclear mantle and a CFL color
superconducting core, there is a large density discontinuity at the
interface~\cite{ARRW}.  The two sharp density edges (at the core
radius and at the star radius) could create features at two distinct
time scales in the gravitational wave profile emitted during the
inspiral and merger of a pair of compact stars of this type.  The
first feature would occur when the less dense nuclear mantles of the stars begin
to deform each other; the second feature would occur only somewhat later
when the denser cores begin to deform.

The phase transition at which color superconductivity sets in as a hot
proto-neutron star cools may yield a detectable signature in the
neutrinos received from a supernova~\cite{CarterReddy}.  At the onset
of color superconductivity the quark quasiparticles acquire gaps and
the density of these quasiparticles is then suppressed by a Boltzmann
factor $\exp(-\Delta/T)$.  As a result the mean free path for neutrino
transport suddenly becomes very long.  All of the neutrinos previously
trapped in the core of the star are able to escape in a sudden
burst that may be detectable as a bump in the time distribution of
neutrinos arriving at an earth detector.

Color superconductivity has a large effect on cooling and transport
processes in quark matter~\cite{Prakash,BlaschkeCool}.  In quark matter, the
neutrino emissivity is dominated by quasiquark modes that have
energies smaller than the temperature $T$.  These modes can rapidly
radiate neutrinos by direct URCA reactions ($d \rightarrow u + e +
\bar\nu$, $u \rightarrow d + e^+ + \nu$, etc.) which then dominate the
cooling history of the star as a whole.  In the CFL phase, all of the
quarks have a gap $\Delta \gg T$; the neutrino emissivity is
suppressed by a Boltzmann factor $\exp(-\Delta/T)$ and the CFL state
does not contribute to cooling.  In a compact star with a CFL core and
a nuclear mantle, the cooling will occur only by neutrino emission from 
the mantle.

This conclusion is revised for non-CFL phases of quark matter.  Both
the crystalline color superconductor and the breached-pair color
superconductor have gapless quasiquarks for momenta at the edges of
``blocking regions'', as discussed in section \ref{sec:unlocking}.
These gapless modes could accommodate direct URCA reactions and
conceivably dominate the entire cooling of the
star~\cite{Prakash,BlaschkeCool}.  A similar effect could occur in the
single-flavor spin-one color superconductor, which can have gapless
quasiquarks at the poles or at the equator of the Fermi surface (in
its non-color-spin-locked versions)~\cite{ABCC}.  Just how these
special gapless modes could affect emissivity rates is unknown and is
worthy of investigation.  The crystalline and single-flavor phases
also have collective modes that will contribute to the heat capacity
(phonons~\cite{LOFFphonon} and spin waves, respectively).

%In a strange star made entirely of CFL matter, the physics of the
%instability to r-mode oscillations is dramatically affected by color
%superconductivity~\cite{Madsen}, although this is not the case for
%neutron stars with quark matter present only in their
%cores~\cite{BildstenUshomirsky,Madsen}.

Recent work suggests that the observation of long-period (of order one
year) precession in isolated pulsars might constrain the possible
behavior of magnetic fields in the core of a compact star~\cite{link}.
Rotating compact stars with superfluid interiors will be threaded with
a regular array of rotational vortices that are aligned along the axis
of rotation.  At the same time, if the core is a type II
superconductor then it will also be threaded with an array of magnetic
flux tubes that are aligned along the magnetic axis of the star.  If
the vortex and flux tube arrays coexist, they prevent any rotational
precession because a precession would entangle the interwoven arrays.

Remarkably, the observed precession therefore might rule out the
standard model of a nuclear core containing coexisting neutron and
proton superfluids, with the proton component forming a type II
superconductor.  But color superconducting interiors {\it can} accommodate
the observed precession: magnetic flux tubes do not occur in either
the CFL phase (which is not an electromagnetic
superconductor~\cite{ABRmag}) or the single-flavor phase (which is a
type I superconductor~\cite{Spin1Meissner}).  

Finally, in this thesis we wish to investigate the possibility that
crystalline quark matter could be a locus for glitch phenomena in
pulsars~\cite{BowersLOFF}.  As the rotation of a pulsar gradually
slows, the array of rotational vortices that fills the interior of the
star should gradually spread apart; thus the star sheds its vortices
and loses angular momentum.  But if a crystalline phase occurs within
the star, the rotational vortices may be pinned in place by features
of the crystal structure.  This impedes the smooth outward motion of
the vortices.  The vortex array could remain rigid until an
accumulated stress exceeds the pinning force.  Then, a macroscopic
movement of vortices will occur, leading to an observed glitch in the
rotational frequency of the pulsar.  In chapter 5 we address the
feasibility of this proposed glitch mechanism.  With the crystal
structure known, a calculation of the vortex pinning force can
proceed.  The first step is the explicit construction of a vortex
state in the crystalline phase, and we discuss efforts in this
direction.  The task is a challenge by virtue of the interesting fact
that the LOFF state is simultaneously a superfluid and a crystal.

%, such a shell may be a locus for glitch
%phenomena~\cite{BowersLOFF}.
%As a function of increasing depth in a compact star,
%$\mu$ increases, $m_s$ decreases, and $\Delta_0$ changes also.
%This means that in some shell within
%the quark matter core of a neutron star (or within a strange
%quark star),  $m_s^2/\mu\Delta_0$ may lie within the
%appropriate window where
%crystalline color superconductivity is favored.
%Because this phase is a (crystalline) superfluid,
%it will be threaded with vortices in
%a rotating compact star.  

\subsection{Atomic physics}
\label{subsec:atomicphysics}

In section \ref{sec:crystalline} we investigated crystalline color
superconductivity with a two-flavor NJL model, i.e.~a toy model with
two species of fermion and a pointlike four-fermi interaction.  This
toy model may turn out to be a better model for the analysis of LOFF
pairing in atomic systems. (There, the phenomenon could be called
``crystalline superfluidity''.)  Recently, ultracold gases of
fermionic atoms such as lithium-6 have been cooled down to the
degenerate regime, with temperatures less than the Fermi
temperature~\cite{ColdFermions,Duke,AtomReview}.  In these atomic systems, a
magnetically-tunable Feshbach resonance can provide an attractive
$s$-wave interaction between two different atomic hyperfine
states~\cite{Feshbach}.  This interaction is short-range but the
scattering length can be quite long, so the system may be
strongly-interacting.  The attractive interaction renders the system
unstable to BCS superfluidity below some critical temperature, and it
seems possible to reach this temperature (perhaps by increasing the
atom-atom interaction, thereby increasing $T_c$, rather than by
further reducing the temperature)~\cite{AtomicBCS,AtomReview}.  In these systems
there really are only two species of fermion (two different atomic
hyperfine states) that pair with each other, whereas in QCD our model
is a toy model for a system with nine quarks.  The atomic interaction
will be short-range and $s$-wave dominated, whereas in QCD it remains
to be seen if this is a good approximation at accessible densities. In
the atomic systems, experimentalists can control the densities of the
two different atoms that pair, and in particular can tune their
density difference. This means that experimentalists wishing to search
for crystalline superfluidity have the ability to dial the most
relevant control parameter~\cite{AtomicLOFF,CombescotMora}.  In QCD,
in contrast, $\delta\mu$ is controlled by $m_s^2/\mu$, meaning that it
is up to nature whether, and if so at what depth in a compact star,
crystalline color superconductivity occurs.
 
Indirect observations of crystalline color superconductivity in the
interior of a distant compact star are formidably difficult.  But the
atomic system provides a terrestrial setting in which the
predictions of this thesis can be directly tested.  In chapter 5 we
will further discuss the prospect of creating and observing the
crystalline state in an atomic gas.  Because the spatial variation of
the gap parameter can create a density modulation in the gas, it may
be possible to literally see the crystal structure.

%% file: chap2.tex
\chapter{Crystalline Superconductivity:  Single Plane Wave}

\section{Overview}

In this chapter we study the simplest example of a crystalline color
superconductor: a condensate that varies like a single plane wave in
position space~\cite{BowersLOFF}.  Equivalently, each Cooper pair in the condensate has
the same total momentum $2\vq$.  We will use a variational method
similar to that originally employed by Fulde and Ferrell~\cite{FF} and
described in more detail by Takada and Izuyama~\cite{Takada1}.  In
section \ref{sec:planewaveLOFF} we will describe the variational
ansatz for the plane wave LOFF state.  We note in particular that,
unlike in the original LOFF context, there is pairing both in $J=0$
and $J=1$ channels.  In section \ref{sec:planewavegap}, we derive the
gap equation for the LOFF state for a model Hamiltonian in which the
full QCD interaction is replaced by a four-fermion interaction with
the quantum numbers of single gluon exchange.  In section
\ref{sec:planewaveresults}, we use the gap equation to evaluate the
range of $\dm$ within which the LOFF state arises. We will see that at
low $\dm$ the translationally invariant BCS state, with gap $\De_0$,
is favored. At $\dm_1$ there is a first order transition to the LOFF
paired state, which breaks translational symmetry. At $\dm_2$ all
pairing disappears, and translational symmetry is restored at a phase
transition which is second order in mean field theory.  In the
weak-coupling limit, in which $\Delta_0\ll \mu$, we find values of
$\dm_1$ and $\dm_2$ which are in quantitative agreement with those
obtained by LOFF. This agreement occurs only because we have chosen an
interaction which is neither attractive nor repulsive in the $J=1$
channel, making the $J=1$ component of our LOFF condensate irrelevant
in the gap equation.  In section \ref{sec:planewaveham}, we consider a
more general Hamiltonian in which the couplings corresponding to
electric and magnetic gluon exchange can be separately tuned.  This
leads to interactions in both $J=0$ and $J=1$ channels, and we show
how it affects the range of $\dm$ within which the LOFF state arises.
In section \ref{sec:planewaveconclusions}, we summarize the results
for the plane wave crystal.

\section{The LOFF plane wave ansatz}
\label{sec:planewaveLOFF}

\begin{figure}%[thb]
\begin{center}
\includegraphics[width=3.0in]{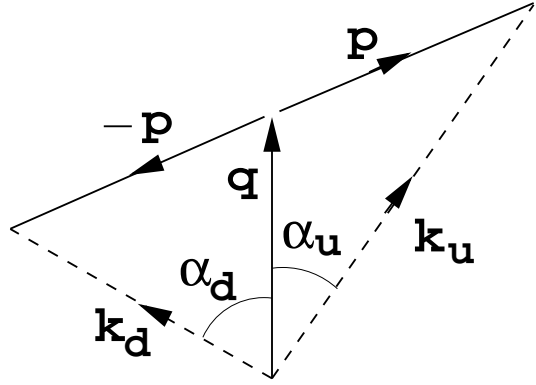}
\end{center}
\caption{
The momenta $\vk_u$ and $\vk_d$
of the two members of a LOFF-state Cooper pair.  We choose
the vector $\vq$, common to all Cooper pairs,
 to coincide with the $z$-axis.  The angles $\alpha_u(\vp)$ and 
$\alpha_d(\vp)$ indicate the polar angles of $\vk_u$ and $\vk_d$,
respectively.
}
\label{fig:angles}
\end{figure}

We begin our analysis of a LOFF state for quark matter by constructing
a variational ansatz for the LOFF wavefunction.  We consider Cooper
pairs which consist of an up quark and a down quark with respective
momenta 
\beql{LOFF:mom} 
\vk_u = \vq+\vp,\hspace{0.3in} \vk_d =
\vq-\vp, 
\eeql 
so that $\vp$ identifies a particular quark pair, and every quark pair
in the condensate has the same nonzero total momentum $2\vq$.  This
arrangement is shown in Figure~\ref{fig:angles}.  The helicity and
color structure are obtained by analogy with the ``2SC'' state as
described in previous work~\cite{ARW2,RappEtc}: the quark pairs will
be color $\mathbf{\bar 3}$ antitriplets, and in our ansatz we consider
only pairing between quarks of the same helicity.

With this in mind, here is a suitable trial wavefunction for the
LOFF state with wavevector $\vq$~\cite{LO,FF,Takada1}:
\beql{LOFF:ansatz}
\ba{rcl}
|\Psi_{\vq}\> &=& B^\ad_L B^\ad_R |0\>,\\[1ex]
B^\ad_L &=& 
  \dsp\prod_{\vp \in {\cal P}, \alpha, \beta} 
    \left( \cos \theta_L(\vp) 
    + \epsilon^{\alpha\beta3} \e^{i \xi_L(\vp)} 
    \sin \theta_L(\vp)\, a^\ad_{Lu\alpha}(\vq\!+\!\vp) 
    \, a^\ad_{Ld\beta}(\vq\!-\!\vp) \right) 
   \\[4ex]
 & \times & 
  \dsp\prod_{\vp \in {\cal B}_u, \alpha} a^\ad_{Lu\alpha}(\vq\!+\!\vp) 
  \times 
  \dsp\prod_{\vp \in {\cal B}_d, \beta} a^\ad_{Ld\beta}(\vq\!-\!\vp) 
  , \\[3ex]
B^\ad_R &=& \mbox{as above,~} L \to R,
\ea
\eeql
where $\alpha$, $\beta$ are color indices, $u$, $d$ and $L$, $R$ are
the usual flavor and helicity labels, and $a^\ad$ 
is the particle creation operator 
(for example, $a^\ad_{Ld\al}$ creates a left-handed down quark with color
$\al$).  The $\theta$'s and $\xi$'s are
the variational parameters of our ansatz: they are 
to be chosen to minimize
the free energy of the LOFF state, as described in the next section.  The
first product in equation \eqn{LOFF:ansatz} creates quark pairs within a
restricted region ${\cal P}$ of the total phase space. This allowed
``pairing region'' will be discussed below. 
The next product
fills a ``blocking region'' ${\cal B}_u$ with unpaired up quarks:
these are up quarks with momenta $\vq+\vp$ for which there are no
corresponding down quarks with momenta $\vq-\vp$.  The final
product fills the blocking region ${\cal B}_d$ with unpaired down
quarks.
The ansatz does not contain a term that would create antiparticle
pairs: we have checked the effect of such a term and found that
it has no qualitative effect on our results.
\begin{figure}[t]
\begin{center}
\includegraphics[width=5.5in]{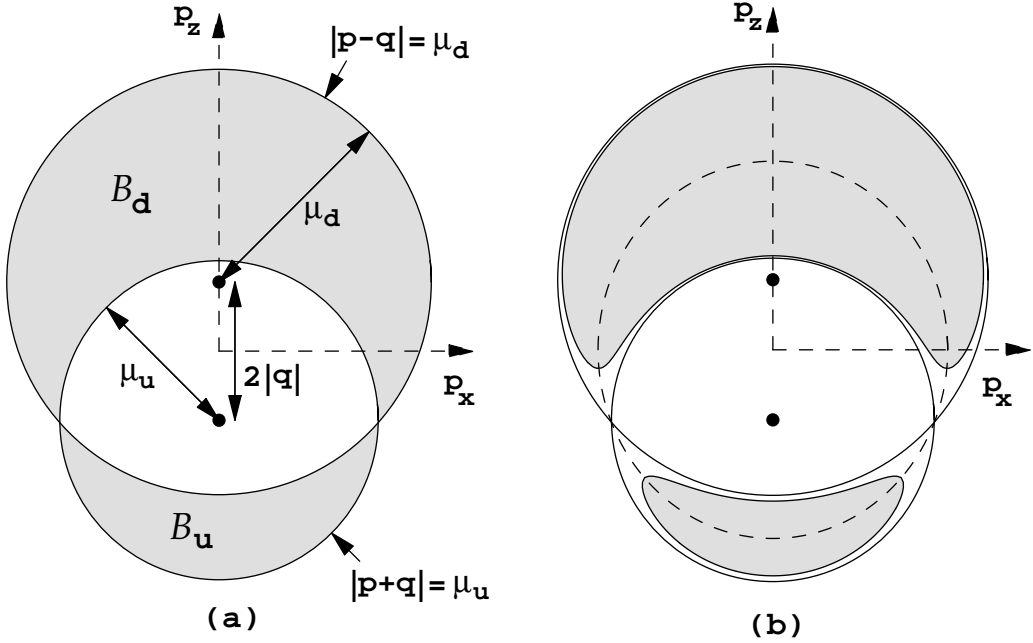}
\end{center}
\caption{
The LOFF phase space, as a function of $\vp$
(\Eqn{LOFF:mom}). We show the $p_y=0$ plane.
(a) The phase space in
the limit of arbitrarily weak interactions. In the shaded
blocking regions ${\cal B}_u$
and ${\cal B}_d$, no pairing is possible. In the inner unshaded region,
an interaction can induce hole-hole pairs. In the outer unshaded
region, an interaction can induce particle-particle pairs.
The region ${\cal P}$ (Eq.~\eqn{LOFF:ansatz})
is the whole unshaded area.
(b)~When the effects of interactions and the formation of the LOFF state
are taken into account, the blocking regions shrink.
The BCS singularity occurs on the dashed ellipse,
defined by $\eps_u+\eps_d=\mu_u+\mu_d$, where making a Cooper pair
costs no free energy in the free case.
}
\label{fig:regions}
\end{figure}

To complete the specification of 
our ansatz we need to describe the allowed pairing and
blocking regions in phase space.  These regions are largely determined
by Pauli blocking as a result of populated Fermi seas.  In the absence
of pairing interactions, the system is in the ``normal'' state and up
and down quarks are distributed in Fermi seas with Fermi momenta
$p_F^u = \mu_u$ and $p_F^d = \mu_d$, respectively (recall that we
consider massless quarks only, so the single particle energy of a
quark with momentum $\vk$ is $\epsilon(\vk) = |\vk|$).  An up quark
carries momentum $\vk_u=\vp + \vq$; in $\vp$-space, therefore, the Fermi sea
of up quarks corresponds to a sphere of radius $\mu_u = \mubar-\dm$
centered at $-\vq$.  Similarly, a down quark carries momentum 
$\vk_d=-\vp +\vq$, giving a 
sphere in $\vp$-space of radius $\mu_d = \mubar+\dm$
centered at $+\vq$.  The two offset spheres are shown in
Figure~\ref{fig:regions}a (we have drawn the 
case $|\vq| > \dm$ so that
the two Fermi surfaces intersect in $\vp$-space).  In the limit of
arbitrarily weak interactions,
the blocking region ${\cal B}_u$ corresponds to the lower shaded area
in the figure: pairing does not occur here since the region is inside
the Fermi sea of up quarks, but outside the Fermi sea of down quarks.
Similarly the upper shaded area is the blocking region ${\cal B}_d$.
The entire unshaded area is the pairing region ${\cal P}$: it includes
the region inside both spheres, where hole-hole pairing can occur, and
the region outside both spheres, where particle-particle pairing can
occur.  

We can now explain how the LOFF wavefunction ansatz can describe
the normal state with no condensate: we choose $\theta_L(\vp) =
\theta_R(\vp) = \pi/2$ for $\vp$ inside both Fermi spheres, and
otherwise all the $\theta$'s are zero.  
With this
choice the first term in \Eqn{LOFF:ansatz} fills
that part of 
each Fermi sea corresponding to the inner unshaded region of 
Figure~\ref{fig:regions}a.  The ${\cal B}_u$ and ${\cal B}_d$ terms fill
out the remainder of each Fermi sea to obtain the normal state.
Note that in the absence of pairing, the normal state can be
described with any choice of $\vq$.  The most convenient choice
is $\vq=0$, in which case $\vk_u=\vk_d=\vp$, 
${\cal B}_u$ vanishes, and ${\cal B}_d$ is a spherical shell.
Other choices of $\vq$ correspond to choosing different
origins of  $\vk_u$-space and $\vk_d$-space, but in the 
absence of any interactions this has no consequence.
Once we turn on interactions and allow pairing, we expect
a particular $|\vq|$ to be favored.

The phase space picture changes slightly when pairing interactions are
included: the blocking regions are smaller when a LOFF condensate is
present, as indicated in Figure~\ref{fig:regions}b.  We will account
for this effect in the next section.  With smaller blocking regions, a
larger portion of the phase space becomes available for LOFF pairing.
Such pairing is guaranteed to be
energetically favorable when it costs zero free 
energy to create an up quark and a down quark, since these quarks 
can then pair to obtain a negative interaction energy.  The zero free
energy condition is 
\beql{LOFF:sing}
\epsilon(\vk_u)+\epsilon(\vk_d) = \mu_u + \mu_d = 2 \bar\mu
\eeql
where $\epsilon(\vk)$ is the single particle energy of a quark with
momentum $\vk$.  For massless quarks, we obtain
$|\vq+\vp|+|\vq-\vp| = 2 \bar\mu$, which describes an ellipsoidal
surface in $\vp$-space.  This surface is indicated by the ellipse
shown in Figure~\ref{fig:regions}b; notice that the ellipsoid and
the two Fermi surfaces all intersect at a circle.

If the interaction is weak, we expect LOFF pairing to be favored in a
thin layer of phase space around this ellipsoid.  This is manifest in
the gap equation derived in the next section (Eq.~(\ref{gap:LOFF})) in
which, as in BCS theory, we find a divergent integrand on this
ellipsoid in the absence of pairing.  Pairing smooths the divergence.
As the interaction gets stronger, the layer of favored pairing gets
thicker.  If there were no blocking regions, we could use the entire
ellipsoid, just as BCS pairs condense over the entire spherical
surface $|\vp| = \mu$ in the symmetric, $\dm=|\vq|=0$ case.  However,
as shown in Figure~\ref{fig:regions}b, the blocking regions exclude
pairing over most of the ellipsoid, leaving a ribbon of unsuppressed
LOFF pairing in the vicinity of the circle where the Fermi surfaces
intersect.  This agrees with our expectation for the particle
distribution in the LOFF state: it is as in the normal state, except
that there is a restricted region (around the aforementioned ribbon)
where each quark in a pair can be near its Fermi surface and where
pairing therefore occurs.

Although the constant single-particle energy 
contours for noninteracting up and down
quarks cross in $\vp$-space (see Figure~\ref{fig:regions}a), we
emphasize that the Fermi surfaces of up and down quarks 
do not cross in momentum ($\vk_u$- and $\vk_d$-)
space.  The $\vp$-space ribbon of unsuppressed pairing
corresponds to unsuppressed pairing between up and down quarks 
with momenta around $\vk$-space ribbons near their 
respective (disjoint) Fermi surfaces.

In the limit of arbitrarily weak interactions, the ribbon in
momentum space along which pairing is unsuppressed shrinks, as
the blocking regions grow to exclude all of the ellipsoid
except the one-dimensional circle at which the two spheres in 
Figure~\ref{fig:regions} intersect.  This circle has
insufficient phase space to lead to a singularity
in the gap equation: the integrand is singular on this circle,
but the integral does not diverge.  Therefore, the LOFF
state is not guaranteed to occur if one takes the weak coupling
limit at fixed $\dm$. In this respect, the LOFF state
is like the BCS state at nonzero $\dm$: for weak coupling,
$\Delta_0\rightarrow 0$ and because the BCS state can only
exist if it has $\Delta_0>\sqrt{2}\dm$, it must vanish
for couplings weaker than some threshold.  
We shall see, however, that at any fixed weak coupling, the LOFF
state, like the BCS state, is guaranteed to occur at some $\dm$: the
BCS state arises if $\dm<\dm_1$ and the LOFF state arises if
$\dm_1<\dm<\dm_2$.

One of the most striking features of the LOFF state is the spin
structure of the condensate.  The familiar ``2SC'' state pairs quarks
of the same helicity and opposite momentum, so the spins are
antiparallel and the quarks are arranged in an antisymmetric
combination to form spin singlet Cooper pairs.  The LOFF state also
pairs quarks of the same helicity, but now the quark momenta are no
longer antiparallel, as can be seen from Figure~\ref{fig:angles}.
Therefore the LOFF Cooper pairs cannot be spin singlets: they are 
superpositions of both spin zero and spin one.  This is revealed
explicitly by evaluating the nonzero $\< \psi\psi \>$ expectation
values in the LOFF state:
\beql{LOFF:cond}
\ba{rcl}
-\< \Psi_{\vq} | \epsilon_{ij} \epsilon_{\alpha \beta 3} 
  \,\psi^{i\alpha}(\vr) C L \,\psi^{j\beta}(\vr) 
  | \Psi_{\vq} \> &=& 2\Gamma_A^L \e^{i 2\vq\cdot\vr} \\[1ex] 
i\< \Psi_{\vq} | (\sigma_1)_{ij} \epsilon_{\alpha \beta 3} 
  \,\psi^{i\alpha}(\vr) C L \sigma^{03} \,\psi^{j\beta}(\vr) 
  | \Psi_{\vq} \> &=& 2\Gamma_B^L \e^{i 2\vq\cdot\vr}  
\ea
\eeql
where $i$, $j$ are flavor indices (1 = up, 2 = down), $\alpha$,
$\beta$ are color indices, $C = i \gamma^0 \gamma^2$, $L =
(1-\gamma_5)/2$ is the usual left-handed projection operator, and
$\sigma_{\mu\nu} = (i/2)[\gamma_\mu,\gamma_\nu]$.  The constants
$\Gamma_A^L$ and $\Gamma_B^L$ are left-handed $J=0$ and $J=1$ 
condensates, respectively.  $\Gamma_A^R$ and $\Gamma_B^R$ 
are defined analogously. The
$\Gamma$'s can be expressed in terms of the variational parameters 
of the LOFF wavefunction:
\beql{LOFF:gammas}
\ba{rcl}
\Gamma_A^L & = & \dsp\frac{4}{V} \dsp\sum_{\vp \in {\cal P}} 
  \sin \theta_L(\vp) \cos \theta_L(\vp) \e^{i\xi_L(\vp)} \,
  \sin \Bigl(\frac{\al_u(\vp)+\al_d(\vp)}{2}\Bigr)\e^{-i\phi(\vp)} \\[1ex]
\Gamma_B^L & = &  \dsp\frac{4}{V} \dsp\sum_{\vp \in {\cal P}} 
  \sin \theta_L(\vp) \cos \theta_L(\vp) \e^{i\xi_L(\vp)} \,
  \sin \Bigl(\frac{\al_u(\vp)-\al_d(\vp)}{2}\Bigr)\e^{-i\phi(\vp)}
\ea
\eeql
Here $V$ is the spatial
volume of the system, $\alpha(\vp)$ are the
polar angles of the quark momenta, as in Figure~\ref{fig:angles},
and the dependence on the azimuthal angle $\phi$ follows from
our use of the spinor conventions described in 
Refs.~~\cite{BailinLove,ARW2,ARW3}.  
The expressions for $\Gamma_A^R$ and $\Gamma_B^R$  are the same
as those in (\ref{LOFF:gammas}) except
that $\phi(\vp)$ is replaced by $\pi-\phi(\vp)$.  
In Eq.~(\ref{LOFF:gammas})
and throughout, $(1/V)\sum_\vp$ becomes $\int d^3p/(2\pi)^3$
in an infinite system.

Once we have
derived a gap equation by minimizing the free energy
with respect to these variational parameters, we expect the condensates to 
be simply related to gap parameters occurring in the gap equation.
We will see explicitly how $\Gamma_A$ and $\Gamma_B$ are determined
in the next section.  

Notice that the condensates of \Eqn{LOFF:cond} are plane
waves in position space by virtue of the nonzero momentum $2\vq$ of a
Cooper pair.  $\Gamma_A$ describes pairing which is antisymmetric in
color, spin, and flavor, while $\Gamma_B$ describes pairing which is
antisymmetric in color but symmetric in spin and flavor (in each case,
Pauli statistics are obeyed).  
In the original LOFF condensate of electrons
there can be no $\Ga_B$, since electrons have no color or flavor,
so that only the spin antisymmetric pairing is possible.

The $J=0$ condensates $\< \psi C L \psi\>$, 
$\< \psi C R \psi \>$ are Lorentz scalars (mixed under parity), while
the $J=1$ condensates $\< \psi C L \sigma^{03} \psi \>$, 
$\< \psi C R \sigma^{03} \psi \>$ are 
{\bf 3}-vectors (also mixed under parity) which point in
the $z$-direction, parallel to
the spontaneously chosen direction $\hat\vq$ of the
LOFF state.
Because the ansatz contains a $J=1$
component, it would be interesting to generalize it
to include the possibility of $LR$ pairing, in addition to
$LL$ and $RR$ pairing. We discuss this further in Section~\ref{sec:planewaveham}.

\section{The gap equation and free energy}
\label{sec:planewavegap}

Having presented a trial wavefunction for the LOFF state, we now
proceed 
to minimize the expectation value of the free energy
$\< F \>$ with respect to the variational parameters
of the wavefunction (the $\theta$'s and $\xi$'s of equation
(\ref{LOFF:ansatz})) to obtain a LOFF gap equation.  The free energy is
$F = H - \mu_u N_u - \mu_d N_d$, where $H$ is the Hamiltonian, and
$N_u$ and $N_d$ are the number operators for up and down quarks,
respectively.  We choose
a model Hamiltonian which has a free quark term $H_0$ and an
interaction term $H_I$, and write the free energy as $F = F_0 +
H_I$, where $F_0 = H_0 - \mu_u N_u - \mu_d N_d$ is the free energy for
noninteracting quarks.  To describe the pairing interaction between
quarks, we use an NJL model consisting of a four-fermion interaction
with the color and flavor structure of one-gluon exchange:
\beql{gap:ham}
H_I = \frac{3}{8} \int d^3 x \left[ G_E (\bar\psi \gamma^0 T^A \psi) 
(\bar\psi \gamma^0 T^A \psi) -  G_M (\bar\psi \gamma^i T^A \psi)  
(\bar\psi \gamma^i T^A \psi) \right]
\eeql
where the $T^A$ are the color $SU(3)$ generators,
normalized so that $\tr(T^AT^B)=2\de^{AB}$.  
Notice that we have
relaxed some constraints on the spin structure of one-gluon exchange:
we allow for the possibility of independent couplings $G_E$ and $G_M$
for electric and magnetic gluons, respectively.  This spoils Lorentz
boost invariance but there is no reason to insist on boost invariance
in a finite-density system.  Indeed, in high density quark matter we
expect screening of electric gluons but only Landau damping of
magnetic gluons, and we might choose to model these effects by setting
$G_E \ll G_M$.  We postpone a discussion of these issues and their
implications for the LOFF state until Section~\ref{sec:planewaveham}.  For now,
we restrict ourselves to the case of Lorentz invariant single gluon
exchange, by letting $G_E = G_M = G > 0$.

We need to
evaluate the expectation value of $F$ in the LOFF state to obtain 
an expression for the free energy of the system in terms of the variational
parameters of the ansatz.  The noninteracting part of the free energy
is simply
\beql{gap:F0}
\ba{rcl}
\< F_0 \> &=&  \dsp\sum_{\vp \in {\cal B}_u} 2 (|\vq + \vp | - \mu_u) + \sum_{\vp \in {\cal B}_d} 2 (|\vq - \vp | - \mu_d) \\
 & & + \dsp\sum_{\vp \in {\cal P}} 2 (|\vq+\vp| + |\vq-\vp| - \mu_u - \mu_d) \sin^2 \theta_L(\vp) \\
 & & + (\mbox{same, with } L \to R).
\ea
\eeql
The first and second terms represent the contributions of the unpaired
left-handed up and down quarks, respectively.  The third term gives
the (noninteracting) free energy of the left-handed quark pairs.  The
three terms are all repeated with $L$ replaced by $R$ to include the
free energy for the right-handed quarks.  The factors of two in
equation (\ref{gap:F0}) appear because there are two quark colors
(``red'' and ``green'') involved in the the condensate.  The ``blue''
quarks do not participate in the pairing interaction and instead
behave as free particles: the blue up and down quarks fill Fermi seas
with Fermi momenta $p_F^u = \mu_u$ and $p_F^d = \mu_d$,
respectively.  Below, we will
want to compare the free energy of the LOFF, BCS and normal 
states.  Since at any given $\mu_u$ and $\mu_d$ 
the free energy of the spectator quarks
is the same in all three states, we 
can neglect these blue quarks in the 
remainder of our analysis even though they do contribute
to the total free energy.  

%Therefore the free energy of these spectator quarks is
%neglected in the above expression and the blue quarks are ignored in the
%remainder of our analysis.  

The expectation value of $H_I$ gives the total binding energy 
of the pairing interaction:
\beql{gap:HI}
\< H_I \> = -\half GV \left( |\Gamma^L_A|^2 + |\Gamma^R_A|^2 \right)
\eeql
where the $\Gamma_A$'s are the $J=0$ LOFF condensates defined
in equations (\ref{LOFF:gammas}).  
These condensates 
are simply related to $J=0$ LOFF gap parameters 
defined as 
\beql{gap:DeltaDefn}
\Delta^{\{L,R\}}_A = G \Gamma_A^{\{L,R\}}\ .
\eeql
The gap parameters $\Delta_A$ correspond to 1PI Green's functions
and are the quantities which will appear in the
quasiparticle dispersion relations 
and for which we will derive the self-consistency conditions
conventionally called gap equations.
We see from Eq.~(\ref{gap:HI}) that with $G > 0$ the interaction
is attractive in the $J=0$ channel and is neither attractive
nor repulsive in the $J=1$ channel.  

Our ansatz breaks rotational invariance, so once $J=0$ pairing
occurs ($\Ga_A\neq 0$) we expect that 
there will also be $J=1$ pairing
($\Ga_B\neq 0$). As we have seen, this arises
even in the absence of any interaction in the $J=1$ channel
as a consequence of the fact that the momenta of
two quarks in a Cooper pair are not anti-parallel if $\vq\neq 0$. 
Because $\langle H \rangle$
is independent of $\Gamma_B$, the quasiparticle 
dispersion relations must also be independent of $\Gamma_B$.  
That is, the $J=1$ gap parameter 
must vanish: $\De_B=0$.
In Section~\ref{sec:planewaveham},
we shall see by direct calculation that
$\Delta_B$ is proportional to $(G_E-G_M)\Gamma_B$. In the
present analysis with $G_E=G_M$, therefore, $\Delta_B=0$ while
$\Gamma_B\neq 0$.

The $\xi$'s  are chosen
to cancel the azimuthal phases $\phi(\vp)$ in equations
(\ref{LOFF:gammas}). 
By this choice we obtain maximum coherence
in the sums over $\vp$, giving the largest possible magnitudes for 
the condensates and gap parameters.  We have
\beql{gap:phases}
\xi_L(\vp) = \phi(\vp) + \varphi_L, \hspace{0.3in} 
\xi_R(\vp) = \pi-\phi(\vp) + \varphi_R
\eeql
where $\varphi_L$ and $\varphi_R$ are arbitrary $\vp$-independent
angles.  
These constant phases do not affect the free 
energy --- they correspond to the Goldstone bosons for the
broken left-handed and right-handed baryon number symmetries --- and
are therefore not constrained by the variational procedure.  
For convenience, we set $\varphi_L = \varphi_R = 0$ 
and obtain condensates
and gap parameters that are purely real.  

The relative phase $\varphi_L - \varphi_R$ 
determines how the LOFF condensate transforms under a parity
transformation. Its value determines whether
the $J=0$ condensate is scalar, pseudoscalar,
or an arbitrary combination of the two and whether
the $J=1$ condensate is vector, pseudovector, or an arbitrary combination.
Because single gluon exchange cannot change the handedness of a
massless quark, the left- and right-handed 
condensates in the LOFF phase are not coupled in the 
free energy of Eq.~(\ref{gap:HI}.)
Our choice of 
interaction Hamiltonian
therefore allows an arbitrary choice of $\varphi_L - \varphi_R$.
A global $U(1)_A$ transformation changes $\varphi_L - \varphi_R$,
and indeed this is a symmetry of our toy model.
If we included $U(1)_A$-breaking interactions in
our Hamiltonian, to obtain a more complete description of QCD,
we would find that the free energy depends
on $\varphi_L - \varphi_R$, and thus selects a 
preferred value. For example, had we taken $H_I$ to be the
two-flavor instanton interaction as in Ref.~\cite{ARW2,RappEtc}, 
the interaction energy would appear as 
$\Gamma^{L*}\Gamma^R + \Gamma^L \Gamma^{R*}$ instead
of as in (\ref{gap:HI}). This would
enforce a fixed phase relation $\varphi_L-\varphi_R=0$,
favoring condensates which are parity conserving~\cite{ARW2,RappEtc}.

%Because single gluon exchange cannot change the handedness of a
%massless quark, the left- and right-handed condensates in the LOFF
%state are independent.  This is manifest in Eq.~(\ref{gap:HI}) by
%the fact that the interaction energy separates into left and right
%handed components, with no cross terms.  The interaction Hamiltonian
%therefore allows an arbitrary relative phase between left and right
%condensates. 
%as determined by the phase angles $\xi(\vp)$ in the
%variational wavefunction.\footnote{By contrast, if we had chosen to
%model the pairing interaction with a two-flavor instanton vertex as in
%~\cite{ARW2}, the interaction energy would appear as $\Delta_L^*
%\Delta_R + \Delta_L \Delta_R^*$ yielding a fixed phase relation
%between left and right handed condensates.  }  

We now apply the variational method to determine the angles
$\theta(\vp)$ in our trial wavefunction, by requiring that the free
energy is minimized: $\partial \< F \> / \partial \theta(\vp) = 0$.
This is complicated by the fact that the pairing region ${\cal P}$ and
the blocking regions ${\cal B}_u$ and ${\cal B}_d$ are themselves
implicitly dependent on the $\theta$ angles: these angles determine
the extent of the LOFF pairing, and the phase space regions ${\cal P}$,
${\cal B}_u$ and ${\cal B}_d$
change
when a condensate is present, as mentioned in Section~\ref{sec:planewaveLOFF}.
For now we simply ignore any $\theta$-dependence of the phase space
regions; our result will nevertheless turn out to be correct.  Everything is
the same for left and right condensates so we hereafter drop the $L$
and $R$ labels.  Upon variation with respect to $\theta(\vp)$,
we obtain
\beql{gap:tan2th}
\tan 2\theta(\vp) = \frac{ 2 \Delta_A \sin (\beta_A(\vp)/2) }{|\vq+\vp| 
+ |\vq-\vp| - \mu_u - \mu_d}
\eeql
where $\beta_A(\vp) = \alpha_u(\vp) + \alpha_d(\vp)$ is the angle
between the two quark momenta in a LOFF pair, as shown in 
Figure~\ref{fig:angles}.  Notice that the denominator on the right hand side
of the above expression vanishes along the ellipsoidal surface of
optimal LOFF pairing described in Section~\ref{sec:planewaveLOFF}.  When $\vq = 0$,
the quark momenta are antiparallel so $\beta_A(\vp) = \pi$ and 
Eq.~(\ref{gap:tan2th}) reduces to the simple BCS result:
$\tan 2\theta = \Delta_A/(|\vp|-\mubar)$.  

With the $\theta$ angles now expressed in terms of a gap parameter
$\Delta_A$, we turn to the LOFF
quasiparticle dispersion relations. They can be obtained
by taking the absolute value of the expressions
\beql{gap:quasi}
\ba{rcrcl}
E_1(\vp) &=& \dm &+& \half( |\vq+\vp| - |\vq-\vp|) \\[1ex]
 && & +& \half \sqrt{ (|\vq+\vp| + |\vq-\vp| - 2\bar\mu)^2 
+ 4 \Delta_A^2 \sin^2(\half\beta_A(\vp))} \\[2ex]
E_2(\vp) &=& -\dm & -& \half( |\vq+\vp| - |\vq-\vp|) \\[1ex]
 && & +& \half \sqrt{ (|\vq+\vp| + |\vq-\vp| - 2\bar\mu)^2 
+ 4 \Delta_A^2 \sin^2(\half\beta_A(\vp))}\ , \\
\ea
\eeql
whose meaning we now describe.
For regions of $\vp$-space which are well outside both Fermi surfaces,
$E_1$ ($E_2$) is the free energy cost of removing
a LOFF pair and adding an up quark with momentum 
$\vq+\vp$ (a down quark with momentum $\vq-\vp$).  
For regions of $\vp$-space which are well inside both Fermi surfaces,
$E_1$ ($E_2$) is the free energy cost of removing
a LOFF hole pair and adding a down hole with momentum 
$\vq-\vp$ (an up hole with momentum $\vq+\vp$).  
Where the Fermi surfaces cross in $\vp$-space and pairing
is maximal, both quasiparticles are equal superpositions of
up and down.  
In the region of $\vp$-space which is well inside the up Fermi
surface but well outside the down Fermi surface, 
$E_1$ is negative, corresponding
to a domain in which it is energetically favorable to have an unpaired
up quark with momentum $\vq+\vp$ rather than a $(\vq+\vp,\vq-\vp)$
quark pair.  Similarly, $E_2$ is negative where it is favorable
to have an unpaired down quark with momentum $\vq-\vp$ rather than a LOFF pair.
Equations (\ref{gap:quasi}) allow us to finally 
complete our description of the LOFF phase by specifying the 
definitions of the phase space regions ${\cal P}$,
${\cal B}_u$ and ${\cal B}_d$.
The blocking region ${\cal B}_u$ 
is the region where $E_1(\vp)$ is negative, and unpaired up quarks
are favored over LOFF pairs.
Similarly ${\cal B}_d$ is the region where $E_2(\vp)$ is
negative. 
The regions $E_1 < 0$ and $E_2 < 0$ are shown as the shaded
areas in Figure~\ref{fig:regions}a for $\Delta_A = 0$, and in 
Figure~\ref{fig:regions}b for $\Delta_A \neq 0$.  LOFF pairing occurs in
the region where $E_1$ and $E_2$ are both positive:
\beql{gap:pregion}
{\cal P} = \{ \vp | E_1(\vp) > 0 \mbox{ and } E_2(\vp) > 0 \}
\eeql
corresponding to the entire unshaded regions of Figure~\ref{fig:regions}.
The actual quasiparticle dispersion functions are $|E_1(\vp)|$ and
$|E_2(\vp)|$: they are nonnegative everywhere, since they represent
energies of 
perturbations of the LOFF state which is the presumed ground state of
the system.\footnote{
Since the LOFF condensate contains pairs with momentum $2\vq$,
the momentum of its quasiparticle excitations is only defined
modulo $2\vq$. The momentum, modulo $2\vq$, 
of a quasiparticle of energy $|E_1(\vp)|$ is $\vp{\rm ~mod~} 2\vq$.}
In the blocking regions, elementary excitations are
created by replacing an unpaired quark with a quark pair, and vice
versa in the pairing region.
When $\vq=0$, Eqs.~(\ref{gap:quasi}) reduce to the more familiar BCS
result: $E_{\{1,2\}}(\vp) = \pm \dm + \sqrt{(|\vp|-\bar\mu)^2 + \Delta_A^2}$.

With the boundaries of the blocking regions specified, one
can verify 
by explicit calculation
that the variation of these boundaries 
upon variation of the $\theta$'s does not change the free energy.
This can be understood as follows.
Notice that 
because we can create zero-energy quasiparticles on
the boundaries of the blocking regions, 
there is no actual energy gap in the
excitation spectrum of the LOFF state.
The change in $\< F \>$ 
due to variation of the boundaries of the 
blocking regions is zero because this
variation simply creates zero-free-energy 
quasiparticles on these boundaries.
This justifies our neglect of the 
$\theta$-dependence of the phase space
regions in the derivation of Eq.~(\ref{gap:tan2th}).

Substituting the expression (\ref{gap:tan2th}) for the $\theta$ angles 
into the expression (\ref{LOFF:gammas}) for the $\Gamma_A$ condensate,
and using the relation $\Delta_A = G
\Gamma_A$, we obtain a self-consistency equation for the gap parameter
$\Delta_A$:
\beql{gap:LOFF}
1 = \frac{2G}{V}\sum_{\vp\in {\cal P}} 
\frac{2\sin^2(\half\beta_A(\vp))}{\sqrt{
  (|\vq+\vp|+|\vq-\vp|-2\bar\mu)^2 + 4\Delta_A^2\sin^2(\half\be_A(\vp))
  }
}.
\eeql
This can be compared to the BCS gap equation, obtained upon setting
$\vq=0$ and eliminating the blocking regions:
\beql{gap:BCSgap}
1 = \frac{2G}{V}\sum_\vp \frac{1}{\sqrt{(|\vp|-\bar\mu)^2+\Delta_0^2}}\ .
\eeql
Note that in the LOFF gap equation (\ref{gap:LOFF}), the gap parameter
appears on the right hand side both explicitly in the denominator and
also implicitly in the definition of the pairing region ${\cal P}$, as
given in (\ref{gap:pregion}).  This means that if the $\vq\rightarrow
0$ limit is taken at fixed $\dm$, the LOFF gap equation will only
become the BCS gap equation if the blocking regions vanish in this
limit. This happens if, as $\vq\rightarrow 0$, $\De_A$ tends to a
limiting value which is greater than $\dm$.  For $\Delta_A < \dm$, the
blocking region $\mathcal{B}_d$ does {\it not} disappear but instead
becomes a spherical shell, as we can see by taking $\vq \rightarrow 0$
in figure~\ref{fig:regions}.  Pairing is blocked in the region 
\begin{equation}
\bar\mu - \sqrt{\dm^2 - \Delta_A^2} < |\vp| < \bar\mu + \sqrt{\dm^2 - \Delta_A^2}.
\end{equation}
This is precisely the ``breached pair'' color superconductor discussed
in chapter 1~\cite{Liu1,Liu2}; the breach is just the blocking region
$\mathcal{B}_d$.  This second solution to the $\vq = 0$ gap equation
was first discovered by Sarma~\cite{Sarma}.  In the present
calculation such states will always have higher free energy than both
the unpaired state ($\Delta_A = 0$) and the BCS state obtained by
solving the gap equation (\ref{gap:BCSgap}) with no blocking
regions..  But there may exist other scenarios in which
the Sarma solution is favored: when one species is very heavy and the
other very light, for example~\cite{Liu2}.

In the next section we will solve the LOFF gap equation 
(\ref{gap:LOFF}) and
determine the circumstances in which the LOFF state is the true ground
state of the system.
Once we have obtained a solution to the gap equation (\ref{gap:LOFF}) for 
$\Delta_A$, the condensates
are given by $\Gamma_A = \Delta_A/G$ 
and 
\beql{gap:GammaB}
\Gamma_B = \dsp\frac{2}{V} \sum_{\vp \in {\cal P}} \frac{ 2 \Delta_A 
\sin(\half\beta_A(\vp))\sin (\half\beta_B(\vp))}{ \sqrt{ ( |\vq+\vp| +
|\vq-\vp| - 2\bar\mu)^2 + 4 \Delta_A^2 \sin^2(\half\beta_A(\vp)) }}
\eeql
where $\beta_B(\vp)=\alpha_u(\vp)-\alpha_d(\vp)$. 
(See Figure~\ref{fig:angles}.)    
We now see explicitly that if the interaction is 
attractive in the $J=0$ channel, creating a nonzero
$\Gamma_A$ and $\Delta_A$, a nonzero $J=1$
condensate $\Gamma_B$ is induced regardless of the
fact that there is no interaction in the $J=1$ channel.
As a check, note that
if $\vq=0$, $\sin(\half\beta_A(\vp))=1$ and 
$\sin(\half\beta_B(\vp))$ is given by the cosine of the polar
angle of $\vp$.  The right hand side of (\ref{gap:GammaB}) 
therefore vanishes upon integration, and $\Gamma_B$ vanishes when $\vq=0$
as it should.
It is now apparent that two features contribute to a nonzero $\Ga_B$.
The first is that the momenta in a quark pair are not antiparallel,
which leads to the factors of $\sin(\half\be_A(\vp))$ in \Eqn{gap:GammaB}.
The second is that the pairing region is anisotropic, since if it were
not the factor of $\sin(\half\be_B(\vp))$ would ensure that the
right-hand side of \eqn{gap:GammaB} vanishes upon integration.

As written, the gap equations (\ref{gap:LOFF})
and (\ref{gap:BCSgap}) are
ultraviolet divergent.  In QCD,
of course, asymptotic freedom implies that the interaction between
quarks decreases at large momentum transfer and we have
not yet represented this fact in our toy model.   
In previous work~\cite{ARW2,ARW3,BergesRajagopal}, 
we chose to mimic the effects of asymptotic
freedom (and to render the right hand side of the gap equation finite)
by introducing a form factor associated with each fermion leg
in the four-fermion interaction.  This is not a good strategy
when $\vq\neq 0$.  The two incident quarks carry
momenta $\vq+\vp$ and $\vq-\vp$ while the 
outgoing quarks carry momenta $\vq+\vp'$ and $\vq-\vp'$.
Were we to cut off these four momenta with form
factors on each leg, we would have a
cutoff which depends explicitly on $\vq$.  
This is not a good representation
of what happens in full QCD, in which the condition
for when the interaction becomes weak is determined by
the momentum $\vp-\vp'$ transferred through the gluon  and
has nothing to do with $\vq$.
For simplicity, 
we choose to introduce a hard cutoff in our NJL model, rather
than a smooth form factor, and choose simply to cut off
the momentum $\vp$.  This is not equivalent to cutting
off the momentum transfer, but has the desired feature
of being a  $\vq$-independent cutoff.
That is, we limit the integration region
to $|\vp|<\Lambda$ in the BCS gap equation (\ref{gap:BCSgap})  
and to $\{\vp \in {\cal P}$ and $|\vp|<\Lambda\}$ in
the LOFF gap equation (\ref{gap:LOFF}).  In the BCS case,
this criterion is equivalent to cutting off the momentum
of each fermion leg. In the LOFF case, it is not equivalent
and is more appropriate.  The choice we have made is not
the only $\vq$-independent cutoff one might try.
For example, 
we have also obtained results upon cutting off
momenta outside a large ellipsoid in $\vp$-space, confocal
with the centers of the two Fermi spheres in Figure~\ref{fig:regions},
but have found that this makes little difference relative
to the simpler choice of the large sphere $|\vp|<\Lambda$.

%To mimic the effects of asymptotic freedom (and to render finite
%the right hand side of the above expression) we shall introduce a cutoff
%to neglect modes above a momentum scale $\Lambda$.  For simplicity we
%choose a hard cutoff which can be implemented by simply limiting the 
%extent of the pairing region ${\cal P}$.  To this effect the pairing 
%region is restricted to $|\vp| < \Lambda$.  [...]

\section{Results}
\label{sec:planewaveresults}
We solve the gap equation \eqn{gap:LOFF} numerically
(and analytically in the limit $\De_A \ll \dm,q,\De_0$) and
calculate the LOFF state free energy as a function of $\dm$ and $q$, for given
coupling $G$, average chemical potential $\mubar$, and cutoff $\La$.
We vary $q$ to minimize the LOFF free energy, and
compare it with that for the standard BCS
pairing \eqn{gap:BCSgap} to see which is favored.
In this way we can map out the phase diagram for the three
phases of pairing between the two species of quark:
BCS, LOFF, and unpaired.

Note that the solution to the gap equation, the LOFF gap parameter
$\De_A$, is not a gap in the spectrum of excitations.  
The quasiparticle dispersion 
relations (\ref{gap:quasi}) vary
with the direction of the momentum, yielding gaps that vary from zero
(for momenta on the edge of the blocking regions in phase space) up to
a maximum of $\De_A$.

We will first discuss the range of $\dm$ in which
there exists a LOFF state as a local energy minimum.
Later we will go on to
study the competition between LOFF and BCS, and see in what range
of $\dm$ the LOFF state
is the global minimum.
We expect the BCS state to be preferred when the mismatch $\dm$
between the Fermi energies of the two species is small.
When the mismatch is comparable to the BCS gap
($\dm\sim\De_0$) we expect a transition to LOFF, and
at larger $\dm$ we expect all pairing to cease.
These expectations are largely borne out.

In general we fix $\La=1~\GeV$ and $\mubar=0.4~\GeV$, and study
different coupling strengths $G$ which we parameterize by
the physical quantity $\De_0$, the BCS gap of Eq.~\eqn{gap:BCSgap}
which increases monotonically with increasing $G$.
When we wish to study the dependence on the cutoff, we
vary $\La$ while at the same time 
varying the coupling $G$ such that $\De_0$ is kept fixed.  
(This is in the same spirit as using a renormalization
condition on a physical quantity---$\De_0$---to fix
the ``bare'' coupling---$G$.)  We expect that the relation between
other physical quantities
and $\De_0$ will be reasonably insensitive to variation
of the cutoff $\La$.

\begin{figure}[t]
\begin{center}
\includegraphics[width=2.1in,angle=-90]{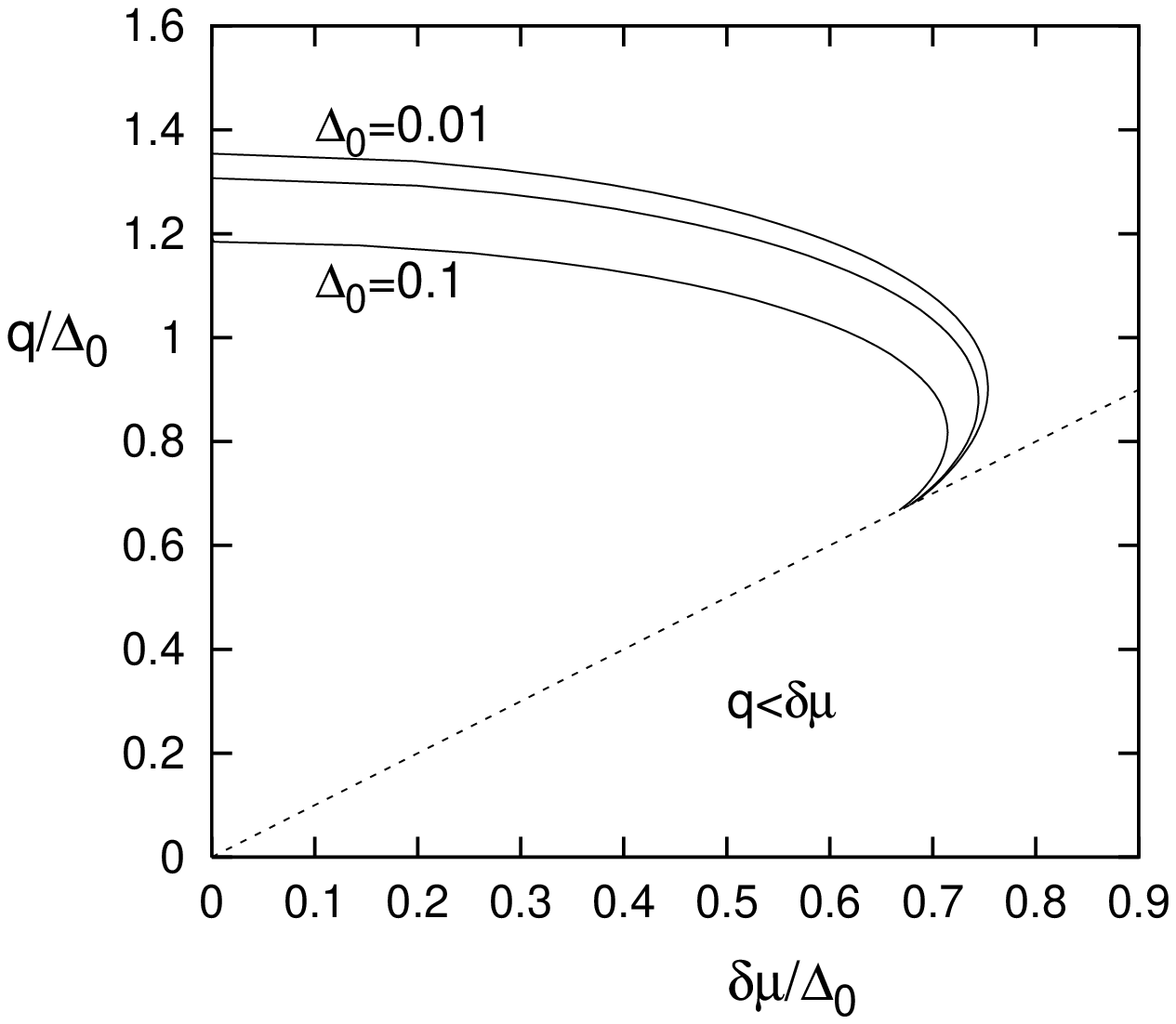}
\phantom{XX}
\includegraphics[width=2.1in,angle=-90]{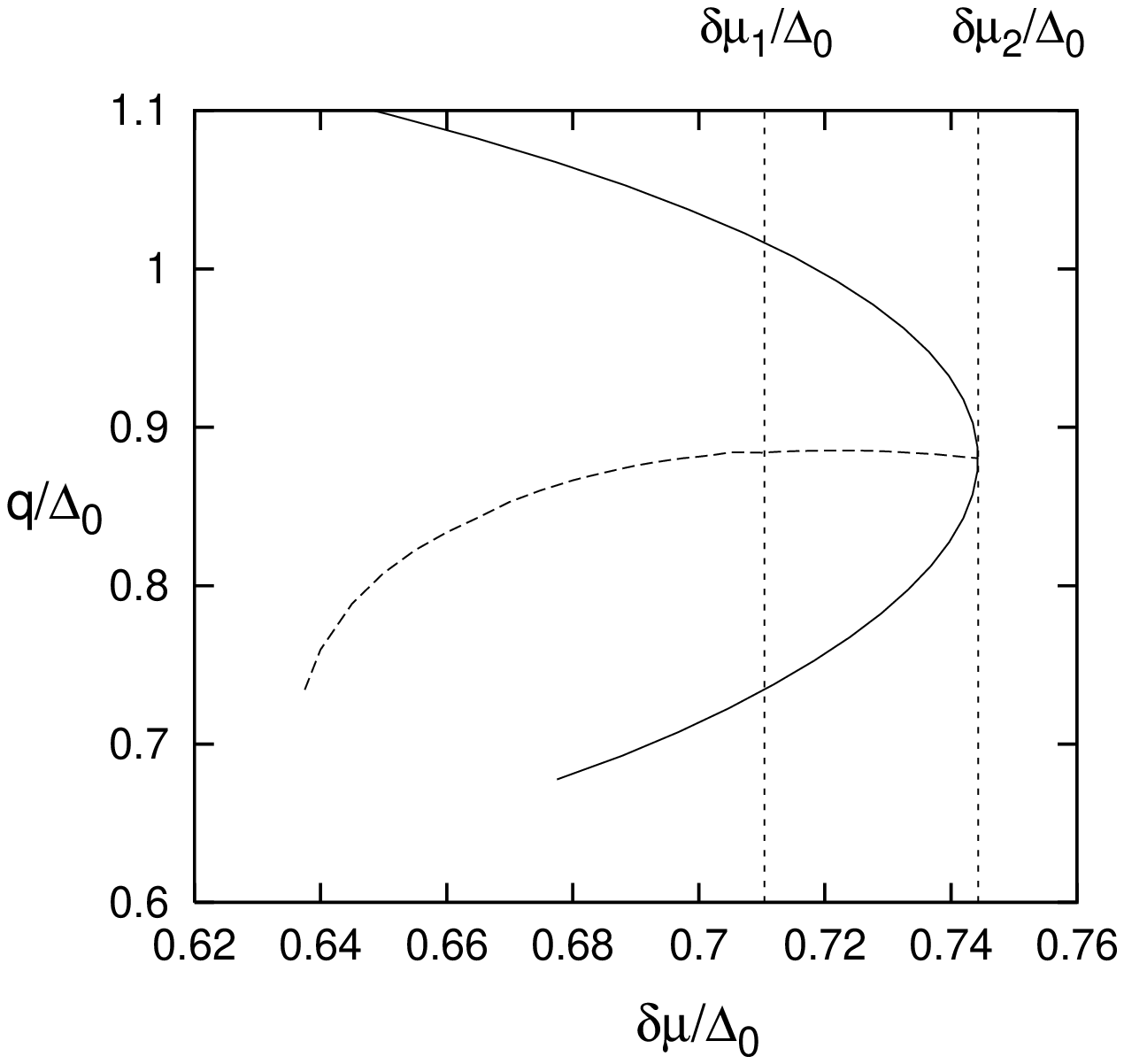}
\end{center}
\caption{
(a)~The zero-gap curves for the LOFF state. 
To the right of a solid curve, there is no solution to the LOFF
gap equation, to the left of the curve there is a solution,
and on the curve the gap parameter is zero.
The three curves are (from strongest
to weakest coupling): 
$\De_0=0.1, 0.04, 0.01~\GeV$.
The region $q<\dm$ is complicated to describe~\cite{FF},
and solutions found in this region never give the
lowest free energy state at a given $\dm$.
(b)~Here, we choose $\De_0=0.04~\GeV$
and focus on the region near $\dmmax$, the maximum value
of $\dm$ at which the LOFF state exists.   The dashed
curve shows the value of $|\vq|$ which minimizes the free energy of the LOFF
state at a given $\dm$.  $\dmmin$, discussed below, is also indicated.
}
\label{fig:zero_gap}
\end{figure}

%\subsection{LOFF pairing versus no pairing}

We wish to determine $\dm_2$, the boundary separating
the LOFF phase and the normal phase.
The LOFF to unpaired phase transition is second order,
so it occurs where the solution $\De_A$ to the LOFF gap equation
\eqn{gap:LOFF} is zero.
Setting $\Delta_A=0$ in the gap equation (\ref{gap:BCSgap}) 
yields an analytical expression relating
$\dm$ and $q$, for any given $G$ and
$\La$.  In Figure~\ref{fig:zero_gap}a we show
the $\De_A=0$ curve
for three couplings corresponding to $\De_0=0.1~\GeV$ 
(strong coupling), $\De_0=0.04~\GeV$  
and $\De_0=0.01~\GeV$ (weak coupling).
We have only drawn the zero-gap curve in the region 
where $q\geq\dm$.
We expect this to be the region of interest for LOFF pairing
because when $q\geq\dm$ the two spheres of Figure~\ref{fig:regions}
do in fact intersect.  We have verified that, as described in some
detail in Ref.~\cite{FF}, there are regions of Figure~\ref{fig:zero_gap}a
with $q<\dm$ within which the LOFF gap equation (\ref{gap:LOFF})
has (one or even two) nonzero solutions, but these solutions
all correspond to phases whose free energy is either 
greater than that of the normal phase or greater than
that of the BCS phase or both.
Figure~\ref{fig:zero_gap} shows that 
for a given coupling strength, parameterized by $\De_0$, there is
a maximum $\dm$ for which the LOFF state exists: we call it
$\dmmax$. For $\dm>\dmmax$, the mismatch of chemical
potentials is too great for the LOFF phase to exist.

We see from Figure~\ref{fig:zero_gap}a that as the
coupling gets weaker, $\dmmax/\De_0$ gets gradually
larger. (Of course, $\dmmax$ itself gets smaller: 
the quantities plotted are $\dm/\De_0$ and $q/\De_0$.)
Note that in the $\De_0\rightarrow 0$ limit, the zero gap
curve  
is essentially that shown in the figure for $\De_0=0.01~\GeV$,
in agreement with the curve obtained at weak coupling
by Fulde and Ferrell~\cite{FF}.  The fact that this curve
ceases to move in the $\De_0\rightarrow 0$ limit means that
$\dmmax \rightarrow 0$ while $\dmmax/\De_0 \rightarrow {\rm const}$
in this limit.

For $\dm\rightarrow\dmmax$ from below, we see from 
Figure~\ref{fig:zero_gap} that
there is a solution to the LOFF gap equation only at a single
value of $q$.  For example, 
at $\De_0=0.04~\GeV$
we find $q=0.880\De_0=1.183\,\dmmax$ at $\dmmax=0.744\De_0$.
(In agreement with Refs.~\cite{LO,FF}, 
in the weak coupling limit we find
$q=0.906\De_0=1.20\,\dmmax$ at $\dmmax=0.754\De_0$.)
For any value of $\dm<\dmmax$,
solutions to the LOFF gap equation exist for a range of $|\vq|$.
We must now find the value of $|\vq|$ for which the free energy
of the LOFF state is minimized.   We obtain the
free energy of the LOFF state at a point in Figure~\ref{fig:zero_gap}
by first solving the gap equation (\ref{gap:LOFF}) numerically to obtain
$\Delta_A$, and then using (\ref{gap:DeltaDefn}) and (\ref{gap:tan2th})
to evaluate $\langle F_0 + H_I \rangle$ given in (\ref{gap:F0})
and (\ref{gap:HI}).  For
each value of $\dm<\dmmax$ we can now
determine which choice of $q$ yields the
lowest free energy.  The resulting ``best-$q$ curve'' curve  is shown in 
Figure~\ref{fig:zero_gap}b for $\De_0=0.04~\GeV$.\footnote{As a check
on our determination of the best $q$,  we have confirmed that
the total momentum of the LOFF state with the best $q$
is zero, as must be the case for the ground state of the
system at a given $\dm$ (by a theorem attributed to Bloch \cite{London}).   This is a powerful check, 
because it requires the 
net momentum of the unpaired quarks in the blocking regions
(which is in the negative $z$ direction; see Figure~\ref{fig:regions})
to be cancelled by the net momentum carried by the LOFF condensates.
When, in future work, our ansatz is extended to describe
a LOFF crystal rather than a single plane wave, this check
will no longer be powerful. Once we go from 
$\Gamma\sim\exp(2i\vq\cdot\vr)$ to $\Gamma\sim\cos(2\vq\cdot\vr)$
or to a more involved crystalline pattern, the total momentum
of the condensates and of the unpaired quarks will each be
zero.}

\begin{figure}[t]
\begin{center}
\includegraphics[width=2.38in,angle=-90]{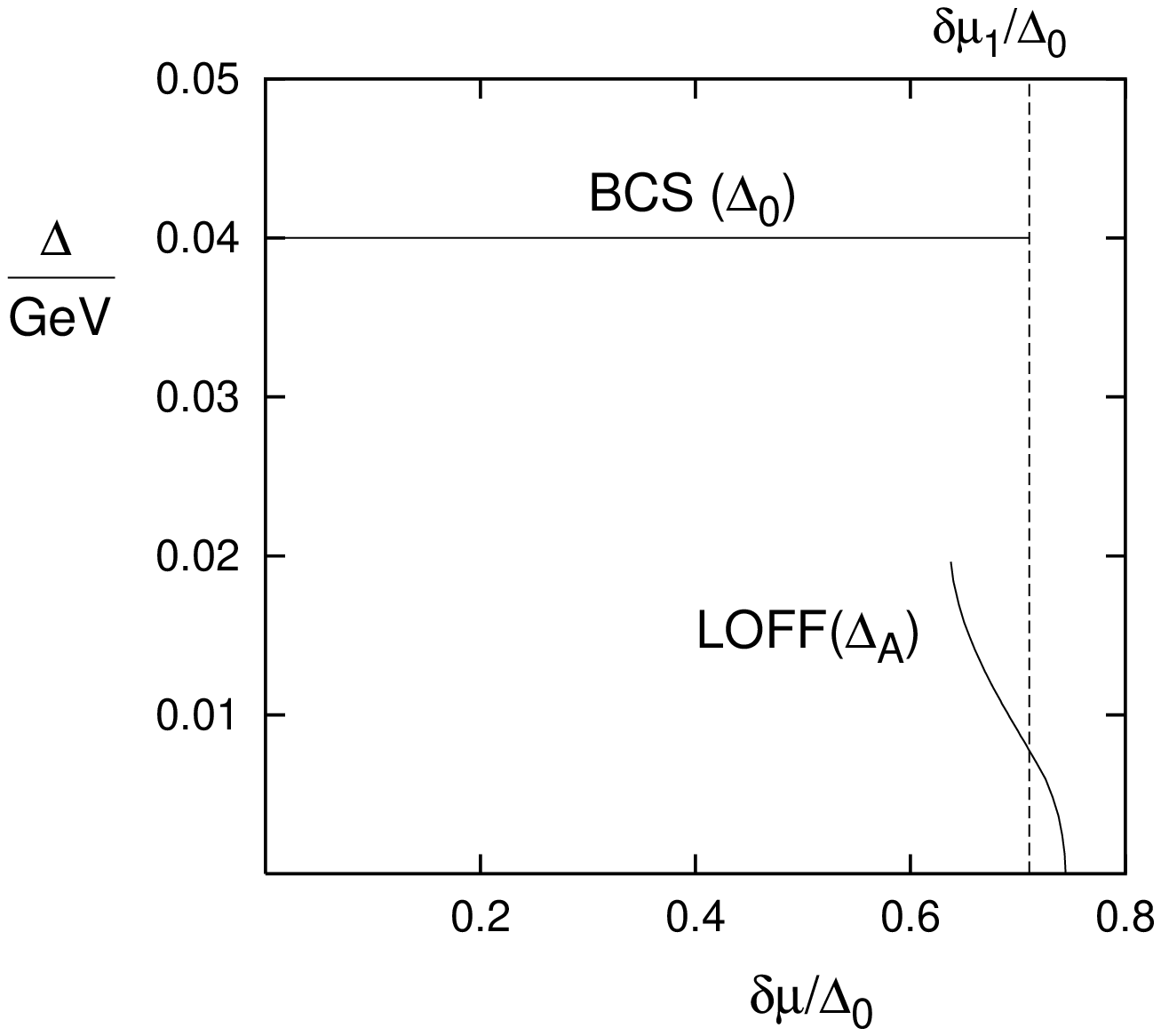}
\includegraphics[width=2.52in,angle=-90]{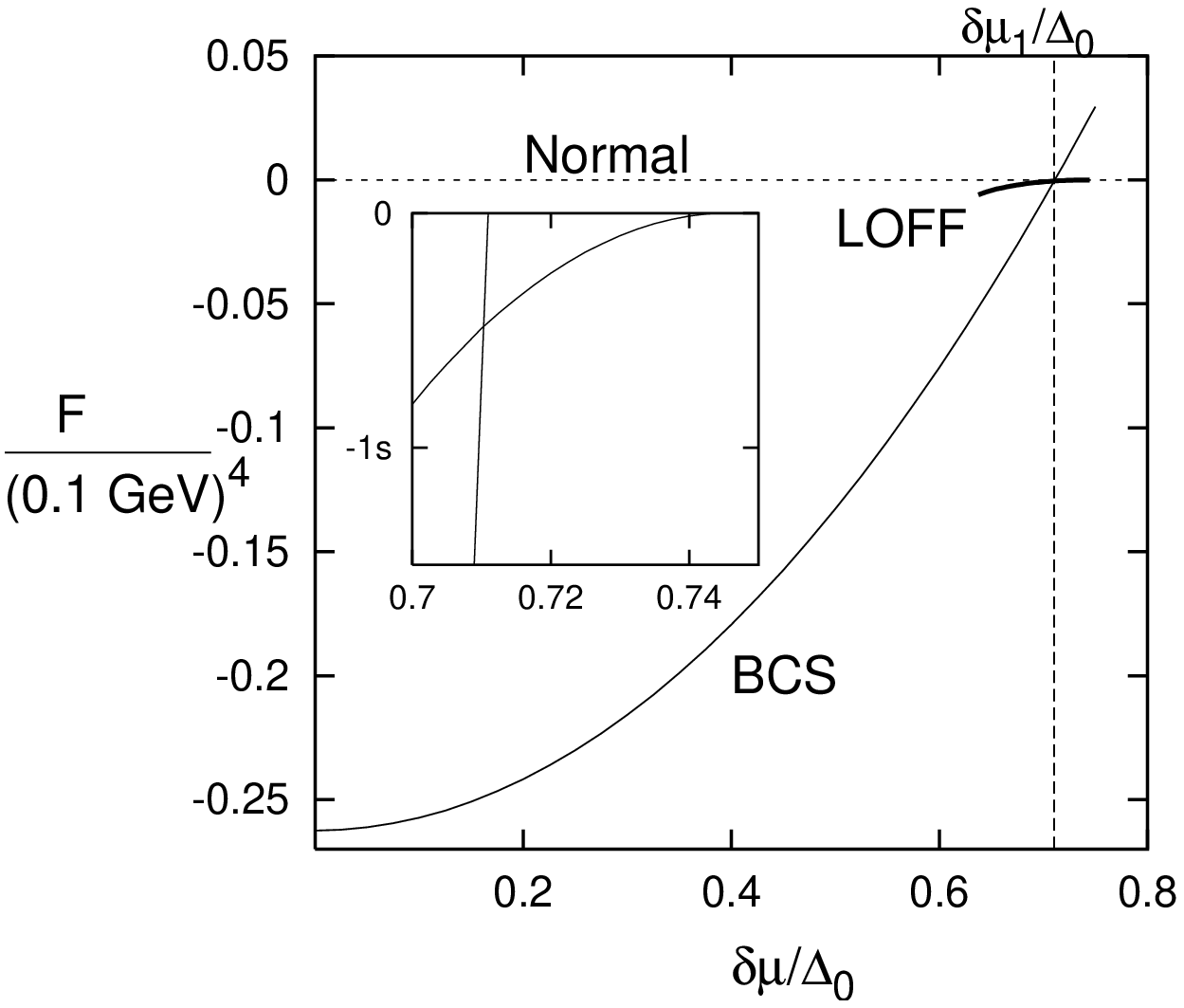}
\end{center}
\caption{LOFF and BCS gaps and free energies
as a function of $\dm$, with coupling chosen so 
that $\De_0=40{\rm ~MeV}$ and with $\mubar=0.4~\GeV,
\La=1~\GeV$. 
Free energies are measured relative to the normal state.
At each $\dm$ we have
varied $q$ to find the best LOFF state.
The vertical dashed line marks $\dm=\dm_1$, the value of $\dm$ above which
the LOFF state has lower free energy than BCS.
The expanded inset (wherein $s=10^{-7}~\GeV^4$) focuses on the region
$\dm_1<\dm<\dm_2$ where the LOFF state has the lowest free energy.
This figure should be compared with the sketch in figure \ref{omegadeltafig}.  
}
\label{fig:F_plot}
\end{figure}
% You need to edit F_vs_dm.eps and F_vs_dm_inset.eps by hand to make
% their bounding boxes just the right size to include labels but no
% extraneous margins. Then edit delta_vs_dm.eps to agree with F_vs_dm.eps.
% gv tells you the coords of the cursor, so it is
% easy to work out what the BBox should be. 
% Then include F_vs_dm.eps and F_vs_dm_inset.eps
% into xfig as pictures, with their original aspect ratio.
% Fiddle the size and posn of the inset using xfig Move and Scale
% commands.

Finally, for each point on the best-$q$ curve  
we ask whether the LOFF free energy at that $\dm$ and (best) $q$
is more or less than the
free energy of the BCS state at the same $\dm$. 
In this way, we find $\dmmin$ at which
a first order phase transition between the LOFF and BCS states
occurs.
In Figure~\ref{fig:F_plot} we show the competition between the
BCS and LOFF states as a function of the Fermi surface mismatch
$\dm$, for a fixed coupling corresponding to $\De_0=40$ MeV.  
The LOFF state exists for
$\dm<\dmmax=0.744\De_0$.  At each $\dm<\dmmax$, we plot the gap
parameter and free energy characterizing the LOFF state 
with the best $q$ for that $\dm$.
Although the BCS gap $\De_0$ is larger than the LOFF gap $\De_A$,
as $\dm$ increases we see from Eq.~(\ref{BCSFreeEnergy}) that
the BCS state pays a steadily increasing free-energetic price for maintaining
$p_F^u=p_F^d$, whereas the LOFF state pays no such price.
We now see that the LOFF state has lower free energy than the BCS state
for $\dm>\dmmin$, in this case $\dmmin=0.7104\De_0$.
At $\dm=\dmmin$, the gap parameter is $\De_A=0.0078~\GeV =0.195\De_0$. 
(Had we calculated $\dm_1$ by comparing the BCS free energy
with that of the unpaired state instead of with
that of the LOFF state, we would have obtained
$\dmmin=0.711\De_0$. 
As the inset to Figure~\ref{fig:F_plot} confirms,
the BCS free energy varies so rapidly that this makes an
almost imperceptible difference.  In later figures,
we therefore obtain $\dmmin$ via the simpler route of comparing
BCS vs. normal.)
At the coupling corresponding to $\De_0=40$ MeV, 
we have found that the LOFF state is favored over both the BCS state
and the normal state in a ``LOFF window''
$0.710 < \dm/\De_0 < 0.744$.

\begin{figure}%[thb]
\begin{center}
\includegraphics[width=3in,angle=-90]{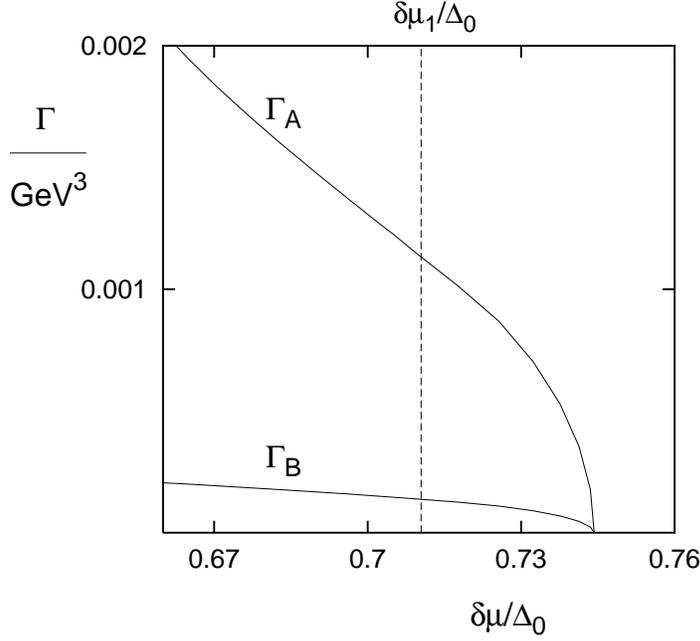}
\end{center} 
\caption{
The two LOFF condensates $\Gamma_A$ $(J=0)$
and $\Gamma_B$ $(J=1)$ for the same choice
of parameters as in Figure~\ref{fig:F_plot}. We focus on the region
$\dmmin<\dm<\dmmax$.   For reference, 
in the BCS phase $\Gamma_A=\De_0/G=0.00583~\GeV^3$ 
and $\Gamma_B=0$.
}
\label{fig:GammaBandA}
\end{figure}

With solutions to the gap equation in hand, we can obtain
the $J=0$ condensate $\Gamma_A=G\De_A$ and the $J=1$ condensate
$\Gamma_B$ given in Eq.~(\ref{gap:GammaB}). In Figure~\ref{fig:GammaBandA},
we show both condensates
within the LOFF window $\dmmin<\dm<\dmmax$.
We see first of all that $\Gamma_B\neq 0$, as advertised.
For the choice of parameters in Figs.~\ref{fig:F_plot} 
and \ref{fig:GammaBandA} we find 
$\Gamma_B/\Gamma_A$
essentially constant over the
whole LOFF window, varying from 0.121 at $\dmmin$ to 0.133 at $\dmmax$.
Increasing $\De_0$ tends
to increase $\Gamma_B/\Gamma_A$, as does decreasing $\La$.  
Second of all, we see that the phase transition at $\dm=\dmmax$,
between the LOFF and normal phases, is second order in the mean-field
approximation we employ throughout.

\begin{figure}[t]
\begin{center}
\includegraphics[width=3in,angle=-90]{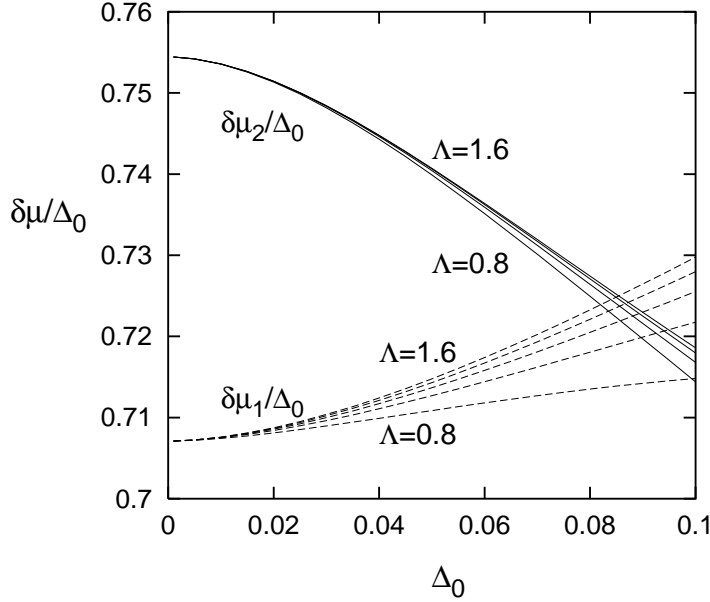}
\end{center}
\caption{The interval of $\dm$ within which the LOFF state occurs,
as a function of the coupling (parameterized as usual by the
BCS gap $\De_0$).
Below the solid line, there is a LOFF state. Below the dashed line,
the BCS state is favored. The different lines of each type correspond to
different cutoffs $\La=0.8~\GeV$ to $1.6~\GeV$. $\dmmin/\De_0$ 
and $\dmmax/\De_0$ show little
cutoff-dependence, and the cutoff-dependence disappears completely
as $\De_0,\dm\to 0$.
}
\label{fig:fish}
\end{figure}

It is interesting to explore how the width of the LOFF window
depends on the strength of the coupling, 
and to confirm that it is insensitive to the cutoff.
We do this in Figure~\ref{fig:fish}, where we plot $\dmmax/\De_0$
(solid lines) and $\dmmin/\De_0$ (dashed lines).
The LOFF state is favored
for $\dmmin/\De_0< \dm/\De_0 < \dmmax/\De_0$, i.e.~between the
solid and dashed curves in Figure~\ref{fig:fish}.
In the weak coupling limit, the LOFF window tends to 
$0.707 < \dm/\De_0 < 0.754$ and $\De_A$ at $\dmmin$ 
tends to $0.23\De_0$, as in Refs.~\cite{LO,FF}.  
Note that if one takes the weak-coupling limit $\De_0\rightarrow 0$
at fixed $\dm$, neither BCS nor LOFF pairing survives because
$\dm/\De_0\rightarrow\infty$.
However, for any arbitrarily
small but nonzero coupling, the LOFF phase
is favored within a range of $\dm$. 
Figure~\ref{fig:fish} thus demonstrates that in an analysis of the LOFF state 
in the weak-coupling limit, it is convenient to
keep $\dm/\De_0$ fixed while taking $\De_0\rightarrow 0$.
We see from Figure~\ref{fig:fish} 
that strong coupling helps the BCS state more than it
helps the LOFF state.
When the coupling gets strong enough, there is no longer
any window of Fermi surface mismatch $\dm$ in which the LOFF
state occurs: the BCS state is always preferred.

The different lines of each type in Figure~\ref{fig:fish} 
are for different cutoffs
and show that there is in fact little sensitivity to the cutoff.
The $\La$ dependence of $\dmmin/\De_0$ and $\dmmax/\De_0$ is mild
for all values of $\De_0$ which are of interest, 
and is weakest for $\De_0\rightarrow 0$. 
This is because in that limit pairing can only occur
very close to the unblocked ribbon of the ellipsoid of 
Fig. 2b, along which the integrand in the gap equation 
is singular and pairing is allowed. Thus
most of the pairing region ${\cal P}$,
and in particular the region near $\La$, become irrelevant in this limit.

The one physical quantity which we have explored which does
turn out to depend qualitatively on $\La$ is the ratio $\Gamma_B/\Gamma_A$.
Those quarks with 
momenta as large as $\La$ which pair have momenta which are 
almost antiparallel, and so contribute much less to $\Gamma_B$ than
to $\Gamma_A$.  For this reason, the ratio $\Gamma_B/\Gamma_A$
is sensitive to the number of Cooper pairs formed at very large $\vp$,
and hence to the choice of $\La$.  As discussed above, pairing
far from the favored ribbon in phase space becomes irrelevant
for $\De_0\rightarrow 0$, and indeed in this limit we find that the
$\La$ dependence of $\Gamma_B/\Gamma_A$ decreases.  However,
for $\De_0=40$ MeV we find that changing $\La$ from $1.2~\GeV$
to $0.8~\GeV$ increases $\Gamma_B/\Gamma_A$ by more than 50\%.

We chose to show results for $\De_0=40$ MeV in Figure~\ref{fig:F_plot}
because with this choice, the LOFF window occurs at values of $\dm$
that are reasonable for quark matter in the interiors of compact
stars.  Consider $\mu = $ 400 MeV and $m_s = $ 300 MeV (recall that
$m_s$ is a density-dependent effective mass which is significantly
larger than the current quark mass).  Substituting these numbers in
equations (\ref{UnpairedFermiMomenta}) describing unpaired neutral
quark matter, we find that the baryon number density is 4 times
nuclear matter density and $\dm=\half(\mu_d-\mu_u) = 28$ MeV.  (Had we
chosen $m_s=200$ MeV, we would have obtained $\dm=12$ MeV and a baryon
number density that is 5 times nuclear matter density.)  Of course,
neither $\dm$ nor the value of $\De_0$ are accurately known for the
quark matter which may exist within a compact star.  Still, it seems
possible that their ratio could be appropriate for the quark matter to
be in the LOFF phase.  If there is a range of radii within a compact
star in which quark matter occurs with $\dmmin < \dm < \dmmax$, this
quark matter will be a crystalline color superconductor.

In Figure~\ref{fig:F_plot}, the LOFF gap parameter $\De_A$ is $7.8$
MeV at $\dm=\dmmin$.  It remains larger than typical neutron star
temperatures $T_{\rm ns}\sim 1$ keV until very close to $\dm=\dmmax$.
Similarly, the LOFF free energy, which is $4.8\times 10^{-8}~\GeV^4 =
4.8\times(10~\MeV)^4$ at $\dm=\dmmin$, is much larger than $T_{\rm
ns}^4$ throughout the LOFF window except very close to $\dm=\dmmax$.
The LOFF gap and free energy are likely to be much larger for a
condensate of multiple plane waves, as we will see in the next
chapter. Furthermore, we will see in chapter 5 that the free energy of
the LOFF state is of the right order to lead to interesting glitch
phenomena.

\section{More general Hamiltonian and ansatz}
\label{sec:planewaveham}

In Section~\ref{sec:planewavegap}, we introduced the four-fermion
interaction Hamiltonian $H_I$ of Eq.~(\ref{gap:ham}) with independent
couplings $G_E$ and $G_M$ for the interactions which model
the exchange of electric and magnetic gluons.
It proves convenient to use the linear combinations
\beql{ham:GAandGB}
\ba{rcl}
G_A &=& \frac{1}{4}(G_E + 3 G_M)\\[0.5ex]
G_B &=& \frac{1}{4}(G_E - G_M)\ ,
\ea
\eeql
of the coupling constants
in terms of which the expectation value of $H_I$ 
in the LOFF state (\ref{LOFF:ansatz}) becomes
\beql{ham:HI}
\langle H_I \rangle = - \half 
G_A V \left( |\Gamma^L_A|^2 + |\Gamma^R_A|^2 \right)
- \half G_B V \left( |\Gamma^L_B|^2 + |\Gamma^R_B|^2 \right)\ .
\eeql
Thus, a positive coupling $G_A$ describes an attractive interaction
which induces a $J=0$ condensate $\Gamma_A$.  As we have seen,
in the LOFF state this is necessarily accompanied by a $J=1$ condensate
$\Gamma_B$.  
In our analysis to this point, we have set $G_A=G>0$ and $G_B=0$.
We now discuss the general case, in which $G_B\neq 0$.

Before beginning, let us consider how to choose $G_B/G_A$ in order 
for our model Hamiltonian to be a reasonable toy model for
QCD at nonzero baryon density.  At zero density, of course,
Lorentz invariance requires $G_B=0$.  At high densities, on
the other hand, electric gluons are screened while 
static magnetic gluons are not.  (Magnetic gluons with nonzero
frequency are damped.)  We now know~\cite{Son} that at asymptotically
high densities it is in fact the exchange of magnetic gluons 
which dominates the pairing interaction.  This suggests
the choice $G_E=0$, corresponding to $G_B/G_A = -1/3$.   
At the accessible densities of interest to us, it is presumably
not appropriate to neglect $G_E$ completely.  Note also that
the four-fermion interaction induced by instantons in QCD
only yields interactions in flavor-antisymmetric channels.
It results in an attractive interaction in the $J=0$ channel and no
interaction in the 
$J=1$ channel.  Thus, although the instanton interaction cannot
be written in the form (\ref{gap:ham}), for our purposes it
can be thought of as adding a contribution to $G_A$, but none
to $G_B$.  
Hence our model is likely to best represent high density QCD
for a ratio of couplings lying somewhere in the range
\beql{ham:range}
-\frac{1}{3} < \frac{G_B}{G_A} < 0 \ .
\eeql 
We plot our results over a wider range of couplings below.

Once $G_B\neq 0$ and there is an interaction in the $J=1$ channel,
we expect, in addition to the $J=1$
condensate $\Gamma_B$, a $J=1$ gap parameter $\Delta_B$.
The quasiparticle dispersion relations
are then determined by $\Delta_A$ and $\Delta_B$, 
which are defined as
\beql{ham:Deltas}
\ba{rcl}
\Delta_A &=& G_A \Gamma_A\\
\Delta_B &=& G_B \Gamma_B \ .
\ea
\eeql
Following through the variational calculation as in
Section~\ref{sec:planewavegap} leads to the coupled gap equations:
\beql{ham:coupledgapeqs}
\ba{rcl}
\Delta_A &=& \dsp\frac{2G_A}{V} \sum_{\vp \in {\cal P}} 
\frac{ 2 S_A 
(\Delta_A S_A + \Delta_B S_B) }
{ \sqrt{ ( |\vq+\vp| + 
|\vq-\vp| - 2\bar\mu)^2 + 4 (\Delta_A S_A 
+ \Delta_B S_B)^2 }} \\[2ex]
\Delta_B &=& \dsp\frac{2G_B}{V} \sum_{\vp \in {\cal P}} 
\frac{ 2 S_B
(\Delta_A S_A + \Delta_B S_B) }
{ \sqrt{ ( |\vq+\vp| + 
|\vq-\vp| - 2\bar\mu)^2 + 4 (\Delta_A S_A 
+ \Delta_B S_B)^2 }} \\[3ex]
S_A &=& \sin(\half\beta_A(\vp)) \\[1ex]
S_B &=& \sin(\half\beta_B(\vp))
\ea
\eeql
with $\beta_A(\vp) = \alpha_u(\vp) + 
\alpha_d(\vp)$, $\beta_B(\vp) = \alpha_u(\vp) - \alpha_d(\vp)$
defined in terms of the angles in Figure~\ref{fig:angles}.
The pairing region ${\cal P}$ is still defined 
by \eqn{gap:pregion} but with new 
quasiparticle dispersion relations obtained from
Eqs.~(\ref{gap:quasi}) with $\De_A^2S_A^2$ replaced by
$(\De_A S_A + \De_B S_B)^2$.

For $G_B=0$, the coupled equations \eqn{ham:coupledgapeqs}
reduce to Eqs.~(\ref{gap:LOFF}) and
(\ref{gap:GammaB}).  Note that if, instead, $G_B>0$ and $G_A=0$, 
we find an attractive interaction
in the $J=1$ channel in Eq.~(\ref{ham:GAandGB}) and
no interaction in the $J=0$ channel.  Analysis
of  Eqs.~(\ref{ham:coupledgapeqs}) in this case yields a nonzero
value of $\Delta_B$, while $\Delta_A=0$ even though $\Gamma_A\neq 0$.
The geometry of the LOFF pairs  
requires $\Gamma_A\neq 0$ when $\Gamma_B \neq 0$.

Rather than describing how every Figure in Section~\ref{sec:planewaveresults}
changes when $G_B\neq 0$, we choose to focus on the question
of how the interval of $\dm$ within which the LOFF state occurs
(the LOFF window)
changes as a function of $G_B/G_A$.  To further simplify the
presentation, we specialize to the weak-coupling limit
in which $\Delta_0\rightarrow 0$. This means that, as in 
Figure~\ref{fig:fish}, the LOFF window is independent of the
cutoff $\Lambda$.  

\begin{figure}[t]
\begin{center}
\includegraphics[width=3in,angle=-90]{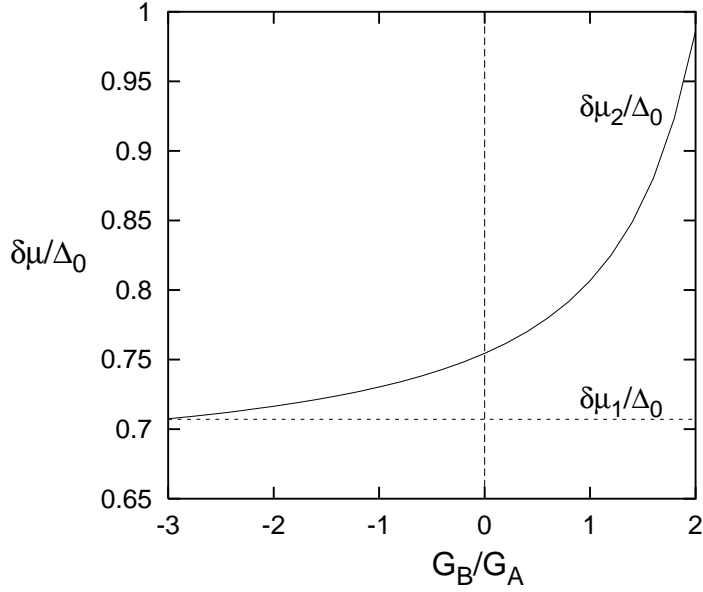}
\end{center} 
\caption{
The interval of $\dm$ in which the LOFF state 
is favored at weak coupling, as a function of the
ratio of couplings $G_B/G_A$. 
Below the solid line, there is  a LOFF state.
Below the dashed line, the ordinary BCS state is favored.
$G_B=0$ corresponds to the Lorentz-invariant interaction with $G_E=G_M$.
QCD at high density is likely best described by a coupling
in the range $-\third < G_B/G_A < 0$.
}
\label{fig:dm_vs_ba}
\end{figure}

We show the dependence of the LOFF window on $G_B/G_A$ in
Figure~\ref{fig:dm_vs_ba}. The lower boundary $\dm=\dm_1$ is, as in
Section~\ref{sec:planewaveresults}, the same (up to a very small correction) as
the $\dm$ at which the BCS and normal states have equal free energies.
We find the upper boundary $\dm=\dm_2$
by first dividing Eqs.~(\ref{ham:coupledgapeqs}) by $\Delta_A$
and then looking for a value of $\dm$ at which $\Delta_A\rightarrow 0$
and $\Delta_B\rightarrow 0$ but $\Delta_A/\Delta_B$ remains nonzero.
As before, this defines a zero-gap curve, and $\dmmax$ is the
maximum value of $\dm$ reached by this curve.

% As in Section \ref{sec:planewaveresults}, the lower boundary $\dm_1$
% of the LOFF window is (up to a very small correction) the
% same as the $\dm$ at which the BCS and normal states have
% equal free energies.  Because the BCS state is purely $J=0$,
% it is completely unaffected by the value of $G_B$.  
% 
% Therefore,
% in the weak coupling limit $\dm_1=0.707$ independent of $G_B/G_A$.
% In contrast, since the LOFF state requires a $J=1$ condensate
% we expect it to be fortified by $G_B>0$ and penalized by $G_B<0$.
% 
% This is borne out in the fact that we find that $\dm_2$, the
% upper boundary of the LOFF window, increases with increasing
% $G_B$ as shown in Figure~\ref{fig:dm_vs_ba}.  

We find that the lower boundary $\dm_1$ is completely unaffected by
the value of $G_B$, since the BCS state is purely $J=0$. So in the
weak-coupling limit we obtain the result of Section~\ref{sec:planewaveresults},
$\dm_1/\De_0=0.707$, independent of $G_B/G_A$.  In contrast, $\dm_2$,
the upper boundary of the LOFF window, increases with increasing
$G_B$. This is understandable: the LOFF state always produces a $J=1$
condensate, so we expect it to be fortified by $G_B>0$ and penalized
by $G_B<0$.  
There is no analogue of this behavior in an electron
superconductor~\cite{LO,FF}, where there can
be no $J=1$ condensate.  Our $J=1$ condensate
affects the gap equation and free energy 
only if $G_B\neq 0$;  for this reason, our weak coupling results
are in agreement
with those of LOFF~\cite{LO,FF} only if $G_B=0$, as in 
Section~\ref{sec:planewaveresults}.
The effect of a coupling $G_B$
in the physically interesting range (\ref{ham:range}) is to reduce the
LOFF window, but only slightly.

%This
%difference only arises if $G_B\neq 0$, and this is the reason why 
%our results in  Section \ref{sec:planewaveresults} were in agreement
%with those of LOFF in the weak-coupling limit
%employed by LOFF.
%we only found deviations from the original
%LOFF results away from weak coupling.  

% We find $\dm_2$ by first dividing 
% Eqs.~(\ref{ham:coupledgapeqs}) by $\Delta_A$
% and then looking for a value of $\dm$ at which 
% $\Delta_A\rightarrow 0$ and $\Delta_B\rightarrow 0$
% but $\Delta_A/\Delta_B$ remains nonzero.  The results
% shown in Figure~\ref{fig:dm_vs_ba} confirm that because
% the LOFF state in QCD includes a $J=1$ condensate while 
% that in an electron superconductor does not, our 
% results differ in the weak coupling limit 
% from those obtained by LOFF.~\cite{LO,FF}  This difference only
% arises if $G_B\neq 0$, and this is the reason why
% in Section \ref{sec:planewaveresults} we only found deviations
% from the original LOFF results away from weak coupling.
% The effect of a coupling $G_B$ in the range
% (\ref{ham:range}) is to reduce the LOFF window,
% but only slightly.

In the vicinity of $\dm_2$ we should evaluate the competition between
LOFF pairing and the formation of single-flavor $\langle uu \rangle$
and $\langle dd \rangle$ condensates.  These are spatially-uniform
spin-one condensates, as introduced in section \ref{sec:singleflavor}
and described in more detail in the NJL survey of chapter 4.  We defer
a complete discussion of these condensates until chapter 4, but here
review some aspects that are relevant in the current context.
Considering only the up quarks, the single-flavor spin-one condensates
can involve LL/RR pairing, with a condensate structure $\langle u C
\sigma_{03} u \rangle$, and LR pairing, with a condensate structure
$\langle u C \gamma_3 u \rangle$.  In the LL/RR channel, the $G_A$
term in our model Hamiltonian gives no interaction, and the $G_B$
term gives a repulsive interaction for $G_B < 0$.  Thus, for
reasonable Fermi couplings that are likely to best represent high
density QCD (as in equation (\ref{ham:range})), our model Hamiltonian
does not predict any pairing in this channel\footnote{ This conclusion
is modified at asymptotically high density, where it has been shown by
Sch\"afer \cite{TS1flav} that long-range single-gluon exchange does in
fact induce pairing in the LL/RR channel.  (The long-range interaction
emphasizes near-collinear scattering which is attractive for both
electric and magnetic gluons.)  Simultaneously, the long-range gluon
interaction considerably widens the window in which the LOFF plane
wave phase can occur~\cite{pertloff,Giannakis}.  Both effects must be
taken into account in the comparison between LOFF and single-flavor 
phases in the high density limit}.  In the LR channel,
the model Hamiltonian gives an attractive interaction if $G_E + G_M >
0$ and is independent of the linear combination of couplings $G_E -3
G_M$.  If we solve the gap equation for the $\langle u C \gamma_3 u
\rangle$ (LR) condensate with $G_E=G_M=G$, $\De_0=40~\MeV$,
$\mu_u=0.4~\GeV$, and $\La=1~\GeV$, we find a gap of 8 keV and a free
energy which is about five orders of magnitude smaller than that of
the LOFF phase.  (If we choose $G_E=0$ and $G_M>0$, the interaction is
still attractive but the gap is even smaller.)  Therefore, even though
for $\dm>\dmmax$ we expect single-flavor pairing and consequent
$\langle u C \gamma_3 u \rangle$ and $\langle d C \gamma_3 d \rangle$
condensates, the resulting condensation energy is so small that it is
a good approximation to neglect these condensates in the evaluation of
$\dmmax$, as we have done.

\section{Conclusions}
\label{sec:planewaveconclusions}

We have studied the formation of a plane-wave LOFF state involving
pairing between two flavors of quark whose chemical potentials differ
by $2\dm$.  This state is characterized by a gap parameter and a
diquark condensate, but not by an energy gap in the dispersion
relation.  In the LOFF state, each Cooper pair carries momentum $2\vq$
with $|\vq| \approx 1.2\dm$.  The condensate and gap parameter vary in
space like a plane wave with wavelength $\pi/|\vq|$.

We focused primarily on an NJL-type four-fermion interaction with the
quantum numbers of single gluon exchange.  In the limit of weak
coupling (BCS gap $\De_0\ll \mu$) the LOFF state is favored for values
of $\dm$ which satisfy $\dm_1 < \dm < \dm_2$, where
$\dm_1/\De_0=0.707$ and $\dm_2/\De_0=0.754$.  The LOFF gap parameter
decreases from $0.23 \De_0$ at $\dm=\dm_1$ to zero at $\dm=\dm_2$.
These are the same results found by LOFF in their original analysis.
Except for very close to $\dm_2$, the critical temperature above which
the LOFF state melts will be much higher than typical neutron star
temperatures.  At stronger coupling the LOFF gap parameter decreases
relative to $\De_0$ and the window of $\dm/\De_0$ within which the
LOFF state is favored shrinks. The window grows if the interaction is
changed to weight electric gluon exchange more heavily than magnetic
gluon exchange.

Because it violates rotational invariance by involving Cooper pairs
whose momenta are not antiparallel, the quark matter LOFF state
necessarily features nonzero condensates in both the $J=0$ and $J=1$
channels.  Both condensates are present even if there is no
interaction in the $J=1$ channel.  In this case, however, the $J=1$
condensate does not affect the quasiparticle dispersion relations;
that is, the $J=1$ gap parameter vanishes.  If there is an attraction
in the $J=1$ channel (as, for example, if the strength of the electric
gluon interaction is increased) the size of the LOFF window increases.

The single plane wave state is a rather feeble state, as we can see
from figure \ref{fig:F_plot}: it only prevails as the favored ground
state in a tiny window of $\dm$, and its gap and free energy are small
compared to those of the BCS phase.  But our results for the
single-plane-wave state are just the starting point for an exploration
of much more complicated and much more robust LOFF crystals involving
multiple plane waves.  We embark on this exploration in the next
chapter.

%% file: chap3.tex
\chapter{Crystalline Superconductivity:  Multiple Plane Waves}

%------------------------------------------------------------------------

\section{Overview}

In this chapter we explore LOFF crystalline states that are
superpositions of multiple plane waves in position space~\cite{BowersCrystal}.  In momentum
space, each plane wave corresponds to another ``pairing ring'' on each
Fermi surface, as in figure \ref{ringsfig}.  These multiple-plane-wave
states are much more robust than the single-plane-wave state that we
studied in chapter 2, for the simple reason that adding more plane
waves utilizes more of the Fermi surface for pairing, with a
corresponding gain in condensation energy.

In studying the multiple-plane-wave states, we do not attempt to write
down variational wave functions as we did for the single-plane-wave
state in chapter 2.  Rather, in section \ref{sec:multiplewavemethod},
we use a Nambu-Gorkov formalism to directly obtain an infinite set of
coupled gap equations for the gap parameters $\{ \Delta_{\vq} \}$.  In
principle, finding the best LOFF crystal requires exhaustively
searching the infinite-dimensional parameter space of these crystal
order parameters, to find the unique solution of the coupled gap
equations that is a global minimum of the free energy.  This is an
infinite task.  To organize this search, we construct a
Ginzburg-Landau free energy function $\Omega(\{\Delta_{\vq}\})$ which
is valid in the vicinity of the LOFF second-order critical point
$\dm_2$.  In section \ref{sec:multiplewavemethod} we show how the
Nambu-Gorkov method can be used to systematically calculate the
coefficients in the Ginzburg-Landau potential.  Then in section
\ref{sec:multiplewaveresults} we use this Ginzburg-Landau potential
to study a large catalog of candidate crystal structures.  It turns
out that the potential is unstable and therefore predicts a strong
first-order phase transition, rather than a second-order transition as
obtained for the single-plane-wave case, and is therefore not under
quantitative control.  Nevertheless, the Ginzburg-Landau approach is
useful and indeed quite powerful because it organizes the calculation
in such a way that simple qualitative lessons emerge which tell us
what features make a particular crystal structure energetically
favorable or unfavorable.  This narrows our search, remarkably, to a
uniquely favored crystal structure.  In section
\ref{sec:multiplewaveconclusions} we discuss this structure, an FCC
crystal consisting of eight plane wave condensates with wave vectors
pointing towards the eight corners of a cube.  We predict that the
phase is quite robust, with gaps comparable in magnitude to the BCS
gap that would form if the Fermi momenta were degenerate.

\section{Methods}
\label{sec:multiplewavemethod}

\subsection{The gap equation}

We study the crystalline superconducting phase in a toy model
for QCD that has two massless flavors of quarks and a pointlike
interaction.  
%Keep in mind that although our calculations proceed in
%this particular context, our results are quite generic and apply to
%any binary system where two species of fermion have a weak,
%attractive, pointlike interaction.  
The Lagrange function 
%for our toy model 
is
\begin{equation}
\mathcal{L} = \bar \psi (i \delslash + \muslash) \psi - \frac{3}{8} \lambda 
(\bar\psi \Gamma^A \psi)( \bar\psi \Gamma_A \psi)
\end{equation}
where $\muslash = \gamma^0 (\mubar - \tau_3 \dm)$.  The $\tau$'s are
Pauli matrices in flavor space, so the up and down quarks have
chemical potentials as in (\ref{UnpairedFermiMomenta}).            
The vertex is $\Gamma^A =
\gamma^\mu T^a$ so that our pointlike interaction mimics the spin,
color, and flavor structure of one-gluon exchange. (The $T^a$ are color
$SU(3)$ generators normalized so that $\tr (T^a T^b) = 2
\delta^{ab}$.) We denote the coupling
constant in the model by $\lambda$.

\begin{figure}
\centering
\begin{psfrags}
\psfrag{deltalabel}[cc][lc]{$\mathbf{\Delta}$}
\parbox{1.5in}{\includegraphics[width=1.25in]{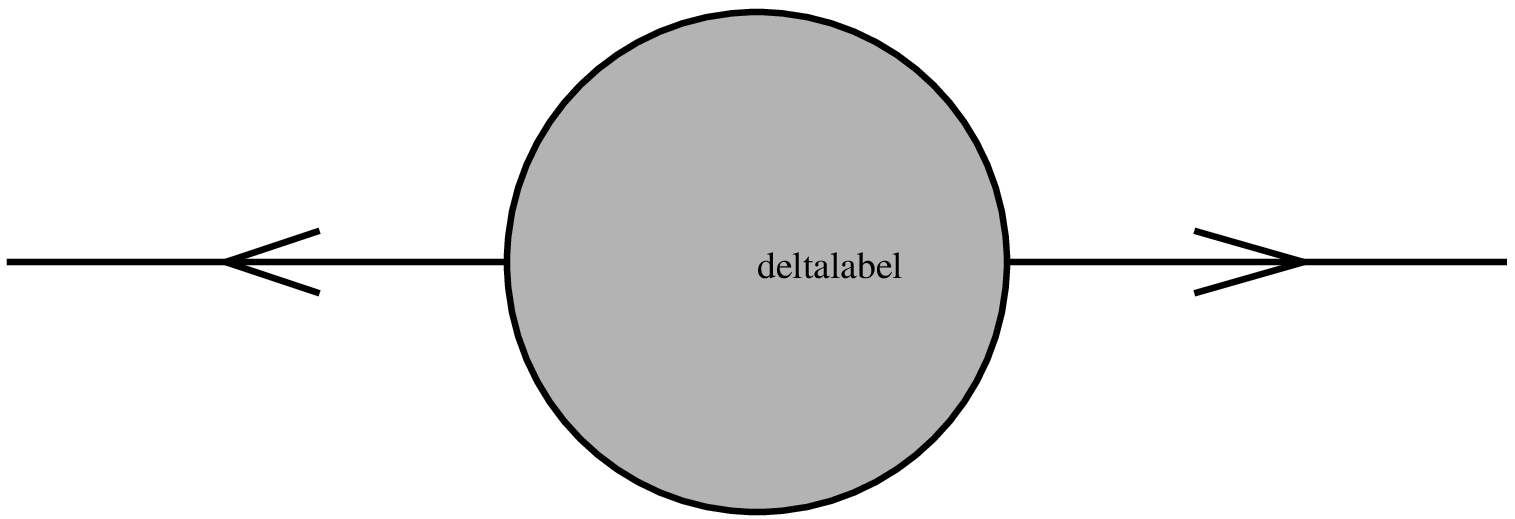}}
\end{psfrags}
$=$ \hspace{0.25in}
\parbox{1.25in}{\includegraphics[width=1in]{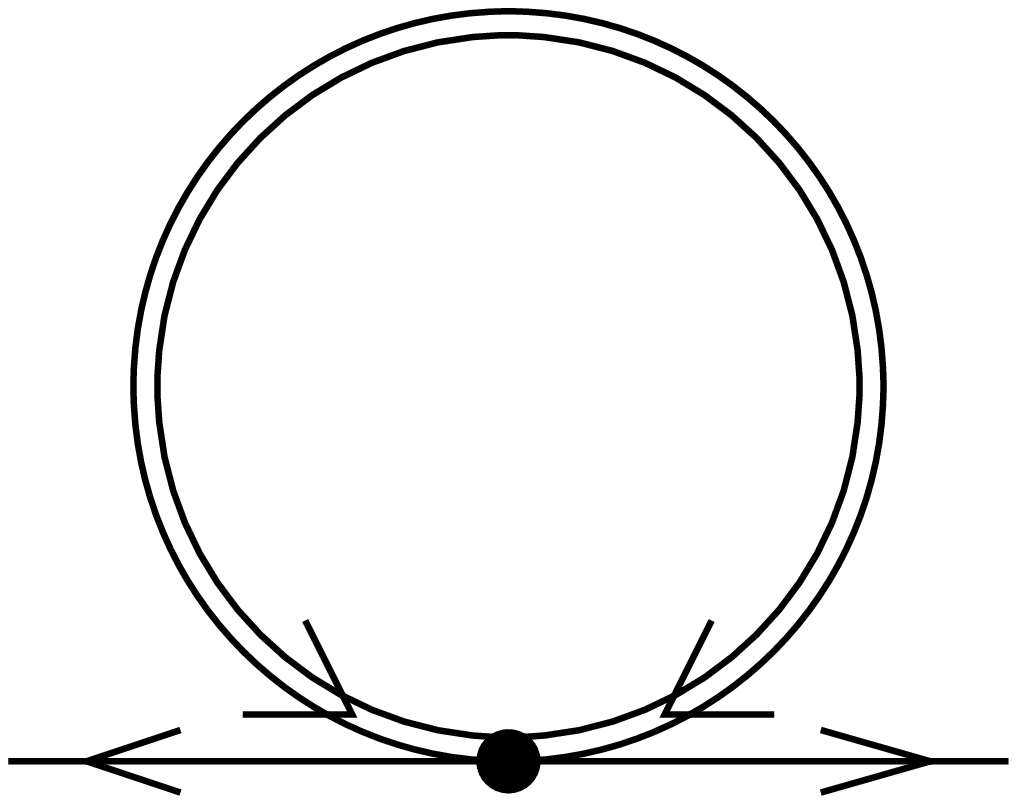}}
\vspace{0.2in}
\caption{
\label{sdeqnfig}
The Schwinger-Dyson graph for the LOFF gap parameter
$\mathbf{\Delta}$.  The black dot is the pointlike interaction vertex and the
double line represents the full anomalous propagator $F$, which
is given in terms of $\mathbf{\Delta}$ in Fig.~\ref{expansionfig}.  }
\end{figure}

It is convenient to use a Nambu-Gorkov diagrammatic method to obtain
the gap equation for the crystalline phase.  Since we are
investigating a phase with spatial inhomogeneity, we begin in position
space.  We introduce the two-component spinor $\Psi(x) = (\psi(x),
\bar\psi^T(x))$ and the quark propagator $i S(x,x') = \langle \Psi(x)
\bar\Psi(x') \rangle$, which has ``normal'' and ``anomalous''
components $G$ and $F$, respectively:
\begin{equation}
i S(x,x') = \left( \begin{array}{cc} iG(x,x') & i F(x,x') \\ i \bar F(x,x') & i\bar G(x,x') \end{array} \right) = \left( \begin{array}{cc} \langle \psi(x) \bar\psi(x') \rangle & \langle \psi(x) \psi^T(x') \rangle \\ \langle \bar\psi^T(x) \bar\psi(x') \rangle & \langle \bar\psi^T(x) \psi^T(x') \rangle \end{array} \right).
\end{equation}
The conjugate propagators $\bar F$ and $\bar G$ satisfy 
\bea
\label{barprops}
i \bar G(x,x') & = & \gamma^0 (i G(x',x))^\dag \gamma^0 \\  
i \bar F(x,x') & = & \gamma^0 (i F(x',x))^\dag \gamma^0.
\eea
The gap parameter $\mathbf\Delta(x)$ that describes the
diquark condensate
is related to the 
anomalous propagator $F$ 
%is related to the gap parameter
%$\mathbf\Delta(x)$ of the diquark condensate 
by a Schwinger-Dyson equation
\begin{equation}
\mathbf\Delta(x) = i \frac{3}{4} \lambda \Gamma^A F(x,x) \Gamma_A^T
\label{sdeqn}
\end{equation}
illustrated diagrammatically in Fig.~\ref{sdeqnfig}.
In our toy model, we are neglecting quark masses and
thus the normal part of the one-particle-irreducible
self-energy is zero; the anomalous part of the 1PI self energy
is just $\mathbf\Delta(x)$.
The crystal order parameter $\mathbf\Delta(x)$ defined
by (\ref{sdeqn}) is a 
matrix in spin, flavor and color space.  In the mean-field
approximation, we can use the equations of motion for $\Psi(x)$ to
obtain a set of coupled equations that determine the propagator
functions in the presence of the diquark condensate 
characterized by $\mathbf\Delta(x)$: 
\begin{equation}
\left(
\begin{array}{cc} 
i \delslash + \muslash & \mathbf\Delta(x) \\
 \bar\mathbf\Delta(x) & (i \delslash - \muslash)^T 
\end{array}
\right) 
\left(
\begin{array}{cc}
G(x,x') & F(x,x') \\  \bar F(x,x') & \bar G(x,x') 
\end{array}
\right) = 
%S(x,x') = 
\left( 
\begin{array}{cc}
1 & 0 \\
0 & 1 
\end{array}
\right)
\delta^{(4)}(x-x')
\label{ngeqns}
\end{equation}
where $\bar\mathbf\Delta(x) = \gamma^0 \mathbf\Delta(x)^\dag
\gamma^0$.  Any function $\mathbf\Delta(x)$ that solves equations
(\ref{sdeqn}) and (\ref{ngeqns}) is a stationary point of the free energy
functional $\Omega [ \mathbf\Delta(x) ]$; of these stationary points,
the one with the lowest $\Omega$ describes the 
ground state of the system.  
Our task, then, is to invert (\ref{ngeqns}), obtaining
$F$ in terms of $\mathbf\Delta(x)$, substitute in (\ref{sdeqn}),
find 
solutions for $\mathbf\Delta(x)$, and then evaluate $\Omega$
for all solutions we find.

There are some instances where analytic solutions to equations
(\ref{sdeqn}) and (\ref{ngeqns}) can be found.  The simplest case is that
of a spatially uniform condensate.  Translational invariance then
implies that the propagators are diagonal in momentum space: $S(p,p')
= S(p) (2 \pi)^4 \delta^{(4)}(p-p')$.  In this case, 
Eqs.~(\ref{ngeqns}) immediately yield
\begin{equation}
S(p)^{-1} = 
\left( 
\begin{array}{cc}
\pslash + \muslash & \mathbf\Delta \\
\bar\mathbf\Delta & (\pslash - \muslash)^T 
\end{array}
\right),
\end{equation}
which is easily inverted to obtain $S$, which can then
be substituted on the right-hand side of Eq.
(\ref{sdeqn}) to obtain a self-consistency equation ({\it i.e.}~a gap
equation) for $\mathbf\Delta$.  The solution of this gap equation 
describes the
familiar ``2SC'' phase~\cite{Barrois,BailinLove,ARW2,RappEtc,Reviews}, 
a two-flavor, two-color BCS condensate,
\begin{equation}
\mathbf\Delta = T^2 \tau_2 C \gamma_5 \Delta_0,
\label{2SCchannel}
\end{equation}
where $T^2$, $\tau_2$, and $C \gamma_5$ indicate that the condensate
is a color antitriplet, flavor singlet, and Lorentz scalar,
respectively.\footnote{The QCD interaction, and thus the
interaction in our toy model,
is attractive in the color
and flavor antisymmetric channel and this dictates the
color-flavor pattern of (\ref{2SCchannel}).  Our toy model
interaction  does not distinguish
between the Lorentz scalar (\ref{2SCchannel}) 
and the pseudoscalar possibility.
But, the instanton interaction in QCD favors the scalar condensate.}
The remaining factor $\Delta_0$, which without loss of generality
can be taken to be real,  gives the magnitude of
the condensate.  In order to solve the resulting gap equation
for $\Delta_0$, we must complete the specification of our
toy model by introducing a cutoff.  In previous
work~\cite{ARW2,RappEtc,ARW3,Reviews},  it has been shown that
if, for a given cutoff, the coupling
$\lambda$ is chosen so that the model describes a reasonable
vacuum chiral condensate, then at $\mu\sim 400-500$ MeV
the model describes a diquark
condensate which has $\Delta_0$ of order 
tens to 100 MeV.
Ratios between physical observables
depend only weakly on the cutoff, meaning that when $\lambda$ 
is taken to vary with the cutoff such that one observable is
held fixed, others depend only weakly on the cutoff.
For this reason, we are free to 
make a convenient choice of cutoff so long as we then choose
the value of $\lambda$ that yields the ``correct'' $\Delta_0$.
Since we do not really know the correct value of $\Delta_0$
and since this is after all only a toy model, we simply think of
$\Delta_0$ as the single free parameter in the model,
specifying the strength of the interaction and thus the 
size of the BCS condensate.
Because the quarks near the Fermi surface 
contribute most to pairing, it is convenient 
to introduce a cutoff $\omega$ defined so as to 
restrict the gap integral to 
momentum modes near the Fermi surface ($||\vp| - \bar\mu | \leq \omega$).
In the weak coupling (small $\lambda$) 
limit, the explicit solution to the gap equation is then
\begin{equation}
\label{bcsgapeqn}
\Delta_0 = 2 \omega e^{-\pi^2/2 \lambda \bar\mu^2}\ .
\end{equation}
This is just the familiar BCS result for the gap. (Observe
that the density of states at the Fermi surface is
$N_0 = 2\bar\mu^2/\pi^2$.)  We denote the gap 
for this BCS solution by $\Delta_0$,
reserving the symbol $\Delta$ for the gap parameter
in the crystalline phase.
We shall see explicitly below that when
we express our results for $\Delta$
relative to $\Delta_0$, they are completely 
independent the cutoff $\omega$ 
as long as $\Delta_0/\mu$ is small.

The BCS phase, with $\Delta_0$ given by (\ref{bcsgapeqn}),
has a lower free energy than unpaired quark matter as
long as $\dm<\dm_1= \Delta_0/\sqrt{2}$~\cite{Clogston}.
The first-order unpairing transition at $\dm=\dm_1$
is the analogue in our two-flavor toy model of the unlocking
transition in QCD.
For $\dm>\dm_1$, the free energy of any crystalline
solution we find below must be compared to that of unpaired
quark matter; for $\dm<\dm_1$, crystalline solutions should be
compared to the BCS phase.  We shall work at $\dm>\dm_1$.

The simplest example of a LOFF condensate is one that
varies like a
plane wave: $\mathbf\Delta(x) = \mathbf\Delta \exp(- i 2 q \cdot x)$.
The condensate is static, meaning that $q = (0, \vq)$.  
We shall denote $|\vq|$ by $q_0$. In this condensate, the
momenta of two quarks in a Cooper pair is $(\vp+\vq,-\vp+\vq)$
for some $\vp$, meaning that the total momentum of each and every
pair is $2\vq$.
See Refs.~\cite{ngloff,massloff} for an analysis of this condensate
using the Nambu-Gorkov formalism. Here, we sketch the results.
If we shift the definition of $\Psi$ in momentum space to
$\Psi_q(p)\equiv(\psi(p+q),\bar\psi(-p+q))$, then in
this shifted basis the propagator is diagonal:
\begin{equation}
i S_q(p,p')= \langle\Psi_q(p)\bar\Psi_q(p')\rangle = i S_q(p)
\delta^4(p-p')
\label{momentumshift}
\end{equation}
and the inverse propagator is simply
\begin{equation}
\label{planewave2}
S_q(p)^{-1} = 
\left( 
\begin{array}{cc}
\pslash + \qslash + \muslash & \mathbf\Delta \\
\bar\mathbf\Delta & (\pslash - \qslash - \muslash)^T 
\end{array}
\right) \ .
\end{equation}
See Refs.~\cite{ngloff,massloff}
for details and to see
how this equation can be inverted and
substituted into Eq.~(\ref{sdeqn}) to obtain a gap equation for $\Delta$.   
This gap equation has nonzero solutions for 
$\dm<\dm_2\simeq 0.7544 \Delta_0$, and has a second-order
phase transition at $\dm=\dm_2$ with $\Delta \sim (\dm_2-\dm)^{1/2}$.
We rederive these results below.

\begin{figure}[t]
\centering
\begin{psfrags}
\psfrag{deltalabel}[cc][lc]{$\mathbf{\Delta}_{\vq_2}$}
\psfrag{deltabarlabel}[cc][lc]{$\bar\mathbf{\Delta}_{\vq_1}$}
\psfrag{p1}{$\vp$}
\psfrag{p2}{$-\vp + 2\vq_1$}
\psfrag{p3}{$\vp - 2\vq_1 + 2\vq_2$}
\parbox{3.25in}{\includegraphics[width=3.25in]{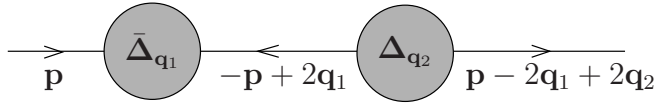}}
\end{psfrags}
\vspace{0.2in}
\caption{
\label{scatterfig}
A process whereby a quark with momentum $\vp$ scatters by interactions
with two plane-wave condensates and acquires a momentum $\vp - 2\vq_1 + 2\vq_2$.
}
\end{figure}

If the system is unstable to the formation of a single
plane-wave condensate, we might expect that a condensate of multiple
plane waves is still more favorable.  
Again our goal is to find gap
parameters $\mathbf\Delta(x)$ that are self-consistent solutions of
Eqs.~(\ref{sdeqn}) and (\ref{ngeqns}).  We use an ansatz that
retains the Lorentz, flavor, and color structure of the 2SC phase:
\begin{equation}
\mathbf\Delta(x) = T^2 \tau_2 C \gamma_5 \Delta(x)
\end{equation}
but now $\Delta(x)$ is a scalar function that characterizes the
spatial structure of the crystal.  We write this function as a
superposition of plane waves:
\begin{equation}
\Delta(x) = \sum_{\vq} \Delta_\vq e^{-i 2 q \cdot x} 
\label{spatialansatz}
\end{equation}
where, as before,  $q = (0,\vq)$.  
The $\{ \Delta_\vq \}$ constitute a set
of order parameters for the crystalline phase.  Our task is to
determine for which set of $\vq$'s the $\Delta_\vq$'s are nonzero.
Physically, for each
$\Delta_\vq\neq 0$ the condensate includes some Cooper pairs for which the
total momentum of a pair is $2 \vq$.  This is indicated by the
structure of the anomalous propagator $F$ in momentum space: 
Eqs.~(\ref{sdeqn}) and (\ref{spatialansatz}) together imply that
\begin{equation}
%F(p,p') = -i \langle \bar\psi^T(-p) \bar\psi(p') \rangle  = \sum_\vq  F_\vq(p) (2\pi)^4 \delta^{(4)}(p-p'+2q)
F(p,p') = -i \langle \psi(p) \psi^T(-p') \rangle  = \sum_\vq  F_\vq(p) (2\pi)^4 \delta^{(4)}(p-p'-2q)
\label{pspaceanomprop}
\end{equation}
and 
\begin{equation}
\mathbf{\Delta}_\vq = i \frac{3}{4} \lambda \int \frac{d^4 p}{(2\pi)^4} \Gamma^A F_\vq(p) \Gamma_A^T
\label{psdeqn}
\end{equation}
where $\mathbf{\Delta}_\vq = T^2 \tau_2 C \gamma_5 \Delta_\vq$.
Eq.~(\ref{psdeqn}) yields an infinite set of coupled gap equations, one
for each $\vq$.  (Note that each $F_\vq$ depends on all 
the $\Delta_\vq$'s.)
It is not consistent to choose only a 
finite set of $\Delta_\vq$ to be nonzero
because when multiple plane-wave condensates are present, these
condensates induce an infinite ``tower'' (or lattice) of 
higher momentum condensates.  This is easily understood by noting that
a quark with momentum $\vp$ can acquire an additional
momentum $2 \vq_2 - 2 \vq_1$ by 
interacting with two different plane-wave condensates as
it propagates through the medium, as shown in Fig.~\ref{scatterfig}.  
Note that this
process cannot occur when there is only a single plane-wave
condensate.  The analysis of the single plane-wave condensate
closes with only a single nonzero $\Delta_\vq$, and is therefore
much easier than the analysis of a generic crystal structure.
Another way that this difficulty manifests itself is that 
once we move beyond the single plane-wave
solution to a more generically nonuniform condensate, 
it is no longer possible to diagonalize the propagator
in momentum space by a shift, as was possible 
in Eqs.~(\ref{momentumshift}, \ref{planewave2}).
 
\subsection{The Ginzburg-Landau approximation}

The infinite system of equations 
(\ref{psdeqn}) has been solved analytically only in one
dimension, 
where it turns out that the gap parameter can be
expressed as a Jacobi elliptic function that, as promised, is composed
of an infinite number of plane waves~\cite{1D}.  
In three dimensions, the crystal structure of the LOFF
state remains unresolved~\cite{BuzdinKachkachi,Houzet,CombescotMora}.
In the vicinity of the second-order transition at $\dm_2$,
however, we can simplify the calculation considerably by
utilizing the smallness of $\Delta$ to make a controlled
Ginzburg-Landau approximation.  
This has the advantage of providing a controlled 
truncation of the
infinite series of plane waves, because near $\dm_2$ the system is
unstable to the formation of plane-wave condensates only for $\vq$'s
that fall on a sphere of a certain radius $q_0$, as we shall see
below.  This was in fact the
technique employed by Larkin and Ovchinnikov in their
original paper~\cite{LO}, and it has been further developed
in Refs.~\cite{BuzdinKachkachi,Houzet,CombescotMora}.
As far as we know, though, no previous authors
have done as complete a study of possible crystal structures
in three dimensions as we attempt. Most have limited
their attention to, at most, structures 1, 2, 5 and 9 from
the 23 structures we describe in Fig.~\ref{stereographicfig}
and Table~\ref{structures} below. As
far as we know, no previous authors have investigated
the crystal structure that we find to be most favorable.

The authors of Refs.~\cite{BuzdinKachkachi,Houzet,CombescotMora}
have focused on using the Ginzburg-Landau
approximation at nonzero temperature, near the critical
temperature at which the LOFF condensate vanishes.
Motivated by our interest in compact stars, 
we follow Larkin and Ovchinnikov in staying at $T=0$
while using the fact that, for
a single plane-wave condensate, $\Delta\rightarrow 0$ for
$\dm\rightarrow \dm_2$ to motivate the Ginzburg-Landau
approximation.  
The down side of this is that,
in agreement with previous authors,
we find that the $T=0$ phase transition becomes first order
when we generalize beyond a single
plane wave.
In the end, therefore, the lessons of our Ginzburg-Landau
approximation must be taken qualitatively.  We nevertheless
learn much that is of value.
\begin{figure}[t]
\psfrag{deltalabel}[cc][lc]{$\mathbf{\Delta}$}
\psfrag{deltabarlabel}[cc][lc]{$\bar\mathbf{\Delta}$}
\psfrag{delta}[cc][lc]{$\mathbf{\Delta}$}
\psfrag{deltabar}[cc][lc]{$\bar\mathbf{\Delta}$}
\begin{eqnarray}
\parbox{.8in}{\includegraphics[width=.8in]{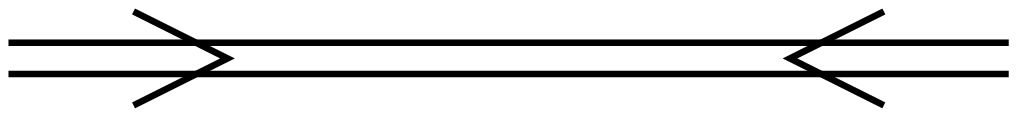}}
& = & 
-
\hspace{0.1in}
\parbox{1.2in}{\includegraphics[width=1.2in]{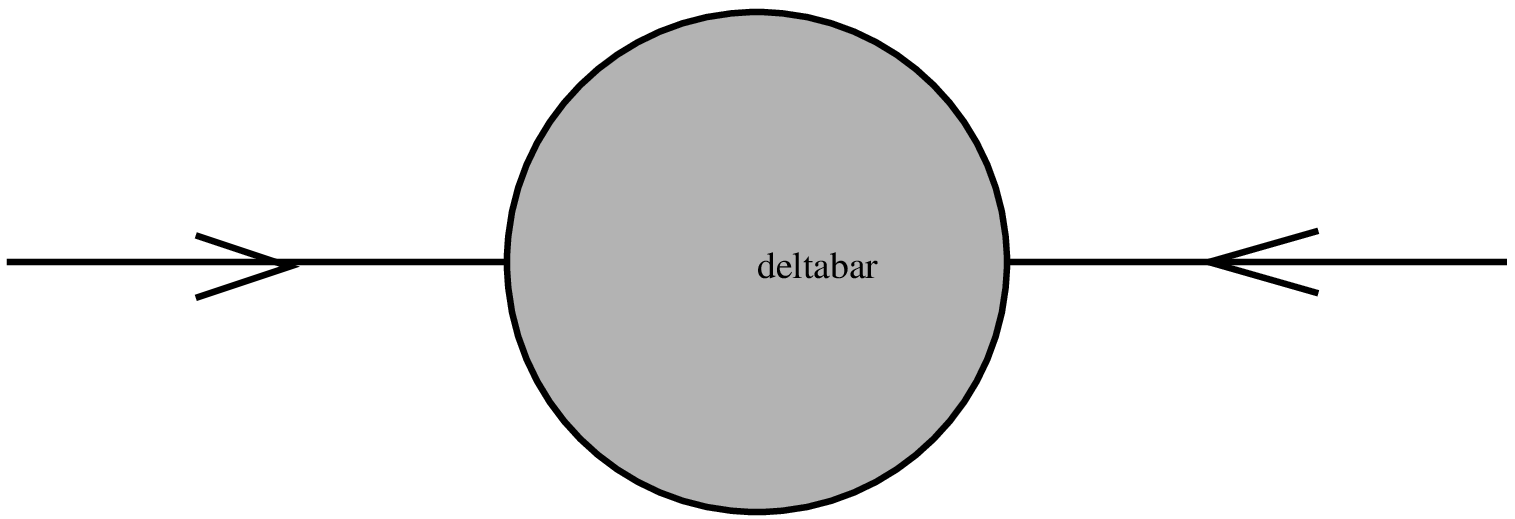}} 
\hspace{0.1in}
+
\hspace{0.1in}
\parbox{2.4in}{\includegraphics[width=2.4in]{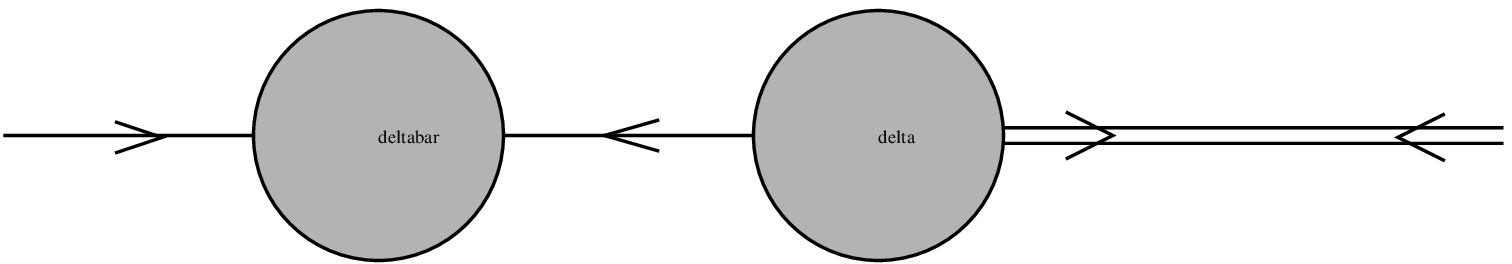}} \nonumber \\
 & & \parbox{3in}{} \nonumber \\
& = & 
-
\hspace{0.1in}
\parbox{1.2in}{\includegraphics[width=1.2in]{deltabar.eps}} 
\hspace{0.1in}
-
\hspace{0.1in}
\parbox{2.8in}{\includegraphics[width=2.8in]{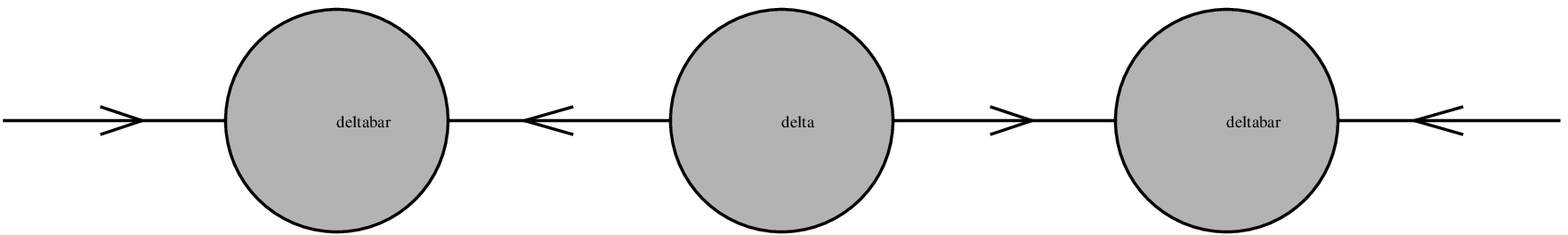}} \nonumber \\
& &
- 
\hspace{0.1in}
\parbox{4.4in}{\includegraphics[width=4.4in]{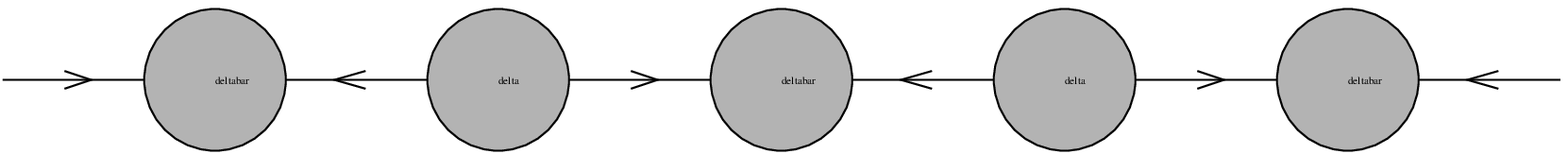}} \nonumber \\
& & 
- 
\hspace{0.1in}
\cdots \nonumber
\end{eqnarray}
\caption{
\label{expansionfig}
The diagrammatic expression for the full anomalous propagator $\bar F$, and the
first three terms in the series expansion in powers of $\Delta$.  }
\end{figure}

To proceed with the Ginzburg-Landau expansion, we first integrate
equations (\ref{ngeqns}) to obtain
\bea
\label{intngeqns}
G(x,x') & = & G^{(0)}(x,x') - \int d^4 z \ G^{(0)}(x,z) \mathbf\Delta(z) \bar F(z,x') \\
\bar F(x,x') & = & -\int d^4 z \ \bar G^{(0)}(x,z) \bar\mathbf\Delta(z) G(z,x')
\eea
where $G^{(0)} = (i \delslash + \muslash)^{-1}$, $\bar G^{(0)} = ((i
\delslash - \muslash)^T)^{-1}$.  Then we expand these equations in
powers of the gap function $\mathbf\Delta(x)$.  For $\bar F(x,x')$ we
find (suppressing the various spatial
coordinates and integrals for notational simplicity)
\begin{equation} 
\ba{rcl} 
\bar F & = & -\bar G^{(0)} \bar\mathbf\Delta G^{(0)} - \bar G^{(0)}
\bar\mathbf\Delta G^{(0)} \mathbf\Delta \bar G^{(0)} \bar\mathbf\Delta
G^{(0)} \\ & & - \bar G^{(0)} \bar\mathbf\Delta G^{(0)} \mathbf\Delta
\bar G^{(0)} \bar\mathbf\Delta G^{(0)} \mathbf\Delta \bar G^{(0)}
\bar\mathbf\Delta G^{(0)} + \mathcal{O}(\Delta^7) 
\ea
\end{equation} 
as expressed diagrammatically in Fig.~\ref{expansionfig}.  We then
substitute this expression for $\bar F$
into the right-hand side of the Schwinger-Dyson 
equation (actually the conjugate of equation (\ref{psdeqn})).  After some spin,
color, and flavor matrix manipulation, the result in momentum space is
%\vfill\eject

\begin{eqnarray}
\label{fullgapeqn}
\Delta_\vq^* & = & - \frac{2 \lambda \bar\mu^2}{\pi^2}\Pi(\vq) \Delta_\vq^* - \frac{2 \lambda \bar\mu^2}{\pi^2} \sum_{\vq_1,\vq_2,\vq_3} J(\vq_1\vq_2\vq_3\vq) \Delta_{\vq_1}^* \Delta_{\vq_2} \Delta_{\vq_3}^* \delta_{\vq_1-\vq_2+\vq_3-\vq} \nonumber \\
 & & - \frac{2 \lambda \bar\mu^2}{\pi^2} \sum_{\vq_1,\vq_2,\vq_3,\vq_4,\vq_5}K(\vq_1\vq_2\vq_3\vq_4\vq_5\vq) \Delta_{\vq_1}^* \Delta_{\vq_2} \Delta_{\vq_3}^* \Delta_{\vq_4} \Delta_{\vq_5}^* \delta_{\vq_1-\vq_2+\vq_3-\vq_4+\vq_5-\vq} \nonumber \\
 & & + \mathcal{O}(\Delta^7) 
\end{eqnarray}

\begin{figure}[t]
\Large
\psfrag{deltabar}[cc][lc]{\small $\Delta_\vq^*$}
\psfrag{delta}[cc][lc]{\small $\Delta_\vq^*$}
\psfrag{delta1}[cc][lc]{\small $\Delta_{\vq_1}^*$}
\psfrag{delta2}[cc][lc]{\small $\Delta_{\vq_2}$}
\psfrag{delta3}[cc][lc]{\small $\Delta_{\vq_3}^*$}
\psfrag{delta4}[cc][lc]{\small $\Delta_{\vq_4}$}
\psfrag{delta5}[cc][lc]{\small $\Delta_{\vq_5}^*$}
\begin{eqnarray}
\parbox{1in}{\includegraphics[width=1in]{deltabar.eps}} & = & - \parbox{0.9in}{\includegraphics[width=0.9in]{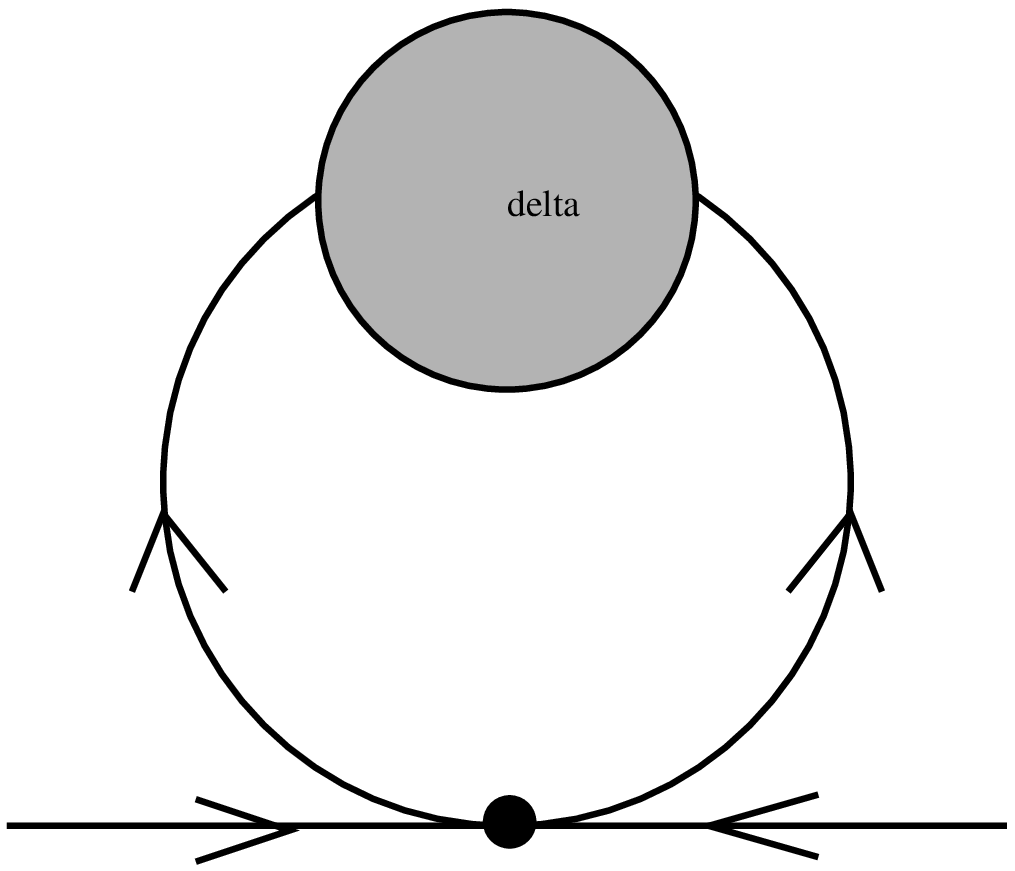}} -  \sum_{\mbox{\small $\vq_1 - \vq_2 + \vq_3 = \vq$}} \parbox{1.25in}{\includegraphics[width=1.25in]{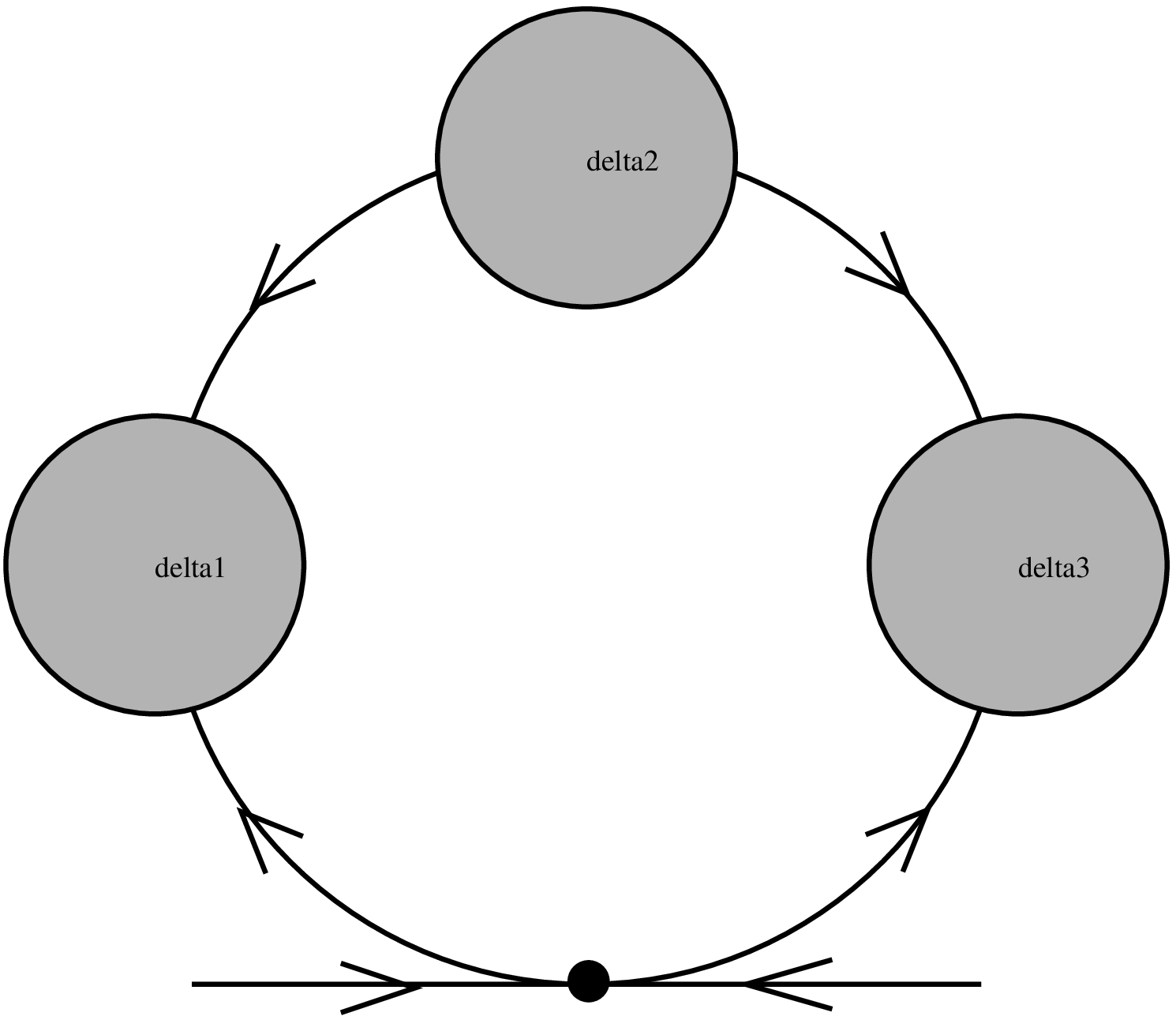}} \nonumber \\
 & & - \sum_{\mbox{\small $\vq_1 - \vq_2 + \vq_3 - \vq_4 + \vq_5 = \vq$}} \parbox{1.4in}{\includegraphics[width=1.4in]{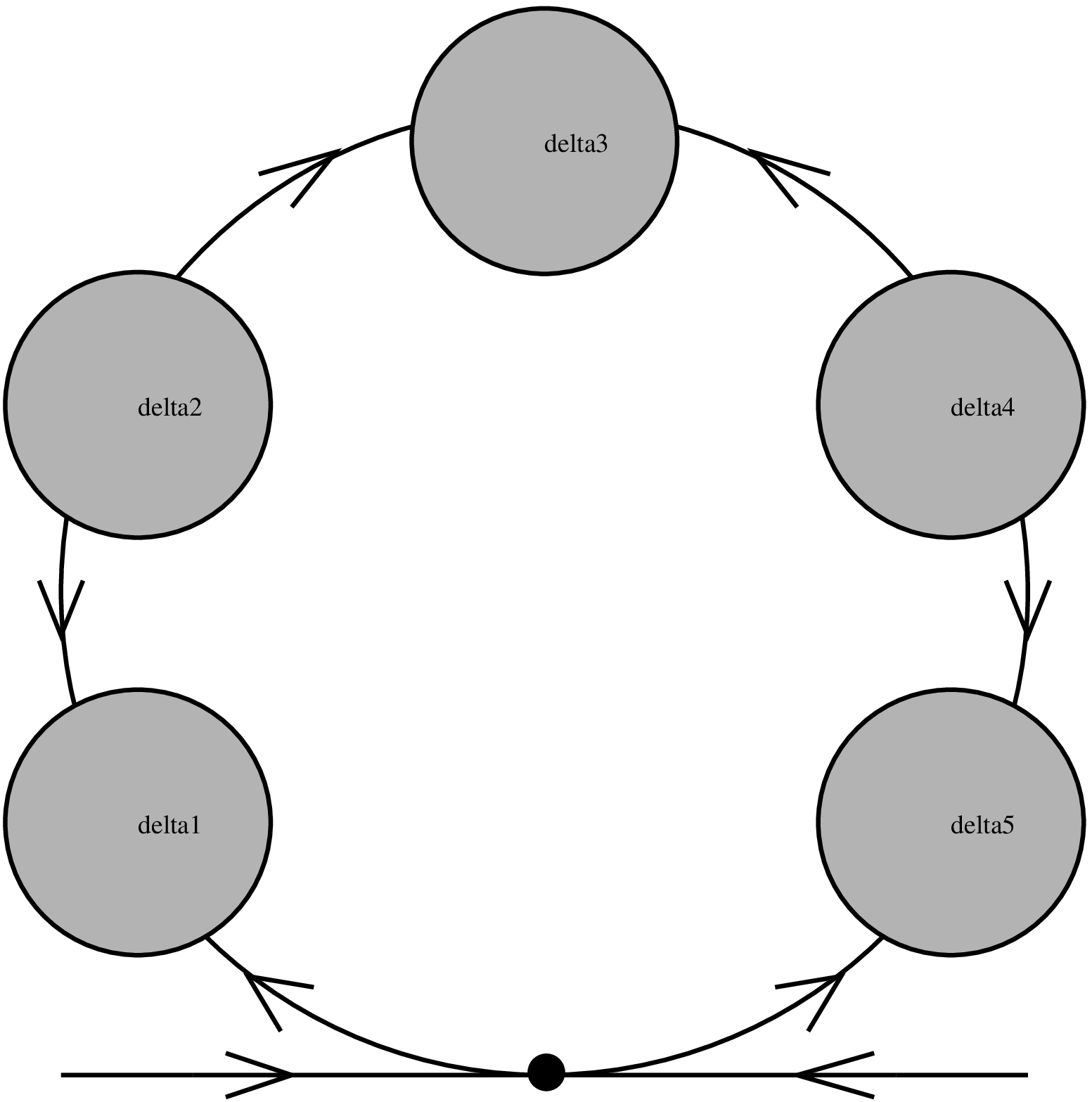}} + \cdots \nonumber
\end{eqnarray}
\caption{
\label{gapeqnfig}
The series expansion corresponding to Eqs.~(\ref{fullgapeqn})
and (\ref{integrals}).  
This diagrammatic equation is obtained 
by substituting the series expansion of Fig.~\ref{expansionfig} 
into the Schwinger-Dyson equation of Fig.~\ref{sdeqnfig}.  
}
\end{figure}

\noindent
as shown in Fig.~\ref{gapeqnfig}.  The prefactors have been chosen
for later convenience.  The functions $\Pi$, $J$, and $K$
corresponding to the three graphs in Fig.~\ref{gapeqnfig}
are given by:
\begin{eqnarray}
\label{integrals}
\Pi(\vq) & = & -i \frac{\pi^2}{\bar\mu^2} \int \frac{d^4 p}{(2\pi)^4} \gamma_\mu (\pslash - \muslash_d)^{-1} (\pslash + 2 \qslash + \muslash_u)^{-1} \gamma^\mu \nonumber \\
J(\vq_1\vq_2\vq_3\vq_4) & = & -i\frac{\pi^2}{\bar\mu^2} \int \frac{d^4 p}{(2\pi)^4} \gamma_\mu (\pslash - \muslash_d)^{-1} (\pslash + 2 \qslash_1 + \muslash_u)^{-1} \nonumber \\
 & & \times (\pslash + 2 \qslash_1 - 2\qslash_2 - \muslash_d)^{-1} (\pslash + 2 \qslash_1 - 2\qslash_2 + 2\qslash_3 + \muslash_u)^{-1} \gamma^\mu \nonumber \\ 
K(\vq_1\vq_2\vq_3\vq_4\vq_5\vq_6) & = & -i\frac{\pi^2}{\bar\mu^2} \int \frac{d^4 p}{(2\pi)^4} \gamma_\mu (\pslash - \muslash_d)^{-1} (\pslash + 2 \qslash_1 + \muslash_u)^{-1} \nonumber \\
 & & \times (\pslash + 2 \qslash_1 - 2\qslash_2 - \muslash_d)^{-1} (\pslash + 2 \qslash_1 - 2\qslash_2 + 2\qslash_3 + \muslash_u)^{-1}  \nonumber \\
 & & \times  (\pslash + 2 \qslash_1 - 2\qslash_2 +2\qslash_3 - 2\qslash_4 - \muslash_d)^{-1} \nonumber \\
 & & \times (\pslash + 2 \qslash_1 - 2\qslash_2 + 2\qslash_3 - 2\qslash_4 + 2\qslash_5 + \muslash_u)^{-1} \gamma^\mu. \nonumber \\
\end{eqnarray}

We shall see that $\dm$ and $|\vq|$ are both of order $\Delta$
which in turn is of order $\Delta_0$.  This means that all these
quantities are much less than
$\bar\mu$ in the weak coupling limit.
Thus, in the weak coupling limit we can choose the cutoff $\omega$
such that $\dm, |\vq| \ll \omega \ll \bar\mu$. In this
limit, $J$ and $K$ are independent of the cutoff $\omega$,
as we shall see in appendix \ref{JKappendix} where we present their
explicit evaluation.  In this limit, 
\begin{eqnarray}
\label{Piandalpha}
\Pi(\vq) &=& 
\left[-1+ \frac{\dm}{2|\vq|} \log\left( \frac{|\vq| + \dm}{|\vq| - \dm}\right)
- \frac{1}{2}\log \left( \frac{\omega^2}{\vq^2 - \dm^2} \right) \right]
\nonumber\\
&=& -\frac{\pi^2}{2\lambda\bar\mu^2} +  
\left[-1+ \frac{\dm}{2|\vq|} \log\left( \frac{|\vq| + \dm}{|\vq| - \dm}\right)
- \frac{1}{2}\log \left( \frac{\Delta_0^2}{4(\vq^2 - \dm^2)} \right) \right]
\nonumber\\
&=& -\frac{\pi^2}{2\lambda\bar\mu^2} +
\alpha\left(\frac{|\vq|}{\Delta_0},\frac{\dm}{\Delta_0}\right)\ ,
\end{eqnarray}
where we have used the explicit solution to the
BCS gap equation (\ref{bcsgapeqn}) to eliminate the cutoff $\omega$
in favor of the BCS gap $\Delta_0$, and where the last equation
serves to define $\alpha$.  Note that $\alpha$
depends on the cutoff $\omega$ only through $\Delta_0$,
and depends only on the
ratios $|\vq|/\Delta_0$ and $\dm/\Delta_0$.

It will prove convenient to use the definition of $\alpha$ to
rewrite the Ginzburg-Landau
equation 
Eq.~(\ref{fullgapeqn}) as
\begin{eqnarray}
\label{newfullgapeqn}
0 & = & \alpha(|\vq|)\Delta_\vq^* \,+
\sum_{\vq_1,\vq_2,\vq_3} J(\vq_1\vq_2\vq_3\vq) 
\Delta_{\vq_1}^* \Delta_{\vq_2} \Delta_{\vq_3}^* 
\delta_{\vq_1-\vq_2+\vq_3-\vq} 
\nonumber \\   
&  & + \sum_{\vq_1,\vq_2,\vq_3,\vq_4,\vq_5}K(\vq_1\vq_2\vq_3\vq_4\vq_5\vq) 
\Delta_{\vq_1}^* \Delta_{\vq_2} \Delta_{\vq_3}^* 
\Delta_{\vq_4} \Delta_{\vq_5}^* \delta_{\vq_1-\vq_2+\vq_3-\vq_4+\vq_5-\vq} 
\nonumber \\
&  & + \mathcal{O}(\Delta^7).
\end{eqnarray}

\begin{figure}[t]
\centering
\psfrag{0.2}[tc][tc]{\small 0.2}
\psfrag{0.4}[tc][tc]{\small 0.4}
\psfrag{0.6}[tc][tc]{\small 0.6}
\psfrag{0.8}[tc][tc]{\small 0.8}
\psfrag{1}[tc][tc]{\small 1.0}
\psfrag{1.2}[tc][tc]{\small 1.2}
\psfrag{1.4}[tc][tc]{\small 1.4}
\psfrag{1.6}[tc][tc]{\small 1.6}
\psfrag{y1}[rc][rc]{\small 0.2}
\psfrag{y2}[rc][rc]{\small 0.4}
\psfrag{y3}[rc][rc]{\small 0.6}
\psfrag{y4}[rc][rc]{\small 0.8}
\psfrag{y5}[rc][rc]{\small 1.0}
\psfrag{xlabel}{$|\vq|/\Delta_0$}
\psfrag{ylabel}{$\dm/\Delta_0$}
\psfrag{ag0}{$\alpha>0$}
\psfrag{ae0}{$\alpha=0$}
\psfrag{al0}{$\alpha<0$}
\psfrag{qline}{$|\vq|=1.200\dm$}
\psfrag{0.754}[rc][rc]{0.754}
\psfrag{0.905}[tc][tc]{0.905}
\includegraphics[width=4.5in]{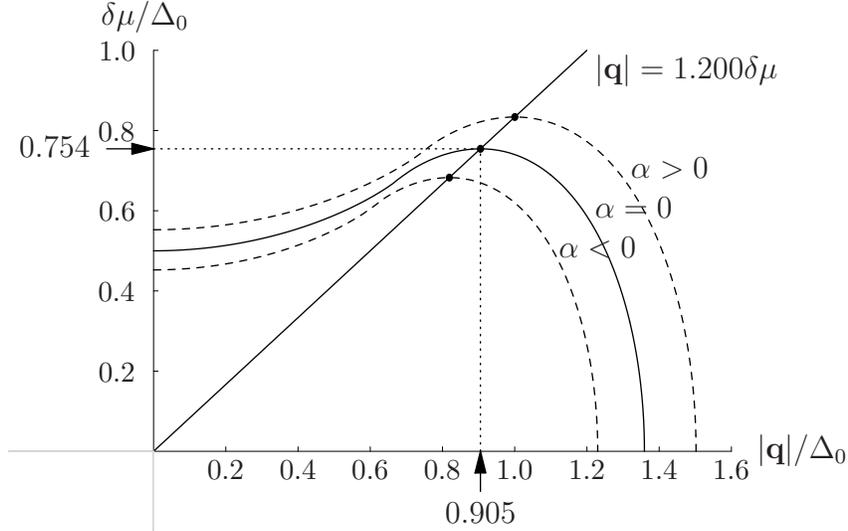}
\caption{
\label{zerogapfig}
Along the solid curve,
$\alpha(|\vq|,\dm)=0$.  The maximum
$\dm$ reached by this curve is $\dm=0.754\Delta_0\equiv\dm_2$,
which occurs at $|\vq|=0.9051\Delta_0=1.1997\dm_2$.  Along the
upper and lower dashed curves, $\alpha=+0.1$  and $\alpha=-0.1$,
respectively.
}
\end{figure}

To learn how to interpret $\alpha$, consider 
the single plane-wave condensate in which $\Delta_\vq\neq 0$
only for a single $\vq$.
If we divide equation (\ref{fullgapeqn}) by $\Delta_\vq^*$, 
we see that the equation
$\Pi = -\pi^2/2\lambda\bar\mu^2$, 
which is to say $\alpha=0$, defines a curve
in the space of $(|\vq|,\dm)$ 
where we can find a solution to the gap equation with
$\Delta_\vq\rightarrow 0$, with $|\vq|$ on the curve and
for any $\hat{\bf q}$. 
This curve is shown in Fig.~\ref{zerogapfig}.  
We shall see below that when only one $\Delta_\vq$ is nonzero,
the ``$J$ sum'' and ``$K$ sum'' in (\ref{fullgapeqn}) are both
positive. This means that wherever $\alpha<0$, {\it i.e.} below
the solid curve in Fig.~\ref{zerogapfig}, there are solutions with 
$\Delta_\vq\neq 0$ for these values of $|\vq|$,
and wherever $\alpha>0$, {\it i.e.}~above the solid
curve, there are no single
plane-wave solutions to the gap equation.  
The solid curve in Fig.~\ref{zerogapfig} therefore 
marks the boundary of the instability towards the formation
of a single plane-wave condensate.
The highest point on this curve is special, as it denotes
the maximum value of $\dm$ for which a single plane-wave
LOFF condensate can arise. This second-order critical point
occurs at
$(|\vq|,\dm) = (q_0, \dm_2)$ with $\dm_2
\simeq 0.7544 \Delta_0$ and $q_0/\dm_2 \simeq 1.1997$, where
$\Delta_0$ is the BCS gap of Eq.~(\ref{bcsgapeqn}).  

As $\dm\rightarrow\dm_2$ from above, only those plane waves
lying on a sphere in momentum space with $|\vq|=q_0$ are becoming 
unstable to condensation.  If we analyze them one by one,
all these plane waves are equally unstable. That is, 
in the
vicinity of the critical point $\dm_2$, 
the LOFF gap equation admits plane-wave
condensates with $\Delta_\vq\neq 0$ for a single $\vq$ lying
somewhere on the sphere
$|\vq| = q_0$. For each such plane wave, the
paired quarks occupy a ring with opening angle   
$\psi_0 = 2 \cos^{-1}(\dm/|\vq|) \simeq 67.1^\circ$
on each Fermi surface, as shown
in Fig.~\ref{ringsfig}.

\subsection{The free energy}

In order
to compare different crystal structures,
with (\ref{newfullgapeqn}) in hand, 
we can now derive a 
Ginzburg-Landau free energy functional
$\Omega[\Delta(x)]$ which characterizes the system in the vicinity of
$\dm_2$, where $\Delta\rightarrow 0$. 
This is most readily obtained by noting that the 
gap equations (\ref{newfullgapeqn}) must be equivalent to 
\begin{equation}
\frac{\partial\Omega}{\partial\Delta_{\vq}} = 0
\end{equation}
because solutions to the gap equations are stationary points
of the free energy.  This determines the free energy
up to an overall multiplicative constant, which can be found by
comparison with the single plane-wave solution previously known.  
The
result is
\begin{eqnarray} 
\label{freeenergy}
\frac{\Omega}{N_0} & = &   \alpha(q_0) 
\sum_{\vq,\ |\vq|=q_0} \Delta_\vq^* \Delta_\vq + \frac{1}{2} \sum_{\vq_1\cdots\vq_4, \ |\vq_i|=q_0} J(\vq_1\vq_2\vq_3\vq_4) \Delta_{\vq_1}^* \Delta_{\vq_2} \Delta_{\vq_3}^* \Delta_{\vq_4} \delta_{\vq_1-\vq_2+\vq_3-\vq_4} \nonumber \\
 & & + \frac{1}{3} \sum_{\vq_1\cdots\vq_6, \ |\vq_i|=q_0} K(\vq_1\vq_2\vq_3\vq_4\vq_5\vq_6) \Delta_{\vq_1}^* \Delta_{\vq_2} \Delta_{\vq_3}^* \Delta_{\vq_4} \Delta_{\vq_5}^* \Delta_{\vq_6} \delta_{\vq_1-\vq_2+\vq_3-\vq_4+\vq_5-\vq_6} \nonumber \\
 & & + \mathcal{O}(\Delta^8) 
\end{eqnarray}
where $N_0 = 2\bar\mu^2/\pi^2$ and where we have restricted
our attention to modes with $|\vq|=q_0$, as we now explain.
Note that in the vicinity of $|\vq|=q_0$  and $\dm=\dm_2$,
\begin{equation}
\alpha \approx \left( \frac{\dm - \dm_2}{\dm_2} \right)\ .
\end{equation}
We see that for $\alpha>0$ (that is, for $\dm>\dm_2$) $\Delta_\vq=0$
is stable whereas for $\alpha<0$ (that is, $\dm<\dm_2$), the LOFF
instability sets in.  In the limit $\alpha\rightarrow 0^-$, only those
plane waves on the sphere $|\vq| = q_0$ are unstable.  For this reason
we only include these plane waves in the expression (\ref{freeenergy})
for the free energy.  Notice that equation (\ref{freeenergy}), which
we have derived starting from the gap equations (\ref{psdeqn}),
is the same as equation (\ref{GLpotential}) which was obtained from
generic arguments of translational and rotational symmetry.  The added
power of our gap equation derivation is that it enables us to
calculate the Ginzburg-Landau coefficients $\alpha$, $J$, and $K$ from
the microscopic theory.

We shall do most of our analysis in the vicinity of
$\dm=\dm_2$, where we choose $|\vq|=q_0=1.1997\dm_2$ 
as just described.
However, we shall also
want to apply our results at $\dm>\dm_2$.
At these values of
$\dm$, we shall choose $|\vq|$ in such a way as to minimize 
$\alpha(|\vq|)$, because this minimizes the quadratic term in 
the free energy $\Omega$ and thus minimizes the free
energy in the vicinity of $\Delta\rightarrow 0$, which
is where the Ginzburg-Landau analysis is reliable.
As Fig.~\ref{zerogapfig} indicates, for any given $\dm$
the minimum value of $\alpha$ is to be found at $|\vq|=1.1997 \dm$.
Therefore, when we apply Eq.~(\ref{freeenergy}) 
away from $\dm_2$, we shall set $q_0=1.1997 \dm$, just as at 
$\dm=\dm_2$. As a consequence, the opening angle of the
pairing rings, $\psi_0=2\cos^{-1}(\dm/|\vq|)$, is unchanged
when we move away from $\dm=\dm_2$.

%------------------------------------------------------------------------

\section{Results}
\label{sec:multiplewaveresults}

\subsection{Generalities}

All of the modes on the sphere
$|\vq|=q_0$ become unstable
at $\dm=\dm_2$.  The quadratic term in the free energy
includes no interaction between modes with different $\vq$'s,
and so predicts that $\Delta_\vq\neq 0$ for all modes
on the sphere. 
Each plane-wave mode corresponds to a ring of paired
quarks on each Fermi surface, so we would obtain
a cacophony of multiple overlapping rings,
favored by the quadratic term because this allows more 
and more of the quarks
near their respective Fermi surfaces to pair. 
Moving beyond lowest order,
our task is to evaluate the quartic
and sextic terms in the free energy.
These higher order terms 
characterize the effects of interactions
between $\Delta_\vq$'s with differing $\vq$'s
(between the different pairing rings) 
and thus determine how condensation in one mode
enhances or deters condensation in other modes.
The results we shall present rely on our ability
to evaluate $J$ and $K$, defined in Eqs.~(\ref{integrals}).
We describe the methods we use to
evaluate these expressions in appendix \ref{JKappendix}
and focus here on 
describing and understanding the
results.
We shall see, for example, that although the quadratic
term favors adding
more rings, the higher order terms strongly disfavor
configurations in which $\Delta_\vq$'s corresponding
to rings that intersect          are nonzero. 
Evaluating the quartic and sextic terms in the free energy will
enable us to evaluate the free energy of 
condensates with various configurations of several plane waves and thereby
discriminate between candidate crystal structures.

A given crystal structure can be described by a set of vectors
$\mathcal{Q} = \{ \vq_a, \vq_b, \cdots \}$, specifying which plane
wave modes are present in the condensate, and a set of gap parameters
$\{ \Delta_{\vq_a}, \Delta_{\vq_b}, \cdots \}$, indicating the
amplitude of condensation in each of the modes.  Let us define
$\mathcal{G}$ as the group of proper and improper rotations that
preserve the set $\mathcal{Q}$.  We make the assumption that
$\mathcal{G}$ is also the point group of the crystal itself; this
implies that $\Delta_{\vq} = \Delta_{\vq'}$ if $\vq'$ is in the orbit
of $\vq$ under the group action.  For most (but not all) of the
structures we investigate, $\mathcal{Q}$ has only one orbit and
therefore all of the $\Delta_\vq$'s are equal.

\begin{figure}
\begin{center}
\psfrag{q1}{$\vq_1$}
\psfrag{q2}{$\vq_2$}
\psfrag{q3}{$\vq_3$}
\psfrag{q4}{$\vq_4$}
\psfrag{q5}{$\vq_5$}
\psfrag{q6}{$\vq_6$}
\psfrag{psi}{$\psi$}
\psfrag{chi}{$\chi$}
%\parbox{1.75in}{(a)} \hspace{0.75in} \parbox{2in}{(b)}
\parbox{1.75in}{\includegraphics[width=1.75in]{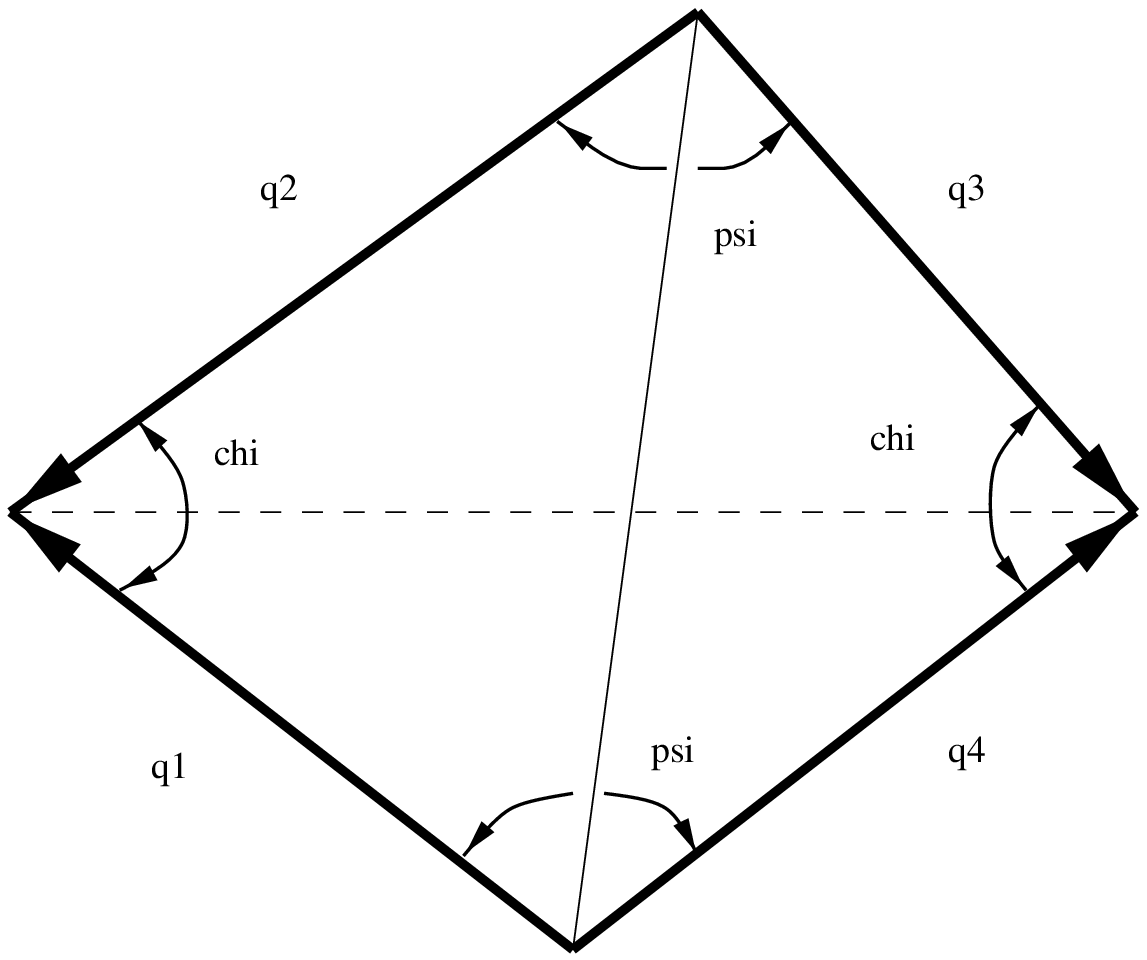}} 
\hspace{0.75in}
\parbox{2in}{\includegraphics[width=2in]{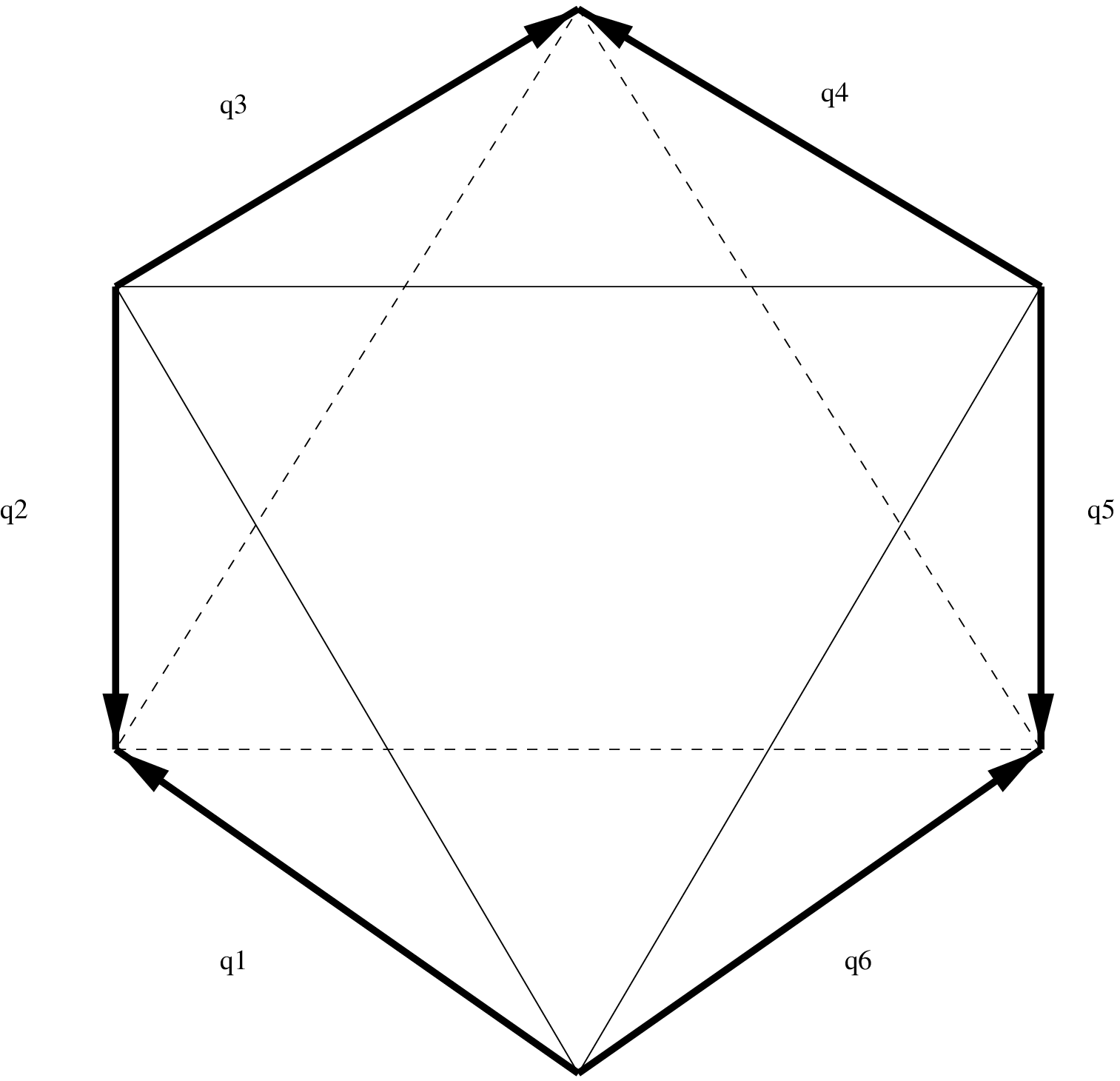}} 
%\parbox{5in}{(c)} \\
%\parbox{3.5in}{\includegraphics[width=4in]{twowaveshapes.eps}}
%\includegraphics[width=2.8in]{rhombus.eps}
%\parbox{4.5in}{\includegraphics[width=3in]{gammafig.eps}} 
%\includegraphics[width=4in]{gammafig.eps} 
\end{center}
\caption{
\label{rhombushexagonfig}
Rhombic and hexagonal combinations of $\vq$'s.  On the left is a rhombus
with $\vq_1-\vq_2+\vq_3-\vq_4 = 0$.  On the right is a hexagon
with $\vq_1-\vq_2+\vq_3-\vq_4+\vq_5-\vq_6 = 0$.  The edges have equal
lengths ($|\vq_i| = q_0$). The shapes are in general nonplanar.
}
\end{figure}

For a given set $\mathcal{Q}$, the quartic term in the free energy
(\ref{freeenergy}) is a sum over all combinations of four 
$\vq$'s that
form closed ``rhombuses'', as shown in Fig.~\ref{rhombushexagonfig}.  
The four $\vq$'s are chosen from the set $\mathcal{Q}$
and they need not be distinct.
By a rhombus we mean a closed figure composed of four equal
length vectors which will in general be nonplanar.
A rhombus  is therefore characterized by two
internal angles $(\psi,\chi)$ with the constraint $0 \leq \psi + \chi
\leq \pi$.  Each shape corresponds to a value of the $J$ function (as
defined in equations (\ref{integrals})); the rotational invariance of
the $J$ function implies that congruent shapes give the same value and
therefore $J(\vq_1\vq_2\vq_3\vq_4) = J(\psi,\chi)$.  So, each unique
rhombic combination of $\vq$'s in the set $\mathcal{Q}$
that characterizes a given crystal structure
yields a unique contribution to the quartic coefficient in
the Ginzburg-Landau 
free energy of that crystal
structure.  The continuation to next order is
straightforward: the sextic term in the free
energy (\ref{freeenergy}) is a sum
over all combinations of six $\vq$'s that form closed ``hexagons'', as
shown in Fig.~\ref{rhombushexagonfig}.  Again these shapes are
generally nonplanar and each unique hexagonal combination of $\vq$'s
yields a unique value of the $K$ function and a unique 
contribution to the sextic
coefficient in the Ginzburg-Landau free energy of the crystal.

As we previewed in chapter 1, when all of the $\Delta_\vq$'s are
equal, we can evaluate aggregate quartic and sextic coefficients
$\beta$ and $\gamma$, respectively, as sums over all rhombic and
hexagonal combinations of the $\vq$'s in the set $\mathcal{Q}$:
\begin{equation}
\label{squarehexagonsums}
\beta = \sum_\square J(\square), \ \gamma =  \sum_{\hexagon}  K(\hexagon).
\end{equation}
Then, for a crystal with $P$ plane waves, the free energy has the simple
form
\begin{equation}
\frac{\Omega(\Delta)}{N_0} =  P \alpha \Delta^2 + \frac{1}{2} \beta \Delta^4 + \frac{1}{3} \gamma \Delta^6 + \mathcal{O}(\Delta^8)
\end{equation}
and we can analyze a candidate crystal structure by calculating the
coefficients $\beta$ and $\gamma$ and studying the resultant form of
the free energy function.  

If $\beta$ and $\gamma$ are both positive,
a second-order phase transition occurs at $\alpha = 0$; near the
critical point the value of $\gamma$ is irrelevant and the
minimum energy solution is
\begin{equation}
\Delta = \left( \frac{P |\alpha|}{\beta} \right)^{\frac{1}{2}}, \ \ \ \ 
\frac{\Omega}{N_0} = -\frac{P^2 \alpha^2}{2 \beta}\ , 
\end{equation}
for $\dm\leq \dm_2$ ({\it i.e.}~$\alpha\leq 0$). 

If $\beta$ is negative and $\gamma$ is positive, the phase transition
is in fact
first order and occurs at a new critical point defined by
\begin{equation}
\alpha =\alpha_* = \frac{3\beta^2}{16 P \gamma}\ .
\label{alphastar}
\end{equation}
In order to find the $\dm_*$ corresponding to $\alpha_*$,
we need to solve 
\begin{equation}
\alpha\left(|\vq|,\dm_*\right)=\alpha\left(1.1997\dm_*,\dm_*\right)
=\alpha_*= \frac{3\beta^2}{16 P \gamma}\ .
\end{equation}
Since $\alpha_*$ is positive, the critical point
$\dm_*$ at which the first-order phase
transition occurs
is larger than $\dm_2$. 
If $\alpha_*$ is small, then $\dm_*\simeq (1+\alpha_*)\dm_2$.
Thus, a crystalline
color superconducting state whose crystal structure
yields a negative $\beta$ and positive $\gamma$
persists as a possible ground state even above $\dm_2$,
the maximum $\dm$ at which the plane-wave
state is possible.  At the first-order critical
point (\ref{alphastar}),
the free energy has degenerate minima at 
\begin{equation}
\Delta = 0\ ,\ \ \ \ \Delta = \left(\frac{3|\beta|}{4\gamma}\right)
^{1/2}\ .
\end{equation}
If we reduce $\dm$ below $\dm_*$, the minimum with $\Delta\neq 0$
deepens.  Once $\dm$ is reduced to the point at which
the single plane wave would just be starting to form with
a free energy infinitesimally below zero, the free energy
of the crystal structure with negative $\beta$ and positive
$\gamma$ has
\begin{equation}
\Delta = \left( \frac{|\beta|}{\gamma} \right)^{\frac{1}{2}}\ ,\ \ \ \ 
\frac{\Omega}{N_0} = - \frac{|\beta|^3 }{6 \gamma^2} 
\label{DeltaOmegaatdm2}
\end{equation}
at $\dm=\dm_2$.

Finally, if $\gamma$ is negative, the order $\Delta^6$ 
Ginzburg-Landau free energy is unbounded from below. 
In this circumstance, we know that we have found a first
order phase transition but we do not know at what $\dm_*$
it occurs, because the stabilization of the Ginzburg-Landau
free energy at large $\Delta$ must come about at order
$\Delta^8$ or higher.

\subsection{One wave}

With these general considerations in mind we now proceed to look at 
specific examples of crystal structures.  We begin with
the
single plane-wave condensate ($P=1$).  The quartic coefficient
of the free energy is 
\begin{equation}
\label{J0eqn}
\beta = J_0 = J(0,0) = \frac{1}{4} \frac{1}{\vq^2-\dm^2} \simeq 
+ \frac{0.569}{\dm^2},
\end{equation}
and the sextic coefficient is 
\begin{equation}
\label{K0eqn}
\gamma = K_0 = K(\vq\vq\vq\vq\vq\vq) = \frac{1}{32} \frac{\vq^2 + 3 \dm^2}{(\vq^2 - \dm^2)^3} \simeq + \frac{1.637}{\dm^4},
\end{equation}
yielding a second-order phase transition at $\alpha = 0$.
These coefficients agree with those 
obtained by expanding the all-orders-in-$\Delta$
solution for the single plane wave which can be obtained
by variational methods~\cite{FF,Takada1,BowersLOFF}
or by starting from 
(\ref{planewave2}), as in Refs.~\cite{ngloff,massloff}.
The coefficient $\beta$ in (\ref{J0eqn}) was 
first found by Larkin and Ovchinnikov~\cite{LO}.

\subsection{Two waves}

\begin{figure}[t]
\centering
\psfrag{psi}{$\psi$}
\parbox{3.5in}{\includegraphics[width=4in]{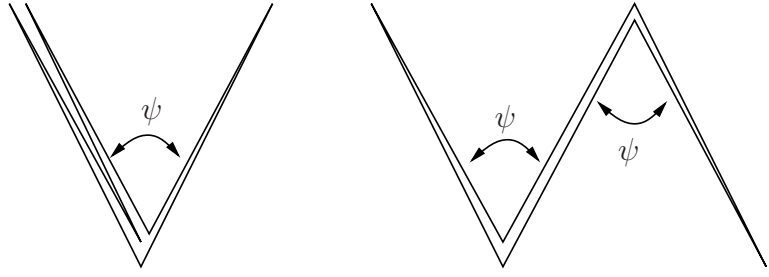}}
\caption{
\label{k1k2fig}
Two different ``hexagonal'' shapes (as in Fig.~\ref{rhombushexagonfig}) 
that can be constructed from
two vectors $\vq_a$ and $\vq_b$. These shapes correspond to the
functions $K_1(\psi)$ and $K_2(\psi)$ in Eq.~(\ref{gammaeqn}).  
}
\end{figure}

\begin{figure}[t]
\centering
\psfrag{xlabela}{$\psi$}
\psfrag{ylabela}{$\beta(\psi)\dm^2$}
\psfrag{xlabelb}{$\psi$}
\psfrag{ylabelb}{$\gamma(\psi)\dm^4$}
\psfrag{30}[tc][tc]{\small $30^{\circ}$}
\psfrag{60}[tc][tc]{\small $60^{\circ}$}
\psfrag{90}[tc][tc]{\small $90^{\circ}$}
\psfrag{120}[tc][tc]{\small $120^{\circ}$}
\psfrag{150}[tc][tc]{\small $150^{\circ}$}
\psfrag{180}[tc][tc]{\small $180^{\circ}$}
\psfrag{y1}[rc][rc]{\small $-2$}
\psfrag{y2}[rc][rc]{\small $2$}
\psfrag{y3}[rc][rc]{\small $4$}
\psfrag{y4}[rc][rc]{\small $6$}
\psfrag{y5}[rc][rc]{\small $8$}
\psfrag{y6}[rc][rc]{\small $10$}
\psfrag{y7}[rc][rc]{\small $10$}
\psfrag{y8}[rc][rc]{\small $20$}
\psfrag{y9}[rc][rc]{\small $30$}
\psfrag{y10}[rc][rc]{\small $40$}
\psfrag{y11}[rc][rc]{\small $50$}
\psfrag{psi0labela}{$\psi = \psi_0 \simeq 67.1^{\circ}$}
\psfrag{psi0labelb}{$\psi = \psi_0$}
\psfrag{arrowlabela}{$\beta(\psi_0) \simeq -1.138/\dm^2$}
\psfrag{arrowlabel}{$\gamma(\psi_0) \simeq +0.249/\dm^4$}
%\parbox{4.5in}{\includegraphics[width=3in]{betafig.eps}} 
\parbox{4in}{\includegraphics[width=4in]{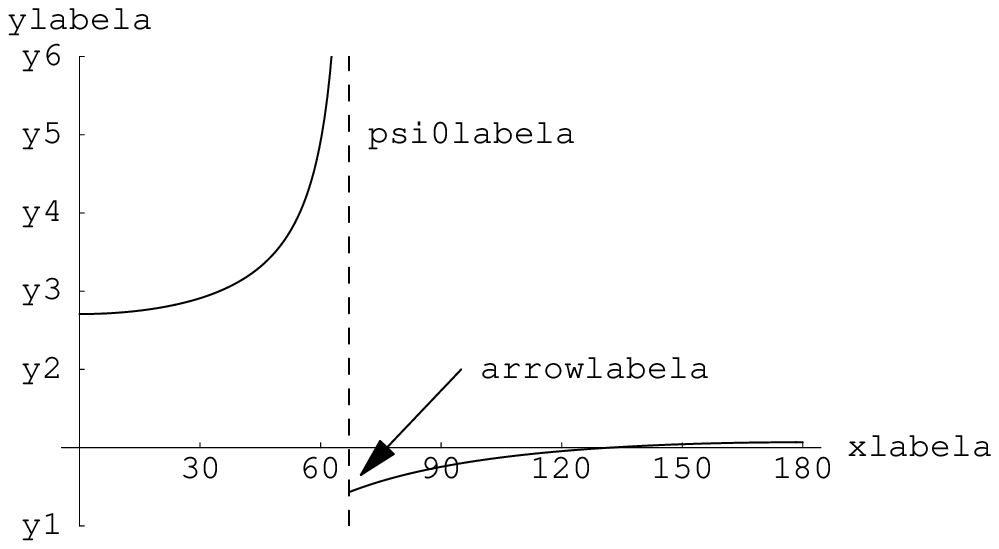}} \\
\parbox{4in}{\phantom{spacer}} \\
%\hspace{0.25in}
%\parbox{4.5in}{\includegraphics[width=3in]{gammafig.eps}} 
\parbox{4in}{\includegraphics[width=4in]{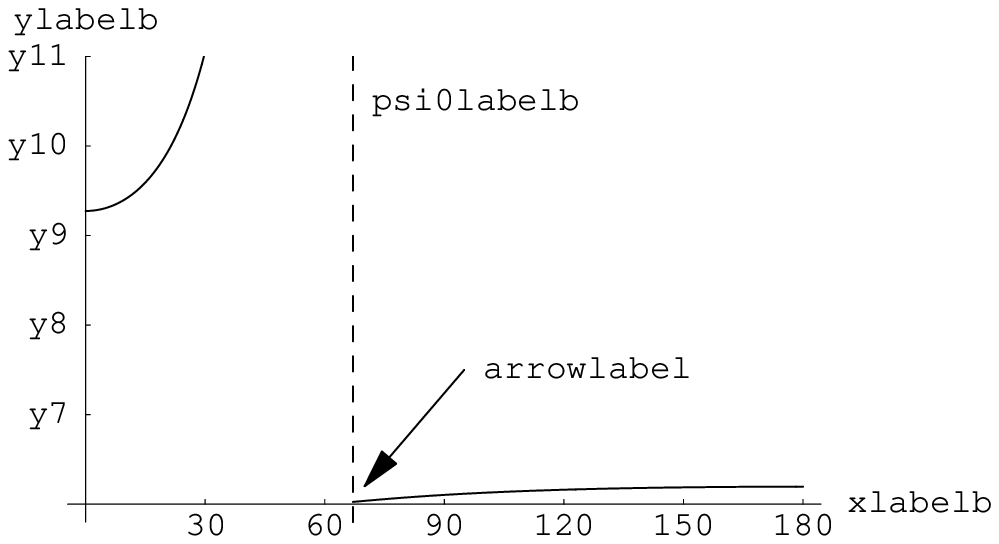} }
\caption{
\label{betagammafig}
$\beta(\psi)$
and $\gamma(\psi)$, the quartic and sextic coefficients in 
the Ginzburg-Landau free energy 
for a condensate consisting of two plane waves whose wave
vectors define an angle $\psi$.  
%The minimum value of $\gamma$ is $\gamma(\psi_0) \simeq 0.249/\dm_2^4$.
}
\end{figure}

Our next example is a condensate of two plane waves ($P=2$) with wave vectors
$\vq_a$ and $\vq_b$ and equal gaps $\Delta_{\vq_a} = \Delta_{\vq_b} =
\Delta$.  The most symmetrical arrangement is an antipodal pair
($\vq_b = -\vq_a$), which yields a cosine spatial variation
$\Delta(\vx) \sim \cos(2\vq_a\cdot\vx)$.  We will find it useful, however,
to study the generic case where $\vq_a$ and $\vq_b$ have the
same magnitude but define an
arbitrary angle $\psi$.  We find that the quartic coefficient
is
\begin{equation}
\beta(\psi) = 2 J_0 + 4 J(\psi,0) 
\end{equation}
and the sextic coefficient is 
\begin{equation}
\label{gammaeqn}
\gamma(\psi) = 2 K_0 + 12 K_1(\psi) + 6 K_2(\psi) 
\end{equation}
where $K_1(\psi) = K(\vq_a\vq_a\vq_a\vq_a\vq_b\vq_b)$ and $K_2(\psi) =
K(\vq_a\vq_a\vq_b\vq_a\vq_a\vq_b)$. ($K_1$
and $K_2$  arise from the ``hexagonal''
shapes shown in Fig.~\ref{k1k2fig}.)  The functions
$\beta(\psi)$ and $\gamma(\psi)$ are plotted in 
Fig.~\ref{betagammafig}.  
These functions manifest a number of interesting
features.  Notice that the functions are singular and discontinuous at
a critical angle $\psi = \psi_0 \simeq 67.1^\circ$, where $\psi_0$ is
the opening angle of a LOFF pairing ring on the Fermi surface.  For
the two-wave condensate we have two such rings, and the two rings are
mutually tangent when $\psi = \psi_0$.  For $\psi < \psi_0$, both
$\beta$ and $\gamma$ are large and positive, implying that an
intersecting ring configuration is energetically unfavorable.  For
$\psi > \psi_0$ the functions are relatively flat and small,
indicating some indifference towards any particular arrangement of the
nonintersecting rings.  There is a range of angles for which $\beta$
is negative and a first-order transition occurs (note that $\gamma$ is
always positive).  The favored arrangement is a pair of adjacent rings
that nearly intersect ($\psi = \psi_0 + \epsilon$).

It is unusual to find coefficients in a Ginzburg-Landau free energy
that behave discontinuously as a function of parameters describing the
state, as seen in Fig.~\ref{betagammafig}.  These discontinuities
arise because, as we described in the caption of Fig.~\ref{ringsfig},
we are taking two limits.  We first take a double scaling weak
coupling limit (as discussed on page \pageref{doublescaling}) in
which $\Delta_0$,~$\Delta$,~$\dm$,~$|\vq| \ll \omega\ll \bar\mu$ while
$\dm/\Delta_0$, $|\vq|/\Delta_0$ and $\Delta/\Delta_0$ (and thus the
angular width of the pairing bands) are held fixed.  Then, we take the
Ginzburg-Landau limit in which $\dm/\Delta_0 \rightarrow
\dm_2/\Delta_0$ and $\Delta/\Delta_0\rightarrow 0$ and the pairing
bands shrink to rings of zero angular width.  In the Ginzburg-Landau
limit, there is a sharp distinction between $\psi< 67.1^\circ$ where
the rings intersect and $\psi>67.1^\circ$ where they do not.  Without
taking the weak coupling limit, the plots of $\beta(\psi)$ and
$\gamma(\psi)$ would nevertheless look like smoothed versions of those
in Fig.~\ref{betagammafig}, smoothed on angular scales of order
$\dm^2/\omega^2$ (for $\beta$) and $\dm^4/\omega^4$ (for $\gamma$).
However, in the weak coupling limit these small angular scales are
taken to zero.  Thus, the double scaling limit sharpens what would
otherwise be distinctive but continuous features of the coefficients
in the Ginzburg-Landau free energy into discontinuities.

\subsection{Crystals}

From our analysis of the two-wave condensate, we can infer that for a
general multiple-wave condensate it is unfavorable to allow the
pairing rings to intersect on the Fermi surface.  For nonintersecting
rings, the free energy should be relatively insensitive to how the
rings are arranged on the Fermi surface.  However, 
Eqs.~(\ref{squarehexagonsums}) suggest that a combinatorial advantage is
obtained for exceptional structures that permit a large number of
rhombic and hexagonal combinations of wave vectors.  
That is, if there are many ways of picking four 
(not necessarily different) wave vectors
from the set of wave vectors that specify the crystal
structure for which $\vq_1-\vq_2+\vq_3-\vq_4 = 0$,
or if there are many ways of picking six wave vectors
for which $\vq_1 -\vq_2 +\vq_3 -
\vq_4 + \vq_5 - \vq_6 = 0$, such a crystal structure
enjoys a combinatorial advantage that will tend to make
the magnitudes of $\beta$ or $\gamma$ large. 
For a rhombic
combination $\vq_1-\vq_2+\vq_3-\vq_4 = 0$, the four $\vq$'s must be
the four vertices of a rectangle that is inscribed in a circle on the
sphere $|\vq| = q_0$. (The circle need not be a great circle, and the
rectangle can degenerate to a line or a point if the four $\vq$'s are
not distinct).  For a hexagonal combination $\vq_1 -\vq_2 +\vq_3 -
\vq_4 + \vq_5 - \vq_6 = 0$, the triplets $(\vq_1\vq_3\vq_5)$ and
$(\vq_2\vq_4\vq_6)$ are vertices of two inscribed triangles that have
a common centroid. In the degenerate case where only four of the six
$\vq$'s are distinct, the four distinct $\vq$'s must be the vertices of an
inscribed rectangle or an inscribed isosceles trapezoid for which one
parallel edge is twice the length of the other.  When five of the six
$\vq$'s are distinct, they can be arranged as a rectangle plus any fifth
point, or as five vertices of an inscribed cuboid
arranged as one antipodal pair plus the three corners adjacent to
one of the antipodes.

\begin{figure}[p]
\caption{
\label{stereographicfig}
Stereographic projections of the candidate crystal structures. 
The points ($\newmoon$) and circles ($\fullmoon$) are projections 
of $\vq$'s that are 
respectively above and below the equatorial plane of the 
sphere $|\vq| = q_0$.  
}
\vspace{0.1in}
\centering
\parbox{1.1in}{\centering point}
\parbox{1.1in}{\centering antipodal pair}
\parbox{1.1in}{\centering triangle}
\parbox{1.1in}{\centering tetrahedron}
\parbox{1.1in}{\centering square}
\parbox{1.1in}{\includegraphics[width=1.1in]{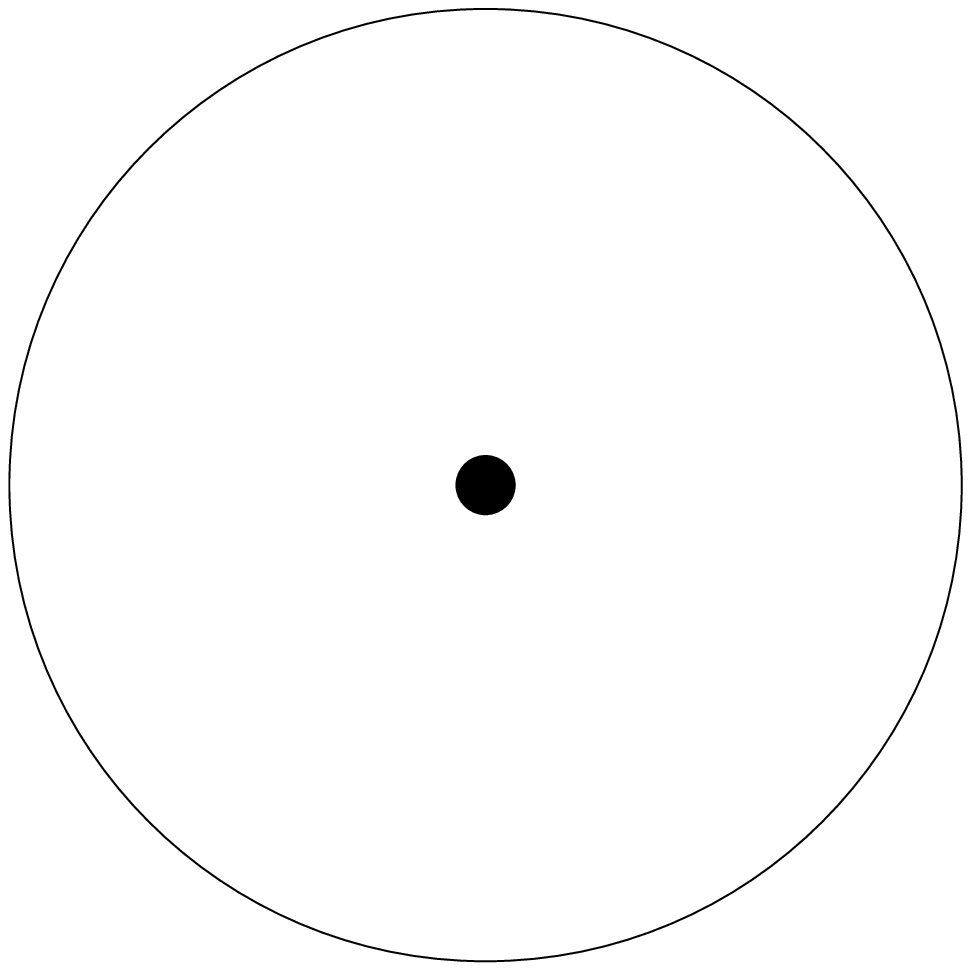}} 
\parbox{1.1in}{\includegraphics[width=1.1in]{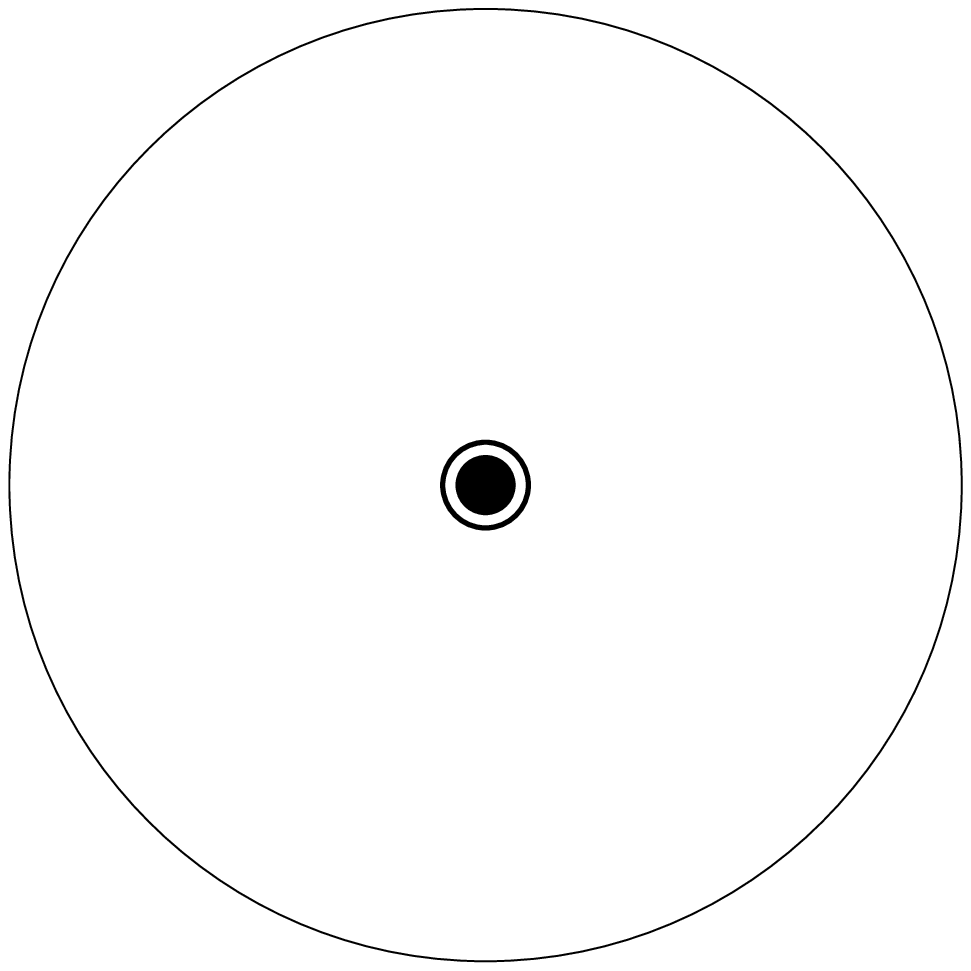}} 
\parbox{1.1in}{\includegraphics[width=1.1in]{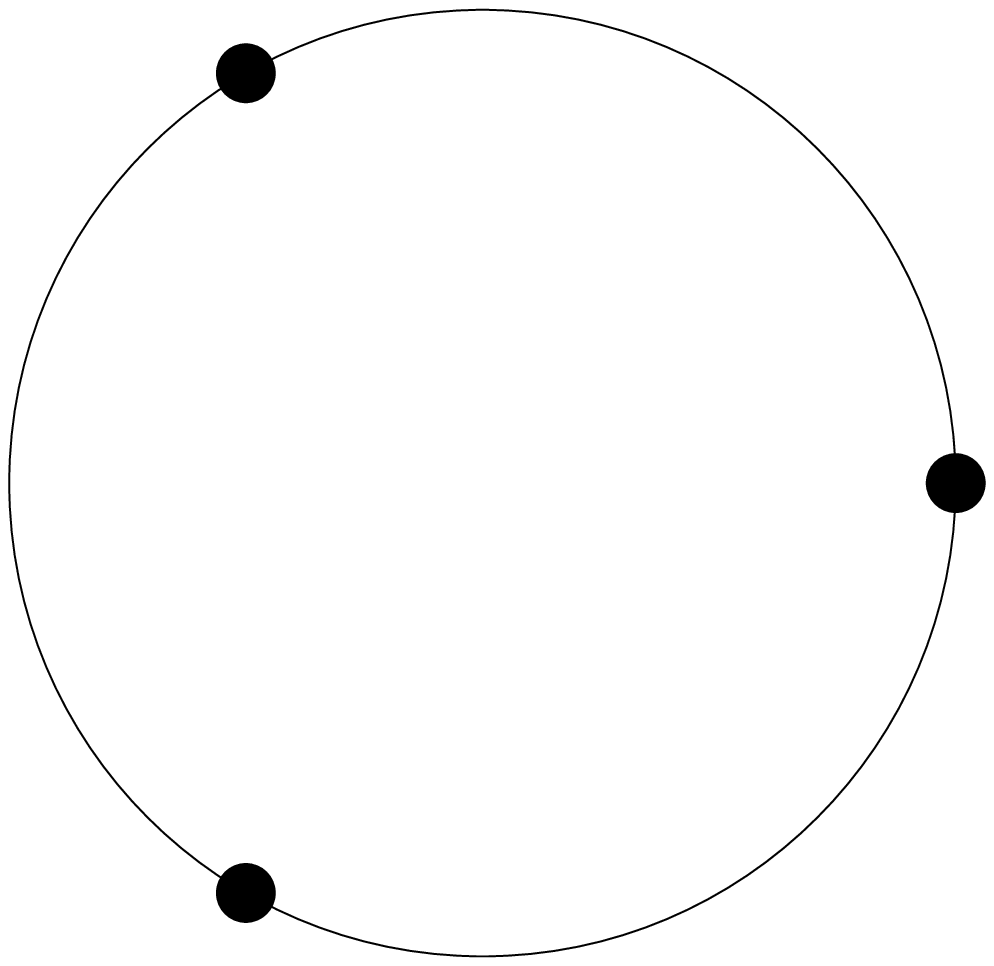}} 
\parbox{1.1in}{\includegraphics[width=1.1in]{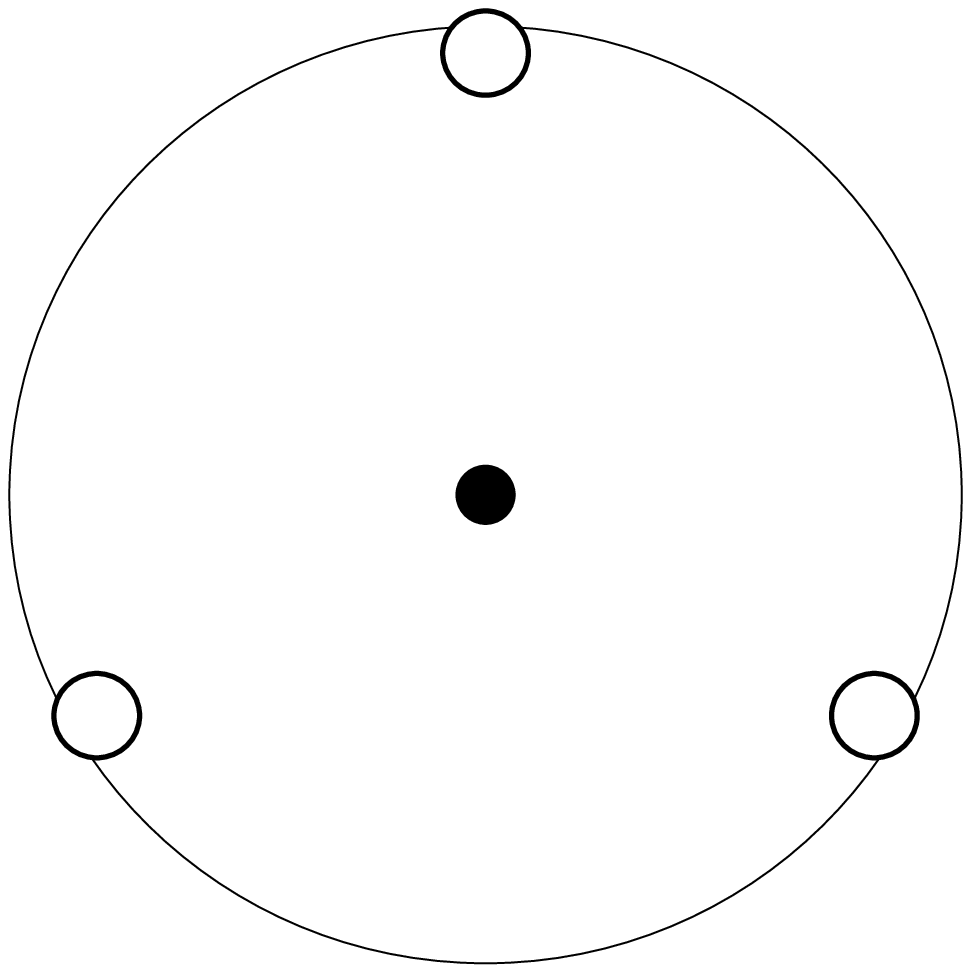}} 
\parbox{1.1in}{\includegraphics[width=1.1in]{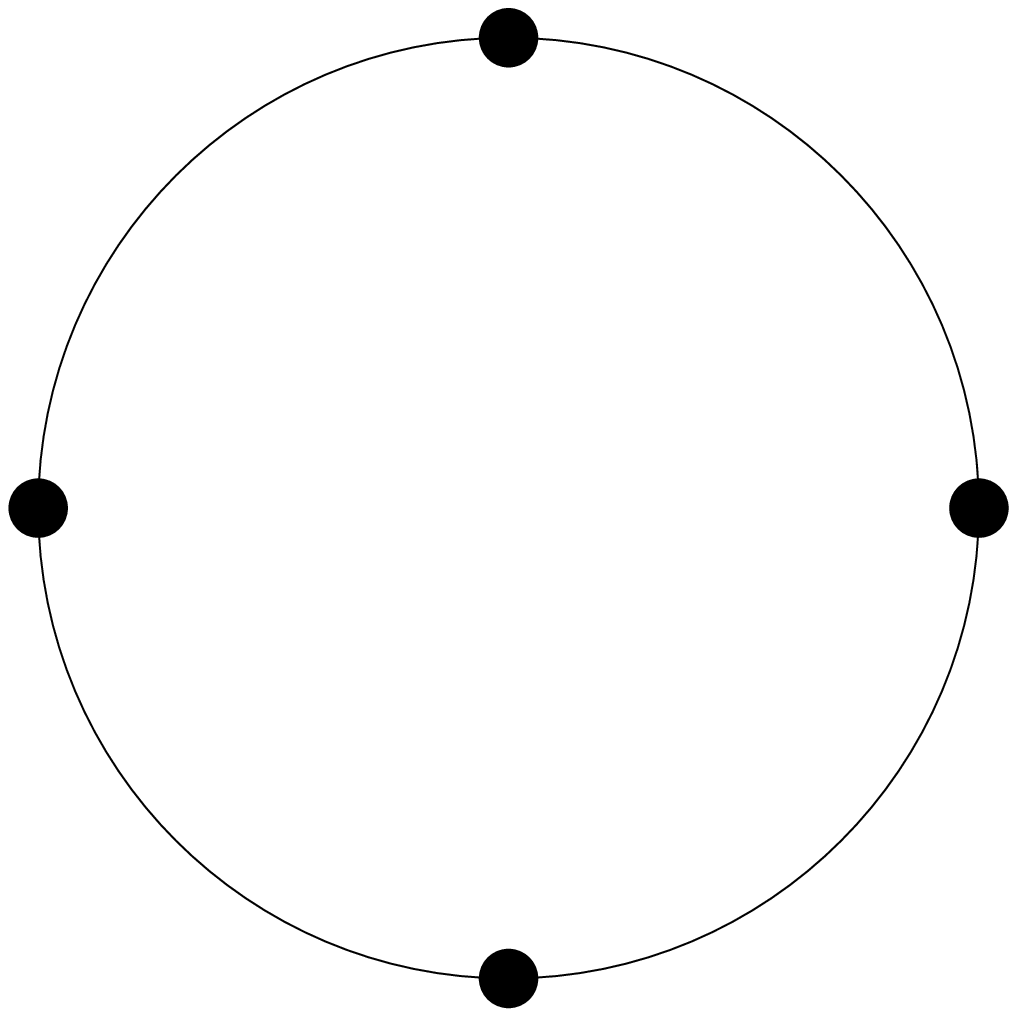}} \\
\parbox{5in}{\phantom{blah}} \\ 
\parbox{1.1in}{\centering \phantom{spacer}  pentagon}
\parbox{1.1in}{\centering trigonal bipyramid}
\parbox{1.1in}{\centering \phantom{spacer}  square pyramid}
\parbox{1.1in}{\centering \phantom{spacer}  octahedron}
\parbox{1.1in}{\centering \phantom{spacer}  trigonal prism}
\parbox{1.1in}{\includegraphics[width=1.1in]{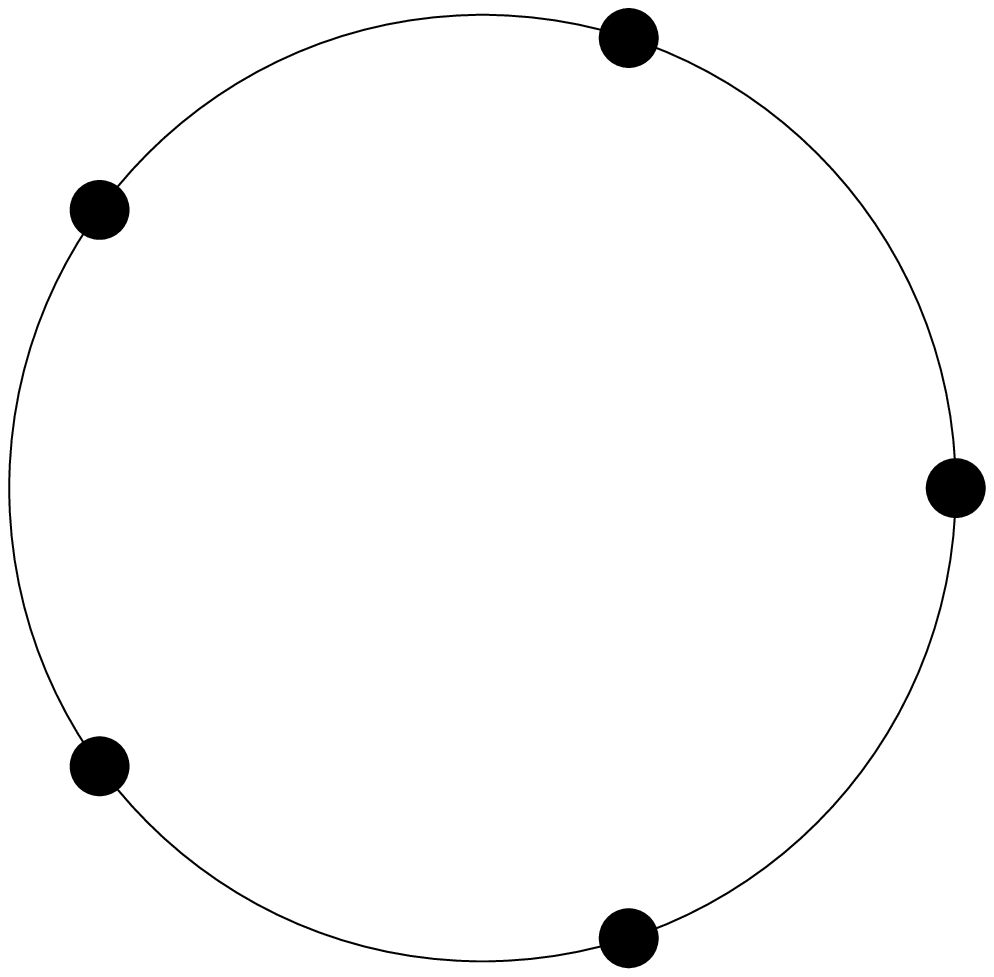}} 
\parbox{1.1in}{\includegraphics[width=1.1in]{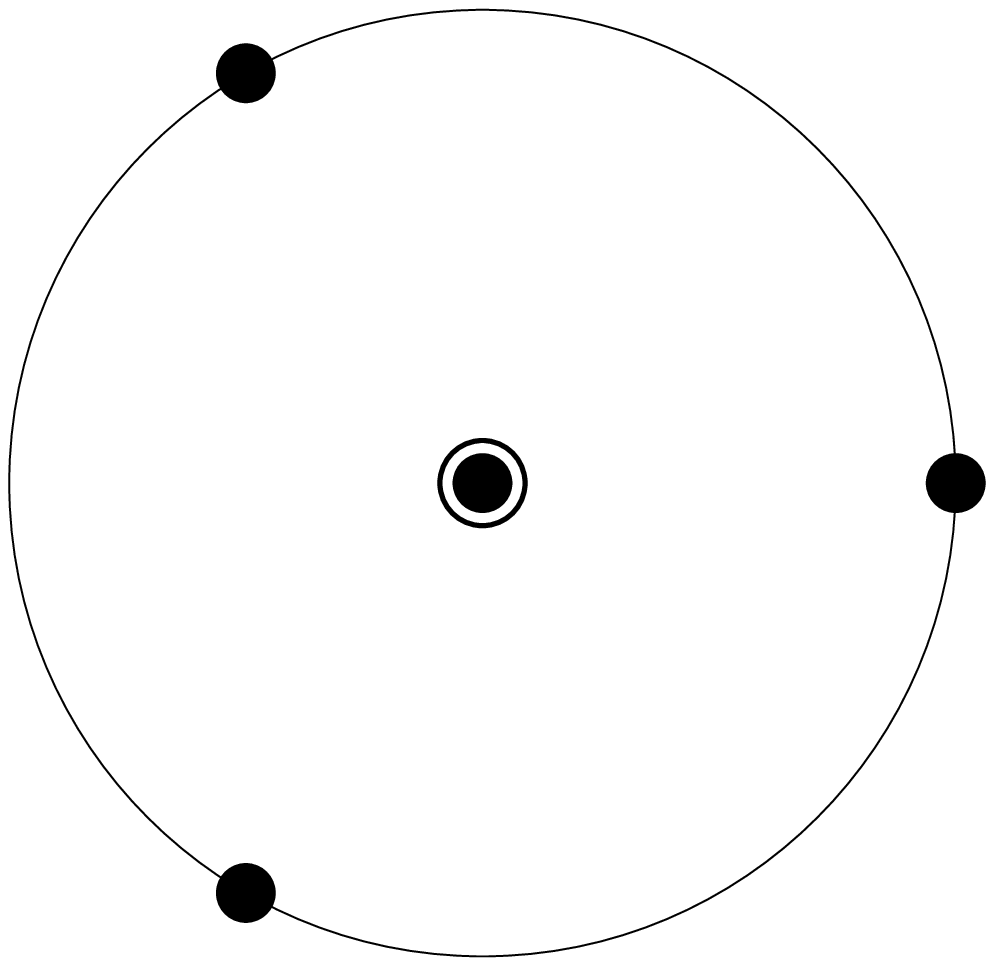}} 
\parbox{1.1in}{\includegraphics[width=1.1in]{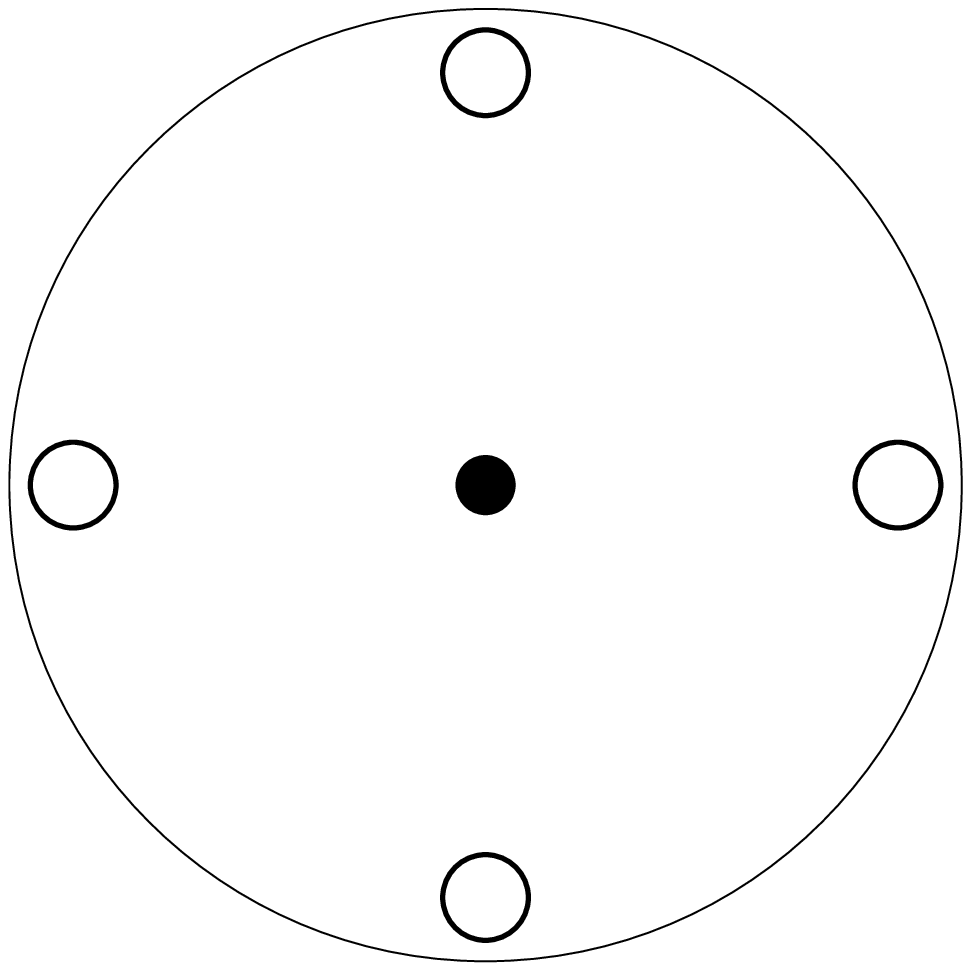}} 
\parbox{1.1in}{\includegraphics[width=1.1in]{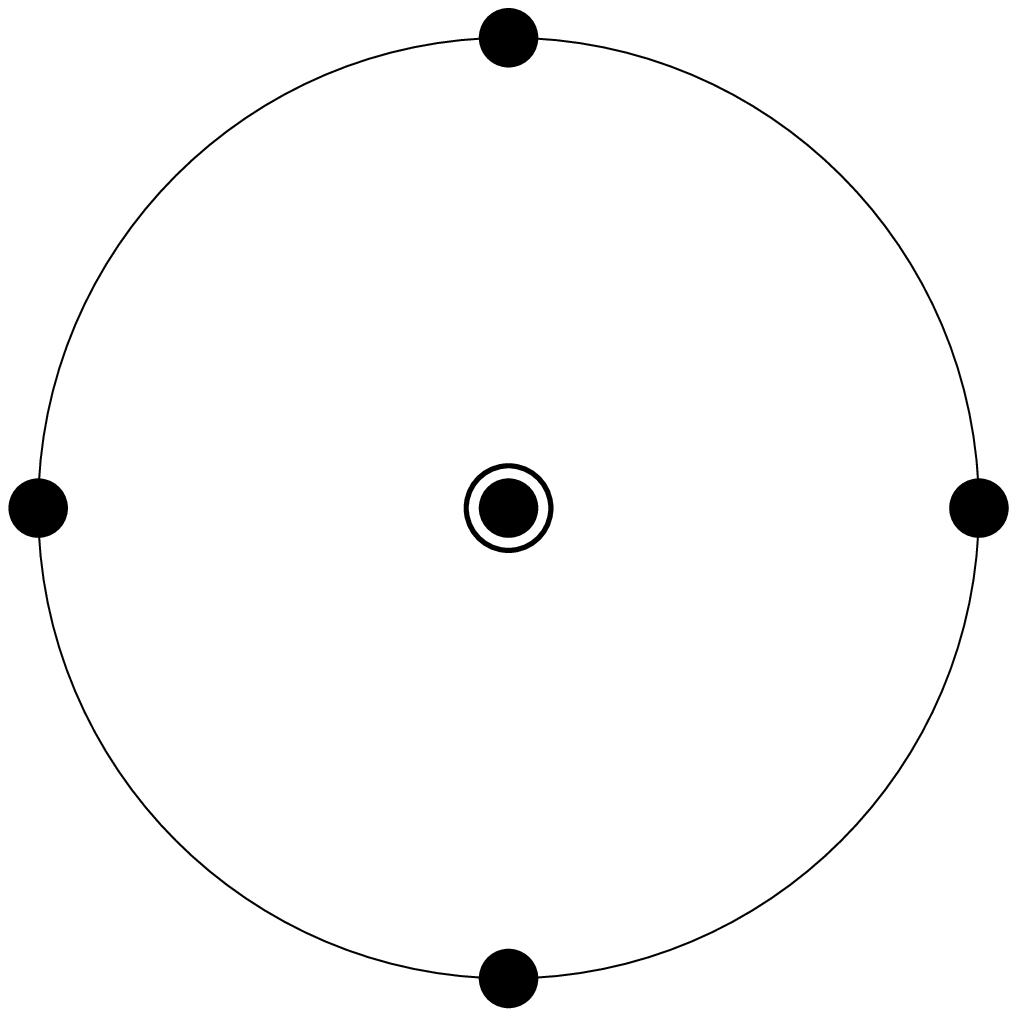}} 
\parbox{1.1in}{\includegraphics[width=1.1in]{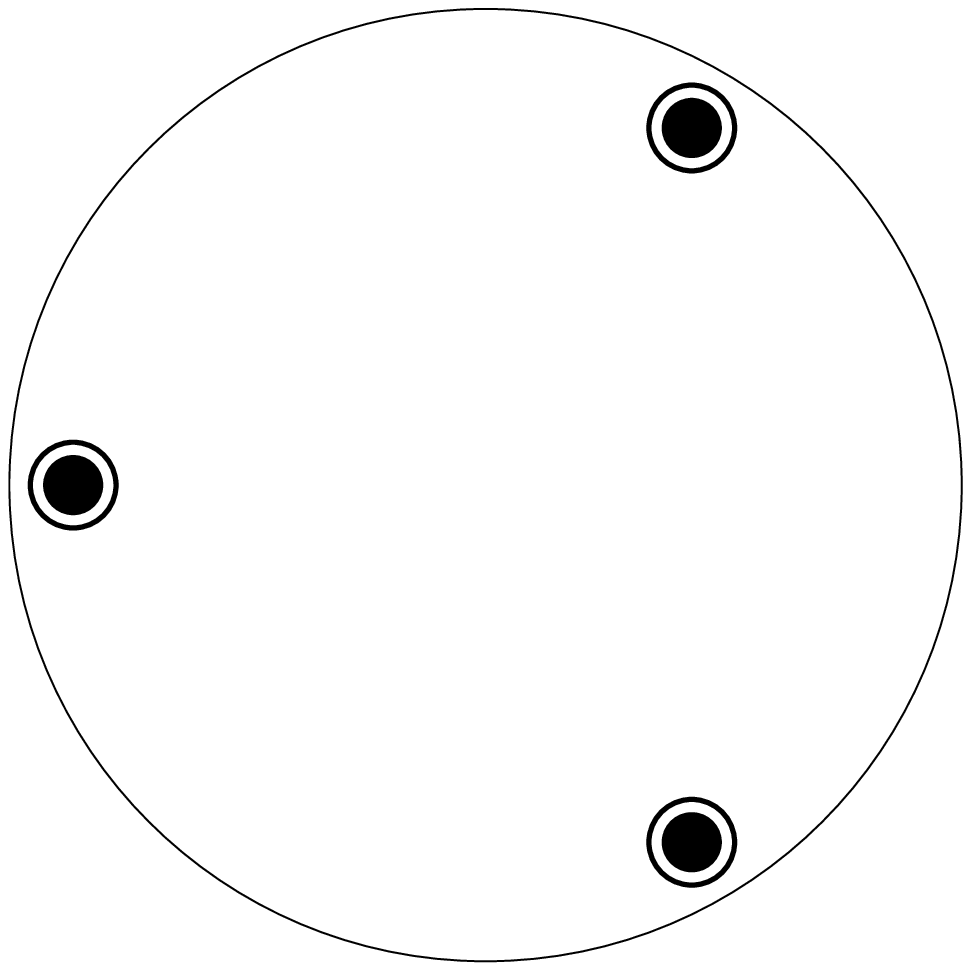}} \\
\parbox{5in}{\phantom{blah}} \\ 
\parbox{1.1in}{\centering \phantom{spacer}  hexagon}
\parbox{1.1in}{\centering pentagonal bipyramid}
\parbox{1.1in}{\centering capped trigonal antiprism}
\parbox{1.1in}{\centering \phantom{spacer}  cube}
\parbox{1.1in}{\centering square antiprism}
\parbox{1.1in}{\includegraphics[width=1.1in]{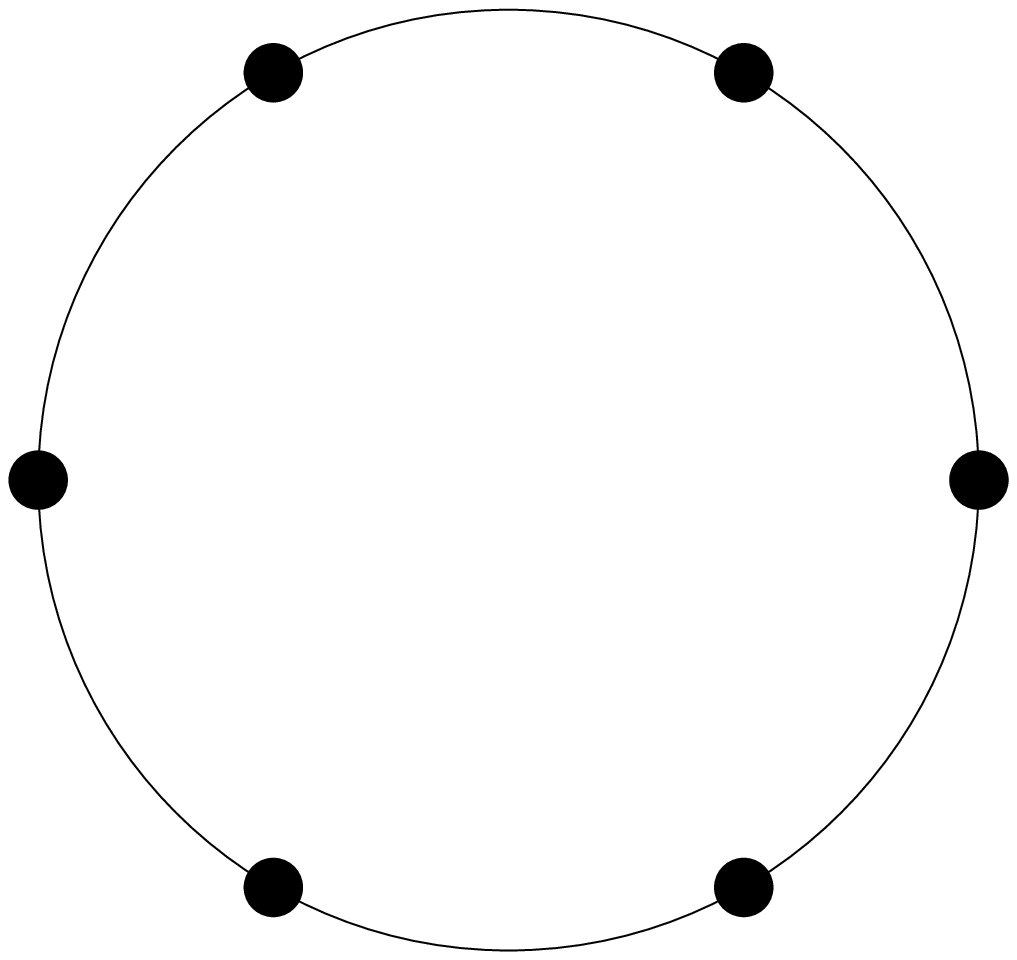}} 
\parbox{1.1in}{\includegraphics[width=1.1in]{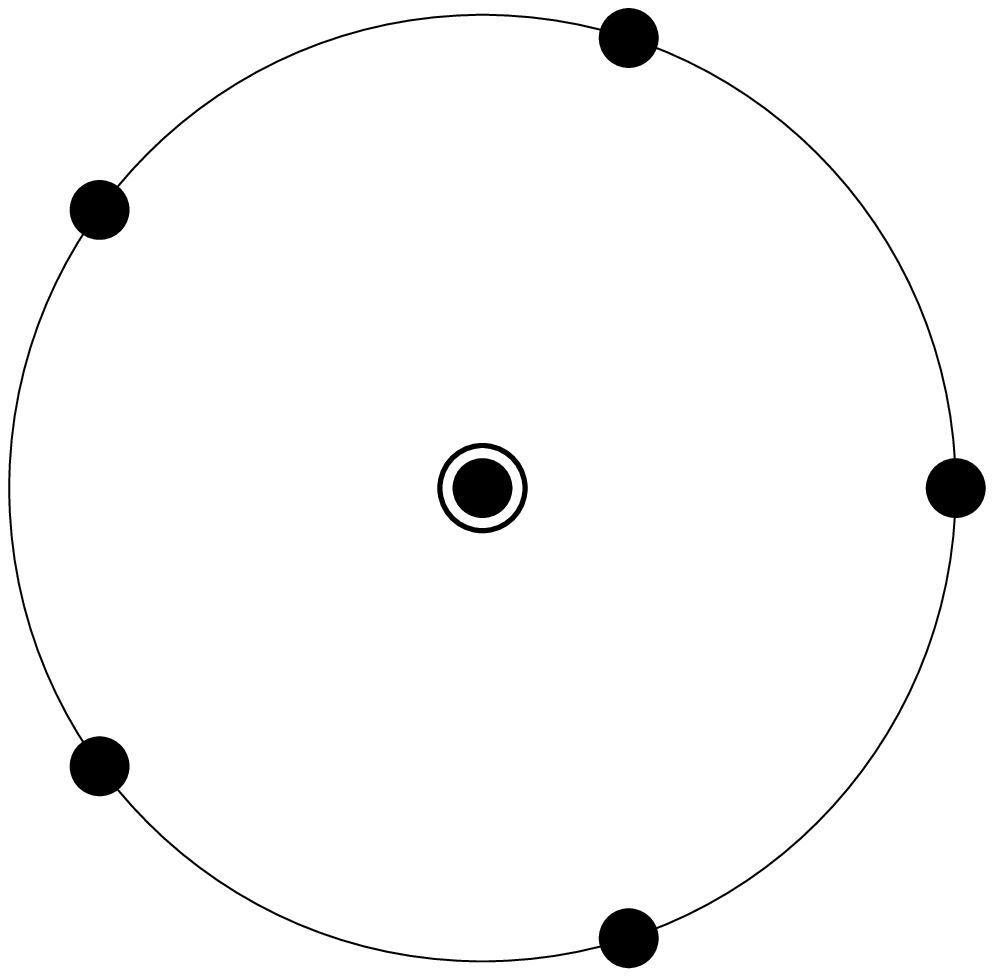}} 
\parbox{1.1in}{\includegraphics[width=1.1in]{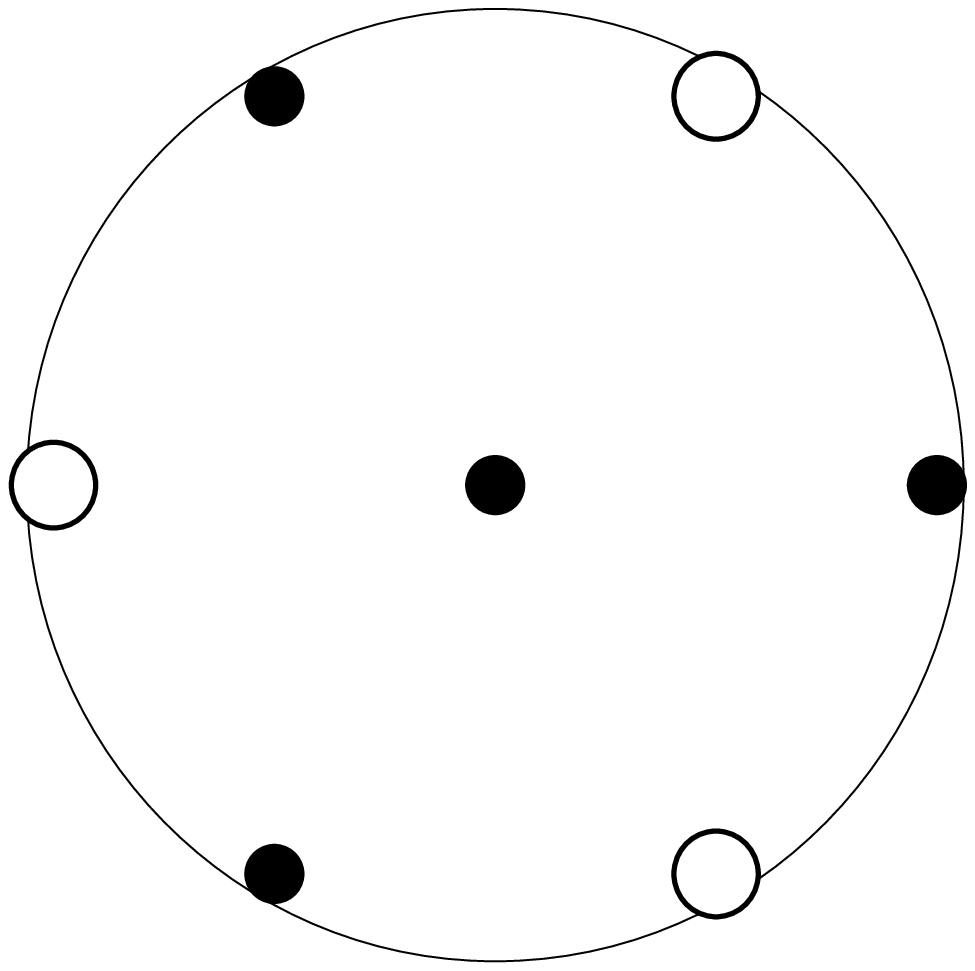}} 
\parbox{1.1in}{\includegraphics[width=1.1in]{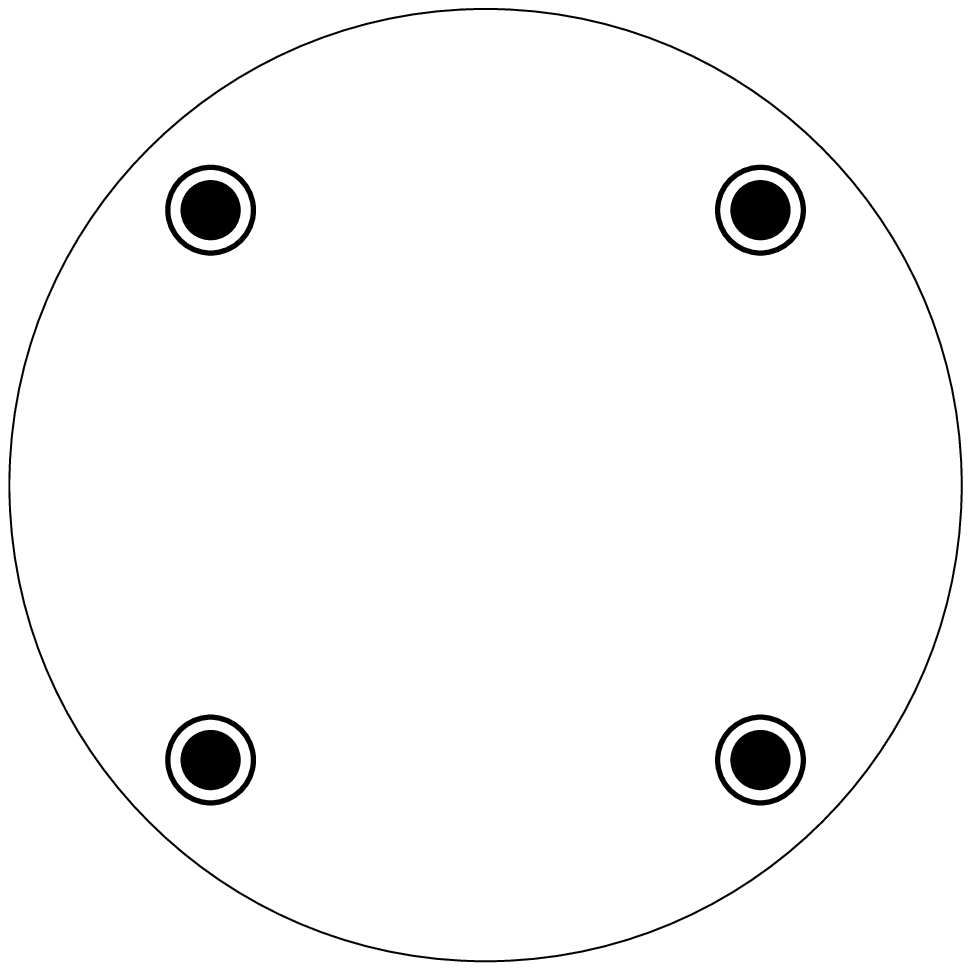}} 
\parbox{1.1in}{\includegraphics[width=1.1in]{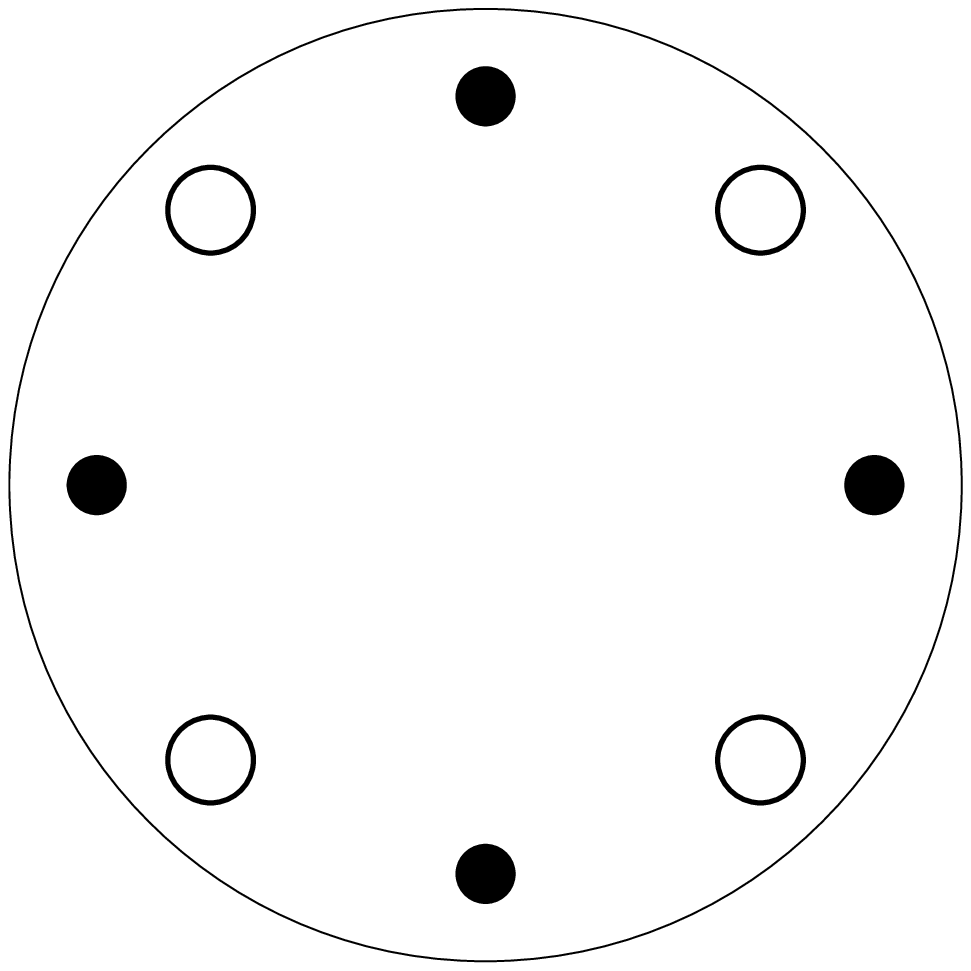}} \\
\parbox{5in}{\phantom{blah}} \\ 
\parbox{1.1in}{\centering \phantom{spacer} hexagonal bipyramid}
\parbox{1.1in}{\centering \phantom{spacer}  augmented trigonal prism}
\parbox{1.1in}{\centering \phantom{spacer}  capped square prism}
\parbox{1.1in}{\centering \phantom{spacer}  capped square antiprism}
\parbox{1.1in}{\centering bicapped square antiprism}
\parbox{1.1in}{\includegraphics[width=1.1in]{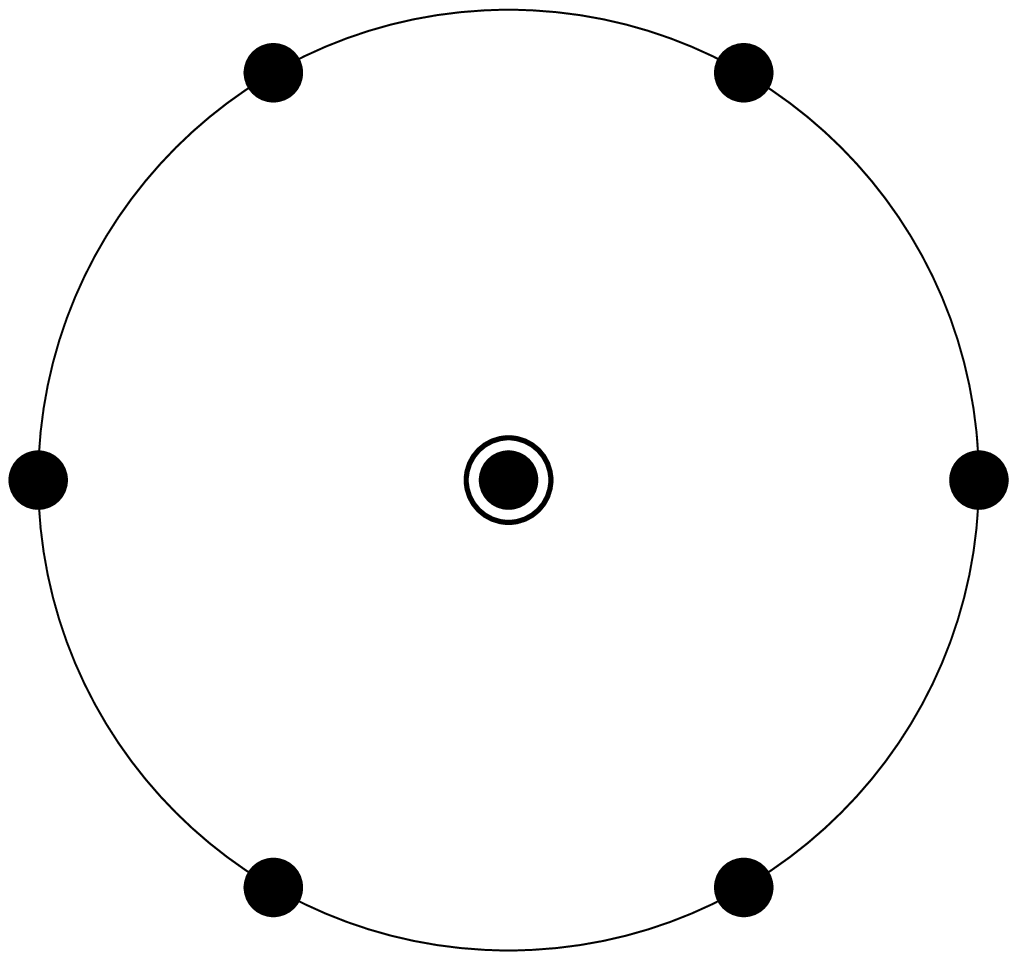}} 
\parbox{1.1in}{\includegraphics[width=1.1in]{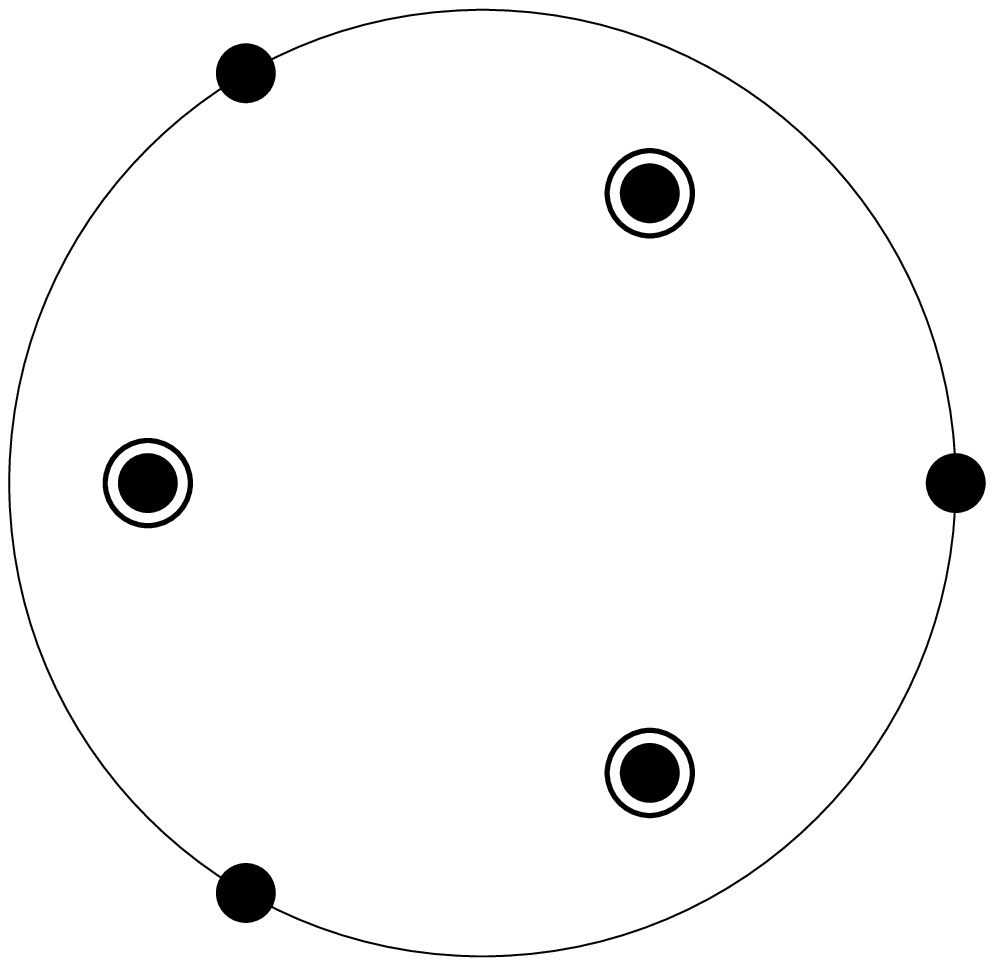}} 
\parbox{1.1in}{\includegraphics[width=1.1in]{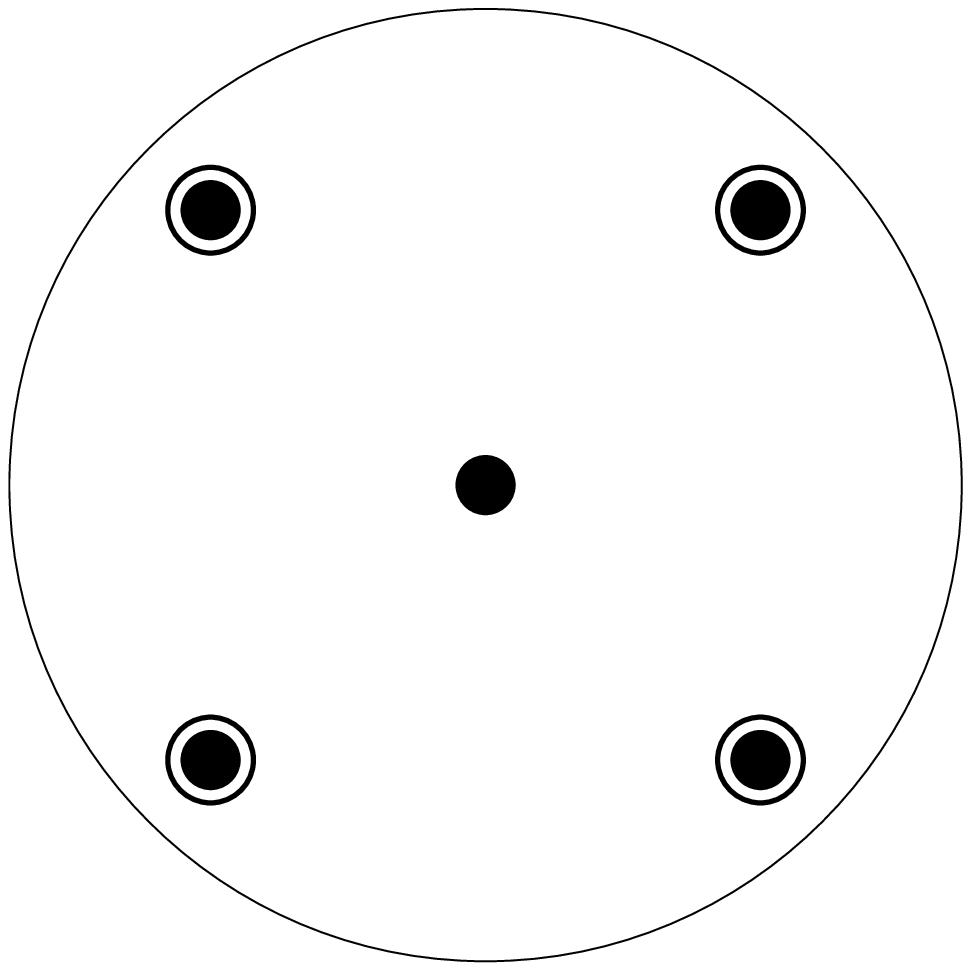}} 
\parbox{1.1in}{\includegraphics[width=1.1in]{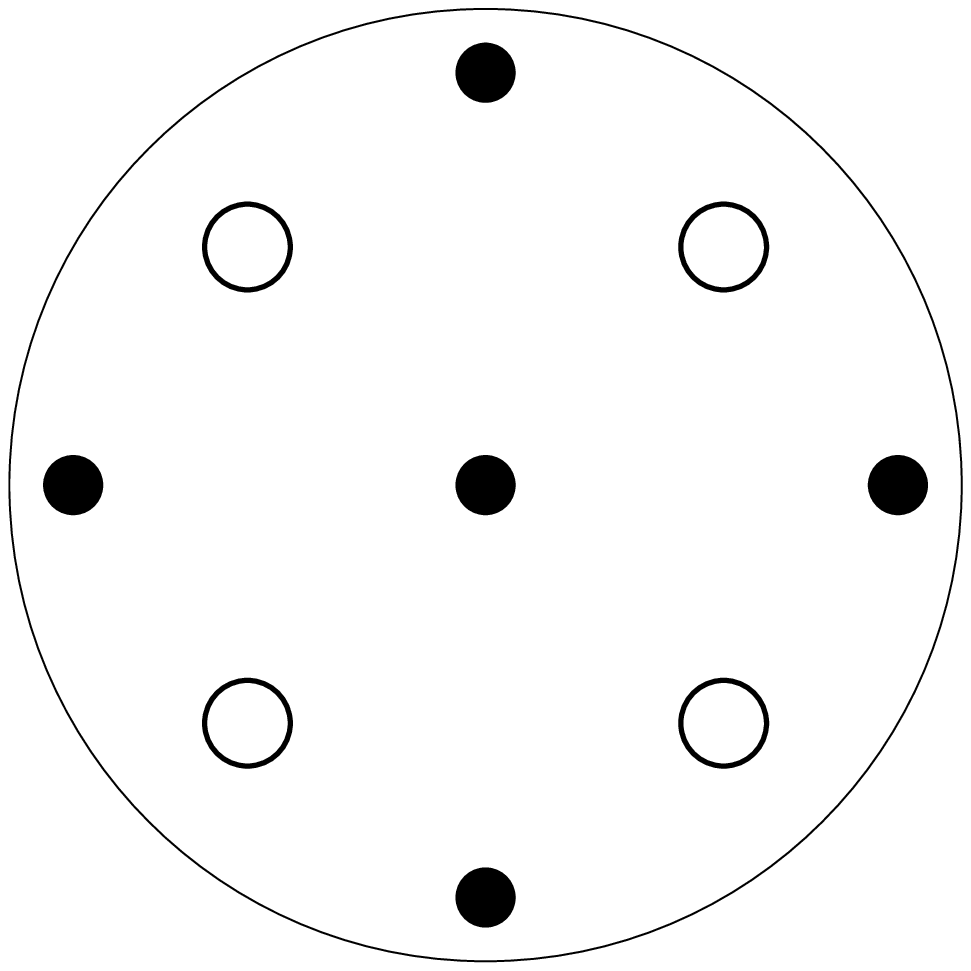}} 
\parbox{1.1in}{\includegraphics[width=1.1in]{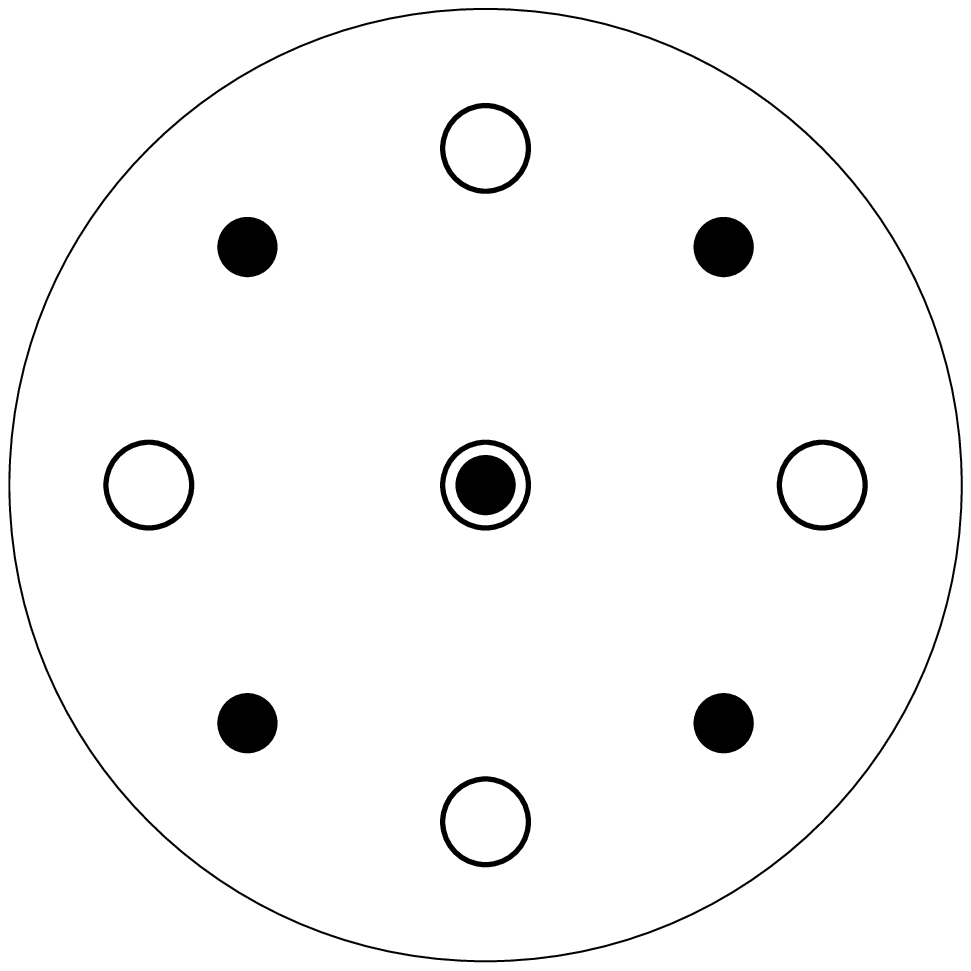}} \\
\parbox{5in}{\phantom{blah}} \\ 
\parbox{1.1in}{\centering icosahedron} 
\parbox{1.1in}{\centering cuboctahedron} 
\parbox{1.1in}{\centering dodecahedron} \\
\parbox{1.1in}{\includegraphics[width=1.1in]{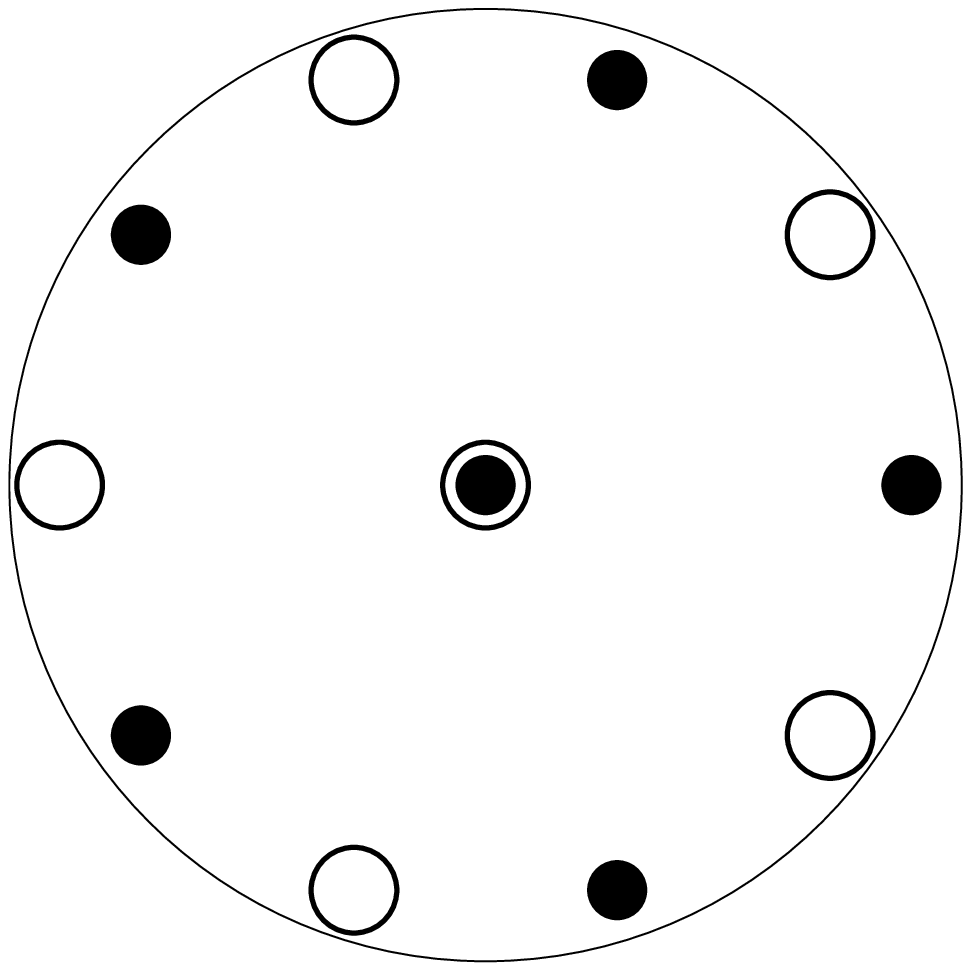}} 
\parbox{1.1in}{\includegraphics[width=1.1in]{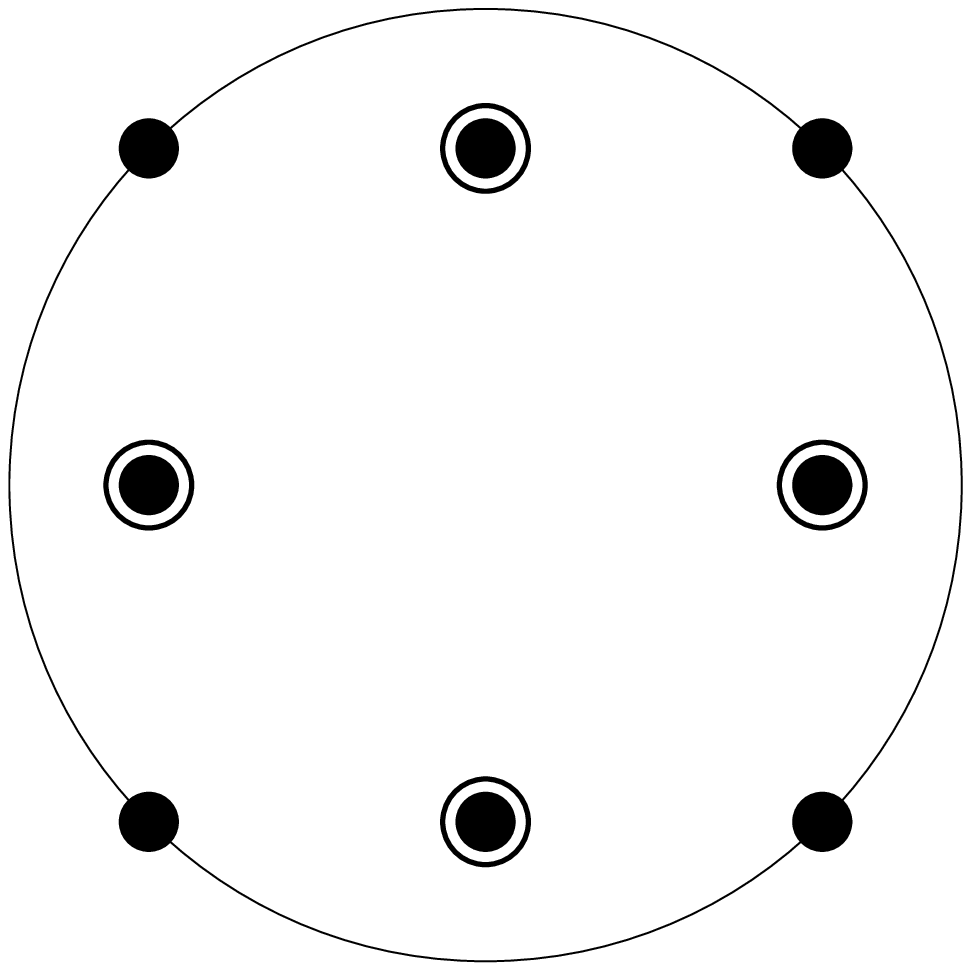}} 
\parbox{1.1in}{\includegraphics[width=1.1in]{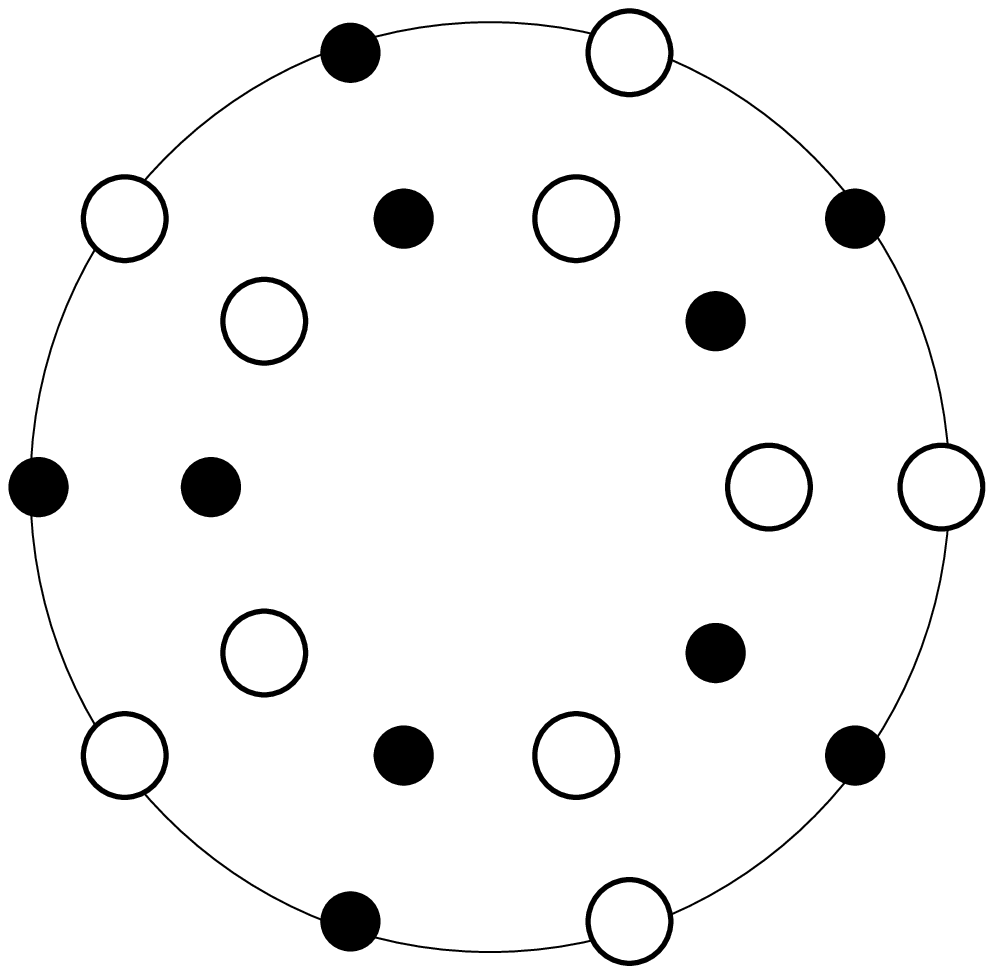}} 
\end{figure}

We have investigated a large number of different multiple-wave
configurations depicted in Fig.~\ref{stereographicfig}
and the results are compiled in Table~\ref{structures}.  The name of
each configuration is the name of a polygon or polyhedron that is
inscribed in a sphere of radius $q_0$; the $P$ vertices of the given
polygon or polyhedron 
then correspond to the $P$ wave vectors in the set
$\mathcal{Q}$.  
%The ``cube'' structure, for example, is the set of
%eight vectors that point from the center of a cube to its eight
%vertices.  
With this choice of nomenclature, keep in mind that what we call the
``cube'' has a different meaning than in much of the previous
literature.  We refer to an eight plane-wave configuration with
the eight wave vectors directed at the eight corners of a cube.
Because this is equivalent to eight vectors directed at
the eight faces of an octahedron --- the cube
and the octahedron are dual polyhedra --- in the nomenclature
of previous literature this eight-wave crystal would
have been called an octahedron, rather than a cube.
Similarly, the crystal that we call the ``octahedron'' 
(six plane waves whose wave vectors point at the six
corners of an octahedron) is the structure that has been
called a cube in the previous literature, because its wave
vectors point at the faces of a cube.

\begin{table}[p]
\caption{ 
\label{structures}
Candidate crystal structures with $P$ plane waves,
specified by their symmetry group $\mathcal{G}$ and F\"{o}ppl configuration.
Bars denote dimensionless equivalents: $\bar\beta = \beta\,\dm^2$,
$\bar\gamma = \gamma\,\dm^4$, $\bar\Omega = \Omega/(\dm_2^2 N_0)$
with $N_0 = 2 \bar\mu^2/\pi^2$.  $\bar\Omega_{\min}$ is the 
(dimensionless) minimum
free energy at $\dm = \dm_2$,
obtained from (\ref{DeltaOmegaatdm2}). 
The phase transition (first order for $\bar\beta<0$ and $\bar\gamma>0$, 
second order
for $\bar\beta>0$ and  $\bar\gamma>0$) occurs at $\dm_*$.
}
\vspace{0.1in}
\begin{minipage}{6.5in}
\small
%\begin{center}
\begin{tabular}{rl|c|c|r|r|c|c}
& \multicolumn{1}{l|}{Structure}                   & P  & \multicolumn{1}{c|}{$\mathcal{G}$(F\"{o}ppl)} & \multicolumn{1}{c|}{$\bar\beta$} & \multicolumn{1}{c|}{$\bar\gamma$} & \multicolumn{1}{c|}{$\bar\Omega_{\min}$} & \multicolumn{1}{c}{$\dm_*/\Delta_0$} \\ \hline
%& \multicolumn{1}{l|}{Structure}                   & P  & \multicolumn{1}{c|}{$\mathcal{G}$(F\"{o}ppl)} & \multicolumn{1}{c|}{$\beta \dm_2^2$} & \multicolumn{1}{c|}{$\gamma \dm_2^4$} & \multicolumn{1}{c|}{$\Omega_{\min}/(\dm_2^2 N_0)$} & \multicolumn{1}{c}{$\dm_2^*/\Delta_0$} \\ \hline
1 & point 		   & 1  & $C_{\infty v}$(1) & 0.569 & 1.637 & 0 & 0.754 \\ 
2 &antipodal pair 	   & 2  & $D_{\infty v}$(11) & 0.138 & 1.952 & 0 & 0.754 \\ 
3 &  triangle 		   & 3  & $D_{3 h}$(3)      & -1.976 & 1.687 & -0.452 & 0.872 \\ 
4 &  tetrahedron 	   & 4  & $T_d$(13)    	    & -5.727 & 4.350 & -1.655 & 1.074 \\ 
5 & square 		   & 4  & $D_{4h}$(4)       & -10.350 & -1.538 & --  & -- \\ 
6 & pentagon  		   & 5  & $D_{5h}$(5)       & -13.004 & 8.386 & -5.211 & 1.607 \\ 
7 & trigonal bipyramid     & 5  & $D_{3h}$(131)       & -11.613 & 13.913 & -1.348 & 1.085 \\ 
8 & square pyramid\footnote{Minimum $\gamma$ and $\Omega_{\min}$ obtained for $\theta_2 \simeq 51.4^{\circ}$ (where $\theta_i$ is the polar angle of the $i$th F\"{o}ppl plane).}
         & 5  & $C_{4v}$(14)    & -22.014 & -70.442 & -- & -- \\
9 & octahedron  	   & 6  & $O_h$(141)          & -31.466 & 19.711 & -13.365 & 3.625 \\ 
10 & trigonal prism\footnote{Minimum $\gamma$ 
at $\theta_1=\pi-\theta_2 \simeq 43.9^{\circ}$. }
        & 6  & $D_{3h}$(33)       & -35.018 & -35.202 & -- & -- \\ 
11& hexagon  		   & 6  & $D_{6h}$(6)       & 23.669 & 6009.225 & 0 & 0.754 \\ 
12& pentagonal             & 7  & $D_{5h}$(151)      & -29.158 & 54.822 & -1.375 & 1.143 \\ 
  &  \ bipyramid 	   &    & &  &  & &\\ 
13& capped trigonal	   & 7  & $C_{3v}$($13\bar3$) & -65.112 & -195.592 & -- & -- \\
  & antiprism\footnote{Minimum $\gamma$ at $\theta_2 = \pi-\theta_3 \simeq 70.5^{\circ}$ (a cube with one 
vertex removed).}             & & & & & & \\
14 & cube  		   & 8  & $O_h$(44)          & -110.757 & -459.242 & -- & -- \\ 
%14 & square prism 	   & 8  & $D_{4h}$(44)	& & & & \\
15 & square antiprism\footnote{Minimum $\gamma$ 
at $\theta_1=\pi-\theta_2 \simeq 52.1^{\circ}$.}
  	   & 8  & $D_{4d}$($4 \bar 4$)       & -57.363 & -6.866 & -- & -- \\ 
16 & hexagonal             & 8  & $D_{6h}$(161)       & -8.074 & 5595.528 & $-2.8\times 10^{-6}$ & 0.755 \\
   & \ bipyramid 	   &    & &  &  & &\\ 
17 & augmented  	   & 9  & $D_{3h}$($3\bar 3\bar 3$)       & -69.857  & 129.259 & -3.401 & 1.656 \\ 
   & \ trigonal prism\footnote{Minimum $\gamma$ and $\Omega_{\min}$ at $\theta_1 =\pi-\theta_3 \simeq 43.9^{\circ}$. }  	   &                     &        &  & & & \\
18 & capped                & 9  & $C_{4v}$(144)    & -95.529 &  7771.152  &  -0.0024  & 0.773  \\
   & \ square prism\footnote{Best configuration is degenerate ($\theta_2 = \theta_3$, a square 
pyramid).   Result shown is for $\theta_2 \simeq 54.7^{\circ}$, $\theta_3 \simeq 125.3^{\circ}$ (a capped cube).} & & & & & & \\ 
19 & capped  		   & 9  & $C_{4v}$($14\bar 4$)       & -68.025 & 106.362 & -4.637 & 1.867 \\
   & \ square antiprism\footnote{Minimum $\gamma$ and $\Omega_{\min}$ at $\theta_2 \simeq 72.8^{\circ}$, $\theta_3 \simeq 128.4^{\circ}$.} 	   &    & &  &  & &\\ 
20 & bicapped  	   	   & 10 & $D_{4d}$($14\bar 41$)       & -14.298 & 7318.885 & $-9.1\times 10^{-6}$ & 0.755 \\ 
   & \ square antiprism\footnote{Best configuration is degenerate ($\theta_2 = 0$, an antipodal
pair).  Result shown is for $\theta_2 \simeq 72.8^{\circ}$.} 	   &    & &  &  & & \\ 
%20 & capped augmented   & 10 & $C_{3v}$($13\bar 3\bar 3$) & & & & \\
%   & \ trigonal prism   & & & & & & \\
21 & icosahedron  	   & 12 & $I_h$($15\bar 5 1$)          & 204.873 & 145076.754 & 0& 0.754 \\ 
22 & cuboctahedron  	   & 12 & $O_h$($4\bar 4\bar 4$)          & -5.296 & 97086.514 & $-2.6\times 10^{-9}$ & 0.754 \\
23 & dodecahedron  	   & 20 & $I_h$(5555)          & -527.357 & 114166.566 & -0.0019 & 0.772 \\
\end{tabular}
%\end{center}
\end{minipage}
\end{table} 

Because the LOFF pairing rings have an opening angle $\psi_0 \simeq
67.1^\circ$, no more than nine rings can be arranged on the Fermi
surface without any intersection~\cite{extremal,tammes}.  For this reason we
have focussed on crystal structures with nine or fewer waves,
but we have included several structures with more waves
in order to verify that such structures are not favored.
We have tried to analyze a fairly exhaustive list of candidate structures.
All five Platonic solids are included in Table~\ref{structures}, as is the
simplest Archimedean solid, the cuboctahedron. (All other Archimedean
solids have even more vertices.) We have analyzed many
dihedral polyhedra and
polygons: regular polygons, bipyramids, 
prisms,\footnote{A trigonal prism is two triangles,
one above the other. A cube is an example of a square
prism.} antiprisms,\footnote{An antiprism is a prism
with a twist. For example, a square antiprism is two
squares, one above the other, rotated relative to 
each other by $45^\circ$. The octahedron is an example
of a trigonal antiprism.}
and various capped 
or augmented polyhedra\footnote{Capping a polyhedron
adds a single vertex on the principal symmetry axis of the 
polyhedron (or polygon). Thus, a capped square is a square pyramid
and an octahedron could be called a bicapped square.
Augmenting a polyhedron means adding vertices on the equatorial
plane, centered outside each vertical facet.  Thus, augmenting
a trigonal prism adds three new vertices.}.
For each crystal
structure we list the crystal point group $\mathcal{G}$ and the
F\"{o}ppl configuration of the polyhedron or
polygon.  The F\"{o}ppl configuration
is a list of the number of vertices on circles formed by intersections
of the sphere with consecutive planes perpendicular to the principal
symmetry axis of the polyhedron or polygon.  
We use a modified notation where $a$
or $\bar a$ indicates that the points on a given circle are
respectively eclipsed or staggered relative to the circle 
above.
Note that polyhedra with several different principal symmetry axes,
namely those with $T$, $O$, or $I$ 
symmetry, have several different F\"{o}ppl descriptions: for example,
a cube is $(44)$ along a fourfold symmetry axis or $(1 3 \bar 3 1)$ along a 
threefold symmetry axis.  (That is, the cube
can equally be described as a square prism or
a bicapped trigonal antiprism. 
This should make clear
that the singly capped trigonal antiprism of Fig.~\ref{stereographicfig}
and Table~\ref{structures} is a cube with one vertex removed.)

We do not claim to have analyzed all possible crystal structures,
since that is an infinite task.  However, there are several classic
mathematical problems regarding extremal arrangements of points
on a sphere and, although we do not know that our problem
is related to one of these, we have
made sure to include  
solutions to these problems. 
For example, 
many of the structures that we have evaluated correspond to solutions
of Thomson's problem~\cite{thomson, extremal} (lowest energy arrangement of 
$P$ point charges on the surface of a sphere) or Tammes's 
problem~\cite{tammes, extremal} (best packing of $P$ equal circles 
on the surface of a sphere without any overlap). In fact, we include
all solutions to the Thomson and Tammes problems for $P \leq 9$.  Our 
list also includes 
all ``balanced'' configurations~\cite{balanced} that are possible for 
nonintersecting rings: a balanced configuration is a set $\mathcal{Q}$
with a rotational symmetry about every $\vq \in \mathcal{Q}$; this 
corresponds to an arrangement of particles on a sphere for which 
the particles are in equilibrium for {\em any} two-particle force law.

For each crystal structure, we have calculated the quartic and sextic
coefficients $\beta$ and $\gamma$ according to 
Eqs.~(\ref{squarehexagonsums}), using methods described
in appendix \ref{JKappendix} to calculate all the $J$ and $K$ integrals.
To further discriminate among the various
candidate structures, we also list the minimum free energy
$\Omega_{\min}$ evaluated at the plane-wave instability point $\dm =\dm_2$
where $\alpha=0$.  
To set the scale, note that the BCS state at $\dm=0$ has 
$\Omega_{\rm BCS}=-\mu^2\Delta_0^2/\pi^2$, corresponding
to $\bar\Omega_{\rm BCS}=-0.879$ in the units of Table~\ref{structures}. 
For those configurations with $\beta>0$ and $\gamma>0$,
$\Omega_{\min} = 0$ at $\dm=\dm_2$  and $\Omega_{\min} < 0$ for
$\dm<\dm_2$, where $\alpha<0$. Thus, we find a second-order
phase transition at $\dm=\dm_2$.
For those configurations with $\beta<0$ and $\gamma>0$,
at $\dm=\dm_2$ the minimum free energy occurs at a
nonzero $\Delta$ with $\Omega_{\min} < 0$.  (The
value of $\Delta$ at which this minimum occurs
can be obtained from (\ref{DeltaOmegaatdm2}).)
Because $\Omega_{\min}<0$ at $\dm=\dm_2$,
if we go to $\dm>\dm_2$, where $\alpha>0$, we lift this minimum
until at some $\dm_*$
it has $\Omega=0$ and becomes degenerate with
the $\Delta=0$ minimum. At $\dm=\dm_*$,
a first-order phase transition occurs. 
For a very weak
first-order phase transition, $\dm_* \simeq \dm_2 \simeq 0.754 \Delta_0$.
For a strong first-order phase transition,
$\dm_* \gg \dm_2$ and the crystalline color superconducting
phase prevails as the favored
ground state over a wider range of $\dm$.

%\subsection{Crystal structures with intersecting rings lose}
\subsection{Crystal structures with intersecting rings}

There are seven configurations in Table~\ref{structures} with very
large positive values for $\gamma$.  These are precisely the seven
configurations that have intersecting pairing rings: the hexagon,
hexagonal prism, capped square prism, bicapped square antiprism,
icosahedron, cuboctahedron, and dodecahedron. The first two
of these include hexagons, and since $\psi_0>60^\circ$ the
rings intersect. The last four
of these crystal structures have more than nine rings, meaning
that intersections between rings  are also inevitable.  
The capped square prism is an example of a nine-wave structure
with intersecting rings. It has a $\gamma$ which is almost
two orders of magnitude larger than that of the
augmented trigonal prism
and the capped square antiprism which, in contrast, are nine
wave structures with no intersecting rings.  
Because of their very large $\gamma$'s 
all the structures with intersecting rings 
have either second-order phase transitions or
very weak first-order phase transitions occurring
at a $\dm_*\simeq \dm_2$.  At $\dm=\dm_2$, all these
crystal structures have $\Omega_{\min}$ very close to zero.
Thus, as our analysis of two plane waves led us to expect,
we conclude that these crowded configurations with
intersecting rings are disfavored.

%\subsection{``Regular'' crystal structures rule, and the cube rules them all}
\subsection{``Regular'' crystal structures}

At the opposite extreme, we see that there are several
structures that have negative values of $\gamma$: 
the square, square
pyramid, square antiprism, trigonal prism, capped
trigonal antiprism, and cube.  
Our analysis demonstrates that the transition to all these
crystal structures (as to those with $\beta<0$ and $\gamma>0$)
is first order.  But, we cannot evaluate $\Omega_{\min}$ or
$\dm_*$ because, to the order we are working, $\Omega$ is
unbounded from below.   For each of these crystal structures,
we could 
formulate a well-posed (but difficult) variational problem in which
we make a variational ansatz corresponding to the structure,
vary, and find $\Omega_{\min}$ without making a 
Ginzburg-Landau approximation.   It is likely, therefore, that within
the Ginzburg-Landau approximation $\Omega$ will be stabilized
at a higher order than the sextic order to which we have
worked.  

Of the sixteen crystal
structures with no intersecting rings, there are seven
that are particularly
favored by the combinatorics of Eqs.~(\ref{squarehexagonsums}).
It turns out that these seven  crystal structures
are precisely the six that we have found with $\gamma<0$,
plus the octahedron, which is the most favored crystal structure
among those with $\gamma>0$.
As discussed earlier, more
terms contribute to the rhombic and hexagonal sums in 
Eqs.~(\ref{squarehexagonsums})
when the $\vq$'s are arranged in such away that
their vertices form rectangles,
trapezoids, and cuboids inscribed in the sphere $|\vq| = q_0$.  Thus
the square itself fares well, as do the square pyramid, square
antiprism, and trigonal prism which contain one, two, and three
rectangular faces, respectively. The octahedron has
three square cross sections.
However, the cube is the outstanding winner
because it has six rectangular faces, six rectangular
cross sections, and also allows the five-corner arrangements
described previously.  The capped trigonal antiprism
in Table~\ref{structures} is a cube with one vertex removed.
This seven-wave crystal has almost as many waves
as the cube, and almost as many combinatorial
advantages as the cube, and it turns out to have the
second most negative $\gamma$.

With eight rings, the cube is close to the maximum
packing for nonintersecting rings with opening angle
$\psi_0\simeq 67.1^\circ$. Although nine rings of this
size can be packed on the sphere,
adding a ninth ring to the cube and deforming the
eight rings into a cuboid, as we have done with the
capped square prism, necessarily results in
intersecting rings
and the ensuing cost
overwhelms the benefits of the
cuboidal structure.  
To form a nine-ring structure with no intersections
requires rearranging the eight rings, spoiling
the favorable regularities of the cuboid.
Therefore a nine-ring arrangement is actually less favorable
than the cuboid,
even though it allows one more plane wave.
We see from Table~\ref{structures} that $\gamma$ for the cube is
much more negative than that
for any of the other combinatorially favored structures.
The cube is our winner, and we understand
why.  

To explore the extent to which the 
cube is favored, we can compare it to the octahedron, which
is the crystal structure with $\gamma>0$ for which we found
the strongest first-order phase transition, with
the largest $\dm_*$ and the deepest $\Omega_{\min}$.
The order $\Delta^6$ 
free energy we have calculated  for the cube is far below
that for the octahedron at all values
of $\Delta$.  To take an extreme example, at
$\dm=\dm_2$ the octahedron has $\bar\Omega_{\min}=-13.365$ 
at  $\Delta=1.263\dm_2=0.953\Delta_0$ whereas for the cube
we find that $\bar\Omega=-2151.5$ at this $\Delta$. 
As another example, suppose that we arbitrarily
add $+\quarter 800\Delta^8/\dm^8$
to the $\bar\Omega$ of the cube.  In this case, at $\dm=\dm_2$ 
we find that the cube has $\bar\Omega_{\min}=-32.5$ 
at $\Delta=0.656\Delta_0$ and is
thus still favored over the octahedron, even though
we have not added any $\Delta^8$ term to the free energy
of the octahedron.
These numerical exercises demonstrate the extraordinary
robustness of the cube, but should not be taken as more
than qualitative.
We do not know at what $\Delta$ and 
at what value $\Omega_{\min}$ 
the true free energy for the cube
finds its minimum.  
However, because the qualitative features
of the cube are so favorable
we expect that 
it will have a deeper $\Omega_{\min}$ and a 
larger $\dm_*$ than the octahedron. Within the Ginzburg-Landau
approximation,
the octahedron already has $\Delta=0.953\Delta_0$ 
and a deep $\bar\Omega_{\min}=-13.365$, about fifteen times
deeper than $\bar\Omega_{\rm BCS}=-0.879$ for the 
BCS state at $\dm=0$   
\footnote{Of course, the BCS state 
isotropically utilizes the entire Fermi surface 
for pairing, so we will not find a crystalline state 
which has a condensation energy greater than that of the 
cube.  The numbers are unreliable here, but can
be taken to indicate that the condensation energy could
be comparable to that of the BCS phase.}.

Even if we were to push the Ginzburg-Landau analysis of
the cube to higher order and find a stable
$\Omega_{\min}$, we would not be able to trust such
a result quantitatively.
Because it predicts a strong first-order phase transition,
the Ginzburg-Landau approximation predicts its own
quantitative demise.
What we have learned from it, however,
is that there are qualitative reasons that make the cube
the most favored crystal structure of them all. And, to
the extent that we can trust the quantitative calculations
qualitatively, they indicate that the first-order phase
transition results in a state with $\Delta$ and $\Omega_{\min}$
comparable to those of the BCS phase,
and occurs at a $\dm_*\gg \dm_2$.

\subsection{Varying continuous degrees of freedom}

None of the
regularities of the cube which make it so favorable
are lost if it is deformed 
continuously into
a cuboid, slightly shorter or taller than it is wide, as long
as it is not deformed so much as to cause rings to cross.
Next, we investigate this and some of the other possible 
continuous degrees of freedom present in a number
of the crystal structures we have described above.

So far we have neglected the fact, mentioned at the start of this
section, that some of the candidate structures have multiple orbits
under the action of the point group $\mathcal{G}$.  These structures
include the square pyramid, the four bipyramids, and the five capped
or augmented structures listed in Table~\ref{structures}; all have two orbits
except for the three singly capped crystal structures, which
have three orbits.
For these multiple-orbit structures each orbit should
have a different gap parameter but in Table~\ref{structures} we have assumed that all
the gaps are equal.  We have, however, analyzed each of these
structures upon assuming different gaps, 
searching for a minimum of the free energy in the
two- or three-dimensional parameter space of gaps.  
In most cases, the deepest minimum
is actually obtained by simply eliminating one of the orbits from the
configuration ({\it i.e.}~let $\Delta = 0$ for that orbit); the resultant
structure with one less orbit appears as another structure in Table~\ref{structures}.  
For example, the bicapped square antiprism has two orbits: the
first is the set of eight $\vq$'s forming a square antiprism, the
second is the antipodal pair of $\vq$'s forming the two ``caps'' of the
structure. Denote the gaps corresponding
to these two orbits as $\Delta_1$ and $\Delta_2$.  
This structure is overcrowded with intersecting rings,
so it is not surprising to find that a lower-energy configuration is
obtained by simply letting $\Delta_2 = 0$, which gives the
``uncapped'' square antiprism.  Configurations with fewer orbits are
generally more favorable, with only three exceptions known to us: the
trigonal bipyramid is favored over the triangle or the antipodal pair;
the square pyramid is favored over the square or the point;
and the capped trigonal antiprism is favored over any of the
structures that can be obtained from it by removing one or
two orbits.  For
these configurations, Table~\ref{structures} lists the results for $\Delta_1 =
\Delta_2\,(=\Delta_3)$; the numbers 
can be slightly improved with $\Delta_1 \neq \Delta_2\,(\neq \Delta_3)$ 
but the difference is unimportant.

For some configurations in Table~\ref{structures} the positions of the points are
completely fixed by symmetry while for others the positions of the
points can be varied continuously, while still maintaining the
point-group symmetry of the structure.  For example, with the square
pyramid we can vary the latitude of the plane that contains the
inscribed pyramid base.  Similarly, with the various polygonal prism
and antiprism structures (and associated cappings and augmentations),
we can vary the latitudes of the inscribed polygons (equivalently, we
can vary the heights of these structures along the principal symmetry
axis).  For each structure that has such degrees of freedom, we have
scanned the allowed continuous 
parameter space to find the favored configuration.
Table~\ref{structures} then shows the results for this
favored configuration, and the latitude angles describing
the favored configuration are given as footnotes.
However, if the structure always has overlapping rings
regardless of its deformation, then either no favorite configuration exists
or the favorite configuration is a degenerate one
that removes the overlaps by changing the structure.  There are two
instances where this occurs: the capped square prism can be deformed
into a square pyramid by shrinking the height of the square prism to
zero, and the bicapped square antiprism can be deformed into an
antipodal pair by moving the top and bottom square faces of the
antiprism to the north and south poles, respectively.  For these
structures, Table~\ref{structures} just lists results for an arbitrarily chosen
nondegenerate configuration.

\begin{figure}
\centering
\psfrag{x1}[tc][tc]{\small $10^{\circ}$}
\psfrag{x2}[tc][tc]{\small $20^{\circ}$}
\psfrag{x3}[tc][tc]{\small $30^{\circ}$}
\psfrag{x4}[tc][tc]{\small $40^{\circ}$}
\psfrag{x5}[tc][tc]{\small $50^{\circ}$}
\psfrag{x6}[tc][tc]{\small $60^{\circ}$}
\psfrag{x7}[tc][tc]{\small $70^{\circ}$}
\psfrag{x8}[tc][tc]{\small $80^{\circ}$}
\psfrag{x9}[tc][tc]{\small $90^{\circ}$}
\psfrag{x10}[tc][tc]{\small $50^{\circ}$}
\psfrag{x11}[tc][tc]{\small $55^{\circ}$}
\psfrag{x12}[tc][tc]{\small $60^{\circ}$}
\psfrag{x13}[tc][tc]{\small $65^{\circ}$}
\psfrag{y1}[rc][rc]{\small 3000}
\psfrag{y2}[rc][rc]{\small 6000}
\psfrag{y3}[rc][rc]{\small 9000}
\psfrag{y4}[rc][rc]{\small 12000}
\psfrag{y5}[rc][rc]{\small 15000}
\psfrag{y6}[rc][rc]{\small 30}
\psfrag{y7}[rc][rc]{\small 60}
\psfrag{y8}[rc][rc]{\small 90}
\psfrag{y9}[rc][rc]{\small 120}
\psfrag{y10}[rc][rc]{\small 150}
\psfrag{xlabel}{$\theta$}
\psfrag{ylabel}{$\gamma(\theta) \dm^4$}
\includegraphics[width=6in]{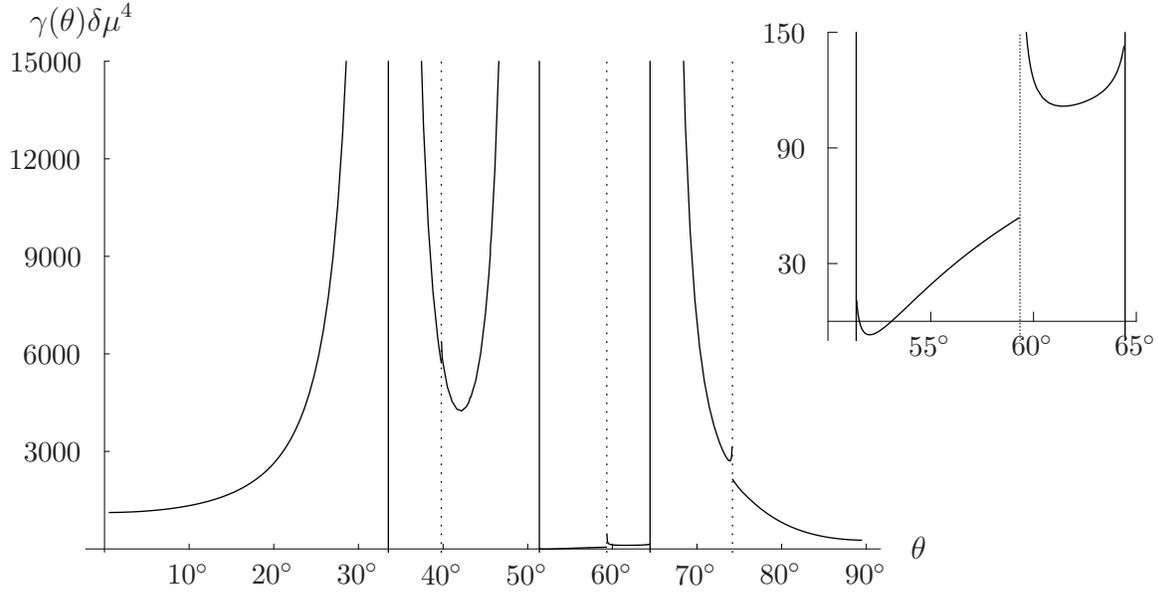}
\caption{
\label{scanfig}
The sextic free energy coefficient $\gamma$ for the square antiprism as a
function of the polar (latitude) 
angle $\theta$ of the top square facet.  (The
polar angle of the bottom square facet is $\pi-\theta$.)
The inset
plot shows the detail in the range of $\theta$ 
where no rings intersect.
Solid and dashed vertical lines indicate the positions of primary and
secondary singularities as discussed in the text (other secondary
singularities occur, but are not discernible on the plot).  }
\end{figure}

A typical parameter scan is shown in Fig,~\ref{scanfig}, where we
have plotted $\gamma$ for the square antiprism as a function of the
polar angle $\theta$ of the top square facet (the polar angle of the
bottom square facet is $\pi-\theta$).  As we expect, $\gamma$ is very
large in regions where any rings intersect, and we search for a
minimum of $\gamma$ in the region where no rings intersect.
The plot has a rather complicated structure of singularities and
discontinuities; these features are analogous to those of 
Fig.~\ref{betagammafig}, and as there they arise as a result
of the double limit we are taking.  
Primary singularities occur at critical angles
where pairing rings are mutually tangent on the Fermi surface.
Secondary singularities occur where rings 
corresponding to harmonic $\vq$'s, 
obtained by taking sums and differences of the fundamental $\vq$'s
that define the crystal structure,
are mutually tangent. Such $\vq$'s arise in the calculation
of $J$ and $K$ because these calculations 
involve momenta corresponding
to various diagonals of the rhombus and hexagon 
in Fig.~\ref{rhombushexagonfig}.

\begin{figure}
\centering
\psfrag{5}[tc][tc]{\small $5^{\circ}$}
\psfrag{10}[tc][tc]{\small $10^{\circ}$}
\psfrag{15}[tc][tc]{\small $15^{\circ}$}
\psfrag{20}[tc][tc]{\small $20^{\circ}$}
\psfrag{25}[tc][tc]{\small $25^{\circ}$}
\psfrag{30}[tc][tc]{\small $30^{\circ}$}
\psfrag{35}[tc][tc]{\small $35^{\circ}$}
\psfrag{40}[tc][tc]{\small $40^{\circ}$}
\psfrag{45}[tc][tc]{\small $45^{\circ}$}
\psfrag{-500}[rc][rc]{\small -500}
\psfrag{-400}[rc][rc]{\small -400}
\psfrag{-300}[rc][rc]{\small -300}
\psfrag{-200}[rc][rc]{\small -200}
\psfrag{-100}[rc][rc]{\small -100}
\psfrag{-100}[rc][rc]{\small -100}
\psfrag{100}[rc][rc]{\small 100}
\psfrag{200}[rc][rc]{\small 200}
\psfrag{xlabel}{$\phi$}
\psfrag{ylabel}{$\gamma(\phi) \dm^4$}
\psfrag{arrowlabel1}[lb][lt]{cube}
\psfrag{arrowlabel2}[tc][rb]{\mbox{square antiprism}}
\includegraphics[width=5in]{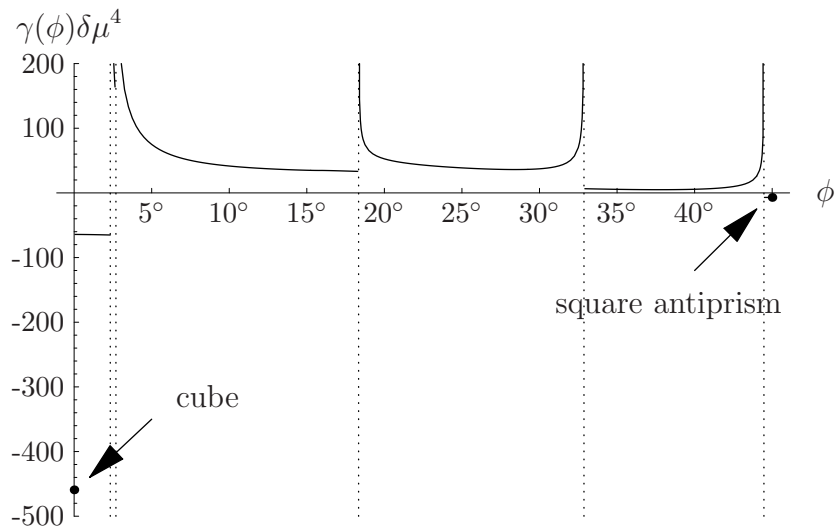}
\caption{
\label{twistfig}
The sextic free energy coefficient $\gamma$ for a scan 
that linearly interpolates from the cube (twist angle
$\phi=0$) to the square antiprism of Table~\ref{structures} (twist
angle $\phi=45^\circ$). Dashed vertical lines indicate
secondary singularities.  
}
\end{figure}

In addition to varying the latitudes of the F\"oppl planes
in various structures, we varied ``twist angles''.  For
example, we explored the continuous degree of
freedom that turns a cube into a square antiprism, 
by twisting the top square relative to the bottom square
by an angle $\phi$ ranging from $0^\circ$ to $45^\circ$.
In Fig.~\ref{twistfig}, we show a parameter scan
in which we simultaneously vary the twist angle $\phi$ and the
latitudes of the square planes in such a way that
the scan interpolates linearly from the cube to the
most favorable square antiprism of Table~\ref{structures}.
In this parameter scan, we find a collection
of secondary singularities and one striking
fact: $\gamma$ is much more negative when 
the twist angle is zero ({\it i.e.} for the cube itself)
than for any nonzero value. For the cube, $\gamma=-459.2/\dm^4$,
whereas the best one can do with a nonzero twist 
is $\gamma=-64.2/\dm^4$, which is the result for
an infinitesimal twist angle. 
Thus, any nonzero twist spoils the regularities
of the cube that contribute to its combinatorial
advantage, and this has    
a dramatic and unfavorable effect on the free energy.

Finally, we have scanned the parameter space of a generic 
cuboid to see how this compares to the special case of a cube.  
That is, we vary the height of the cuboid relative to its
width, without introducing any twist.  This continuous
variation does not reduce the combinatorial advantage
of the  crystal structure.
As expected, therefore, we find that as long as the cuboid
has no intersecting rings, it has a free energy
that is very similar to that of the cube itself.
Any cuboid with intersecting rings is very
unfavorable.  In the restricted parameter space of nonintersecting
cuboidal arrangements, the cuboid with the most negative free
energy
is a square prism
with a polar angle of $51.4^{\circ}$ for the top square face. (For this
polar angle the pairing rings corresponding to the four corners of the
square are almost mutually tangent.)  This prism is slightly
taller than a perfect cube, which has a polar angle of $54.7^{\circ}$.
The free energy coefficients of the best cuboid
are $\beta = -111.563/(\dm)^2$, $\gamma =
-463.878/(\dm)^4$.  These coefficients differ by less than $1\%$ from those
for the cube, given in Table~\ref{structures}.  There is no significant
difference between the cube and this very slightly 
more favorable cuboid:
all the qualitative arguments that favor the cube favor any 
cuboid with no
intersecting rings equally well.  We therefore expect that
if we could determine the exact (rather than Ginzburg-Landau)
free energy, we would find that the favored crystal structure
is a cuboid with a polar angle somewhere between $51.4^\circ$ and
$56.5^\circ$, as this is the range for which no rings
intersect.   We expect no important distinction between the
free energy of whichever cuboid in this narrow range happens to
be favored and that of the cube itself.

%------------------------------------------------------------------------

\section{Conclusions}
\label{sec:multiplewaveconclusions}

We have argued that the cube crystal structure is the favored ground state at
zero temperature near the plane-wave instability point $\dm = \dm_2$.
By the cube we mean
a crystal structure constructed as the sum of eight
plane waves with wave vectors pointing towards the 
corners of a cube.
The qualitative points (which we have demonstrated in explicit
detail via the analysis of many different crystal structures)
that lead us to conclude that the cube is the winner are:
\begin{itemize}
\item
The quadratic term in the Ginzburg-Landau free energy wants
a $|\vq|$ such that 
the pairing associated with any single choice of $\hat\vq$
occurs on a ring with
opening angle $\psi_0\simeq 67.1^\circ$ on each Fermi surface. 
\item
The quadratic term in the Ginzburg-Landau free energy
favors condensation with many different
wave vectors, and thus many different pairing rings on the Fermi
surfaces.
However, the quartic and sextic terms in the free energy
strenuously prohibit the intersection of pairing rings.
No more than nine rings with opening angle $67.1^\circ$
can be placed on the sphere without overlap.
\item
The quartic and sextic terms favor regular crystal structures,
for example those that include
many different sets of wave vectors whose tips form rectangles.
None of the  nine-wave structures with no intersections between
pairing rings are regular in the required sense.
The cube is a very regular eight-wave crystal structure.
\end{itemize}
Quantitatively, we find that a cube (actually, a cuboid that is only
slightly taller than it is wide) has by far the most negative
Ginzburg-Landau free energy, to sextic order, of all the many crystal
structures we have investigated.  For the cube structure, the eight
$\vq$ vectors are the eight shortest vectors in the reciprocal lattice
of a face-centered-cubic crystal.  Therefore, we find that
$\Delta(\vx)$ exhibits face-centered-cubic (FCC) symmetry.  The
explicit form of the gap function $\Delta(\vx)$ is shown in equation
(\ref{fcccrystal}), and a unit cell of the FCC crystal is shown in
figure \ref{unitcellfig}.

Our Ginzburg-Landau analysis predicts a first-order phase transition
to the cubic crystalline color superconductor at
some $\dm=\dm_*$.  The fact that we predict a first-order
phase transition means that the Ginzburg-Landau analysis
cannot be trusted quantitatively.  Furthermore, at order $\Delta^6$,
which is as far as we have gone, the Ginzburg-Landau free energy
for the cube is unbounded from below.  We therefore 
have no quantitative prediction of
$\dm_*$ or the magnitude of $\Delta$.
The best we can do is to note that the cube is significantly
favored over the octahedron, for which the order $\Delta^6$
Ginzburg-Landau analysis predicts $\dm_*\simeq 3.6\Delta_0$
and predicts that at $\dm=\dm_2=0.754\Delta_0$, the gap
is $\Delta\simeq 0.95\Delta_0$ and the condensation energy is
larger than that in the BCS state by a factor of about fifteen.
As we have warned repeatedly, these numbers should not
be trusted quantitatively: because the Ginzburg-Landau approximation
predicts a strong first-order phase transition, it predicts
its own breakdown. We have learned several qualitative
lessons from it, however:  
\begin{itemize}
\item
We have understood the qualitative reasons that make the cube the most
favored crystal structure of them all.  It has the largest number of
pairing rings that can be ``regularly'' arranged on the Fermi surface
without ring overlaps.
\item
The cube structure has an unstable Ginzburg-Landau free energy.  As a
result the Ginzburg-Landau analysis cannot give quantitative results,
but the large instability suggests a robust crystalline phase.  The
gap parameter $\Delta$ for the crystalline phase could be comparable
to the gap $\Delta_0$ for the BCS phase.  Similarly, the condensation
energy by which the crystalline phase is favored over unpaired quark
matter at $\dm\neq 0$ could be comparable to that for the BCS phase at
$\dm=0$.
%(Note that the spatial average
%of $\Delta(\vx)^2$ in (\ref{fcccrystal}) is $8\Delta^2$. It is
%not surprising that $\Delta\sim\Delta_0$ corresponds
%to an $|\Omega_{\min}|$ in the crystalline phase which is
%larger than $|\Omega_{\rm BCS}|$.)
\item
Although we began our analysis in the vicinity of the second-order
plane-wave instability point $\dm_2$, the transition to the
crystalline phase is strongly first-order and occurs at a $\dm_* \gg
\dm_2$.  The emergence of a first-order transition from a study 
of a second-order point has a precedent in the Ginzburg-Landau
analysis for the liquid-solid transition~\cite{Chaikin}.  
\item
We learn that the crystalline color superconductivity window
$\dm_1<\dm<\dm_*$ is large.  Because $\dm_2$ is not much larger than
$\dm_1$, the original LOFF window $\dm_1<\dm<\dm_2$ wherein the single
plane-wave condensate is possible is quite narrow.  We have learned,
however, that $\dm_*\gg\dm_2$.  Furthermore, because the condensation
energy of the crystalline phase is so robust, much greater than that
for the single plane wave and likely comparable to that for the BCS
phase, the value of $\dm_1$, the location of the transition between
the crystalline phase and the BCS phase, will be significantly
depressed.  So the LOFF window is considerably widened in both
directions.
\item
Given the robustness of the FCC crystalline phase in our two 
flavor model, in real QCD with three flavors of quarks we can 
expect that the crystalline phase will occupy a large regions
of the $(T,\mu)$ phase diagram of figure \ref{phasediagram}.  
\end{itemize}

%% file: chap4.tex
\chapter{Single Color and Single Flavor Color Superconductivity}

\section{Overview}
\label{sec:1scoverview}

There are nine types of quark (3 colors, 3 flavors) in dense quark
matter.  In general we expect that all nine quarks will find
attractive channels in which to form Cooper pairs, because any pairing
lowers the free energy by BCS condensation.  The color-flavor-locked
(CFL) phase directly accomplishes this feat, but for non-CFL phases of
color superconductivity that occur at intermediate densities as in
figure \ref{phasediagram}, it is not obvious how the system contrives to
pair all nine quarks.  Here we speculate on some of the contrivances
that might occur.

In the 2SC phase (which seems unlikely to occur in neutral quark
matter), only two colors and two flavors pair, leaving five
``orphaned'' quarks: the blue up and down quarks, and all three colors
of strange quark.  In this context it has been proposed that the blue
up and down quarks could pair together in a single-color
condensate~\cite{ARW2,BHO}, and all the strange quarks could pair
together in a single-flavor condensate which involves all three
colors~\cite{TS1flav,IwaIwa}.  

Single-flavor pairing can also occur in the limit of very large
separation between the Fermi surfaces for the different flavors.  In
this context the system will abandon inter-species pairing and form
$\langle uu \rangle$, $\langle dd \rangle$, and $\langle ss \rangle$
condensates.  Each could involve three colors (e.g.~in a
color-spin-locked (CSL) phase~\cite{TS1flav}) and therefore all nine
quarks are paired.

In the LOFF crystalline phase, so far we have only discussed the
two-flavor situation of pairing between up and down quarks.  This is a
``2SC-LOFF'' phase: just like the spatially-uniform 2SC phase, it
pairs only two colors and two flavors and leaves five orphaned quarks.
In a real three-flavor context, the crystalline phase could involve
$ud$, $us$, and $ds$ pairs, thus involving all nine quarks in a
``CFL-LOFF'' phase.  Alternatively, the three-flavor LOFF state might
involve only $ud$ and $us$ pairs, i.e.~it might only pair quarks with
adjacent Fermi surfaces (recall that in neutral quark matter the Fermi
momenta are split with $p_F^s < p_F^u < p_F^d$).  In this case, the
$ud$ pairing could involve $r$ and $g$ quarks, and the $us$ pairing
could involve $r$ and $b$ quarks, and two quarks are orphaned: the
blue down quark and the green strange quark.  These lonely quarks might
resort to self-pairing if the QCD interaction permits an attraction in
this channel (we will see that it does in some NJL models, if the
quark is heavy enough).  

In this chapter we survey a large catalog of different
color-flavor-spin channels for diquark condensation~\cite{ABCC}.  This catalog
includes the single-color and/or single-flavor pairing channels that
are mentioned in the various scenarios described
above~\cite{IwaIwa,ARW2,TS1flav,BHO}.  We consider only
translationally invariant phases, but many of the condensates in our
survey have non-zero spin, and therefore spontaneously break
rotational invariance.  In section \ref{sec:1scmean} we identify the
attractive channels using NJL models with four-fermion interactions
based on instantons, magnetic gluons, and combined electric and
magnetic gluons.  In section \ref{sec:1scgap} we solve the NJL
mean-field gap equations for the channels that are attractive.  In
section \ref{sec:1scdispersion} we investigate the quasiquark dispersion
relations for the spin-one condensates.  The quasiquark energy gaps are
anisotropic and gapless modes can occur at the poles or at the equator
of the Fermi surface.  Finally in section \ref{sec:1scconclusions},
we summarize the results, suggest directions for further study, and 
speculate about implications for the physics of compact stars.

%This question is of direct relevance to compact star physics,
%where the smallest gap controls transport properties such
%as the specific heat and neutrino emission rate.
%The density of a compact star rises from nuclear density
%near the surface to a much higher but as yet unknown value
%in the core, so unless there is a direct transition from
%nuclear to color-flavor-locked quark matter, there will be regions where
%some sort of single color or flavor pairing is likely.

\section{Mean-field survey of quark pairing channels}
\label{sec:1scmean}

\subsection{Calculation}
To see which channels are attractive we perform a mean-field calculation
of the pairing energy for a wide range of condensation patterns.
We write the NJL Hamiltonian in the form
\beq
\label{hamiltonian}
\ba{rcl}
H &=& H_{\rm free} + H_{\rm interaction} \\[1ex]
H_{\rm free} &=& \psibar (\dslash - \mu\ga_0 + m) \psi \\[1ex]
 H_{\rm interaction} &=& {\psi^\dag}_{\al}^{ia} \psi^{\be b}_{j}
{\psi^\dag}_{\ga}^{kc} \psi^{\de d}_{l} 
\,\,{\cal H}^{\al}_{ia}{\,}_{\be b}^{j}{\,}^{\ga}_{kc}{\,}_{\de d}^{l}\ ,
\ea
\eeq
where color indices are $\al,\be,\ga,\de$, flavor indices are $i,j,k,l$,
spinor indices are $a,b,c,d$. The four-fermion interaction is
supposed to be a plausible model of QCD, so 
in the interaction kernel ${\cal H}$ we include three
terms, with the color-flavor-spinor structure of a two-flavor
instanton, electric gluon exchange, and magnetic gluon exchange,
\beq\label{kernel}
\ba{rcl}
{\cal H} &=& {\cal H}_{\rm elec} + {\cal H}_{\rm mag} + {\cal H}_{\rm inst}
  \\[1ex]
{\cal H}_{\rm elec} &=& \phm\frac{3}{8} G_E \,\,\de_i^j \de_k^l \,\,
  \de_{ab} \de_{cd} \,\,
  \frac{2}{3} (3 \de^\al_\de \de^\ga_\be - \de^\al_\be \de^\ga_\de) \\[1ex]
{\cal H}_{\rm mag} &=& \phm\frac{3}{8} G_M \,\,\de_i^j \de_k^l \,\,
  \sum_{n=1}^3 [\ga_0\ga_n]_{ab} [\ga_0\ga_n]_{cd} \,\,
  \frac{2}{3} (3 \de^\al_\de \de^\ga_\be - \de^\al_\be \de^\ga_\de) \\[1ex]
{\cal H}_{\rm inst} &=& -\frac{3}{4} G_I \,\,\ep_{ik} \ep^{jl} \,\,
  \quarter \Bigl( [\ga_0(1+\ga_5)]_{ab} [\ga_0(1+\ga_5)]_{cd} +
  [\ga_0(1-\ga_5)]_{ab} [\ga_0(1-\ga_5)]_{cd}
  \Bigr) \\
&& \phantom{-\frac{3}{4} G_I} 
   \frac{2}{3} (3 \de^\al_\be \de^\ga_\de - \de^\al_\de \de^\ga_\be)
\ea
\eeq

We consider condensates that factorize into separate color, flavor, and
Dirac tensors (i.e.~that do not show ``locking'')
and calculate their binding energy by contracting them with \eqn{kernel}.

There is no Fierzing ambiguity in this procedure.
For a given pairing pattern $X$, the condensate is
\beq
\label{qqcond}
\<\psi^{\be b}_{j}\psi^{\de d}_{l}\>^{\vphantom{y}}_{1PI} = \De(X)
  \mathfrak{C}_{(X)}^{\be\de} 
  \mathfrak{F}^{\vphantom{\be}}_{(X)jl} 
  \Ga_{(X)}^{bd\vphantom{\be}}~.
\eeq
We can then calculate the interaction (``binding'') energy of the various
condensates,
\beq
\label{qqstrength}
H = -\sum_X \De(X)^2 \Bigl(
   S^{(X)}_{\rm elec} G_E 
 + S^{(X)}_{\rm mag} G_M
 + S^{(X)}_{\rm inst} G_I \Bigr)
\eeq
The binding strengths $S^{(X)}_{\rm interaction}$ give the strength of
the self-interaction of the condensate $X$ due to the specified
part of the interaction Hamiltonian.

\subsection{Properties of the pairing channels}
In Table \ref{tab:channels} we list the the simple (translationally
invariant, factorizable)  channels available for quark pairing.
The meanings of the columns are as follows.
\ben
\item\underline{Color}:  
two quarks either make an antisymmetric color triplet
(which requires quarks of two different colors)  or
a symmetric sextet (which can occur with quarks of two different colors, 
and also if both quarks have the same color). 
For the \thrbarA\ we use 
$\mathfrak{C}^{\be\de}=\ep^{\be\de}$ in Eq.~\eqn{qqcond}. For the
\sixS\ we use a single-color representative
$\mathfrak{C}^{\be\de}=\de^{\be,1}\de^{\de,1}$ in Eq.~\eqn{qqcond}.
\item\underline{Flavor}: 
two quarks either make an antisymmetric flavor singlet
(which requires quarks of two different flavors) or
a symmetric triplet (which can occur with quarks of two different flavors, 
and also if both quarks have the same flavor).
For the \oneA\ we use $\mathfrak{F}_{jl}=
\si^2_{jl}$ and for the \thrS\ we use $\mathfrak{F}_{jl}=\si^1_{jl}$
in \eqn{qqcond}.
\item \underline{Spin, parity}: since the chemical potential explicitly breaks
the Lorentz group down to three-dimensional rotations and
translations, it makes sense to classify condensates by their total
angular momentum quantum number $j$ and parity.  
\item\underline{Dirac}:  This column gives the Dirac matrix structure
$\Ga^{bd}$ used in \eqn{qqcond}, 
so the condensate is  $\psi^T \Ga \psi$.
We also designate
each condensate as ``LL'' (even number of gamma matrices, so pairs
same-chirality quarks) or ``LR'' (odd number of gamma matrices, so
pairs opposite-chirality quarks). 
\item\underline{BCS-enhancement}: Condensates that correspond to pairs
of particles or holes near the Fermi surface have a BCS singularity
in their gap equation that guarantees a solution, no matter how
weak the coupling. To see which condensates have such a BCS enhancement,
we expanded the field operators in terms of creation and annihilation
operators (see Appendix \ref{app:angmom}). The order of the coefficient of the
$a(\vp)a(-\vp)$ and $b^\dag(\vp)b^\dag(-\vp)$ terms is given
in the table. $\O(1)$ means BCS-enhanced,
$0$ means not BCS-enhanced.
In the $C\ga_0\ga_5$ condensate the coefficient goes to zero as the quark
mass goes to zero (hence it is labelled ``$\O(m)$'' in the table)
meaning that the channel loses its BCS enhancement
in the chiral limit. This is discussed further in  Appendix \ref{app:angmom}.
\item\underline{Binding strength}: 
For each channel we show the binding strength for the instanton
interaction, the full (electric plus magnetic) gluon, 
which could reasonably be used at
medium density, and for the magnetic gluon alone,
which is known to dominate at ultra-high density \cite{Barrois,Son}.
Channels with a positive binding strength
and BCS enhancement will always support pairing (the gap equation
always has a solution, however weak the couplings $G_I, G_E, G_M$).
Other things being equal, the pairing with the 
largest binding strength will have the lowest free energy, and
is the one that will actually occur.
\een

It may seem strange that there are entries in the table with
angular momentum $j=1$ and an antisymmetric Dirac structure
($C\ga_3\ga_5$), and with $j=0$ but a symmetric Dirac structure
($C \ga_0$). If all the angular momentum came from spin this
would be impossible. But even though there are
no explicit spatial derivatives in the diquark operators, there
can still be orbital angular momentum. In Appendix \ref{app:angmom}
the angular momentum content of the particle-particle component of the
condensates is analyzed into its
spin and orbital content. We see, for example, that
$C\ga_3\ga_5$ has an antisymmetric space wavefunction
($l=1$) and a symmetric spin wavefunction ($s=1$), combined
to give an antisymmetric $j=1$.

\subsection{Results}

\begin{table}[htb]
\setlength{\tabcolsep}{0.25em}
\newcommand{\attr}{\color{black}}
\newcommand{\rep}{\color{black}}
\newcommand{\st}{\rule[-0.5ex]{0em}{2.8ex}}
\newlength{\instwid}
\newlength{\gluonwid}
\newlength{\magwid}
\settowidth{\instwid}{instanton} %{$S_{\rm inst}$}
\settowidth{\gluonwid}{x+$S_{\rm mag}$x}
\settowidth{\magwid}{mag.~only}
\begin{tabular}{cccccccccccc}
\hline\hline
\multicolumn{6}{c}{Structure of condensate} 
 & \rule[-1ex]{0em}{3.5ex} &
\multicolumn{3}{c}{Binding strength} \\
\cline{1-6}\cline{8-10}
& & & & & & 
  & \rule[-2ex]{0em}{5ex} \makebox[\instwid]{instanton~} 
  & \multicolumn{2}{c}{gluon} \\[-1ex]
\cline{9-10}\\[-4.5ex]
color & flavor & $j$ & parity &
\multicolumn{2}{c}{Dirac} &  
\parbox{4em}{\bc BCS\\[-0.5ex] enhance-\\[-0.5ex] ment\ec}
 & \parbox{\instwid}{\bc ~\\ ~ \\[-0.2ex] $S_{\rm inst}$ \ec}
 & \parbox{\gluonwid}{
      \bc full\\ $S_{\rm elec}$\\[-0.2ex]+$S_{\rm mag}$ \ec}
 & \parbox{\magwid}{\bc mag.~only\\ ~ \\[-0.2ex] $S_{\rm mag}$ \ec } \\[-1ex]
\hline
\thrbarA & \oneA & $0_A$ & $+$& $C\ga_5$ & LL & $\O(1)$ 
  & \attr$+64$ & \attr$+64$ & \attr$+48$\st \\
\thrbarA & \oneA & $0_A$ & $-$& $C$ & LL & $\O(1)$  
  &  \rep$-64$ & \attr$+64$ & \attr$+48$ \\
\thrbarA & \oneA & $0_A$ & $+$& $C\ga_0 \ga_5$ & LR & $\O(m)$ 
  & 0 &  \rep$-32$ &  \rep$-48$  \\
\thrbarA & \oneA & $1_{A}$ &$-$& $C\ga_3 \ga_5$ & LR & $\O(1)$ 
  & 0 & \attr$+32$ & \attr$+16$  \\
\hline
\sixS & \oneA & $1_S$ &$-$& $C\si_{03} \ga_5$ & LL & $\O(1)$ 
  &  \rep$-16$ & 0 & \attr$+4$ \\
\sixS & \oneA & $1_S$ & $+$ & $C\si_{03}$ & LL & $\O(1)$ 
  &  \attr$+16$ & 0 & \attr$+4$ \\
\sixS & \oneA & $0_{S}$ &$-$& $C\ga_0$ & LR & 0  
  & 0 & \attr$+8$ & \attr$+12$ \\
\sixS & \oneA & $1_S$ & $+$ & $C\ga_3$ & LR & $\O(1)$ 
  & 0 &  \rep$-8$ &  \rep$-4$ \\
\hline
\thrbarA & \thrS & $1_S$ &$-$& $C\si_{03} \ga_5$ & LL & $\O(1)$ 
  & 0 & 0 &  \rep$-16$ \st \\
\thrbarA & \thrS & $1_S$ & $+$ & $C\si_{03}$ & LL & $\O(1)$ 
  & 0 & 0 &  \rep$-16$ \\
\thrbarA & \thrS & $0_{S}$ &$-$& $C\ga_0$ & LR & 0 
  & 0 &  \rep$-32$  &  \rep$-48$ \\
\thrbarA & \thrS & $1_S$ & $+$ & $C\ga_3$ & LR & $\O(1)$ 
  & 0 & \attr$+32$ & \attr$+16$ \\
\hline
\sixS & \thrS & $0_A$ & $+$ & $C\ga_5$ & LL & $\O(1)$ 
  & 0 &  \rep$-16$ &  \rep$-12$  \\
\sixS & \thrS & $0_A$ &$-$& $C$ & LL & $\O(1)$ 
  & 0 &  \rep$-16$ &  \rep$-12$  \\
\sixS & \thrS & $0_A$ & $+$ & $C\ga_0 \ga_5$ & LR & $\O(m)$  
  & 0 & \attr$+8$ & \attr$+12$ \\
\sixS & \thrS & $1_{A}$ &$-$& $C\ga_3 \ga_5$ & LR & $\O(1)$ 
  & 0 &  \rep$-8$  &  \rep$-4$  \\
\hline\hline
\end{tabular}
\caption{Binding strengths of diquark channels in NJL models
in the mean-field approximation.
The first 6 columns specify the channels,
and the last 3 columns give their attractiveness in NJL
models with various
types of four-fermion vertex: 2-flavor instanton, single gluon exchange,
single magnetic gluon exchange (expected to dominate at higher density).
See equations \eqn{qqcond} and \eqn{qqstrength} 
and subsequent explanation.
}
\label{tab:channels}
\end{table}

The results of the binding strength calculation are shown in Table
\ref{tab:channels}. The first block is antisymmetric in flavor and
color, and so describes pairing of two flavors and two colors. 
The second
block is for two flavors and one color, the third for one flavor and
two colors, and the final block is for one color and one flavor.

\begin{figure}[hbt]
\begin{center}
\includegraphics[width=0.9\textwidth]{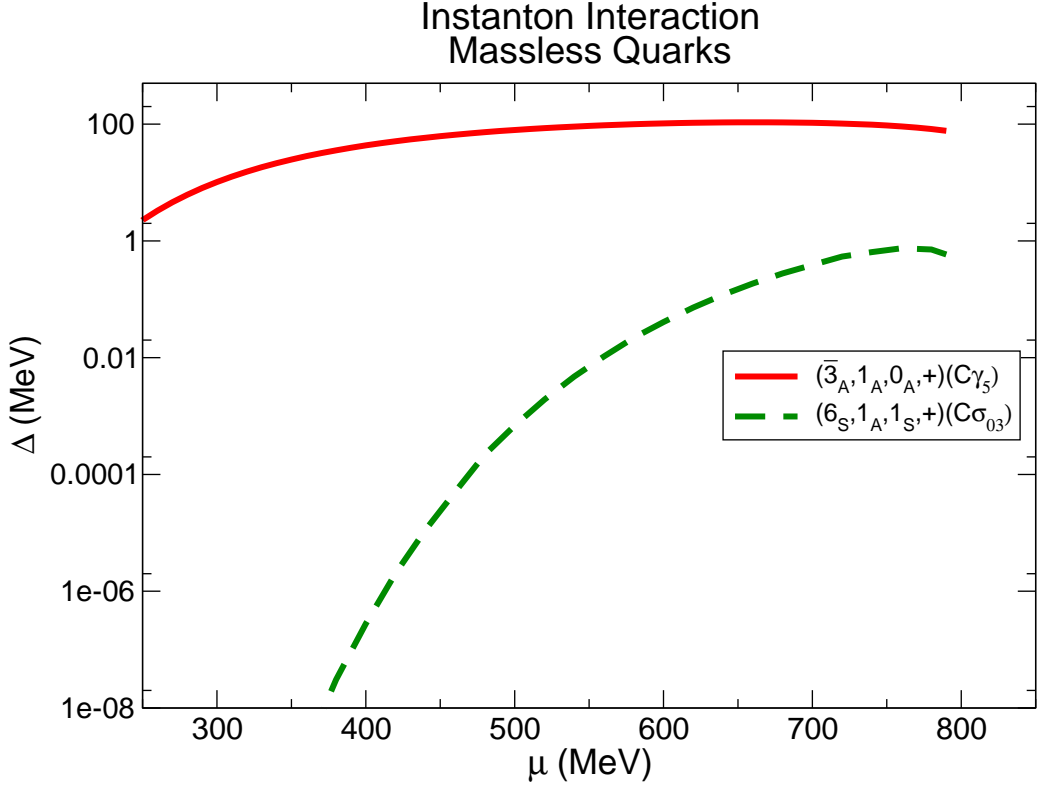}
\end{center}
\caption{Gap parameters in the attractive
channels for an NJL interaction based on the two-flavor instanton.
Since the instanton interaction requires two quark flavors, we
take the quarks to be massless, which is a good approximation for
the $u$ and $d$. The cutoff is $\La=800~\MeV$.}
\label{fig:Inst800}
\end{figure}

\begin{figure}[hbt]
\begin{center}
\includegraphics[width=0.9\textwidth]{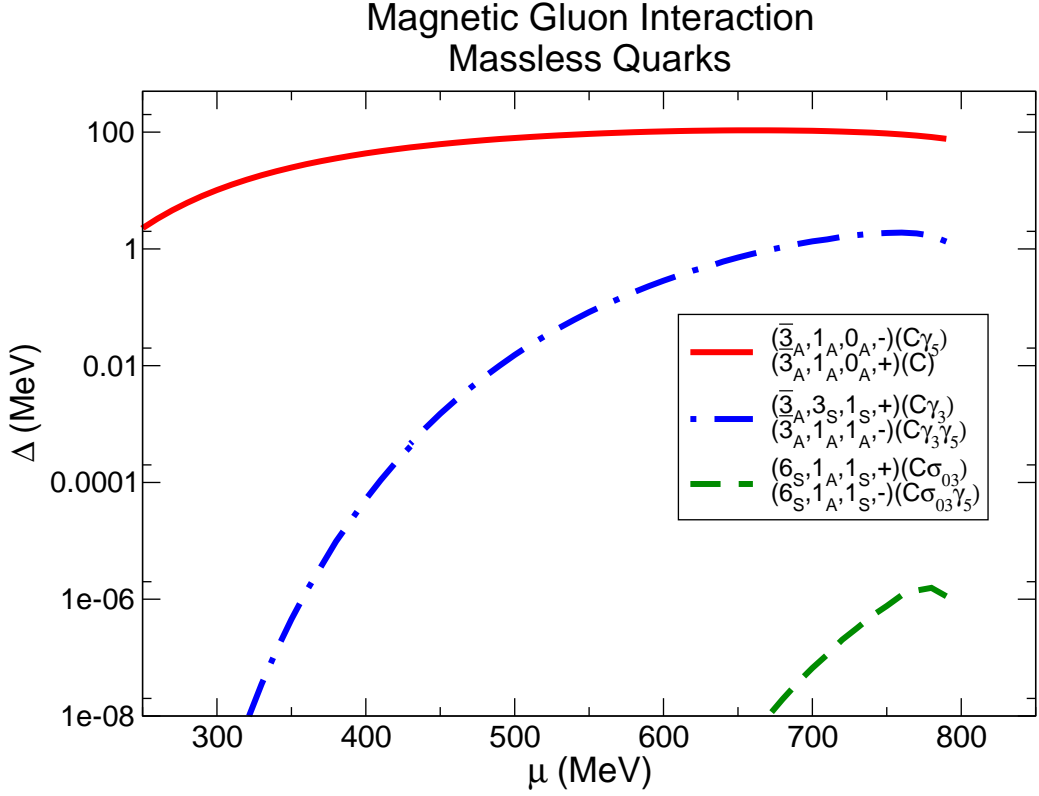}
\end{center}
\caption{Gap parameters in the attractive
channels for an NJL interaction based on 
magnetic-gluon exchange. We show the one-flavor and two-flavor channels,
for massless quarks.
The cutoff is $\La=800~\MeV$.
}
\label{fig:Mag800}
\end{figure}

\begin{figure}[hbt]
\begin{center}
\includegraphics[width=0.9\textwidth]{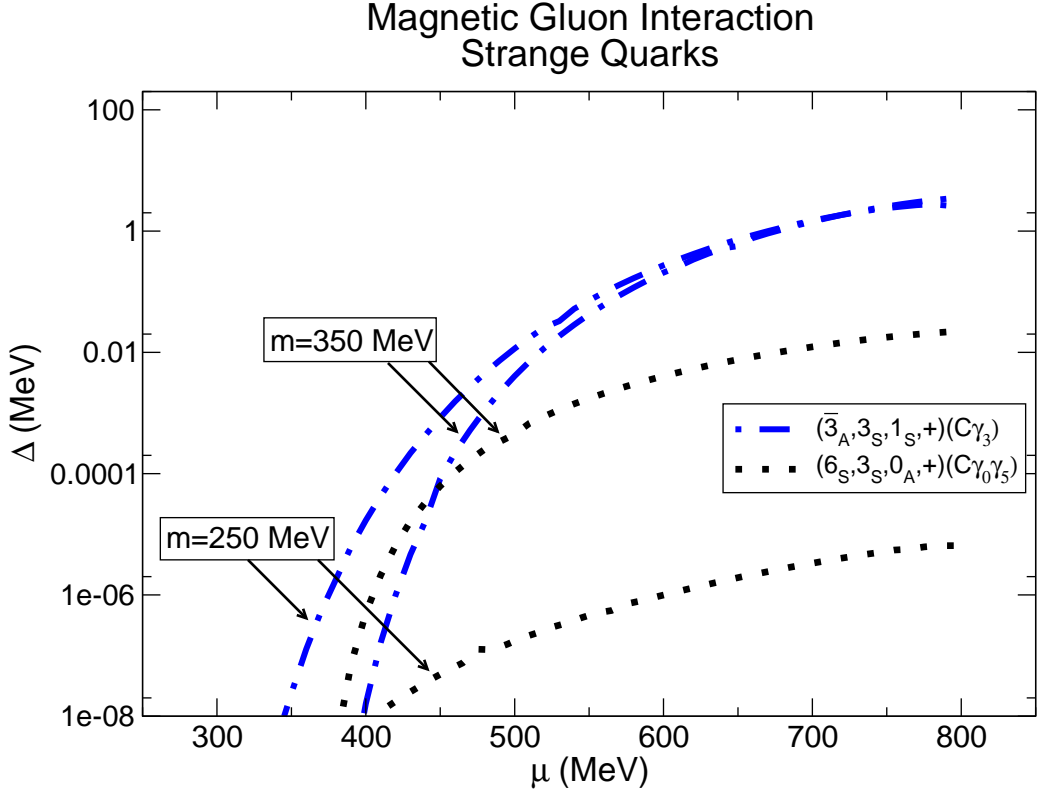}
\end{center}
\caption{Gap parameters in the attractive
channels for an NJL interaction based on magnetic-gluon exchange,
for quarks of mass 250 and 350~\MeV,
a reasonable range of values for the strange quark at medium density.
We show only the single-flavor channels. The cutoff is $\La=800~\MeV$.
}
\label{fig:Mag800strange}
\end{figure}

Certain features can be easily understood: the flavor-symmetric condensates
all have zero instanton binding energy, because the instanton
vertex is flavor-antisymmetric in the incoming quarks. 
The gluonic vertices give the same results for $C\ga_5$ as for $C$,
and for $C \si_{03}\ga_5$ as for $C \si_{03}$, because the gluonic
interaction is invariant under $U(1)_A$ transformations, 
under which the LL condensates
transform into each other ($C\ga_5 \rightleftharpoons C$ and
$C \si_{03}\ga_5 \rightleftharpoons C \si_{03}$) while the LR
condensates are invariant.
We see that there are many attractive channels:
\begin{list}{}{
 \setlength{\topsep}{-0.5\parskip}
 \setlength{\itemsep}{-0.5\parskip}
 }
\item[1)] Two colors and two flavors (\thrbarA,\oneA,$\ldots$).\\
The strongly attractive channel (\thrbarA,\oneA,0,$+$)($C\ga_5$) 
is the 2SC and CFL quark Cooper pairing pattern, and has been
extensively studied. The gap is large enough that even species 
with different masses, whose
Fermi momenta are quite far apart, can pair (hence the CFL phase
which pairs red and green $u$ and $d$, red and blue $u$ and $s$,
and green and blue $d$ and $s$ in this channel).
Its parity partner (\thrbarA,\oneA,0,$-$)($C$) is disfavored by instantons,
and is therefore unlikely to occur at phenomenologically interesting densities.
The additional channel
(\thrbarA,\oneA,1,$-$)($C\ga_3\ga_5$) is more weakly attractive
and also breaks rotational invariance,
and is therefore expected to be even less favored.
This is confirmed by gap equation calculations 
(Fig.~\ref{fig:Mag800}) which show that its gap is smaller 
by a factor of 10 to 100.
\item[2)] One color, two flavors (\sixS,\oneA,$\ldots$). \\
It is generally emphasized that the quark-quark interaction
is attractive in the color-antisymmetric \thrbarA\
channel. But, as we see in
table \ref{tab:channels}, the color-symmetric
(\sixS,\oneA,1,+)($C\si_{03}$) is attractive for instantons and
the magnetic gluon four-fermion interaction.
The instanton gives it a gap
of order 1~\MeV\ (Fig.~\ref{fig:Inst800}), while
the gluon interaction gives a small gap of order $1~\eV$
(Fig.~\ref{fig:Mag800}).
This channel was originally suggested for pairing of the blue
up and down quarks that are left out of 2SC \cite{ARW2},
and is discussed in more detail in Ref.~\cite{BHO}.
Its gap is small, so it could only pair quarks of similar mass,
i.e.~the light quarks, but
in a real-world uniform phase
such pairing will not occur either, because charge neutrality causes 
the up and down chemical potentials (and hence Fermi momenta)
to differ by tens of MeV, which is larger than the gap.
In a non-uniform mixture
of two locally charged phases \cite{Glendenning}, 
however, it is conceivable that
the up and down Fermi momenta could be similar enough to allow
pairing in this channel.
The parity partner (\sixS,\oneA,1,$-$)($C\si_{03}\ga_5$) is
disfavored by instantons. The channel 
(\sixS,\oneA,0,$-$)($C\ga_0$) is attractive, but 
has no particle-particle component, 
and presumably only occurs for sufficiently
strong coupling. Solving
the gap equations for reasonable coupling strength we find
no gap in this channel.
\item[3)] Two colors and one flavor (\thrbarA,\thrS,$\ldots$).\\
The only attractive channel is
(\thrbarA,\thrS,1,$+$)($C\ga_3$). 
This is a pairing option for
red and green strange quarks when the up and down 
quarks are paired in the 2SC state.  .
% Jeff has already done this gap eqn (by variational methods?)
We have solved the relevant gap equation
(Figs.~\ref{fig:Mag800},\ref{fig:Mag800strange}) 
and find gaps of about 1 MeV or less.
If three colors are available then
a competing possibility is to lock the colors to the spin (CSL),
so the condensate is
a linear combination of $C \ga_i$ and $C \si_{0i}$
with a color structure that is correlated with the spatial direction,
e.g.~red and green quarks pair in the $z$ direction, red and blue
in the $y$ direction, green and blue in the $x$ direction.
This leaves an unbroken global $SO(3)$ of spatial rotations
combined with color rotations, so the gap is isotropic, which
helps to lower the free energy \cite{TS1flav,Schmitt:2002sc}.
Note also that the channels
(\thrbarA,\thrS,1,$+$)($C\si_{03}$) 
and (\thrbarA,\thrS,1,$-$)($C\si_{03}\ga_5$)
which are repulsive in the NJL model become
attractive at asymptotic density when 
the gluon propagator provides a form factor that
strongly emphasizes small-angle scattering  \cite{TS1flav}.
\item[4)] One color and one flavor (\sixS,\thrS,$\ldots$).\\ There is
an attractive channel here, the (\sixS,\thrS,0,$+$)($C\ga_0\ga_5$).
It loses its particle-particle component as the quark mass goes to
zero, making it very weak for up and down quarks, but stronger for
strange quarks (Fig.~\ref{fig:Mag800strange}).  It is suitable for the
blue strange quarks when red and green strange quarks have paired in the
(\thrbarA,\thrS,1,$+$)($C\ga_3$) channel.
\end{list}

Many of the attractive channels have repulsive partners
with the
same symmetries, so a condensate in the attractive
channel will automatically generate a small additional one in the
repulsive channel.
For example, the (\thrbarA,\oneA,0,$+$)($C\ga_5$) can
generate (\thrbarA,\oneA,0,$+$)($C\ga_0\ga_5$). This was discussed in
Ref.~\cite{ABR2+1}, where the induced ($C\ga_0\ga_5$) condensate
(there called ``$\ka$'') in 2+1 flavor CFL was calculated and found
to be small. In Ref.~\cite{Fugleberg:2002rk} it was observed that
if all three quarks are massive then this condensate may be important.
In the context of CFL the (\thrbarA,\oneA,0,$+$)($C\ga_5$) can
also generate (\sixS,\thrS,0,+)($C\ga_5$) \cite{ARW3,Pisarski:1999cn}, 
since they both break the full symmetry group down to the same subgroup.

%Similarly, (\sixS,\thrS,0,$+$)($C\ga_0\ga_5$)
%can generate (\sixS,\thrS,0,$+$)($C\ga_5$), and so on.
% and (\thrbarA,\thrS,1,$+$)($C\ga_3$) can generate
% (\thrbarA,\thrS,1,$+$)($C\si_{03}$), and (\sixS,\oneA,1,+)($C\si_{03}$)
% can generate (\sixS,\oneA,1,+)($C\ga_3$).

\section{Gap calculations for the attractive diquark channels}
\label{sec:1scgap}

For the attractive channels we performed uncoupled gap equation calculations,
and obtained the dependence of the quark pairing on $\mu$. The
amount of pairing is given by the gap parameter $\De(\mu)$, which
occurs in the self energy (See Appendix \ref{app:gap}) as
\beq
\Delta^{\alpha\beta ab}_{ij}(p)=\Delta(\mu)\,\,{\cal C}^{\rm \alpha\beta}
  {\cal F}_{ij} \Gamma^{ab}~,
\eeq
with color matrix ${\cal C}$, flavor matrix ${\cal F}$, and Dirac
structure $\Gamma^{ab}$.
Note that $\De(\mu)$ is a gap {\em parameter}, not the gap. It sets
the scale of the gap in the quasiparticle excitation spectrum, but
as we will see in Section~\ref{sec:1scdispersion} the gap itself 
often depends on the direction of the momentum.

The 4-fermion interactions that we use are nonrenormalizable, so our
gap equations (shown explicitly in Appendix \ref{app:gapeqns}) involve
a 3-momentum cutoff $\La$, which represents the decoupling of our
interactions at higher momentum, due to instanton form factors,
effective gluon masses, etc.  The usual procedure for NJL model
calculations is to calibrate the coupling strength for each cutoff
$\La$ by known low-density physics such as the size of the chiral
condensate. However, it is well known that this leads to an
approximately cutoff-independent maximum gap (as a function of $\mu$)
in the $\psi C\ga_5\psi$ channel, so we used that criterion directly
as our calibration condition, setting the maximum gap to $100~\MeV$.

The results of our calculations,
for cutoff $\La=800~\MeV$, are plotted in
Figs.~\ref{fig:Inst800},
 \ref{fig:Mag800}, and \ref{fig:Mag800strange}. For other cutoffs the overall shape
of the curves is very similar. Because we use a sharp cutoff $\La$, the gap
falls to zero when $\mu$ reaches $\La$ (see, e.g., Eq.~\eqn{analyticgap}).
We show gap plots for the instanton interaction (Fig.~\ref{fig:Inst800})
and magnetic gluon interaction. The full electric + magnetic gluon gives
results that are similar to those for the magnetic gluon, but with
no gap in the (\sixS,\oneA,1,+)($C\si_{03}$) channel.

% (3bar,1,1,-) (i.e. NOT (3bar,3,1,+)
For the magnetic gluon, we show a gap plot for massless quarks
(Fig.~\ref{fig:Mag800}) which includes the two-flavor channels
(\thrbarA,\oneA,0,+)($C\ga_5$),  (\thrbarA,\oneA,1,$-$)($C\ga_3\ga_5$)
and (\sixS,\oneA,1,+)($C\si_{03}$)
which could sustain $u$-$d$ pairing, as well as the single-flavor
channel (\thrbarA,\thrS,1,+)($C\ga_3$) which could sustain $u$-$u$
or $d$-$d$ pairing.

% We also show a gap plot for quarks with a mass appropriate to the 
% strange quark (Fig.~\ref{fig:Mag800strange})
% which includes the single-flavor channels that could sustain $s$-$s$
% pairing with two colors (\thrbarA,\thrS,1,+)($C\ga_3$) 
% or one color (\sixS,\thrS,0,+)($C\ga_0\ga_5$).

We also show a gap plot (Fig.~\ref{fig:Mag800strange})
  for a single quark flavor with mass
  of 250 or 350 MeV. This is appropriate for the strange quark, whose 
  effective mass is expected to lie between 150 MeV and about
  400 MeV (see Ref.~\cite{BuballaOertel}, Fig.~1).

The relative sizes of the gaps in the different channels reflect the
pairing strengths given in Table \ref{tab:channels}. We see that the
Lorentz scalar (\thrbarA,\oneA,0,+)($C\ga_5$) (solid line) is dominant.
The $j=1$ channels have much smaller gap parameters.
The (\thrbarA,\thrS,1,+)($C \ga_3$) 
gap parameter (dash-dot line in Figs.~\ref{fig:Mag800} 
and \ref{fig:Mag800strange})
rises to a few \MeV\ with the magnetic gluon interaction.
The (\sixS,\oneA,1,+)($C \si_{03}$) gap parameter
(dashed line) rises to about 1~\MeV\ with an instanton interaction,
but only 1~\eV\ with the magnetic gluon interaction. 
It should be remembered, however, that
the temperature of a compact star can be anything
from tens of MeV at the
time of the supernova to a few eV after millions of years, so
gaps anywhere in this range are of potential phenomenological interest.

The (\sixS,\thrS,0,+)($C\ga_0\ga_5$) channel (dotted line), 
which is the only attractive channel for
a single color and flavor of quark, is highly suppressed for
massless quarks at high density
but reaches about 10~\keV\ for strange
quarks ($m=350~\MeV$, Fig.~\ref{fig:Mag800strange}).
This is because its particle-particle component goes to
zero as $m\to 0$ (Eq.~\eqn{BCSenhancement} and Table~\ref{tab:channels}).

Up to this point we have not mentioned the $j=1, m_j=\pm 1$ channels
(e.g.~$\psi C\ga_\pm\psi \equiv \psi C(\ga_1\pm i\ga_2)\psi$).
We have only discussed the $j=1, m_j=0$
channels (e.g.~$\psi C\ga_3\psi$). That is because rotational invariance
of the interaction Hamiltonian that we are using
guarantees that changing $m_j$ from 0 to $\pm 1$ will not affect
the binding
energy and gap equation. This can be seen by considering the form of
the binding energy. From \Eqn{hamiltonian} it is
\beq
E_B \sim \<\psi\psi\>^{\dag ac} \<\psi\psi\>^{bd} {\cal H}_{abcd}~.
\eeq
Note that it is quadratic in the diquark condensate, with one of the
factors being complex conjugated.
So if we have some 3-vector condensate,
for example $\phi=\sum_i \phi_i \<\psi\ga_i \psi\>$, then its binding
energy is
\beq
E_B \propto |\phi_x|^2 + |\phi_y|^2 + |\phi_z|^2
\eeq
It is clear that the $m_j=0$ condensate $\phi_i=(0,0,1)$ has the same
binding energy as the $m_j=\pm 1$ condensate $\phi_i=(1/\sqrt{2})(1,\pm i,0)$.
We have explicitly solved the gap equations for the $m_j=\pm 1$
condensates, and find their solutions identical to the corresponding
$m_j=0$ condensates.
However, the quasiquark excitations in the two cases are quite different, 
and we proceed to study these in the next section.

\section{Quasiquark dispersion relations}
\label{sec:1scdispersion}

The physical behavior of quark matter will be dominated by its lowest energy
excitations. As well as Goldstone bosons that arise from spontaneous
breaking of global symmetries, there will be fermionic excitations
of the quarks around the Fermi surface.
In the presence of a diquark condensate, the spectrum of quark
excitations is radically altered. Instead of arbitrarily low energy
degrees of freedom, associated with the promotion of a quark from a
state just below the Fermi surface to just above it, there is
a minimum excitation energy (gap), above which the excitation
spectrum is that of free
quasiquarks, which are linear combinations of a particle and a hole.

The dispersion relations of the quasiparticles can be calculated 
straightforwardly by including a condensate of the desired structure
in the inverse propagator $S^{-1}$, shown in Eq.~\eqn{invprop}.
Poles in the propagator correspond to zeros in $S^{-1}$, so the
dispersion relations are obtained by solving 
$\det S^{-1}(p_0,\vp,\mu,\De,m)=0$ for the energy $p_0$
as a function of the 3-momentum $\vp$ of the quasiparticle,
quark chemical potential $\mu$, gap parameter $\De$, and quark mass $m$.

\begin{figure}[hbt]
\begin{center}
\includegraphics[width=0.8\textwidth,angle=-90]{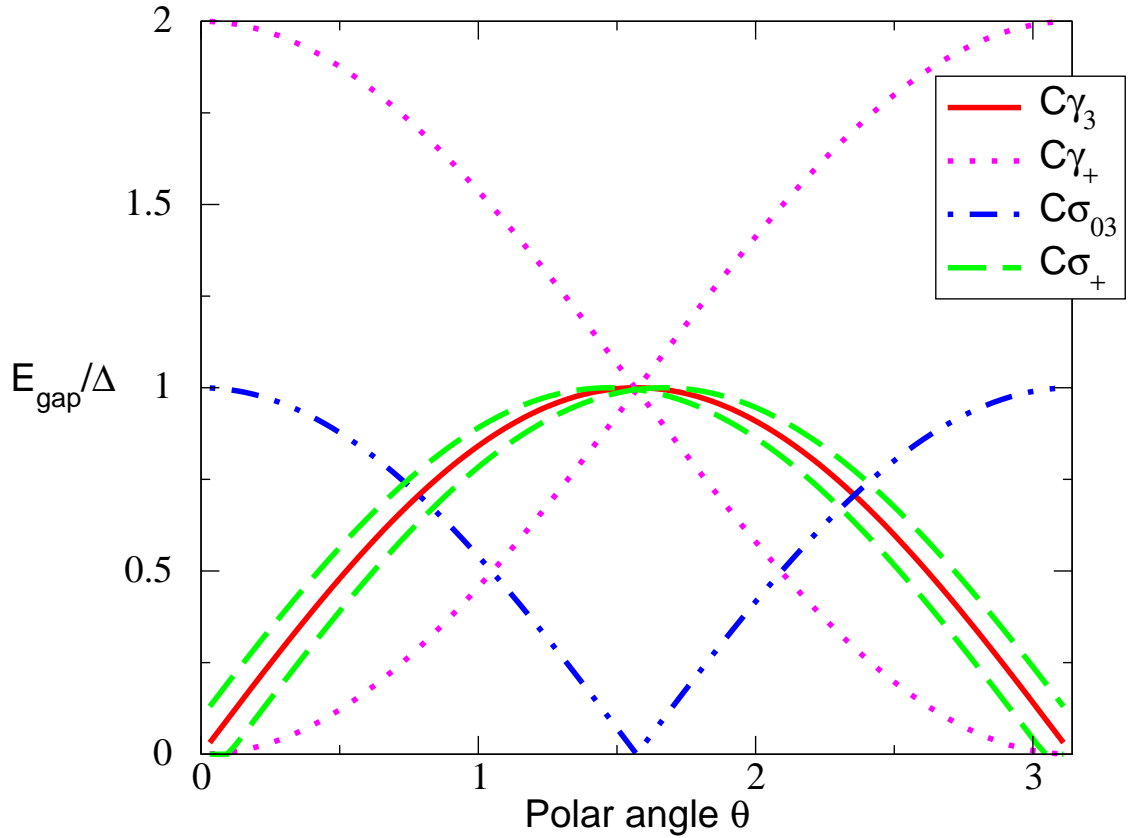}
\end{center}
\caption{Energy gap in units of the gap parameter
as a function of polar angle on the
Fermi surface for 
rotational symmetry breaking phases with
massless quarks, at $\mu=500~\MeV$, gap parameter $\De=50~\MeV$.
$\ga_+ \equiv \ga_1 + i\ga_2$, $\si_+ \equiv \si_{01} + i\si_{02}$.
}
\label{fig:DRmassless}
\end{figure}

\begin{figure}[hbt]
\begin{center}
\includegraphics[width=0.8\textwidth,angle=-90]{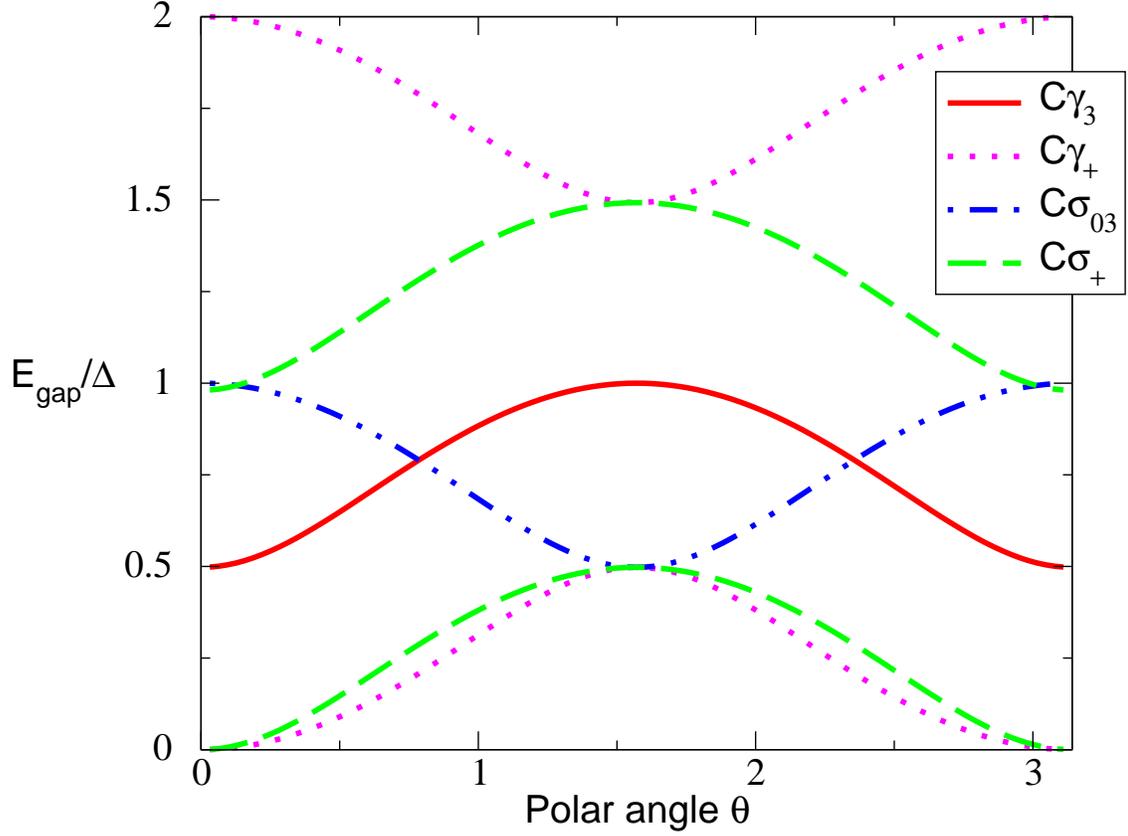}
\end{center}
\caption{Energy gap in units of the gap parameter
as a function of polar angle on the
Fermi surface for
rotational symmetry breaking phases at 
$\mu=500~\MeV$, with gap parameter $\De=50~\MeV$.
The quarks have mass $m=250~\MeV$.
$\ga_+ \equiv \ga_1 + i\ga_2$, $\si_+ \equiv \si_{01} + i\si_{02}$.
}
\label{fig:DRmassive}
\end{figure}

The gap is by definition the energy required to excite
the lowest energy quasiquark mode. Isotropic condensates have a uniform
gap, but one of the most interesting features of $j> 0$ condensates
is that they are not in general fully gapped: the gap goes to zero
for particular values of momentum $\vp$, which correspond to
particular places on the Fermi surface. This means that
transport properties such as viscosities and emissivities, 
which are suppressed by factors of $\exp(-\De/T)$ in phases
with isotropic quark pairing, may not be so 
strongly suppressed by a $j>0$ condensate.
In Figs.~\ref{fig:DRmassless} and \ref{fig:DRmassive} we show the 
variation of the gap over the Fermi surface by plotting
the energy of the lowest excitation as a function of angle, 
\beq
E_{\rm gap}(\th) = \min_{p,i} |E_i(p,\th)|
\eeq
where $E_i(p,\th)$
is the energy of the $i^{\rm th}$ quasiquark excitation
with momentum $(p\sin\th\cos\phi, p\sin\th\sin\phi, p\cos\th)$.
For the plots we take $\mu=500~\MeV$ and $\De=50~\MeV$,
with quark mass $m=0$ (Fig.~\ref{fig:DRmassless}) or
$m=250~\MeV$ (Fig.~\ref{fig:DRmassive}).

\newcommand{\mueff}{\mu_{\rm eff}}
\newcommand{\Deltaeff}{\De_{\rm eff}}

\begin{list}{\arabic{enumi})}{
  \usecounter{enumi}
  \setlength{\itemsep}{-0.5\parsep} % added to \parsep
  \setlength{\labelwidth}{0mm}
  \setlength{\labelsep}{0mm}
  \setlength{\leftmargin}{0 mm} % space for label and sep
 }
\item ~~{$C\ga_3$} condensate: $j=1,m_j=0$.\\
There is one quasiquark excitation with
energy less than the gap parameter $\De$.
\beq
\ba{rcl}
E(p)^2 &=& (\sqrt{p^2 + m^2 \mu^2/\mueff^2}\pm \mueff)^2 + \Deltaeff^2\\[1ex]
\mueff(\th)^2 &=& \mu^2 + \Delta^2 \cos^2(\theta) \\[1ex]
\Deltaeff(\th)^2 &=& \Delta^2 \Bigl( 
  \sin^2(\theta) + m^2/\mueff^2 \cos^2(\theta) \Bigr)
\ea
\eeq
From the expression for $\Deltaeff(\th)$ we see that for massless quarks
the gap goes to zero for momenta parallel to the $z$-axis, i.e.~at the
poles on the Fermi surface (solid curve in Fig.~\ref{fig:DRmassless}). 
Massive quarks retain a small gap of order
$ m\De/\mu$ at the poles (solid curve in Fig.~\ref{fig:DRmassive}).

\item ~~$C(\ga_1\pm i\ga_2)$ condensate: $j=1,m_j=\pm 1$.\\
There are two quasiquark excitations with
energy less than $2\De$,
\beq
\ba{rl}
E(p)^2 = 2\Delta^2 + m^2+ \mu^2 + p^2 \pm &
  \Bigl( 4\De^4 + 4\mu^2(p^2+m^2) +  2\De^2 p^2(1 - \cos(2\th)) \\ 
  & \pm 4\De^2\mu \sqrt{4 m^2 + 2 p^2(1+\cos(2\th))} \Bigr)^\half
\ea
\eeq
For this condensate the effective gap again goes to zero at the poles,
but in this case it remains zero even in the presence of a quark mass
 (dotted curve in
Figs.~\ref{fig:DRmassless},\ref{fig:DRmassive}).

\item ~~$C\si_{03}$ condensate:  $j=1,m_j=0$.\\
There is one quasiquark excitation with energy less than $\De$,
its dispersion relation is \cite{BHO}
\beq
\ba{rcl}
E(p)^2 &=& (\sqrt{p^2 + m^2 \mu^2/\mueff^2}\pm \mueff)^2 + \Deltaeff^2\\[1ex]
\mueff(\th)^2 &=& \mu^2 + \Delta^2 \sin^2(\theta) \\[1ex]
\Deltaeff(\th)^2 &=& \Delta^2 \Bigl( 
  \cos^2(\theta) + m^2/\mueff^2 \sin^2(\theta) \Bigr)
\ea
\eeq
This is related to the dispersion relation for the 
$C\ga_3$ condensate by $\th \to \pi/2-\th$: for massless quarks the
quasiquarks are gapless around the equator of the Fermi sphere
(dash-double-dot curve in Fig.~\ref{fig:DRmassless})
and in the presence of a quark mass they gain a small gap of order
$\De m/\mu$ (dash-double-dot curve in Fig.~\ref{fig:DRmassive}). 
The equator is a larger proportion of the Fermi surface than
the poles, so in this case we might expect a greater effect
on transport properties.

\item ~~$C(\si_{01}\pm i\si_{02})$ condensate: $j=1,m_j=\pm 1$.\\
There are two quasiquark excitations with 
energy less than $\De$. They have
rather complicated dispersion relations. 
Going to the massless
case, and assuming $E,(p-\mu) \ll \mu$, which will be true for the low-energy
quasiquark degrees of freedom that we are interested in, we find
\beq
\ba{rcl}
E(p) &=& (\De^2 \mu + \De^2 p \cos(\th) \pm \eta/\sqrt{2})/(2\mu^2) \\
\eta^2 &=& ( 8\mu^4(\mu - p)^2 + 8\De^2\mu^2(\mu^2 - \cos(\th)^2 p^2) +
  2\De^4(\mu + \cos(\th) p)^2 )
\ea
\eeq
In this case there is a region near the poles, $\th \lesssim \De/\mu$,
where the gap is zero (dashed curve in Fig.~\ref{fig:DRmassless}). 
This is because at those angles $E(p)$ has
zeros at  two values of $p$ close to $\mu$. When $\th\approx \De/\mu$
those two zeros merge and disappear from the real $p$ axis.
The presence of a quark mass $m>\De$ wipes out this effect, but there
is still no gap at the poles on the Fermi surface (dashed curve
in Fig.~\ref{fig:DRmassive}).
\end{list}

In comparing Figs.~\ref{fig:DRmassless} and \ref{fig:DRmassive}
it is interesting to note that introducing a mass for the quark
opens up a gap whenever the  gap lines intersect each other
at a non-zero angle (after one includes the mirror-image negative-energy
gap curves for the quasiholes). This occurs at zero energy at the poles
for $C\ga_3$ and at the equator for $C\si_{03}$. It occurs at
non-zero energy for $C(\ga_1\pm i\ga_2)$. The case of 
$C(\si_{01}\pm i\si_{02})$ is similar, but it is not obvious from the
gap plot for reasons described above.

We see that the $j\neq 0$ phases show a rich variety of
quasiquark dispersion relations. 
For massless quarks they are all gapless in special regions
of the Fermi surface, and for massive quarks the $m_j=\pm 1$
condensates remain gapless for momenta parallel to the spin.
It follows that for these phases the quasiquark excitations will
play an important role in transport properties, even
when the temperature is less than the gap parameter.

Moreover, different condensates ($m_j=\pm 1$ vs.~$m_j=0$) that 
because of rotational invariance of the Hamiltonian have
exactly the same binding energy and gap equation, nevertheless
have completely different energy gaps over the Fermi surface.
They will therefore behave quite differently when exposed
to nonisotropic external influences, such as magnetic fields
or neutrino fluxes, and also in their coupling to external
sources of torque, e.g.~via electron-quasiquark scattering.
All these influences are present in compact stars, and it will
be interesting and complicated to sort out which is favored
under naturally occurring conditions. And it should not
be forgotten that these conditions vary with the age
of the star. 

\section{Conclusions}
\label{sec:1scconclusions}

In this chapter we have completed a mean-field survey of 24 different
channels for diquark condensation, using an NJL model with Fermi
couplings for instantons and pointlike gluons.  Our catalog of
channels is by no means comprehensive, but it does exhaust the
possibilities for structures that can be factorized into separate
tensors for color, flavor, and spin (notably, this excludes interesting
``locking'' phases such as color-flavor-locking or
color-spin-locking).

As we promised in section \ref{sec:1scoverview}, we can identify
single-color and/or single-flavor phases that are useful for pairing
``orphaned'' quarks in the various pairing scenarios described at the
start of this chapter.  In fact our survey reveals five channels that
are attractive and therefore susceptible to BCS condensation in the
NJL model.  In the same order that they are listed in table
\ref{channels}, these channels are:
\begin{center}
\begin{tabular}{ll}
2SC & (\thrbarA,\oneA,0,$+$)($C\ga_5$)  \\
spin-one 2SC & (\thrbarA,\oneA,1,$-$)($C\ga_3\ga_5$) \\
single-color & (\sixS,\oneA,1,+)($C\si_{03}$) \\
single-flavor & (\thrbarA,\thrS,1,$+$)($C\ga_3$) \\ 
single-color and single-flavor & (\sixS,\thrS,0,$+$)($C\ga_0\ga_5$) \\
\end{tabular}
\end{center}
The first is well-known, the second is an obscure (and never-favored)
spin-one cousin of the 2SC phase, and the last three are the channels
that are good for pairing orphaned quarks.  For these attractive
channels we have calculated gaps and the results are shown in
Figs.~\ref{fig:Inst800}, \ref{fig:Mag800}, and
\ref{fig:Mag800strange}.  (Note that Fig.~\ref{fig:Mag800} also shows
two more channels which are parity-partners of the 2SC and
single-color channels.  These channels are attractive for gluons, but
they are parity odd and disfavored by instantons.)

The gap of the 2SC phase has been calculated, by various methods, to
be in the range of 10 to 100 MeV.  In the present work we have {\em
assumed} a maximum 2SC gap of 100 MeV and used this assumption to
calibrate the NJL models.  Then we find, as shown in
Figs.~\ref{fig:Inst800}, \ref{fig:Mag800}, and
\ref{fig:Mag800strange}, that the ``exotic'' phases have gaps that are
no larger than about 1 MeV and can be orders of magnitude smaller than
this.  Notice that while the 2SC gap curve is fairly flat on these
plots, the gap curves for the exotic phases are quite steep.  Because
these are semilog plots this implies that the exotic gaps are {\em
drastically} $\mu$-dependent.  They can change by more than two orders
of magnitude when the chemical potential is increased from 400 MeV to
500 MeV.

Moreover, the gap values are very sensitive to the choice of NJL
model.  For example, the gap for the single-color phase is seen to
differ by {\em six} orders of magnitude when it is calculated with an
instanton interaction (dashed line in Fig.~\ref{fig:Inst800}) versus a
magnetic gluon interaction (dashed line in Fig.~\ref{fig:Mag800}).  In
both cases the NJL model is calibrated to give the same value (100
MeV) for the 2SC gap.  Both the instanton and magnetic gluon
interactions are reasonable models for some physics at moderate
densities, but they yield very different results for the gap in this
single-color phase.

Buballa {\it et al}~\cite{BHO} observed similar difficulties in their
NJL investigation of the single-color state.  The model uncertainties
cannot be resolved by appealing to a perturbative calculation at
asymptotically high densities, because the long-range magnetic gluon
which dominates at these densities~\cite{Son} is always repulsive in
the color-symmetric channel~\cite{TS1flav} and therefore does not
predict a single-color pairing state.

The calculation for the single-flavor phase is also very
model-sensitive.  However, in this case the state can be studied with
a perturbative, model-independent calculation in the high density
limit~\cite{TS1flav,Schmitt:2002sc}.  In this regime we find the
simple result that the spin-one (single-flavor) gap is $\Delta = c
\Delta_0$, where $\Delta_0$ is the spin-zero (2SC) gap and $c$ a
constant (independent of $\mu$) that is somewhere between 0.002 and
0.01 (the constant is known exactly and depends on the particular spin
ansatz, e.g.~polar phase or color-spin-locking phase).  The spin-one
gap therefore varies less drastically with $\mu$ than in NJL models
because it is just a constant fraction of the spin-zero gap. If we
assume (without justification) that this relation holds at moderate
densities, we predict a spin-one gap of 200 keV to 1 MeV, for a
spin-one gap of 100 MeV.  For $\mu$ of 400 to 500 MeV, the NJL
calculation predicts a much smaller gap that ranges from 0.1 to 10 keV
in this $\mu$ interval.

In the high-density limit, the single-flavor condensate also includes
an admixture of the $C \sigma_{03}$ spin channel.  This channel is
repulsive in the NJL model but attractive for long-range magnetic
gluons.  In the high-density limit the color structure alone dictates
the attractiveness of the pairing channel: color-symmetric pairing is
repulsive and color-antisymmetric pairing is attractive, independent
of color or spin~\cite{TS1flav}.  But this simple conclusion is
unlikely to persist at moderate densities when various in-medium
effective interactions (like those of our NJL model) can supersede
this perturbative result.

The preceding discussion is intended to convey that our NJL
calculations should be interpreted conservatively.  They are useful
because they indicate which pairing channels are most likely to be
attractive at moderate densities that are of interest to us.  But they
provide only very rough estimates for the magnitudes of the gaps in
the various channels.  In our summary of pairing channels shown in
table \ref{channels}, we have estimated rough upper limits for the
gaps by taking the peak values of these gaps from figures
Figs.~\ref{fig:Inst800}, \ref{fig:Mag800}, and
\ref{fig:Mag800strange}.  These numbers should be interpreted as
reasonable upper bounds for gaps which could be orders of magnitude
smaller.

Although many of the channels we have studied have very small gaps and
therefore very small critical temperatures, they will be
phenomenologically relevant if they are the best pairing option
available for some of the quarks. Since the temperature of a compact
star falls to tens of eV when its age reaches millions of years,
pairing can occur in such channels late in the star's life.  The
corresponding quasiquark excitations will become massive and this can
suppress their participation in transport processes.  With this in
mind it would be useful to extend our analysis of the exotic
condensates to finite temperature.  Recent work indicates that the
usual BCS relationship between the critical temperature and the gap
parameter may be modified for spin-one condensates
\cite{BHO,Schmitt:2002sc}.

The spin-one condensates are gapless at special 3-momenta (or nearly
so with nonzero quark masses).  Recall that the $(j=1,m_j=\pm 1)$
condensates ($\psi C(\gamma_1\pm i\gamma_2)\psi$ or $\psi
C(\sigma_{01}\pm i \sigma_{02})\psi$) always have gapless quasiquarks
at the poles of the Fermi sphere.  The $j=1,m_j=0$ condensates have
gapless regions for massless quarks, at the poles ($\psi C\ga_3\psi$)
or around the equator ($\psi C\si_{03}\psi$), but if the quarks are
massive then the quasiquarks have a minimum gap of order $m\De/\mu$
(Figs.~\ref{fig:DRmassless}, \ref{fig:DRmassive}).  The gapless or
nearly gapless modes may continue to play a role in transport
processes, even when the temperature is much less than the gap
parameter.  It would be very useful to develop a transport theory for
the spin-one condensates, to determine how the light modes contribute
to neutrino emission/absorption via URCA processes or otherwise, and
how they affect specific heat, viscosity, conductivities, etc.  A
natural first step would be to write down an effective theory, which
would contain the lowest quasiquark modes and Goldstone bosons arising
from the breaking of rotational symmetry (which could be called ``spin
waves'' by analogy with helium-3~\cite{helium3}).

If spin-one color condensates occur in compact stars, they are likely
to have interesting and complicated spatial and rotational textures
like those of the various spin-one helium-3 phases.  The $j=1$
condensates can carry angular momentum simply by aligning themselves
in large domains, without involving any superfluid vortices, but it
seems they will typically occur in conjunction with other phases that
are superfluid.  It would be interesting to investigate how the
angular momentum is carried in this situation.

Finally, we note that since the single-flavor color superconductor has
a small gap and also exhibits an electromagnetic Meissner
effect~\cite{Spin1Meissner} (unlike the CFL phase~\cite{ABRmag}), it
could have an upper critical magnetic field smaller than that of some
pulsars.  We can estimate the critical field by assuming that the
critical magnetic energy density is comparable to the condensation
energy of the color superconductor, i.e.~$B_c^2/8\pi \sim \mu^2
\Delta^2/\pi^2$.  By this estimate, pulsars with typical fields of
$10^{12}-10^{13}$ Gauss could destroy single-flavor pairing if the gap
is smaller than $0.1-1$ keV, and magnetars with fields of order
$10^{15}-10^{16}$ Gauss could destroy condensates with gaps of as
large as $0.1-1$ MeV.

%(4) The various $j=1$ condensates show completely different variation
%of the energy gap over the Fermi surface
%(Section~\ref{sec:1scdispersion}).  It would be useful to know how
%they behave when exposed to the nonisotropic external influences that
%are common in compact stars, such as magnetic fields or neutrino
%fluxes, and also in their coupling to external sources of torque, eg
%via electron-quasiquark scattering.

%% file: chap5.tex
\chapter{Applications and Outlook}

\section{Overview}

In this final chapter we take a closer look at two physical settings in
which the crystalline phase might occur with observable consequences.
The first setting is astrophysical: the quark matter core of a
rotating compact star, where a layer of crystalline quark matter can
pin vortices and cause glitches.  The second setting is terrestrial:
an ultracold trapped gas of fermionic atoms, where a spin imbalance
can induce the formation of a crystalline superfluid.  In the last
section, we close by describing unsolved problems and directions for
future research.

\section{Pulsar glitches}
\label{sec:glitches}

Many pulsars have been observed to glitch.  Glitches are sudden jumps
in rotation frequency $\Omega$ which may be as large as
$\Delta\Omega/\Omega\sim 10^{-6}$, but may also be several orders of
magnitude smaller.  The frequency of observed glitches is
statistically consistent with the hypothesis that all radio pulsars
experience glitches~\cite{AlparHo}.  Glitches are thought to originate
from interactions between the rigid crust, somewhat more than a
kilometer thick in a typical neutron star, and rotational vortices in
the neutron superfluid.  The inner kilometer of the crust consists of
a rigid lattice of nuclei immersed in a neutron
superfluid~\cite{NegeleVautherin}.  Because the pulsar is spinning,
the neutron superfluid (both within the inner crust and deeper inside
the star) is threaded with a regular array of rotational vortices.  As
the pulsar's spin gradually slows due to emission of electromagnetic
radiation, these vortices must gradually move outwards since the
rotation frequency of a superfluid is proportional to the density of
vortices.  Deep within the star, the vortices are free to move
outwards.  In the crust, however, the vortices are pinned by their
interaction with the nuclear lattice.  What happens next varies from
model to model.  Perhaps the vortices exert sufficient force on the
crust to tear it apart, resulting in a sudden breaking and
rearrangement of the crust and a change in the moment of
inertia~\cite{RudermanGlitch}.  Perhaps a large cluster of vortices
within the inner crust builds up enough outward pressure to overcome
the pinning force, suddenly becomes unpinned, and moves
macroscopically
outward~\cite{AndersonItoh,AndersonEtAl,AAPS1,AAPS2,AAPS3,PinesAlpar,Recent}.
This sudden decrease in the angular momentum of the superfluid within
the crust results in a sudden increase in angular momentum of the
rigid crust itself, and hence a glitch.  Perhaps, due to interactions
between neutron vortices and proton flux tubes, the neutron vortices
pile up just inside the inner crust before suddenly coming
unpinned~\cite{SedrakianCordes}.  Although the models differ in
important respects, all agree that the fundamental requirements are
the presence of rotational vortices in a superfluid, and the presence of a
rigid structure which impedes the motion of these vortices (by a
pinning force of suitable magnitude) and which encompasses enough of
the volume of the pulsar to contribute significantly to the total
moment of inertia.\footnote{The first model of glitches which was
proposed~\cite{Starquake} relies on the cracking and settling of the
neutron star crust (``starquakes'') as the neutron star spins down.
This model does not require the presence of rotational vortices.
However, this model fails to explain the magnitude and frequency of
glitches in the Vela pulsar~\cite{PinesAlpar,Recent}.}

Although it is premature to draw quantitative conclusions, it is
interesting to speculate that some glitches may originate not at the
crust, but deep within a pulsar which has a color superconducting
quark matter core.  If some region of the core is in the crystalline
phase, because this phase is a superfluid it will be threaded by an
array of rotational vortices, and these vortices may be pinned in the
crystal by the periodic spatial modulation of the diquark condensate.
The basic reasoning is that because the diquark condensate must vanish
in the core of a rotational vortex, the vortex might prefer to reside
at at a node of the LOFF crystal.  It is interesting to note that
enhanced pinning of magnetic flux tubes has been proposed as an
experimental signature of the LOFF phase in an electron
superconductor~\cite{Modler}.  (However, the analogy may be
misleading, because the underlying pinning mechanisms may differ for
quark matter versus electrons in a solid, as we will discuss.)

Before we delve into an assessment of the feasibility of quark matter
glitching, it is worth noting that perhaps the most interesting
consequence is for compact stars made entirely of quark matter (also
commonly called ``strange quark matter'' because it will contain
strange quarks in addition to up and down quarks).  The work of
Witten~\cite{Witten} and Farhi and Jaffe~\cite{FarhiJaffe} suggests
that strange quark matter may be energetically stable relative to
nuclear matter even at zero pressure.  Then it might occur that
observed compact stars are strange quark stars~\cite{HZS,AFO} rather
than neutron stars.  This has recently been suggested for certain
accreting compact stars~\cite{Bombaci}, although the evidence is far
from unambiguous~\cite{ChakrabartyPsaltis}.\footnote{
Recently, the isolated compact star RXJ185635-3754 has also been
proposed as a candidate strange star from a simple black-body fit to
its observed thermal radiation~\cite{Drake}, but this result is
contradicted: the spectrum deviates from black-body, and when
atmospheric effects are taken into account the inferred radius is
consistent with conventional stellar
models~\cite{PWLPN,WalterLattimer}.
}
A conventional neutron star may feature a core made of strange quark
matter, but strange quark stars are made (almost) entirely of quark
matter with either no hadronic matter content at all or with a thin
crust, of order one hundred meters thick, which contains no neutron
superfluid~\cite{AFO,GlendenningWeber} (the nuclei in this thin crust
are supported above the quark matter by electrostatic forces; these
forces cannot support a neutron fluid).  Because of the absence of
superfluid neutrons, and because of the thinness of the crust, no
successful models of glitches in the crust of a strange quark star
have been proposed.  Since pulsars are observed to glitch, the
apparent lack of a glitch mechanism for strange quark stars has been
the strongest argument that pulsars cannot be strange quark
stars~\cite{Alpar,OldMadsen,Caldwell}.  This conclusion must now be
revisited: the quark matter in a strange star would be a color
superconductor, and glitches may originate from pinning of vortices
within a layer of the strange star which is in a crystalline color
superconducting state\footnote{Madsen's conclusion~\cite{Madsen} that
a strange quark star is prone to r-mode instability due to the absence
of damping must also be revisited, since the relevant fluid
oscillations may be damped within or at the boundary of a region of
crystalline color superconductor.}.  

Crystalline quark matter glitching is more interesting for strange
stars than for other compact stars, because it is the only way for
strange stars to glitch.  Certainly, crystalline quark matter
glitching could also occur in a more conventional neutron star with a
quark matter core, but the core would be one of two potential
locations where glitches may originate.  Any attempt to observe a
quark matter glitch would therefore require glitching models that are
sophisticated enough to enable differentiation between the signatures
of a quark matter core glitch and a nuclear matter crustal glitch.

In this context the reader may be concerned that a glitch deep within
the quark matter core of a neutron star may not be observable.
However, due to electron scattering off vortices, the rotation of the
superfluid interior of the star is coupled to the rotation of the
electron plasma on a very short time scale (on the order of seconds),
and the rotation of the electron plasma is subsequently coupled to the
rotation of the outer crust on a similarly short time scale.
Therefore the crust will rapidly respond to a change in the rotation
of the interior superfluid~\cite{AlparLangerSauls}.  This rapid
coupling of the superfluid to the crust, due to the fact that the
electron plasma penetrates throughout the star, is usually invoked to
explain that the core nucleon superfluid speeds up quickly after a
crustal glitch: the only long relaxation time is that of the vortices
within the inner crust.  Here, we invoke it to explain that the crust
speeds up rapidly after a core glitch has accelerated the superfluid
just outside the LOFF layer.  After a glitch in the LOFF region, the
only long relaxation times are those of the vortices in the LOFF
region and in the inner crust.

Our proposed quark matter glitch mechanism requires that the
crystalline color superconducting phase occupies enough of the interior
of a compact star to appreciably contribute to the star's moment of
inertia.  To address this question we need to complete a realistic
three-flavor analysis of the crystalline state, and map out the
$\mu$~interval that the crystalline state will occupy in the real QCD
phase diagram (figure \ref{phasediagram}).  The extent of this
interval will determine the extent of any volume of crystalline color
superconductivity inside a compact star.  So far, of course, we have
only studied the crystalline state in a two-flavor model.  In this
model, we saw that the plane wave crystalline state is favored only
when $\dm/\Delta_0$ is in a small interval $[ \dm_1/\Delta_0,
\dm_2/\Delta_0] \approx [0.707,0.754]$ (where $\Delta_0$ is the gap in
the homogeneous BCS phase, i.e.~2SC), but when we extended our
analysis to multiple plane wave crystals we found that the crystalline
phase could occur over a much larger interval $[\dm_1^\prime/\Delta_0,
\dm_*/\Delta_0]$ (as sketched in figure \ref{omegadeltafig}).
Extending the analysis to three flavors, we expect that the
$\dm/\Delta_0$ interval of the two-flavor problem will correspond to
an interval of $m_s^2/4\mu\Delta_0$ in the three-flavor problem.  If
the plane wave crystal were favored, it would only occupy a small
sliver in the QCD phase diagram.  The much more robust FCC crystal can
occupy a much larger region, perhaps even the entire interval between
the hadronization and unlocking transitions.  As a function of
increasing depth in a compact star, $\mu$ increases, $m_s$ decreases,
and $\Delta_0$ also changes, and the crystalline phase will occur
wherever $m_s^2/4\mu\Delta_0$ is in the (large) interval where the
crystalline phase is favored.  Therefore while a quantitative analysis
is yet to be done, we have reason to believe that the crystalline
phase could encompass a significant volume of the quark matter core of
a compact star.

The next issue to resolve is whether the crystalline phase will
actually pin vortices.  A real demonstration of pinning, and a real
calculation of the pinning force, will require explicitly constructing
a vortex in the crystalline phase, and then observing how the energy
of the vortex varies as it is moved across the crystal.  As we shall
discuss below, however, it is difficult to construct a vortex solution
on top of the existing modulation of the crystalline condensate.  The
difficulty is related to the fact that the vortex is itself a
deformation of the crystal, because both vortex and crystal are made
of the same diquark condensate.  It is easier to understand pinning in
the nuclear crust, because there are two components involved: the
first component (the lattice of nuclei) is a substrate that causes
pinning in the second component (the neutron superfluid).  It is the
spatial variation of the substrate that pins superfluid vortices, not
a crystalline variation of the superfluid order parameter itself.  In
other words, the superfluid does not ``self-pin''.  In an electron
superconductor, magnetic flux tubes are pinned to a substrate of
random magnetic impurities. Enhancement of pinning in an electronic
LOFF phase could occur by self-pinning of the LOFF crystal; however,  an
alternative proposal is that the the LOFF phase does not self-pin but
just makes it easier for the magnetic flux tubes to be pinned by the
magnetic impurities.\footnote{If the electronic LOFF condensate were
to have a standing wave variation $\cos(2\vq\cdot\vr)$ rather than an
FCC structure, the condensate would vanish on nodal planes that are
perpendicular to the magnetic flux tubes~\cite{Modler,Tachiki2} (the
arrangement is similar to that of the vortex state in a layered
high-$T_c$ cuprate superconductor with a magnetic field perpendicular
to the CuO$_2$ planes~\cite{Tachiki1}).  The appearance of nodal
planes leads to a segmentation of vortices; the individual vortex
segments are more flexible and are more readily pinned to random
impurities in the substrate.  In particular, the vortices would {\it
not} be pinned to nodal sites of the LOFF crystal, in fact there would
be no spatial variation of the LOFF crystal transverse to the flux
tubes.}  There is no ``substrate'' underlying quark matter, so if
pinning does occur in the crystalline color superconductor, it must
occur by self-pinning of superfluid vortices by the crystalline
modulation of the same superfluid.

Supposing that there is pinning, we can make an order-of-magnitude
estimate of the pinning force which sidesteps the complications
involved in explicitly constructing a vortex solution (at the end of
this section we will return to the vortex problem and discuss some
preliminary investigations).  We perform a calculation similar to that
done by Anderson and Itoh~\cite{AndersonItoh} for the pinning of
neutron vortices in the inner crust of a neutron star. In that
context, the pinning calculation has since been made
quantitative~\cite{Alpar77,AAPS3,Recent}.  We will attempt an estimate
for both the plane wave LOFF phase and the more robust multiple-plane
wave state (i.e.~the FCC crystal).  For the former, we can use numbers
for the gap and free energy as in Figure~\ref{fig:F_plot}.  That is,
we assume that at $\bar\mu = 400\ \mbox{MeV}$ we have a BCS gap
$\Delta_0 = 40\ \mbox{MeV}$ and a BCS free energy of $\Omega_0 = 2.6
\times 10^7\ \mbox{MeV}^4$ at $\dm = 0$; then at $\dm_1$ the gap for
the plane-wave LOFF state is $\Delta_{\mbox{\scriptsize pw}} \simeq 8\
\mbox{MeV} \simeq 0.2 \Delta_0$ and the phase is favored over the
normal state by a free energy $\Omega_{\mbox{\scriptsize pw}} \simeq 5
\times 10^4\ \mbox{MeV}^4$.  The periodicity of the crystal is
$b_{\mbox{\scriptsize pw}}=\pi/(2|\vq|)\simeq 9$ fm, and the thickness
of a rotational vortex is given by the correlation length
$\xi_{\mbox{\scriptsize pw}}\sim 1/\Delta_{\mbox{\scriptsize pw}} \sim
25$~fm.  For the FCC state, we cannot yet calculate numbers for the
gap and free energy but we will use $\Delta_{\mbox{\scriptsize fcc}}
\sim \Delta_0/2 = 20\ \mbox{MeV}$ and $\Omega_{\mbox{\scriptsize fcc}}
\sim \Omega_{\mbox{\scriptsize BCS}}/4 = 6.5 \times 10^6\
\mbox{MeV}^4$, reflecting our prediction that the FCC phase is robust.
The node spacing is $b_{\mbox{\scriptsize fcc}}=\sqrt{3}
\pi/(2|\vq|)\simeq 15$ fm (i.e.~half the lattice constant, equation
(\ref{latticeconstant})), and the thickness of a rotational vortex is
$\xi_{\mbox{\scriptsize fcc}}\sim 1/\Delta_{\mbox{\scriptsize fcc}}
\sim 10$ fm.  All these numbers are quite uncertain, but we will use
them for the present.  In the context of crustal neutron superfluid
vortices, there are three distinct length scales: the vortex thickness
$\xi$, the lattice spacing between nuclei $b$, and $R$, the radius of
the individual nuclei.  (The condensate vanishes within regions of
size $R$ separated by spacing $b$.)  In the LOFF phase, the latter two
length scales are comparable: since the condensate varies like a sum
of plane waves it is as if $R\sim b$.  The fact that these length
scales are similar in the LOFF phase will complicate a quantitative
calculation of the pinning energy; it makes our order of magnitude
estimation easier, however.  The pinning energy is the difference
between the energy of a section of vortex of length $b$ which is
centered on a node of the LOFF crystal versus one which is centered on
a maximum of the LOFF crystal. It is of order
\begin{equation}
\label{crude1}
E_p \sim \Omega b^3 \sim \left\{ 
\begin{array}{ll} 
  4 \ \mbox{MeV} \phantom{blah} & \mbox{plane wave} \\
  3 \ \mbox{GeV} & \mbox{FCC crystal}
\end{array}
\right.
\end{equation}
The resulting pinning force per unit length of vortex is of order
\begin{equation}
\label{crude2}
f_p \sim \frac{E_p}{b^2} \sim  \left\{
\begin{array}{ll}
% (4 \ \mbox{MeV})/(80\ \mbox{fm}^2) \phantom{blah} & \mbox{plane wave} \\ 
% (3 \ \mbox{GeV})/(200\ \mbox{fm}^2) & \mbox{FCC crystal}
 (5 \ \mbox{MeV})/(100\ \mbox{fm}^2) \phantom{blah} & \mbox{plane wave} \\ 
 (1 \ \mbox{GeV})/(100\ \mbox{fm}^2) & \mbox{FCC crystal}
\end{array}
\right.
\end{equation}
The fact that $b$ and $\xi$ are comparable length scales will make a
complete pinning force calculation more difficult and is likely to
yield an $f_p$ which is significantly less than that we have obtained
by dimensional analysis~\cite{AAPS3,Recent}.  Therefore these figures
should be interpreted as upper-bound estimates.  Note that our
estimate of $f_p$ is quite uncertain both because it is only based on
dimensional analysis and because the values of $\Delta$, $b$ and
$\Omega$ are known to only within an order of magnitude at best.  For
comparison, we present the corresponding numbers for the pinning of
crustal neutron vortices: the pinning energy of neutron vortices in
the inner crust is~\cite{AAPS3}
\begin{equation}
\label{serious1}
E_p \approx 1-3  {\rm \ MeV}
\end{equation}
and the pinning force per unit length is~\cite{AAPS3,PinesAlpar}
\begin{equation}
\label{serious2}
f_p\sim \frac{E_p}{b\xi}\approx
\frac{1-3 {\rm ~MeV}}{(25-50 {\rm ~fm})(4-20{\rm ~fm})}\ ,
\end{equation}
where the form of this expression is appropriate because $\xi<b$.  Of
course it is premature to compare our crude results (\ref{crude1},
\ref{crude2}) for quark matter pinning to the results (\ref{serious1},
\ref{serious2}) of serious calculations for the pinning of crustal
neutron vortices as in Refs.~\cite{Alpar77,AAPS3,Recent}.
Nevertheless, we observe that the results are comparable for the case
of a plane wave LOFF state, while pinning in the more robust FCC state
could be orders of magnitude stronger than nuclear pinning (again,
however, we are likely overestimating the pinning force so the numbers
should be interpreted as upper bounds).  

The nuclear pinning force (\ref{serious2}), when applied in a stellar
model, does yield crustal glitches in accord with those observed in
pulsars.  It remains to be seen whether a much stronger pinning force,
occurring in a crystalline region of the quark matter core, would also
yield glitches consistent with observation.  If the pinning force is
large, then the restraint of vortices could be limited by the critical
shear stress of the crystal: if the critical shear stress is exceeded,
then the vortex can be released by a crystal dislocation.  A
dimensional analysis, like that which we have done for the pinning
force, can only predict that the pinning force and the critical shear
stress are of comparable magnitude, and no conclusion can be reached
without a more careful investigation. 

We now return to the problem of constructing a vortex in the
crystalline phase.  Vortices are usually constructed beginning with a
Ginzburg-Landau free energy functional written in terms of
$\Delta(\vx)$ and $\boldsymbol{\nabla} \Delta(\vx)$.  Instead, we have constructed
a Ginzburg-Landau potential (equation \ref{freeenergy}) that is
written in terms of the momentum modes $\Delta_\vq$ of the order
parameter .  In principle, this contains the same information, but it
is not well-suited to the analysis of a localized object like a
vortex.  In position space, the Ginzburg-Landau free energy should
look like
\begin{equation}
\label{GLpositionspace}
\Omega[\Delta(\vx)] = \Delta(\vx)^* \left[ \alpha + C(\nabla^2 + 4 q_0^2)^2 \right] \Delta(\vx)
+ \mathcal{O} (\Delta^4)
\end{equation}
where $C$ is a positive constant and $\alpha$ is the same parameter
that appears in equation (\ref{freeenergy}), i.e.~it is negative below
the second-order point $\dm_2$.  The quadratic term successfully
reproduces the plane-wave instability of the LOFF phase: just below
the second-order point, all the modes on the sphere $|\vq| = q_0$
become unstable.  Adding quartic and higher terms to equation
(\ref{GLpositionspace}) is difficult, however, because these terms are
Fourier transforms of the complicated $J$ and $K$ functions defined in
chapter 3.  Terms with high powers of spatial derivatives will be
required to reproduce the effects of $J$ and $K$ and somehow yield a
functional that is minimized by an FCC crystal (equation
\ref{fcccrystal}).  So it is not at all clear that the usual
methodology for constructing a vortex starting from a position-space
Ginzburg-Landau potential is the right approach.  

As always, it should be easier to understand the physics of a vortex
far from its core. The natural expectation is that far from its core,
a vortex will be described simply by multiplying the $\Delta(\vx)$ of
(\ref{fcccrystal}) by $\exp[i\theta(\vx)]$, where $\theta(\vx)$ is a
slowly varying function of $\vx$ that winds once from 0 to $2\pi$ as
you follow a loop encircling the vortex at a large distance.  In a
uniform superfluid, this slowly varying phase describes a
particle-number current flowing around the vortex.  But in the 
crystal, the resulting particle-number current is 
\begin{equation}
\mathbf{J} = (\boldsymbol{\nabla} \theta(\vx)) |\Delta(\vx)|^2
\end{equation}
and the current does not flow in a large loop because it vanishes at
the nodal planes of the crystal (see figure \ref{unitcellfig}).  The
dilemma is that the FCC crystal structure divides all of space into
small cells bounded by intersecting nodal planes, and a supercurrent
cannot flow across a nodal plane.  In the presence of a vortex, the
crystal structure must change to accommodate the rotational
supercurrent flow.

One obvious possibility is that the condensate just reduces to a
one-dimensional standing wave $\cos(2 \vq\cdot\vx)$ with no variation
in the transverse direction.  This condensate is just a stack of
layers separated by nodal planes perpendicular to $\vq$, and vorticity
is easily achieved (the supercurrent just circulates in each
transverse layer).  If this is what happens, then the vortices will
not be pinned because they can move freely in the transverse
directions.  

However, it may not be necessary to change the crystal structure so
drastically.  Suppose we instead consider changing the different
$\Delta_{\vq}$'s by small phases, i.e.~we consider condensates of the
form
\begin{equation}
\label{phaseshifteqn}
\Delta(\vx) = \sum_{\vq} \Delta_\vq e^{i 2\vq\cdot\vx} = \sum_{\vq} \Delta e^{i \phi_q} e^{i 2 \vq\cdot\vx}
\end{equation}
(we assume that all the $\Delta_\vq$'s keep the same magnitude
$\Delta$).  These are small distortions of the FCC ground state but
perhaps these small distortions can change the nodal structure to
accommodate vortices.  For small phases, we can expand the free energy
quadratically:
\begin{equation}
\Omega(\{\phi_\vq\}) = \Omega_0 + \sum_{\vq,\vq^\prime} \phi_\vq M_{\vq \vq^\prime} \phi_{\vq^\prime} + \mathcal{O}(\phi^4)
\end{equation}
Odd powers of $\phi$ are not allowed because the free energy should be
symmetric under $\Delta(\vx) \rightarrow \Delta(\vx)^*$.  The matrix
$M$ can be diagonalized to find eight normal modes for the phase
angles.  Four of these normal modes will have zero eigenvalue.  One
mode is a common phase for all eight $\vq$'s (i.e.~the superfluid
mode).  Recalling that the eight $\vq$'s correspond to eight corners
of a cube, the other three zero modes assign a common phase $e^{i\ph}$
to the four corners of one face of the cube, and an opposite phase
$e^{-i\ph}$ to the four corners of the opposite face.  These are the
three phonon zero modes; they correspond to translations in the $x$,
$y$, or $z$ directions by a displacement $\sqrt{3} \ph/(2|\vq|)$.  There is one
nonzero ``tetrahedral'' mode which assigns a common phase to four
tetrahedral corners of the cube (i.e.~four nonadjacent corners) and
the opposite phase to the other four corners.  Finally, there are
three ``skew'' modes which have the same nonzero eigenvalue of $M$;
these modes assign a common phase to the four corners that are the
endpoints of two opposite edges of the cube, and an opposite phase to
the other four corners.

\begin{figure}
\begin{center}
\includegraphics[width=2.8in]{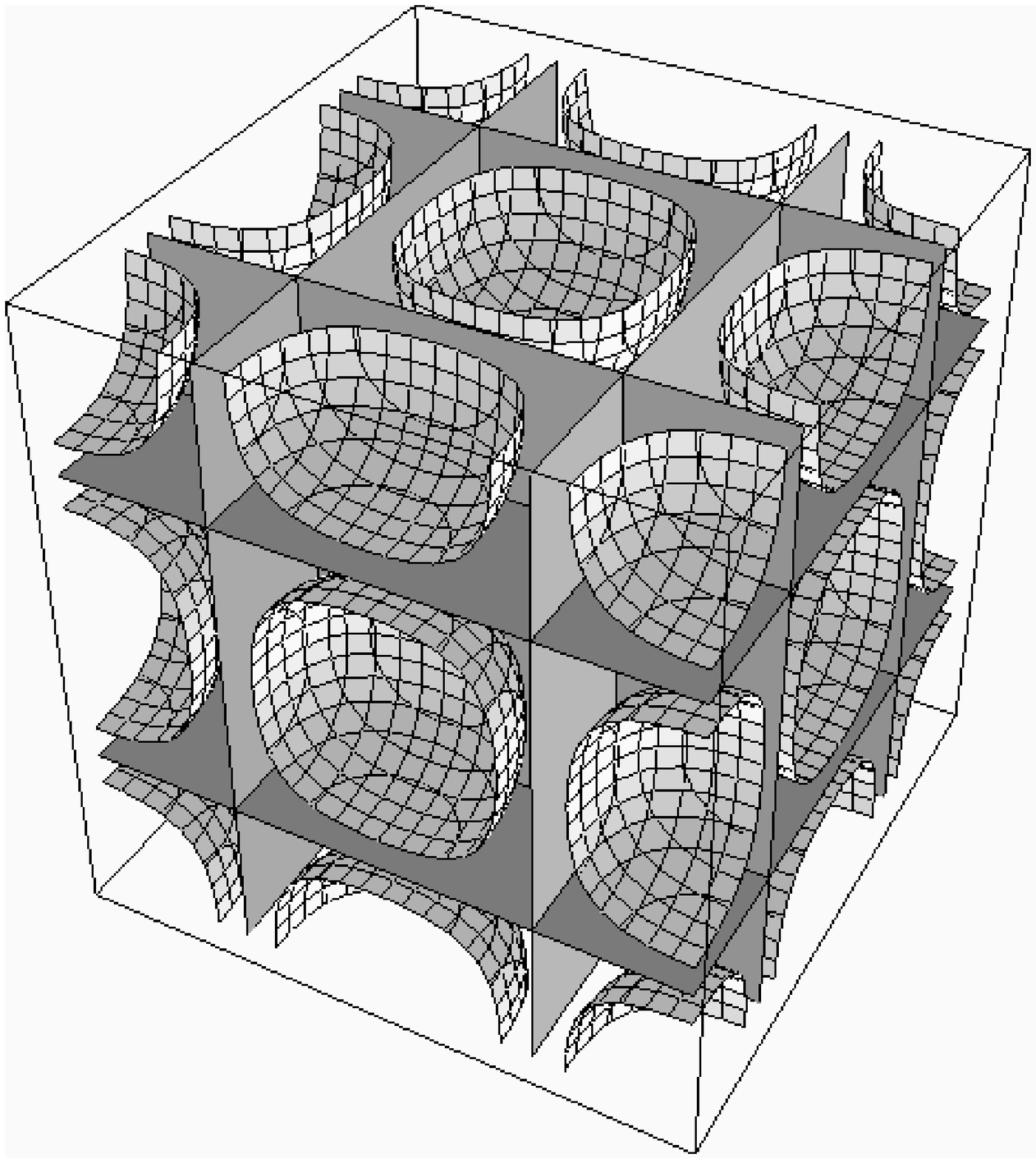}
\hspace{0.2in}
\includegraphics[width=2.8in]{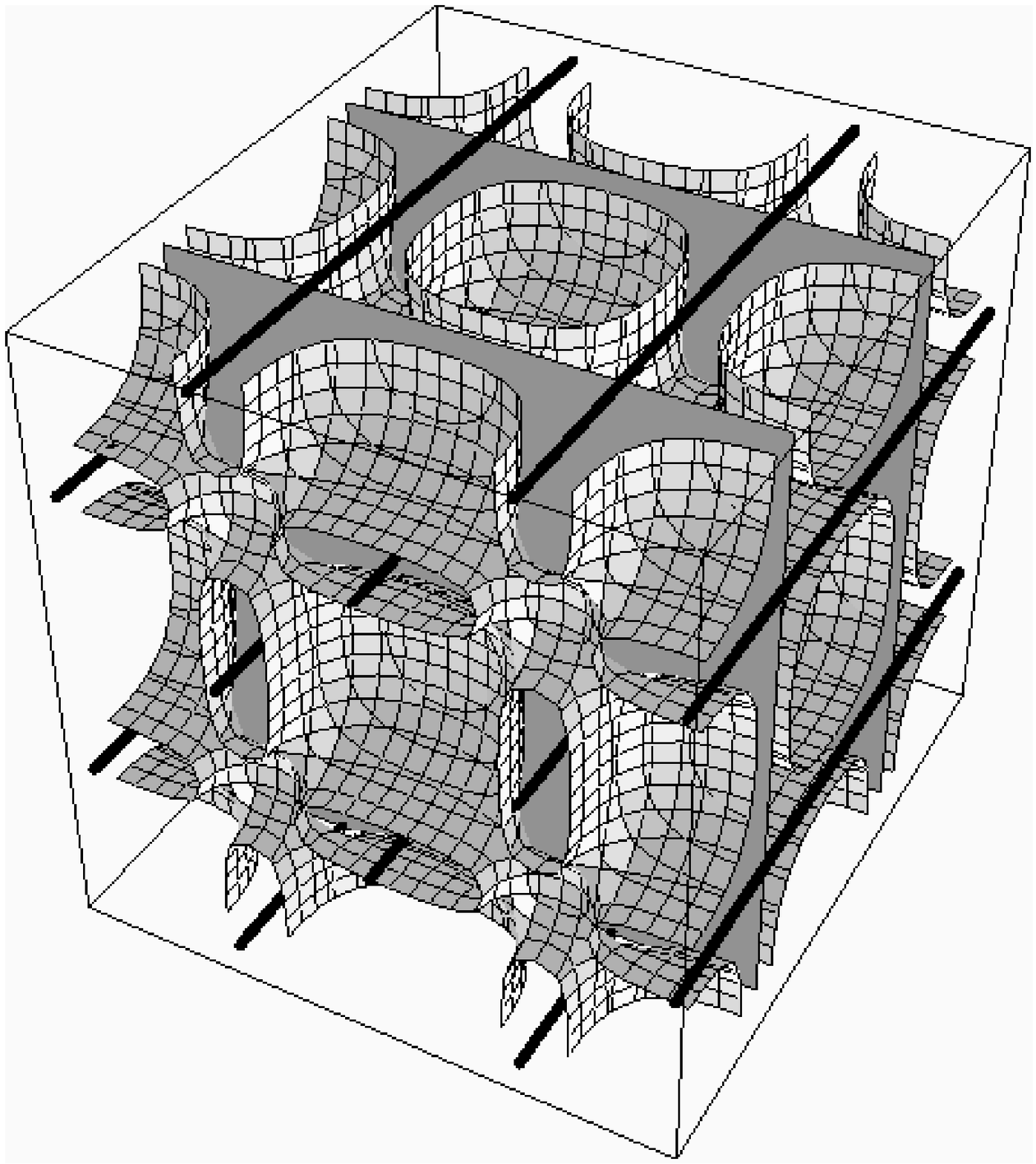}
%\vspace{3in}
\caption{
\label{distortion}
The effect of distorting the crystal by a phase shift, as in equation
(\ref{phaseshifteqn}).  On the left is the FCC crystal (phase shift
0); on the right is the distorted crystal (with a ``skew'' phase shift
of magnitude $\pi/8$).  The gap function $\Delta(\vx)$ vanishes on the
gray nodal planes and on the heavy black nodal lines.  The curved
surfaces are contours where $|\Delta(\vx)| = 2.6 \Delta$ (note that
for the undistorted crystal, $|\Delta(\vx)|$ is simple cubic even
though $\Delta(\vx)$ is FCC).}
\end{center}
\end{figure}

It is these last three ``skew'' modes that might be of interest.  In
figure \ref{distortion} we see what happens to the crystal if a small
phase is applied for one of these modes.  The result is that only one
set of parallel nodal planes remains, while the other nodal planes
vanish and are replaced by nodal lines.  With this deformation it
should be possible to have circulating supercurrents in the layers
between the remaining nodal planes, and the cost of the nodal
deformation that accommodates this vorticity is small when the skew
phase angle is small. Moreover, there is still a transverse variation
of the condensate, and this could cause vortex pinning.  In
particular, the vortices might pin at the new nodal lines.  In the
deformed crystal, a current can be turned on by letting the different
$\Delta_\vq$'s have different magnitudes.  The next step is to slowly
vary these magnitudes so that the current circulates in a large loop.
This investigation is proceeding at the time of this writing.

%All of
%this remains to be worked out, but these tentative explorations of the
%vortex solution suggest that it is possible to pin vortices in the
%crystalline color superconducting phase.

\section{Ultracold Fermi gases}
\label{sec:gases}

One of the most exciting prospects of the LOFF crystalline phase is
the possibility that it could be observed as a ``crystalline
superfluid'' in a trap of ultracold fermionic atoms~\cite{AtomicLOFF}.
This could provide a laboratory test of our prediction that the LOFF
phase has an FCC crystal structure.  Even more significantly, the LOFF
phase should be of interest to experimenters because it might in fact
be easier to observe than the uniform phase: while the onset of
uniform BCS superfluidity in the atomic system could be subtle to
detect~\cite{PitaevskiiStringari}, the onset of crystallization could
have a dramatic observational signature.  Indeed, it might be possible
to literally see the crystal structure, as we will discuss.

In section \ref{subsec:atomicphysics}, we described how the two-flavor
NJL model that we have used to investigate the crystalline color
superconducting phase might be a surprisingly good model for the
analysis of LOFF pairing in an atomic system.  This is why we expect
our prediction of an FCC crystal structure to apply in the atomic
context.  In this context, the two species that pair are the two
hyperfine states of the fermionic isotope that is being trapped
(experimenters have used ${}^6$Li or ${}^{40}$K~\cite{ColdFermions}).
The interaction between atoms, accomplished by a Feshbach
resonance~\cite{Feshbach}, is in the $s$-wave channel and is
well-described as pointlike: the range $R$ of the interatomic
potential is quite small ($R \sim 10-100$ \AA, a typical van der Waals
scale) compared to the atomic spacing achieved in the atom trap (on
the order of 0.1 $\mu$m), so the system is dilute.  On the other hand,
the scattering length $a$ is quite large, of approximately the same
order as the atomic spacing (the Duke group reports $a \simeq 0.5\
\mu$m~\cite{Duke}), which means that the system is strongly
coupled~\cite{Heiselberg}.  The $s$-wave scattering length $a$ is
proportional to the four-fermi coupling constant $g$ by the relation
$g = 4 \pi \hbar^2 a/m$.

The atoms are confined in a harmonic trap that is ``cigar-shaped''
with a very large aspect ratio ($\sim$30 for the CO$_2$ laser trap
used by the Duke group~\cite{Duke}, and $\sim$70 for the magnetic trap
used by the MIT group~\cite{MIT}).  The transverse size of the cloud
is a few to ten microns, while the longitudinal size is several
hundred microns.  The number density in the cloud varies somewhat with
depth in the harmonic potential, but a typical density is $n =
10^{14}$ cm$^{-3}$ (for total number density of atoms in both
hyperfine states).  For now we assume that both states A and B are
equally populated, i.e.~$n_A = n_B = n/2$.  Then both species have a
common Fermi wave vector $k_F ( = p_F/\hbar) = (3 \pi^2 n)^{1/3}
\simeq 1.4 \times 10^5$ cm$^{-1}$, a Fermi energy $E_F \simeq 7 \times
10^{-10}$ eV, and a Fermi temperature $T_F = E_F/k_B \simeq 10\ \mu$K.
The interatomic spacing is $l \sim n^{-1/3} \simeq 0.2\ \mu$m.

Evaporative cooling has been used to obtain a highly-degenerate Fermi
gas with a temperature perhaps as small as $\sim$0.2 $T_F$~\cite{Duke}
(there is some uncertainty in this figure because it is not easy to
measure the temperature in the high-degeneracy regime).  It is
difficult to achieve temperatures very much smaller than $T_F$ because
the evaporative cooling efficiency is diminished by Pauli
blocking~\cite{AtomReview}.  The critical temperature for BCS
superfluidity is~\cite{Gorkov}
\begin{equation}
T_c \simeq 0.28 T_F \exp \left( -\frac{\pi}{2 k_F a} \right)
\end{equation}
but this result is valid only in the weak-coupling regime ($k_F a \ll
1$) where it predicts an exponential suppression of $T_c$ relative to
$T_F$.  The atomic systems are deliberately strongly coupled ($k_F a >
1$) to avoid this exponential suppression.  The strong coupling is
achieved by tuning a magnetic field to the vicinity of a Feshbach
resonance to obtain a very long scattering length (the Duke group
reports $k_F a = 7.3$ for their experimental conditions).  In the
strong coupling regime, the value of $T_c$ is not well known, but it
is expected to be a substantial fraction of the Fermi
temperature~\cite{AtomicBCS,Heiselberg}; a recent
calculation~\cite{HongHsu} predicts that $T_c$ could be as large as
$0.4 T_F$ near the Feshbach resonance.

It was recently proposed~\cite{Stringari} that the superfluid phase
transition can be identified from the free expansion of the
cigar-shaped gas cloud after the trap is switched off.  In particular,
if the gas is in the normal (non-superfluid) state before the trap is
switched off, then it will just expand ballistically and acquire a
spherical shape.  On the other hand, if the gas is in the superfluid
state, then it will expand according to a hydrodynamic equation for
the superfluid mode, leading to a pressure gradient and an enhanced
expansion in the radial direction and an inversion of the cloud shape,
i.e.~the cigar flattens into a pancake.  A recent investigation by the
Duke experimenters seems to produce this behavior, leading them to
speculate that they may have observed the superfluid
phase~\cite{Duke}.  However, they also concede that collisional
hydrodynamics of a non-superfluid gas could create the same inversion
of the cloud shape.  In the strongly-coupled system it is hard to rule
out the influence of collisional hydrodynamics, so this may obscure
any signature for the onset of superfluidity that involves the behavior
of the expanding gas~\cite{PitaevskiiStringari}.

The crystalline superfluid might reveal itself in a more obvious way.
To assess how we might observe the crystalline phase, we first make a
rough estimate of the lattice spacing for the LOFF crystal in the
atomic system.  We write the BCS gap as
\begin{equation}
\label{atomgap}
\Delta_0 = C E_F = C \frac{p_F^2}{2 m}
\end{equation}
where $C$ is a number of order $1/10$ to $1/2$ that is poorly known.
The LOFF phase is likely to occur when the separation of the Fermi
energies of the two hyperfine states (A and B) is comparable to the
value of the BCS gap, i.e.
\begin{equation}
\label{atomunlock}
\delta E_F = \frac{p_{FA}^2}{2m} - \frac{p_{FB}^2}{2m} \approx v_F \delta p_F \approx \Delta_0
\end{equation}
where $v_F$ is the Fermi velocity and $\delta p_F = p_{FA} - p_{FB}$.
To achieve a Fermi surface separation of this magnitude, the
populations of the two hyperfine states A and B should be adjusted
so that there is a fractional population difference 
\begin{equation}
\frac{N_A - N_B}{N_A + N_B} \approx \frac{3}{2} \frac{\delta p_F}{p_F} \approx \frac{3}{4} \frac{\Delta_0}{E_F} = \frac{3}{4} C
\end{equation}
where we have substituted equations (\ref{atomgap}) and
(\ref{atomunlock}).  The momentum of a LOFF Cooper pair is 
\begin{equation} 
q_0 = 0.6 \delta p_F \approx 0.3 C p_F, 
\end{equation}
again substituting equations (\ref{atomgap}) and (\ref{atomunlock}).
Finally, the spacing between nodal planes of the FCC crystal is (see
equation (\ref{latticeconstant}))
\begin{equation}
b = \frac{\sqrt{3} \pi \hbar}{2 q_0} \approx \left( \frac{9}{C} \right) \frac{1}{k_F} \approx \left( \frac{3}{C} \right) l
\end{equation}
where $l = n^{-1/3}$ is the interatomic spacing.  The interatomic
spacing is about $0.2\ \mu$m, so we expect a nodal spacing $b = 1 - 6\
\mu$m, for $C = 1/2 - 1/10$.  

All of these numbers are quite uncertain, first of all because they
are derived from a weak-coupling analysis, whereas the atomic system
is strongly coupled; and second of all because we have obtained our
results at zero temperature, whereas the atomic system is likely to be
observed near its critical temperature.  Taking these numbers as rough
estimates, however, it is interesting to consider how one might
observe the crystalline phase.  Our estimate of the nodal spacing
suggests that the cigar-shaped cloud could have on the order of 1 to
10 nodes along the transverse width, and a few hundred nodes in the
longitudinal direction.  There will be a periodic modulation of the
atom density in the crystalline superfluid: the density is depleted at
the nodal regions of the crystal, in the same way that density is
depleted in the core of a superfluid BCS vortex~\cite{Nygaard} (where
the condensate also vanishes).  The ratio of density modulation to
total density is
\begin{equation}
\label{modulation}
|\delta n(\vx)|/n \sim |\Delta(\vx)|^2/E_F^2 \lesssim C^2 
\end{equation}
where $C$ is the number from equation (\ref{atomgap}).  Notice that
since the modulation varies as $|\Delta|^2$ it will actually be simple
cubic rather than FCC.  The density depletion is relatively modest (of
order 10\% or less).  However, it was recently proposed~\cite{Bulgac}
that density depletion in a vortex core might be enhanced at strong
coupling, beyond the expectation of equation (\ref{modulation}), to
give a density reduction even as large as 70\%.  Although it has not
been worked out, a similarly large depletion might occur at the nodal
sites of the crystalline superfluid, leading to a dramatic density
variation that could be easy to detect.

The nodal spacing is perhaps too small to resolve by direct absorption
imaging of the {\it in situ} trapped gas (although it is worthwhile to
investigate whether the crystal could be detected by Bragg scattering
of infrared light).  However, it might be possible to magnify the
crystal structure by turning off the trapping potential and letting
the gas expand until the structure can be seen.  This is precisely the
method that allowed the striking visualization of the vortex lattice
in a rotating Bose-condensed gas~\cite{vortexexpt}.  In the BEC
context, the expanding BEC cloud with its lattice of vortices is
described by a time-dependent Gross-Pitaevskii
equation~\cite{vortexthy}.  For a cigar-shaped cloud rotated along its
axis, the axial expansion is slower and the radial expansion is just a
dilation of the vortex lattice.  For a pancake-shaped cloud rotated
along a normal axis, the axial expansion is faster and the radial
expansion is not a dilation: the vortex core radius increases faster
than the cloud radius, so as the vortex lattice expands the vortex
holes become proportionally larger.  These results for the expanding
BEC vortex lattice are inspiration for an analogous calculation for
the expanding LOFF crystal, to see how the density profile is deformed
or dilated by the expansion.  This is likely to require solving a
time-dependent Ginzburg-Landau equation for the space-time evolution
of the LOFF order parameter.

If the gas does not expand like a fluid, but instead just expands
ballistically when the trap is turned off, then the expanded gas is
unlikely to bear a resemblance of the original crystal structure.  In
this scenario, the crystalline superfluid could be revealed by a
measurement of the directional distribution of momenta in the expanded
gas.  The distribution should have an anisotropy reflecting the
preferred arrangement of pairing rings on the Fermi surface,
i.e.~reflecting the fact that the FCC LOFF phase chooses to pair atoms
in Cooper pairs with center-of-mass momenta pointing toward the eight
corners of a cube.  While a polar variation in the momentum distribution
could always occur as a result of the initial axial confinement
of the gas, any azimuthal variation would be an unmistakable 
signature of a crystalline phase. 

%This Fermi surface anisotropy could appear as an
%azimuthal variation in the momentum distribution, which would not
%occur except in the crystalline phase (a polar variation, on the other
%hand, could always occur as a result of the initial axial confinement
%of the gas).

\section{Outlook}
\label{sec:outlook}

Our investigation of the crystalline color superconducting phase and
other non-CFL phases of quark matter has, not surprisingly, raised as
many new questions as it has answered old ones.  In this final section
we enumerate some of the interesting directions for future work

\begin{enumerate}

\item %1
Surely the single most important problem to address is the fact that
we have been unable to calculate a gap and free energy for the FCC
crystal state.  Our Ginzburg-Landau analysis is a powerful
method that has taught us what features make a particular crystal
structure energetically favorable or unfavorable, and therefore has
led us to the FCC structure as the likely ground state.  However, the
Ginzburg-Landau potential is unstable and therefore does not predict
values for the gap and free energy.  Given that, and given the
prediction of a strong first-order phase transition, the
Ginzburg-Landau approximation should now be discarded.  What should be
taken from our work is the prediction that the structure
(\ref{fcccrystal}) is favored.  Although $\Delta$ could be estimated
by going to higher order in the Ginzburg-Landau approximation, a much
better strategy is to do the calculation of $\Delta$ upon assuming the
crystal structure (\ref{fcccrystal}) but without requiring $\Delta$ to
be small.  To do this will require constructing the anomalous
propagator $F(p,p')$ (\ref{pspaceanomprop}) by a resummation of the
infinite geometric series shown diagrammatically in figure
\ref{expansionfig}, rather than just keeping the first few terms in
the geometric series as we have done in our Ginzburg-Landau analysis.
The only difficulty is that $F$ is not diagonal in momentum space;
rather, it should be described as a vector of propagators $F_\vq$, one
for each plane wave mode in the crystal, as in equation
(\ref{pspaceanomprop}).  Exactly resumming the geometric series of
figure \ref{expansionfig} will generate an infinite number of nonzero
$\Delta_\vq$'s and corresponding $F_{\vq}$'s, one for each $\vq$ in
the reciprocal lattice of the chosen crystal structure.  A strategy
that will make this calculation tractable is a truncation of this
reciprocal lattice (analogous to the method of Ref.~\cite{RappCrystal}
to study chiral crystal structures), to lowest order just including
the fundamental $\vq$ vectors (i.e.~the eight $\vq$ vectors that
contribute to (\ref{fcccrystal})), to next order including the next
set of $\vq$ vectors at larger radius in the reciprocal lattice space,
etc.  This calculation is proceeding at the time of this writing.

\item %2
If the crystalline color superconducting phase appears in nature, it
will involve all three flavors of quark.  We need to do a three-flavor
analysis of the crystalline phase, including a realistic strange quark
mass and imposing charge neutrality and weak equilibrium.  Possible
pairing strategies for the three flavor system include a CFL-LOFF
phase, involving $ud$, $us$, and $ds$ crystalline condensates, or a
double-2SC-LOFF phase, involving only $ud$ and $ds$ pairings
(i.e.~pairings between species with adjacent Fermi surfaces).  The
three-flavor analysis is necessary to determine whether the
crystalline phase is a superfluid, and it will also determine
electromagnetic properties.  Given the robustness of the crystalline
color superconducting phase in the two-flavor system that we have
studied, and its ability to occur over a large interval in $\dm$, we
expect that crystalline color superconductivity will be a generic
feature of nature's QCD phase diagram, occurring wherever quark matter
that is not color-flavor-locked is to be found.

\item %3
We need to understand how to rotate a chunk of crystalline color
superconducting quark matter, i.e.~we need to determine the structure
of a rotational vortex in the crystalline phase.  A determination of
the vortex structure will resolve whether pinning can occur, and it is
the first step toward a subsequent real calculation of the pinning
force.  As we have seen in section \ref{sec:glitches}, even the
long-distance behavior of the vortex is interesting to work out.  
%And
%there may be a larger issue here related to the fundamental novelty of
%a system that is both a crystal and a superfluid.

\item %4
Also related to pinning, we need to calculate the shear modulus for
the crystal.  Pinning of vortices in a crystalline color
superconductor is likely to require both a nonzero pinning force and a
nonzero shear modulus, because vortices can become unpinned by crystal
dislocation if a critical shear stress is exceeded.  The shear modulus
could be determined from the phonon dispersion relation, which
contains information about the elastic moduli of the
crystal~\cite{LOFFphonon}.

\item %5
The crystalline, single-flavor, and breached-pair color
superconducting phases all have gapless quasiquark modes.  If any of
these phases were to appear in a compact star, these gapless modes
might accommodate direct URCA reactions and play a critical role in the
entire cooling history of the star.  Therefore it is important to work
out just how these special modes might contribute to cooling processes.

\item %6
In the atomic LOFF context, it will be crucial to study the free
expansion of the atomic cloud after the trapping potential is shut
off.  For the scenario of fluid-like expansion, a calculation should
be done in position space to determine how the crystal structure
dilates or deforms as the cloud expands.  For the case of ballistic
expansion, a calculation should be done in momentum space to determine
the anisotropy of the momentum distribution.  It will also be useful
to generalize our analysis to nonzero temperature, because the system
temperature is likely to be near the critical temperature.

\item %7
Our analysis does not apply to QCD at asymptotically high density,
where the QCD coupling becomes weak.  In this regime, quark-quark
scattering is dominated by gluon exchange and because the gluon
propagator is almost unscreened, the scattering is dominated by
forward scattering.  This works in favor of crystalline color
superconductivity~\cite{pertloff}, but it also has the consequence of
reducing $q_0/\dm$ and hence reducing $\psi_0$.  The authors of
Ref.~\cite{pertloff} find $q_0/\dm$ reduced almost to 1, meaning
$\psi_0$ reduced almost to zero.  However, the authors of
Ref.~\cite{Giannakis} find $q_0/\dm\simeq 1.16$ at asymptotically high
density, meaning that $\psi_0\simeq 61^\circ$.  If the opening angle
of the pairing rings on the Fermi surface does become very small at
asymptotic densities, then the crystal structure there is certain to
be qualitatively different from that which we have found.  At present,
the crystal structure at asymptotic densities is unresolved.  This is
worth pursuing, since it should ultimately be possible to begin with
asymptotically free QCD (rather than a model thereof) and calculate
the crystal structure at asymptotic density from first principles.
(At these densities, the strange quark mass is irrelevant and a
suitable $\dm$ would have to be introduced by hand.)  Although such a
first-principles analysis of the crystalline color superconducting
state has a certain appeal, it should be noted that the asymptotic
analysis of the CFL state seems to be quantitatively reliable only at
 densities that are more than fifteen orders of magnitude larger than
those reached in compact stars~\cite{RajagopalShuster}.  At accessible
densities, models like the one we have employed are at least as likely
to be a good starting point.

\end{enumerate}

%% file: appa.tex
\chapter{Evaluating $J$ and $K$ Integrals}
\label{JKappendix}

%-------------------------------------------------------------------

%\renewcommand{\theequation}{A.\arabic{equation}}
% redefine the command that creates the equation no.
%\setcounter{equation}{0}  % reset counter 
%\section*{Appendix}  % use *-form to suppress numbering

In this Appendix, we outline the explicit evaluation 
of the loop integrals
in Eqs.~(\ref{integrals}) that occur in $\Pi$,
$J$ and $K$.  For all loop integrals, the
momentum integration is restricted to modes near the Fermi surface by
a cutoff $\omega \ll \bar\mu$, meaning that the density
of states can be taken as constant within the integration
region:
\begin{equation}
\int d^4 p 
= 
\int_{-\infty}^{+\infty} dp^0 \int_{\bar\mu-\omega}^{\bar\mu+\omega} |\vp|^2 d|\vp| \int_{4\pi} d\hat{\vp}
\approx
\bar\mu^2 \int_{-\infty}^{+\infty} dp^0 \int_{-\omega}^{+\omega} ds \int_{4\pi} d\hat{\vp}
\end{equation}
where $s \equiv |\vp|-\bar\mu$.  Each integral is further simplified by
removing the antiparticle poles from the bare propagators $G^{(0)}$
and $\bar G^{(0)}$ that appear in the integrand. (We can disregard
the antiparticles because their effect on the Fermi surface
physics of interest is suppressed by of order $\Delta/\bar\mu$.)
To see how to remove the antiparticle poles,
consider the propagator $(\pslash + 2 \qslash + \muslash_u)^{-1}$ that
appears in the $\Pi$ integral.  Recall that $\mu_u = \bar\mu - \dm$
and we work in the limit where $|\vq|, \dm \ll \omega \ll |\vp|,
\bar\mu$. We are only interested in the behavior of the propagator in
the vicinity of the particle poles where $p^0 \sim \pm(|\vp|-\bar\mu)
\ll \bar\mu$.  Therefore we can factor the denominator and drop
subleading terms proportional to $p^0$, $\dm$, or $|\vq|$ when they
occur outside of the particle pole:
\begin{eqnarray}
\frac{1}{\pslash + 2\qslash + \muslash_u} & = & 
\frac{(p^0 + \mu_u) \gamma^0 - (\vp + 2 \vq)\cdot\boldsymbol{\gamma}}{(p^0 + \mu_u - |\vp + 2\vq|)(p^0 + \mu_u + |\vp + 2 \vq|)} \nonumber \\
 & \approx & \frac{\bar\mu \gamma^0 - \vp\cdot\boldsymbol{\gamma}}{(p^0 + \bar\mu - \dm - |\vp| - 2\vq\cdot\hat{\vp})(2\bar\mu)} \nonumber \\
 & \approx & \frac{1}{2} \frac{\gamma^0 - \hat{\vp} \cdot \boldsymbol{\gamma}}{(p^0 - s - \dm - 2\vq\cdot\hat{\vp})}
\end{eqnarray}
We simplify all of the propagators in this way.  In the numerator
of each integrand we are then left with terms of the form $\gamma_\mu
\gamma^\alpha \gamma^\beta \cdots \gamma^\mu$.  After evaluating these
products of gamma matrices, the $\Pi$ integral can be written as
\begin{equation}
\Pi(\vq)  =  \int \frac{d p^0}{2\pi i} \ \frac{d\hat{\vp}}{4\pi} \int_{-\omega}^{+\omega} ds \ \left[ (p^0 + s - \dm)(p^0 - s - \dm - 2 \vq\cdot\hat{\vp}) \right]^{-1}.
\end{equation}
This integral is straightforward to evaluate: Wick rotate $p^0
\rightarrow i p_4$, do a contour integration of the $p_4$ integral,
and then do the remaining simple integrals to obtain 
Eq.~(\ref{Piandalpha}).

By power counting, we see that while the $\Pi$ integral has a
logarithmic dependence on the cutoff $\omega$, the $J$ and $K$
integrals have $1/\omega^2$ and $1/\omega^4$ cutoff dependences,
respectively.  We can therefore remove the cutoff dependence in the
$J$ and $K$ integrals by taking the limit $\omega/\dm$,
$\omega/|\vq|\rightarrow \infty$.
Then the $J$ and $K$ integrals depend only on
$\dm$ and the $\vq$'s and take the form
\begin{eqnarray}
\label{jkeqns}
\lefteqn{J(\vq_1 \vq_2 \vq_3 \vq_4) = } \nonumber \\
 & & \int \frac{d p^0}{2\pi i}  \ \frac{d\hat{\vp}}{4\pi} \int_{-\infty}^{+\infty} ds \ \prod_{i=1}^{2} \left[ (p^0 + s - \dm + 2\vk_i\cdot\hat{\vp}) (p^0 - s - \dm - 2\vl_i\cdot\hat{\vp}) \right]^{-1} \nonumber \\
\lefteqn{K(\vq_1 \vq_2 \vq_3 \vq_4 \vq_5 \vq_6)  = } \nonumber \\
 & &  \int \frac{d p^0}{2\pi i}  \ \frac{d\hat{\vp}}{4\pi} \int_{-\infty}^{+\infty} ds \ \prod_{i=1}^{3} \left[ (p^0 + s - \dm + 2\vk_i\cdot\hat{\vp}) (p^0 - s - \dm - 2\vl_i\cdot\hat{\vp}) \right]^{-1} \nonumber \\
%\lefteqn{\Pi(\vq)  =   - \frac{2 \lambda \bar\mu^2}{\pi^2} \int \frac{d p^0}{2\pi i} \ \frac{d\hat{\vp}}{4\pi} \ ds \ \left[ (p^0 + s - \dm)(p^0 - s - \dm - 2 \vq\cdot\hat{\vp}) \right]^{-1}}  \nonumber \\ 
%\lefteqn{J(\vq_1 \vq_2 \vq_3 \vq_4) = } \nonumber \\
% & & \int \frac{d p^0}{2\pi i}  \ \frac{d\hat{\vp}}{4\pi} \ ds \ \left[ (p^0 + s - \dm) (p^0 - s - \dm - 2\vq_1\cdot\hat{\vp}) \right. \nonumber \\
% & & \times \left. (p^0 + s - \dm + 2(\vq_1-\vq_2)\cdot\hat{\vp})(p^0 - s - \dm - 2(\vq_1-\vq_2+\vq_3)\cdot\hat{\vp}) \right]^{-1} \nonumber \\
%\lefteqn{K(\vq_1 \vq_2 \vq_3 \vq_4 \vq_5 \vq_6)  = } \nonumber \\
% & &  \int \frac{d p^0}{2\pi i}  \ \frac{d\hat{\vp}}{4\pi} \ ds \ \left[ (p^0 + s - \dm) (p^0 - s - \dm - 2\vq_1\cdot\hat{\vp}) \right. \nonumber \\
% & & \times  (p^0 + s - \dm + 2(\vq_1-\vq_2)\cdot\hat{\vp})(p^0 - s - \dm - 2(\vq_1 - \vq_2 + \vq_3)\cdot\hat{\vp}) \nonumber \\
% & & \times (p^0 + s - \dm + 2(\vq_1-\vq_2+\vq_3-\vq_4)\cdot\hat{\vp}) \nonumber \\
% & & \times \left. (p^0 - s - \dm - 2(\vq_1 - \vq_2 + \vq_3 - \vq_4 + \vq_5)\cdot\hat{\vp}) \right]^{-1}.
\end{eqnarray}
where we have introduced new vectors
\begin{center}
\begin{tabular}{lll}
$\vk_1 = 0$,     & $\vk_2 = \vq_1-\vq_2$,        & $\vk_3 = \vq_1 - \vq_2 + \vq_3 - \vq_4$  \\
$\vl_1 = \vq_1$, \ & $\vl_2 = \vq_1-\vq_2+\vq_3$, \ & $\vl_3 = \vq_1 - \vq_2 + \vq_3 - \vq_4 + \vq_5$.
\end{tabular}
\end{center}
Notice that these vectors are the coordinates of vertices in the
rhombus and hexagon shapes of Fig.~\ref{rhombushexagonfig}. In
particular, $(\vk_1\vk_2)$ and $(\vl_1\vl_2)$ are the pairs of
endpoints for the solid and dashed diagonals of the rhombus figure,
while $(\vk_1\vk_2\vk_3)$ and $(\vl_1\vl_2\vl_3)$ are the triplets of
vertices of the solid and dashed triangles in the hexagon figure.

We now introduce Feynman parameters to combine the denominator
factors in Eqs.~(\ref{jkeqns}).  Two sets of Feynman parameters
are used, one set for the factors involving $\vk_i$'s and one set for
the factors involving $\vl_i$'s.  For the $J$ integral the result is
\begin{eqnarray}
J & = & \int_0^1 dx_1 \ dx_2 \ \delta(x_1+x_2-1) \int_0^1 dy_1 \ dy_2 \ \delta(y_1+y_2-1) \nonumber \\
 & & \times \int \frac{dp_4}{2\pi} \ \frac{d\hat{\vp}}{4\pi} \ ds \ (s-\dm+ip_4+2\vk\cdot\hat{\vp})^{-2} (s+\dm-ip_4 + 2\vl\cdot\hat{\vp})^{-2}
\label{appJeqn}
\end{eqnarray}
where $\vk = \sum_i x_i \vk_i$, $\vl = \sum_i y_i \vl_i$.  Next, we do
the $s$ integral by contour integration, followed by the $\hat{\vp}$ 
and $p_4$ integrals.  For the $p_4$ integral, noting that the 
$s$ integration introduces a sign factor $\mbox{sgn}(p4)$
and that the integrand in (\ref{appJeqn}) depends only on $ip_4$,
we use
\begin{equation}
\int_{-\infty}^{+\infty} dp_4 \ \mbox{sgn}(p_4) \ (\ \cdots\ ) = 2\  \Re \int_{\epsilon}^{\infty} dp_4 \ (\ \cdots\ )
\end{equation}
where $\epsilon$ is an infinitesimal positive number.  The final result 
is
\begin{equation}
\label{Jinteqn}
J = \frac{1}{4} \ \Re \int_0^1 dx_1\ dx_2 \ \delta(\mbox{$\sum$} x - 1) \int_0^1 dy_1\ dy_2 \ \delta(\mbox{$\sum$} y -1) \ \frac{1}{|\vk-\vl|^2 - \dm_+^2}
\end{equation}
where $\dm_+ = \dm + i\epsilon$.  We include the infinitesimal
$\epsilon$ is so that the integral is well-defined even when
$|\vk-\vl|=\dm$ is encountered in the integration region.  This 
is a ``principal value'' specification for a multidimensional 
integral that is not Riemann-convergent.
A similar analysis for the 
$K$ integral gives the result
\begin{equation}
\label{Kinteqn}
K = \frac{1}{8} \ \Re \int_0^1 dx_1\ dx_2\ dx_3 \ \delta(\mbox{$\sum$} x-1) \int_0^1 dy_1\ dy_2\ dy_3 \ \delta(\mbox{$\sum$} y - 1) \ \frac{|\vk-\vl|^2 + 3\dm^2}{(|\vk-\vl|^2 - \dm_+^2)^3}.
\end{equation}

For the case of a single plane wave,
where all the $\vq_i$'s are equal ($\vq_1 = \vq_2 = \cdots =
\vq$), notice that $\vk_1 = \vk_2 = \vk_3 = 0$ and $\vl_1 = \vl_2 =
\vl_3 = \vq$.  Then the integrands in (\ref{Jinteqn}) and (\ref{Kinteqn})
are constants and we immediately obtain the results of (\ref{J0eqn}) and
(\ref{K0eqn}), respectively.

Finally, we must integrate the Feynmann parameters. For the J
integral, two of the integrals can be done using the delta functions,
a third can be done analytically, and the final integration is done
numerically, using an integration contour that avoids the singularity
at $|\vk-\vl|=\dm$.  For the $K$ integral, we do the $x_3$ and $y_3$
integrals using the delta functions, and then make a linear
transformation of the remaining integration variables $x_1$, $x_2$, $y_1$,
$y_2$, introducing new variables
\begin{equation}
r_i = \sum_j a_{ij} x_j + \sum_j b_{ij} y_j + c_i, \ \ \  i = 1,\ldots,4
\end{equation}
with $a_{ij}$, $b_{ij}$ and $c_i$ chosen such that 
\begin{equation}
|\vk-\vl|^2 = r_1^2 + r_2^2 + r_3^2.
\end{equation}
While such a transformation puts the integrand in a convenient simple
form, it complicates the description of the integration 
region considerably.
Therefore 
we use a Fourier-Motzkin 
elimination procedure~\cite{FourierMotzkin} to express
the four-dimensional integration region as a sum of
subregions, for each of which we have an
iterated integral with
``affine'' limits of integration:
\begin{eqnarray}
\lefteqn{\int_0^1 dx_1 \int_0^{1-x_1} dx_2 \int_0^1 dy_1 \int_0^{1-y_1} dy_2 \ ( \ \cdots \ ) \hspace{1in}} \nonumber \\
 & = & \sum_A \left[ \prod_{i=1}^4 \left( \int_{u_{i0}^{(A)} + \sum_{j<i} u_{ij}^{(A)} r_j}^{v_{j0}^{(A)} + \sum_{j<i} v_{ij}^{(A)} r_j} dr_i \right) \left\| \frac{\partial(x,y)}{\partial r} \right\| ( \ \cdots \ ) \right]
\end{eqnarray}
%\begin{eqnarray}
%\lefteqn{\int_0^1 dx_1 \int_0^{1-x_1} dx_2 \int_0^1 dy_1 \int_0^{1-y_1} dy_2 \ ( \ \cdots \ ) } \nonumber \\
% & = & \sum_A \left[ \int_{x_1}^{x_2} dx \int_{m_1 x + b_1}^{m_2 x + b_2} dy \int_{p_1 x + q_1 y + c_1}^{p_2 x + q_2 y + c_2} dz \int_{s_1 x + t_1 y + u_1 z + v_1}^{s_2 x + t_2 y + u_2 z + v_2} dw \ ( \ \cdots \ ) \right]
%\end{eqnarray}
For each subregion $A$, we can immediately do the $r_4$ integration
since the integrand is independent of $r_4$.  We are left with a
three-dimensional integral over the volume of a polyhedron with six
quadrilateral faces.  We then
apply the divergence theorem to turn the three-dimensional
integral into 
a sum of surface integrals over the faces.  For each surface integral,
we convert to plane polar coordinates $(\rho, \phi)$ so that
$|\vk-\vl|^2 = \rho^2 + d^2$, where $d$ is the distance from the
origin $(r_1,r_2,r_3) = (0,0,0)$ to the plane of integration.  Now the
$\phi$ integration can be done because the integrand is independent of
$\phi$.  Finally, the $\rho$ integration is done numerically, using a
deformed integration contour that avoids the singularity at 
$|\vk-\vl| = \dm$.

%% file: appb.tex
\chapter{Spin-One Calculations}

\section{Calculational details}
\label{app:gap}

To allow the possibility of quark pairing, we use
8-component Nambu-Gorkov spinors,
\beq
\Psi=\left(\begin{tabular}{c}$\psi(p)$ \\ $\bar{\psi}^T(-p)$
 \end{tabular}\right)
\eeq
with
\beq
\bar{\Psi}=\left(\bar{\psi}(p),\psi^T(-p)\right)~.
\eeq
In Minkowski space
the inverse quark propagator for massive fermions takes the form,
\beq
S^{-1}(p)=\Bigg(\begin{tabular}{cc}$\pslash-m+\mu\gamma_0$ & $\bar{\Delta}$\\
				  $\Delta$ & $(\pslash+m-\mu\gamma_0)^T$
	       \end{tabular}\Bigg)
\label{invprop}
\eeq
where
\beq
\bar{\Delta}=\gamma_0\Delta^\dagger\gamma_0~.
\eeq
The gap matrix $\Delta$ is a matrix in color, flavor and Dirac space,
multiplied by a gap parameter also denoted as $\De$,
\beq
\Delta^{\alpha\beta ab}_{ij} = \Delta(\mu) {\cal C}^{\rm \alpha\beta}
  {\cal F}_{ij}\Gamma^{ab}~.
\eeq
The relation between the proper self energy and the full propagator is,
\beq
S^{-1}=S_0^{-1}+\Sigma=\Bigg(\begin{tabular}{cc}$\pslash-m+\mu\gamma_0$ & 0\\
				  0 & $(\pslash+m-\mu\gamma_0)^T$
	       \end{tabular}\Bigg)+
		\Bigg(\begin{tabular}{cc} 0 & $\bar{\Delta}$\\
				  $\Delta$ & 0
	       \end{tabular}\Bigg)
\eeq
where $S_0^{-1}$ is the inverse propagator in the absence of interactions.
The gap is determined by solving a self-consistent Schwinger-Dyson equation for
$\Sigma$. For a 4-fermion interaction modelling single gluon exchange, this
takes the form
\beq
\Sigma=-6iG\int\frac{d^4p}{(2\pi)^4}V^A_\mu S(p)V^{A\mu}
\label{gapeq}
\eeq
where $V^A_\mu$ 
is the interaction vertex in the Nambu-Gorkov basis. 
We study three interactions, the quark-gluon vertex
\beq
V^A_\mu=\Bigg(\begin{tabular}{cc}$\gamma_\mu\lambda^A/2$ & 0\\
				  0 & $-(\gamma_\mu\lambda^A/2)^T$
	       \end{tabular}\Bigg)~,
\eeq
the quark-magnetic gluon vertex
\beq
V^A_i=\Bigg(\begin{tabular}{cc}$\gamma_i\lambda^A/2$ & 0\\
				  0 & $-(\gamma_i\lambda^A/2)^T$
	       \end{tabular}\Bigg)~,
\eeq
and the quark-instanton vertex, for which
\beq
\Sigma_{ik}^{\al\ga}=-6iG\int\frac{d^4p}{(2\pi)^4} \Bigl(
 V_{L\mu}^A S_{jl}^{\be\de}(p)V_L^{A\mu} + V_{R\mu}^A S_{jl}^{\be\de}(p)V_R^{A\mu} \Bigr)\Xi_{ik\be\de}^{jl\al\ga}
\eeq
where
\beq
\Xi_{ik\be\de}^{jl\al\ga}=-\ep_{ik}\ep^{jl}\frac{2}{3} (3 \de^\al_\be \de^\ga_\de - \de^\al_\de \de^\ga_\be)
\eeq
and
\beq
V_L^A=\Bigg(\ba{cc}({\mathbf 1}+\gamma_5) & 0\\
		 0 & ({\mathbf 1}+\gamma_5)^T
	       \ea\Bigg)~,
\quad\hbox{and}\quad
V_R^A=\Bigg(\ba{cc}({\mathbf 1}-\gamma_5) & 0\\
		 0 & ({\mathbf 1}-\gamma_5)^T
	       \ea\Bigg)~.
\eeq
In the case of the $\psi C\ga_5\psi$ condensate for the full gluon interaction
we obtain the gap equation, which after rotation to Euclidean space becomes
\beq
1=16G\int\frac{d{p_0}d^3p}{(2\pi)^4}
  \frac{4(\Delta^2+\mu^2+{p_0}^2+p^2)}{{W}}
\eeq
where
\beq
{W}=\Delta^4+\mu^4+({p_0}^2+p^2)^2+2\Delta^2(\mu^2+{p_0}^2+p^2)-2\mu^2(-{p_0}^2+p^2)~.
\eeq
The $p_0$ integral can be explicitly evaluated,
\beq
1=\frac{2G}{\pi^2}\int^\Lambda_0 dp\left[
   \frac{p^2}{\sqrt{\Delta^2+(p+\mu)^2}}
  +\frac{p^2}{\sqrt{\Delta^2+(p-\mu)^2}}\right]~.
\eeq
The momentum integral can be performed
analytically, giving
\beq
\Delta=2\sqrt{\Lambda^2-\mu^2}\exp\left(
  \frac{\Lambda^2-3\mu^2}{2\mu^2}\right)\exp\left(-\frac{\pi^2}{4\mu^2G}\right)
\label{analyticgap}
\eeq
for $\De\ll\mu$.

\section{Gap equation summary}
\label{app:gapeqns}

Here are the gap equations for the attractive channels. In the
following, positive square roots are implied and we 
define $p_r^2\equiv(p_x)^2+(p_y)^2$.

$$ \int d|p|\equiv \int^\La_{0} d|p|,\quad
  \int dp_r dp_z \equiv \int^\La_0 dp_r \int^{\sqrt{\La^2-p_r^2}}_{-\sqrt{\La^2-p_r^2}} dp_z
$$

\subsection{$C\ga_5$ and $C$ gap equations}

\begin{equation}
1=N\frac{G}{\pi^2}\int d|p|\; \left[
   \frac{|p|^2}{\sqrt{\De^2+(|p|-\mu)^2}}
  +\frac{|p|^2}{\sqrt{\De^2+(|p|+\mu)^2}}
  \right]
\end{equation}
where N is a constant that differs for each interaction.
$$
\begin{array}{lc}
\mbox{Instanton} &\hspace{2cm} N=4 \\
\mbox{Magnetic + Electric Gluon} &\hspace{2cm} N=2 \\
\mbox{Magnetic Gluon} &\hspace{2cm} N=\frac{3}{2} 
\end{array}
$$

The $C$ channel produces an identical gap equation for both the full
gluon and magnetic gluon interactions. The instanton interaction is
not attractive in this channel.

\subsection{$C\sigma_{03}$ and $C\sigma_{03}\ga_5$ gap equations}

\begin{equation}
1=N\frac{G}{\pi^2}\int dp_r \;dp_z \: \left[
  \frac{p_r(\mathcal{E}+p_r^2)}{\mathcal{E} E_+}
 +\frac{p_r(\mathcal{E}-p_r^2)}{\mathcal{E} E_-} 
 \right]
\end{equation}
with
\begin{eqnarray*}
\mathcal{E}^2&=& \De^2 p_r^2 + \mu^2|p|^2 \\
E_{\pm}^2&=&\De^2 + \mu^2 + |p|^{2} \pm 2\mathcal{E} 
\end{eqnarray*}
where N is a constant that differs for each interaction.
$$
\begin{array}{lc}
\mbox{Instanton} &\hspace{2cm} N=1 \\
\mbox{Magnetic Gluon} &\hspace{2cm} N=\frac{1}{8} 
\end{array}
$$

The $C\sigma_{03}\ga_5$ channel produces an identical gap equation for
magnetic gluon interaction. The instanton and the full gluon
interactions are not attractive in this channel.

\subsection{$C(\sigma_{01} \pm i \sigma_{02})$ gap equation}

\begin{equation}
1=N\frac{-iG}{\pi^3}\int dp_r\;dp_z \int^\infty_{-\infty} dp^0\: 
  \frac{p_r (\mu^2-(p^0)^2-p_z^2-2p^0p_z) }{W}
\end{equation}
where
$$W=\mu^4+(-(p^0)^2+|p|^2)^2+2 \De^2(\mu^{2}-(p^0)^2-p_z^2-2p^0p_z)-2 \mu^2((p^0)^2+|p|^2)$$
and N is a constant that differs for each interaction.
$$
\begin{array}{lc}
\mbox{Instanton} &\hspace{2cm} N=2 \\
\mbox{Magnetic Gluon} &\hspace{2cm} N=\frac{1}{4} 
\end{array}
$$

\subsection{$C\ga_{3}$ gap equation}

\begin{equation}
1=N\frac{G}{\pi^2}\int dp_r \;dp_z \: \left[
  \frac{p_r(\mathcal{E}+p_z^2)}{\mathcal{E} E_+}
 +\frac{p_r(\mathcal{E}-p_z^2)}{\mathcal{E} E_-} 
 \right]
\end{equation}
with
\begin{eqnarray*}
\mathcal{E}^2&=& \De^2 p_z^2 + \mu^2(|p|^2+m^2) \\
E_{\pm}^2&=&\De^2 + \mu^2 +m^2 + |p|^{2} \pm 2\mathcal{E} 
\end{eqnarray*}
where N is a constant that differs for each interaction.
\[
\begin{array}{lc}
\mbox{Magnetic + Electric Gluon} &\hspace{2cm} N=\frac{1}{2} \\
\mbox{Magnetic Gluon} &\hspace{2cm} N=\frac{1}{4} 
\end{array}
\]

\subsection{$C\ga_{3}\ga_5$ gap equation}

This channel is not attractive for instantons and its gap equation
with electric or magnetic gluon
interaction is the same as for the massless $C\ga_{3}$ channel, i.e.

\begin{equation}
1=N\frac{G}{\pi^2}\int dp_r \;dp_z \: \left[
  \frac{p_r(\mathcal{E}+p_z^2)}{\mathcal{E} E_+}
 +\frac{p_r(\mathcal{E}-p_z^2)}{\mathcal{E} E_-} 
 \right]
\end{equation}
with
\begin{eqnarray*}
\mathcal{E}^2&=& \De^2 p_z^2 + \mu^2|p|^2 \\
E_{\pm}^2&=&\De^2 + \mu^2 + |p|^{2} \pm 2\mathcal{E} 
\end{eqnarray*}
where
\[
\begin{array}{lc}
\mbox{Magnetic + Electric Gluon} &\hspace{2cm} N=\frac{1}{2} \\
\mbox{Magnetic Gluon} &\hspace{2cm} N=\frac{1}{4} 
\end{array}
\]

\subsection{$C (\ga_1 \pm i \ga_2)$ gap equation}

\begin{equation}
1=N\frac{-iG}{\pi^3}\int dp_r\;dp_z \int^\infty_{-\infty} dp^0\: 
  \frac{p_r (2\De^2 E_{1-} E_{1+}+(m^2+\mu^2-(p^0)^2+p_z^2) E_{2-} E_{2+})}{
    E_{1-} E_{1+}(4\De^4 + 4\De^2(m^2+\mu^2-(p^0)^2+p_z^2) + E_{1-} E_{1+})} 
\end{equation}
with
\begin{eqnarray*}
E_{1\pm}&=&m^2-(\mu\pm p^0)^2+p_z^2 \\
E_{2\pm}&=&m^2-(\mu\pm p^0)^2+|p|^2 
\end{eqnarray*}
where N is a constant that differs for each interaction.
$$
\begin{array}{lc}
\mbox{Magnetic + Electric Gluon} &\hspace{2cm} N=1 \\
\mbox{Magnetic Gluon} &\hspace{2cm} N=\frac{1}{2} 
\end{array}
$$

\subsection{$C\ga_0\ga_5$ gap equation}

\begin{equation}
1=N\frac{G}{\pi^2}\int d|p| \: \left[
  \frac{|p|^2(\mathcal{E}+|p|^2)}{\mathcal{E} E_+}
 +\frac{|p|^2(\mathcal{E}-|p|^2)}{\mathcal{E} E_-} 
 \right]
\end{equation}
with
\begin{eqnarray*}
\mathcal{E}^2&=& \De^2 |p|^2 + \mu^2(|p|^2+m^2) \\
E_{\pm}^2&=&\De^2 + \mu^2 +m^2 + |p|^{2} \pm 2\mathcal{E} 
\end{eqnarray*}
where N is a constant that differs for each interaction.
$$
\begin{array}{lc}
\mbox{Magnetic + Electric Gluon} &\hspace{2cm} N=\frac{1}{2} \\
\mbox{Magnetic Gluon} &\hspace{2cm} N=\frac{3}{4} 
\end{array}
$$
For $m=0$ this reduces to 
\begin{equation}
1=N\frac{2G}{3\pi^2}\frac{\La^3}{\sqrt{\De^2+\mu^2}}
\end{equation}

\section{Orbital/spin content of the condensates}
\label{app:angmom}

In the non-relativistic limit it is meaningful to ask about the
separate contributions of the orbital and spin angular momenta
to the total angular momentum of the diquark condensates.
% We can find out by 
We can identify these by
expanding the
field operators out of which the condensates are built in terms
of creation and annihilation operators,
\beq
\psi^\alpha_i=\sum_{k,s}\left(\frac{m}{VE_k}\right)^\frac{1}{2}
  \left[u^s(k)a^s_{ki\alpha}e^{-ikx}
       +v^s(k)b^{s\dagger}_{ki\alpha}e^{ikx}\right]
\eeq
Inserting the explicit momentum-dependent spinors in any basis allows
the creation/annihilation operator expansions of the condesates to be
calculated.

In the Dirac basis,
\beq
\ba{rcl@{\qquad}rcl}
u_1^D(\vk) &=& A\left(
 \ba{c} 1 \\ 0 \\ Bk_3 \\ B(k_1+ik_2)\ea\right)
&
u_2^D(\vk) &=& A\left(
 \ba{c} 0 \\ 1 \\ B(k_1-ik_2) \\ -Bk_3\ea\right)\\[6ex]
v_1^D(\vk) &=& A\left(
 \ba{c} B(k_1-ik_2) \\ -Bk_3 \\ 0 \\ 1\ea\right)
&
v_2^D(\vk) &=& A\left(
 \ba{c} Bk_3 \\ B(k_1+ik_2) \\ 1 \\ 0\ea\right)
\ea
\eeq
%\textbf{Chiral Spinors}
%\beq
%\begin{tabular}{cc}
%$u_1$(\vk)=$\frac{A}{\sqrt{2}}\left(\begin{tabular}{c} $1-Bk_3$ \\ $-B(k_1+ik_2)$ \\ $1+Bk_3$ \\ $B(k_1+ik_2)$\end{tabular}\right)$
%&
%$u_2$(\vk)=$\frac{A}{\sqrt{2}}\left(\begin{tabular}{c} $-B(k_1-ik_2)$ \\ $1+Bk_3$ \\ $B(k_1-ik_2)$ \\ $1-Bk_3$\end{tabular}\right)$\\\\
%$v_1$(\vk)=$\frac{A}{\sqrt{2}}\left(\begin{tabular}{c} $B(k_1-ik_2)$ \\ $-Bk_3-1$ \\ $B(k_1-ik_2)$ \\ $1-Bk_3$\end{tabular}\right)$
%&
%$v_2$(\vk)=$\frac{A}{\sqrt{2}}\left(\begin{tabular}{c} $Bk_3-1$ \\ $B(k_1+ik_2)$ \\ $1+Bk_3$ \\ $B(k_1+ik_2)$\end{tabular}\right)$
%\end{tabular}
%\eeq
\beq
A=\left(\frac{E+m}{2m}\right)^\half, \qquad B=\frac{1}{E+m}
\eeq
%The Chiral spinors were obtained from the Dirac spinors via the $\gamma$-
%matrix relation,
%\beq
%\gamma^\mu_{chiral}=U\gamma^\mu_{Dirac}U^\dagger
%\eeq
%where,
%\beq
%U=\frac{1}{\sqrt{2}}\left(\begin{tabular}{cc}1 & $-$1 \\ 1 & 1 \end{tabular}\right)
%\eeq

Eq.~\eqn{eqn:C_op_exp} shows the result of performing such a 
calculation for the $\psi C \psi$ condensate,
\bea
\nonumber
\label{eqn:C_op_exp}
\conds{C}&=&\frac{1}{E}\left[(a^2_{pi\alpha}a^2_{-pj\beta}
  +b^{\dagger 1}_{pi\alpha}b^{\dagger 1}_{-pj\beta})(p_1-ip_2)
  \right.\\\nonumber\\\nonumber
&\phantom{=}&-(a^1_{pi\alpha}a^1_{-pj\beta}
  +b^{\dagger 2}_{pi\alpha}b^{\dagger 2}_{-pj\beta})(p_1+ip_2)\\\nonumber\\
&\phantom{=}&\left.+2(a^1_{pi\alpha}a^2_{-pj\beta}
  +b^{\dagger 1}_{pi\alpha}b^{\dagger 2}_{-pj\beta})p_3\right]\cfms
\eea
where $\cfms$ is the
color-flavor matrix which is symmetric under the
interchange $i\rightleftharpoons j$ (flavor) and 
$\alpha\rightleftharpoons\beta$
(color). A sum over momentum $p$ should be performed on the right hand 
side.

Once the operator expansions have been obtained it is a relatively simple
procedure to obtain the angular momentum content by rearranging the terms
and inserting the relevant spherical harmonics. It is important to include 
contributions from momenta $k$ and $-k$ together, since they involve the
same creation/annihilation operators.
For example, for the condensate in \eqn{eqn:C_op_exp},
\bea
p=k&:&\frac{1}{E}\left[a^2_{ki\alpha}a^2_{-kj\beta}(k_1-ik_2)
-a^1_{ki\alpha}a^1_{-kj\beta}(k_1+ik_2)
+2a^1_{ki\alpha}a^2_{-kj\beta}k_3\right]\\
p=-k&:&\frac{1}{E}\left[-a^2_{-ki\alpha}a^2_{kj\beta}(k_1-ik_2)
+a^1_{-ki\alpha}a^1_{kj\beta}(k_1+ik_2)
-2a^1_{-ki\alpha}a^2_{kj\beta}k_3\right]\\\nonumber
&\to&\frac{1}{E}\left[a^2_{ki\alpha}a^2_{-kj\beta}(k_1-ik_2)
-a^1_{ki\alpha}a^1_{-kj\beta}(k_1+ik_2)
+2a^2_{ki\alpha}a^1_{-kj\beta}k_3\right]
\eea
where we have relabelled $k\to -k, i\leftrightarrow j, 
\al \leftrightarrow \be$ in the last line.
The final result is a sum over $k,\al,\be,i,j$ of
\beq
\frac{2}{E}\left[a^2_{ki\alpha}a^2_{-kj\beta}(k_1-ik_2)
-a^1_{ki\alpha}a^1_{-kj\beta}(k_1+ik_2)
+(a^1_{ki\alpha}a^2_{-kj\beta}+a^2_{ki\alpha}a^1_{-kj\beta})k_3\right]
\eeq
Upon inserting the relevant spherical harmonics and using standard
arrow notation for the spins we obtain
\beq
\conds{C}\rightarrow-2\sqrt{8\pi} \frac{p}{E}\left[
  \frac{1}{\sqrt{3}}\uu Y_1^{-1}
 +\frac{1}{\sqrt{3}}\dd Y_1^1
 -\frac{1}{\sqrt{3}}\frac{1}{\sqrt{2}}[\ud+\du]Y_1^0
 \right]
\eeq
which has precisely the correct Clebsch-Gordan structure to be interpreted
as a state with orbital angular momentum $l=1$, which gives an 
antisymmetric spatial wave function, and spin $s=1$, which gives
a symmetric spin wavefunction, combined to give $j=0$. We write this
as $ |l=1_A,s=1_S\rangle$.
Applying this to all the condensates we studied, we can make a table of the
particle-particle (as opposed to particle-hole) content of each of them,
\beq
\label{BCSenhancement}
\ba{c@{\qquad}rl}
\conds{C\gamma_5} & 4\sqrt{2\pi} & | l=0_S,s=0_A\rangle\\[1ex]
\conds{C} & \dsp -2\sqrt{8\pi}\frac{p}{E} & | l=1_A,s=1_S\rangle\\[2ex]
\conds{C\gamma_0\gamma_5} 
  & \dsp 4\sqrt{2\pi}\frac{m}{E} & | l=0_S,s=0_A\rangle\\[2ex]
\conds{C\gamma_3\gamma_5}
  & \dsp 8\sqrt{\frac{\pi}{3}}\frac{p}{E} & | l=1_A,s=1_S\rangle\\[2ex]
\conda{C\sigma_{03}\gamma_5}
& \dsp -4i\sqrt{\frac{2\pi}{3}}\frac{p}{E} & | l=1_A,s=0_A\rangle\\[2ex]
\conda{C\sigma_{03}}
  & \multicolumn{2}{r}{ \left\{
    \ba{r} \dsp -\frac{8}{3}\sqrt{\pi}i\frac{p^2}{E(E+m)} 
    | l=2_S,s=1_S\rangle\\[3ex]
   \dsp + 2i\sqrt{2\pi}\left[\frac{(E+m)^2-\frac{1}{3}p^2}{E(E+m)}\right]
    | l=0_S,s=1_S\rangle
    \ea \right. }  \\[8ex]
\conda{C\gamma_0}& 0 \\[4ex]
\conda{C\gamma_3}
  &  \multicolumn{2}{r}{ \left\{
   \ba{r} \dsp \frac{8}{3}\sqrt{\pi}\frac{p^2}{E(E+m)} 
    | l=2_S,s=1_S\rangle\\[3ex]
    \dsp +2\sqrt{2\pi}\left[\frac{(E+m)^2-\frac{1}{3}p^2}{E(E+m)}\right] 
    |  l=0_S,s=1_S\rangle
    \ea \right. }  \\
\ea
\eeq
These results are summarized in the ``BCS-enhanced'' 
column of Table~\ref{tab:channels}.
We can see explicitly that the $\psi C \ga_0\ga_5 \psi$ condensate
has no particle-particle component in the massless limit, which is why
it cannot occur at high density for the up and down quarks.
This reflects basic physics: the condensate has spin zero, so the two spins
must be oppositely aligned. But it is an ``LR'' condensate 
(see Table~\ref{tab:channels}), so in the massless limit the two quarks,
having opposite momentum and opposite helicity, have parallel spins.

%% file: biblio.tex
% This defines the bibliography file (main.bib) and the bibliography style.
%% If you want to create a bibliography file by hand, change the contents of
%% this file to a `thebibliography' environment.  For more information 
%% see section 4.3 of the LaTeX manual.
%\bibliography{main}
%\bibliographystyle{plain}

\renewcommand{\bibname}{References}